\documentclass[a4paper,11pt]{book}
 \pdfoutput=1

\usepackage[a4paper]{geometry}
\usepackage[latin1]{inputenc}
\usepackage[T1]{fontenc}
\usepackage[authoryear]{natbib}
\usepackage{lipsum} 
\usepackage[table]{xcolor}
\usepackage[english,francais]{babel}

\usepackage{lmodern}
\rmfamily
\DeclareFontShape{T1}{lmr}{b}{sc}{<->ssub*cmr/bx/sc}{}
\DeclareFontShape{T1}{lmr}{bx}{sc}{<->ssub*cmr/bx/sc}{}

\usepackage{amssymb,amsthm,amscd}
\usepackage{amsmath}
\usepackage{mathrsfs}

\usepackage{turnstile}

\usepackage{bm}

\usepackage{multirow}

\usepackage{subfig}
\usepackage{longtable}
\usepackage{tabularx}
\usepackage{calc}
\usepackage{graphicx} 

\usepackage{hyperref}

\usepackage[nottoc]{tocbibind}

\usepackage{tikz}
\usetikzlibrary[patterns]

\usepackage[francais]{minitoc}

\usepackage{diagbox}
\usepackage{booktabs}

\usepackage{bbold}

\newcommand\lam{{\lambda} }

\newcommand\eps{{\varepsilon} }

\newcommand\ab{{\overline a} }

\newcommand\Hb{{\overline H} }
\newcommand\ol{\overline  }

\newcommand\xt{{\tilde x}}
\newcommand\yt{{\tilde y}}
\newcommand\zt{{\tilde z}}
\newcommand\Zt{{\tilde Z}}

\newcommand\cG{{\cal G} }
\newcommand\cW{{\cal W} }
\newcommand\gO{{\cal O} }
\newcommand\cS{{\cal S} }

\newcommand\cF{{\cal F} }

\newcommand\gH{{\cal H} }
\newcommand\cQ{{\cal Q} }
\newcommand\cV{{\cal V} }
\newcommand\cR{{\cal R} }
\newcommand\cM{{\cal M} }
\newcommand\cN{{\cal N} }
\newcommand\cA{{\cal A} }

\newcommand\cP{{\cal P} }
\newcommand\cL{{\cal L} }

\newcommand\Dv{{\Delta \varpi} }
\newcommand\Ddv{{\Delta \dot \varpi} }
\newcommand\Deta{{\Delta \eta} }

\newcommand{\be}{\begin{equation}}
\newcommand{\ee}{\end{equation}}

\newtheorem{rem}{Remarque}

\newcommand{\an}[1]{\textcolor{blue}{#1}}  

\DeclareMathOperator{\e}{e}

\usepackage[final]{pdfpages}

\usepackage{epigraph}

\let\originalepigraph\epigraph 
\renewcommand\epigraph[2]{\originalepigraph{\textit{#1}}{\textsc{#2}}}

\usepackage{makeidx}
\makeindex

\makeatletter
\newskip\@bigflushglue \@bigflushglue = -100pt plus 1fil

\def\bigcentering{\let\\\@centercr\rightskip\@bigflushglue
\leftskip\@bigflushglue
\parindent\z@\parfillskip\z@skip}

\makeatother

\usepackage{fancyhdr}
\makeatletter
\let\ps@plain=\ps@empty
\makeatother

\let\origdoublepage\cleardoublepage
\newcommand{\clearemptydoublepage}{%
  \clearpage
  {\pagestyle{empty}\origdoublepage}%
}
\let\cleardoublepage\clearemptydoublepage

\FrenchFootnotes
\AddThinSpaceBeforeFootnotes

\newcommand*\chapterstar[1]{\chapter*{#1}
  \addcontentsline{toc}{chapter}{#1}
  \markboth{#1}{#1}}

\theoremstyle{plain}

\theoremstyle{definition}

\makeatletter
\newskip\@bigflushglue \@bigflushglue = -100pt plus 1fil

\def\bigcentering{\let\\\@centercr\rightskip\@bigflushglue
\leftskip\@bigflushglue
\parindent\z@\parfillskip\z@skip}

\makeatother

\title{}
\author{Adrien Leleu}

\begin{document}

  \frontmatter

  \pdfbookmark[0]{Page de garde}{garde}

\thispagestyle{empty}

\includegraphics[width=0.25\linewidth]{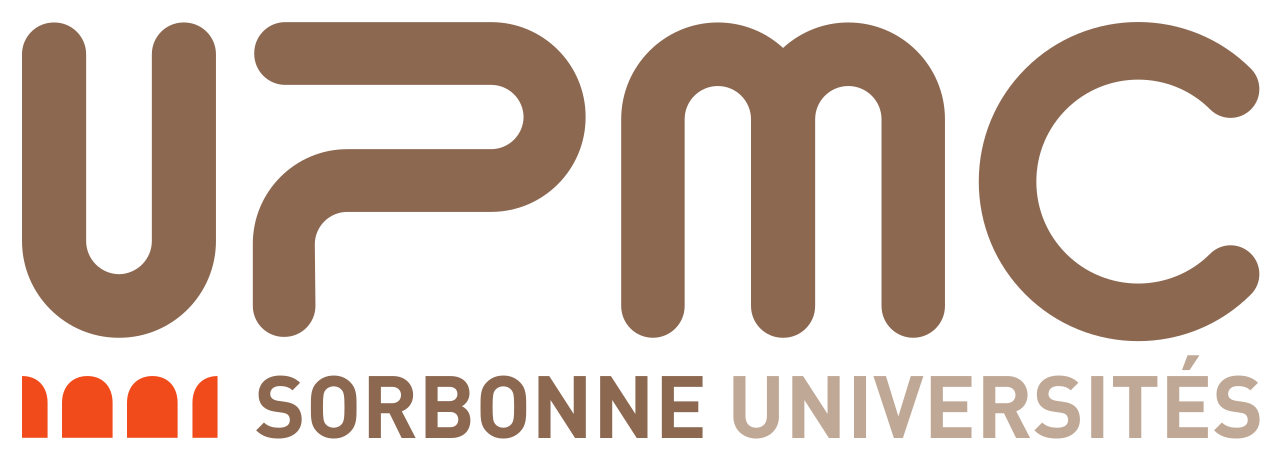} \hfill \includegraphics[width=0.3\linewidth]{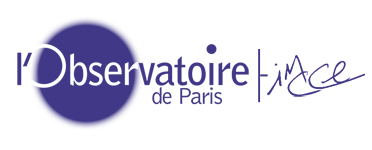}


\begin{center}
  \vspace{\stretch{1}}
  {\large \textbf{École Doctorale d'Astronomie et d'Astrophysique d'île-de-France}}

  \vspace{\stretch{2}}
  {\Huge \textsc{Thèse de Doctorat de l'Université Paris VI}}

  \vspace{\stretch{1}}
  {\LARGE Disciplines : Astronomie et Astrophysique}

  \vspace{\stretch{2.5}}
  {\large présentée par}

  \vspace{\stretch{1}}
  \textbf{{\LARGE Adrien \textsc{Leleu}}}

  \vspace{\stretch{2}}
  \hrule

  \vspace{\stretch{1}}
  {\LARGE \textbf{\textsc{Dynamique des planètes coorbitales}}}

  \vspace{\stretch{1}}
  \hrule

  \vspace{\stretch{2}}
  {\large dirigée par}
  
  \vspace{\stretch{1}}
  	  {\Large
    \begin{tabular}{lcr}
 Philippe \textsc{Robutel} & IMCCE\\
  Alexandre C.M. \textsc{Correia}  & Université d'Aveiro
  \end{tabular}
    }
    
  \vspace{\stretch{4}}
  {\Large Soutenue le 27 septembre 2016 devant le jury composé de :}

  \vspace{\stretch{1}}
  {\Large
  \begin{tabular}{lcr}
    M. Alexandre \textsc{Correia}  & co-directeur\\ 
    M. Bálint \textsc{Érdi}  & rapporteur\\ 
    M. Alain \textsc{Lecavelier des Etangs}  & examinateur\\ 
	M. Alessandro \textsc{Morbidelli}  & examinateur\\ 
	M. David \textsc{Nesvorny}  & rapporteur\\ 
	M. Philippe \textsc{Robutel}  & directeur\\ 
  \end{tabular}
  }

\end{center}


\newpage

\vspace*{\fill}

\noindent
\begin{center}
  \begin{minipage}[t]{(\textwidth-2.5cm)/2}
    IMCCE\\
    Observatoire de Paris\\
    77, avenue Denfert-Rochereau \\
    75 014 Paris\\
    France
  \end{minipage}
  \hspace{1.5cm}
  \begin{minipage}[t]{(\textwidth-2cm)/2}
    UPMC\\
    4 place Jussieu\\
    75005 Paris\\
    France
  \end{minipage}
\end{center}

  \dominitoc
  \dominilof
  \dominilot

  \include{resume}

   \section*{Abstract}

Ce travail porte principalement sur la dynamique et les méthodes de détection des exoplanètes coorbitales. Nous appelons "coorbitale" toute configuration pour laquelle deux planètes orbitent avec le même moyen mouvement moyen autour d'une même étoile. Dans un premier temps, nous revisitons les résultats du cas coplanaire circulaire. Nous rappelons également que les variétés des coorbitaux circulaires et celle des coorbitaux coplanaires sont toutes deux invariantes par le flot du Hamiltonien moyen. Nous nous intéressons donc à ces deux cas particuliers. L'accent est mis sur le cas coplanaire (excentrique), où nous étudions l'évolution de familles d'orbites quasi-périodiques de dimension non maximale en fonction de l'excentricité des planètes. Nous montrons que la géométrie des ces familles dépend fortement de l'excentricité, ce qui entraine des changements de topologie importants dans l'ensemble de l'espace des phases à mesure que celle-ci augmente. Un chapitre est dédié à la détection des exoplanètes coorbitales. On y rappelle les différentes méthodes de détection adaptées au cas coorbital. On développe particulièrement le cas des vitesses radiales, ainsi que leur combinaison avec des mesures de transit. Enfin, on décrit une méthode permettant d'étudier l'effet de perturbations orbitales sur les résonances spin-orbite d'un corps indéformable. Nous appliquons cette méthode dans deux cas: le cas coorbital excentrique, et le cas circumbinaire.\\

This work focuses on the dynamics and the detection methods of co-orbital exoplanets. We call "co-orbital" any configuration in which two planets orbit with the same mean mean-motion around the same star. First, we revisit the results of the circular coplanar case. We also recall that the manifold associated to the coplanar case and the manifold corresponding to the circular case are both invariant by the flow of the averaged Hamiltonian. We hence study these two particular cases. We focus mainly on the coplanar case (eccentric), where we study the evolution of families of non-maximal quasi-periodic orbits parametrized by the eccentricity of the planets. We show that the geometry of these families is highly dependent on the eccentricity, which causes significant topology changes across the space of phases as the latter increases. A chapter is dedicated to the detection of co-orbital exoplanets. We recall the different detection methods adapted to the co-orbital case. We focus on the radial velocity technique, and the combination of radial velocity and transit measurements. Finally, we describe a method to study the effect of orbital perturbations on the spin-orbit resonances for a rigid body. We apply this method in two cases: the eccentric co-orbital case and the circumbinary case.



\thispagestyle{empty}

	  \epigraph{The sounding cataract\\
      Haunted me like a passion: the tall rock,\\
      The mountain, and the deep and gloomy wood,\\
      Their colours and their forms, were then to me\\
      An appetite; a feeling and a love,\\                   
      That had no need of a remoter charm,\\
      By thought supplied, nor any interest\\
      Unborrowed from the eye.}{Wordsworth}

  \setcounter{tocdepth}{2}
  \pdfbookmark[0]{Table des matières}{tablematieres}
  \tableofcontents

  \mainmatter

  \chapterstar{Introduction}

Au début XVII siècle Kepler trouva, en analysant les observations de Tycho Brahe, trois lois décrivant le mouvement des planètes autour du soleil. Quelques décennies plus tard, Newton développa la théorie de la gravitation: les équations du mouvement de corps à symétrie sphérique soumis uniquement à leur attraction gravitationnelle mutuelle. Appliquant sa nouvelle théorie au cas du problème à deux corps, Newton intégra les équations du mouvement et trouva des trajectoires suivant les lois décrites par Kepler. Il fut cependant démuni face au problème à trois corps, comme de nombreux mathématiciens après lui jusqu'à ce que Poincaré montre en 1892 que ce problème n'était pas intégrable.

Si les observations de notre système solaire multi-planétaire ont permis à Kepler d'établir des lois décrivant le problème à deux corps, c'est parce que les perturbations mutuelles des planètes sont suffisamment faibles pour que sur des temps courts les trajectoires ne diffèrent pas significativement de celles de corps isolés. En effet, si les masses des planètes sont faibles par rapport à celle de l'étoile et que leurs distances mutuelles ne deviennent pas trop petites, les équations du mouvement de ce système sont proches d'une somme de problèmes à deux corps. Cette approche, dite perturbative, a été utilisée par Lagrange en 1778 afin de donner des solutions approchées du problème à deux corps perturbé.


Mais si il n'est pas possible d'obtenir une solution analytique générale et exacte du problème à trois corps, cela n'exclut évidemment pas la possibilité d'identifier des solutions particulières de ce problème. Dès 1767, Euler en trouva trois: trois configurations où les corps sont alignés et sont sur une orbite périodique. En 1772, Lagrange porta ce nombre à 5: il ajouta aux configurations d'Euler deux nouvelles configurations où les trois corps sont au sommet d'un triangle équilatéral indéformable tournant autour du centre de gravité du système à une vitesse constante. Soit $m_0$, $m_1$ et $m_2$ les masses de ces trois corps, avec $m_0 \geq m_2 \geq m_1$. Nous notons $L_1$ et $L_2$ les configurations alignées où les corps $1$ et $2$ sont du même côté du corps $0$, $L_3$ la configuration où $m_1$ et $m_2$ sont chacun d'un côté de $m_0$, $L_4$ la configuration équilatérale où $m_1$ est en avance sur $m_2$ et $L_5$ l'autre configuration équilatérale. On appellera également ces 5 orbites les `équilibres de Lagrange', car ce sont des points fixes dans le repère tournant avec chacune de ces configurations. $L_1$, $L_2$ et $L_3$ sont des équilibres instables. En 1843, Gascheau montra que la stabilité (linéaire) des équilibres $L_4$ et $L_5$ dépendait de la répartition de masse entre les trois corps\footnote{$\frac{m_0m_1+m_0m_2+m_1m_2}{(m_0+m_1+m_2)^2}<\frac{1}{27}$}: ces configurations sont stables si $m_1$ et $m_2$ sont suffisamment faibles devant $m_0$. Quand cette condition est vérifiée, il existe un ensemble de conditions initiales non vide pour lesquelles un système de trois corps libre au voisinage de cet équilibre sur des temps longs (où `long' reste ici à définir). Il fallu attendre 1906 pour que le premier corps dans cette configuration, Achille, soit découvert par \citet{Wolf1906} au voisinage du point $L_4$ de Jupiter. Depuis, des milliers d'astéroïdes ont été découverts dans le voisinage des points $L_4$ et $L_5$ des planètes du système solaire, principalement de Jupiter\footnote{http://www.minorplanetcenter.org/}. Par analogie au nom donné aux premiers de ces astéroïdes découverts, nous appellerons cette configuration `troyenne', où les corps librent autour des sommets d'un triangle équilatéral. 

Les 5 orbites particulières précédemment décrites sont périodiques, leur mouvement est donc régi par une unique fréquence fondamentale. Les trois corps ont donc, sur chacune de ces orbites, le même moyen mouvement autour du barycentre du système. Plus généralement on parlera de configuration coorbitale, ou résonance coorbitale par analogie aux autres résonances en moyen mouvement, pour toute orbite où les trois corps auront le même moyen mouvement moyen. Les configurations troyennes précédemment évoquées sont donc des configurations coorbitales, mais ce ne sont pas les seules connues. La topologie globale de la résonance coorbitale à été étudiée en particulier par \cite{Garfinkel1976,Garfinkel1978} et \cite{Ed1977} par approche perturbative. Leurs travaux fournissent un modèle de la résonance coorbitale dans le cadre du problème restreint ($m_1=0$), valable tant que les longitudes des corps $1$ et $2$ ne sont pas trop proches. Dans leur modèle apparaît, en plus des régions troyennes, une région où l'angle résonant $\zeta=\lambda_1 - \lambda_2$ subit de larges librations durant lesquelles l'orbite oscille autour des équilibres $L_3$, $L_4$ et $L_5$. Cette configuration est nommée `fer-à-cheval', d'après la forme que décrit l'orbite de $m_1$ dans le repère tournant avec le moyen mouvement moyen des trois corps. Quelques années plus tard, les premiers corps en configuration fer-à-cheval furent découverts: les satellites de Saturne Janus et Epiméthée \citep{SmiReFoLa1980,SyPeSmiMo1981}.

Récemment, \cite{RoPo2013,RoNi2015} proposèrent un modèle pour la résonance coorbitale coplanaire quasi-circulaire dans le cas planétaire ($m_0 \gg m_2 \geq m_1$). Ces travaux constituent un point de départ pour ce manuscrit et seront rappelés, parmi d'autres résultats concernant la résonance coorbitale circulaire plane, dans le chapitre 1.

Les 5 équilibres de Lagrange sont des configurations coplanaires avec des orbites circulaires. Les domaines des orbites troyennes et en fers à cheval sont également définis dans ce cas particulier. Cependant, les configurations d'Euler et de Lagrange du cas circulaire plan se prolongent en des configurations homothétiques dans le cas excentrique, auxquelles s'ajoutent d'autres familles d'orbites remarquables, telles que les `Anti-Lagranges' \citep{GiuBeMiFe2010}, ou les `quasi-satellites'  \citep{Namouni1999,MiIn2006}. Il existe donc une grande variété de configurations différentes dans la résonance coorbitale. De plus, le cas incliné apporte de nouvelles configurations, telles que les coorbitaux rétrogrades \citep{MoNa2013}. Il est difficile d'explorer cet espace des phases de manière exhaustive. Nous allons cependant développer une approche dans le but d'en apprendre plus sur la dynamique coorbitale, en nous concentrant principalement sur le cas plan excentrique. Les résultats de cette étude seront présentés dans le chapitre 2, avec un intérêt particulier pour les modifications de l'espace des phases à mesure que l'excentricité augmente. Le cas circulaire incliné sera étudié plus brièvement dans le chapitre 3.\\



En se basant sur les connaissances acquises dans les chapitres 1 à 3, on s'intéressera ensuite à la détection des exoplanètes coorbitales. Dans ces chapitres, nous montrons que pour de faibles excentricités et des masses planétaires suffisamment faibles, de grandes régions de l'espace des phases contiennent des orbites stables sur le long terme. Il est donc étonnant que parmi les milliers d'exoplanètes découvertes ces vingt dernières années\footnote{http://www.exoplanets.eu/}, aucune ne soit en configuration coorbitale. Cette question sera abordée dans le chapitre 4. On y discutera deux causes possibles de cette absence: dans une première partie, on rappellera les résultats existants concernant la formation et l'évolution des système coorbitaux. Dans un second temps, on verra pourquoi les techniques "traditionnelles" de détection confondent, au moins au premier ordre, des coorbitaux pour une planète seule. Enfin nous verrons comment des techniques adaptées peuvent lever cette indétermination. 

Dans le dernier chapitre, nous prendrons un peu de recul par rapport à la résonance coorbitale et nous discuterons de l'effet des perturbations orbitales sur la dynamique du spin d'un corps asymétrique indéformable. Nous y présenterons une méthode générale qui sera appliquée dans deux cas: le cas coorbital excentrique, et le cas circumbinaire.\\

\an{
In 1767, Euler found three periodic solutions to the three body problem, where the three bodies are aligned: the configurations $L_1$, $L_2$ and $L_3$. In 1772, Lagrange found two additional periodic solutions $L_4$ and $L_5$, where the three bodies are located at the vertices of an equilateral triangle. $L_1$, $L_2$ and $L_3$ are unstable (hyperbolic) equilibriums. However, \cite{Ga1843} showed the linear stability of $L_4$ and $L_5$ in the planetary case if, noting $m_1$ and $m_2$ the masses of the planets and $m_0$ the mass of the star, $\frac{m_0m_1+m_0m_2+m_1m_2}{(m_0+m_1+m_2)^2}<\frac{1}{27}$.\\
For each of these equilibrium points, the three bodies orbit with the same mean motion around the center of mass of the system. In the planetary case, we will call `co-orbital' configuration, or co-orbital resonance, any configuration where the two planets orbit with the same mean mean-motion around the star. \\
In 1906, Wolf discovered the first co-orbital body in our solar system: the Jupiter's trojan Achilles. This was the first of many, as we know now thousands of bodies located in the vicinity of the $L_4$ and $L_5$ equilibrium points of Jupiter. We call `trojan' the configuration where the bodies librate in the vicinity of the vertices of an equilateral triangle.\\
However, this is not the only possible co-orbital configuration: using perturbative approaches, \cite{Garfinkel1976,Garfinkel1978} and \cite{Ed1977} modelled the circular co-planar co-orbital resonance in the restricted case ($m_0 \gg m_2 \gg m_1$). They found that, in addition to the trojan configuration, there is a domain where the resonant angle $\zeta=\lambda_1-\lambda_2$ librate with a large amplitude, while the orbit encompasses the three equilibriums $L_3$, $L_4$ and $L_5$. This configuration is called `horseshoe', after the shape that the orbits describe in the rotating frame. The first bodies discovered on a horseshoe orbit where the satellites of Saturn, Janus and Epimetheus, found by \citep{SmiReFoLa1980,SyPeSmiMo1981}. Recently, \cite{RoNi2015} extended this model to the planetary case ($m_0 >> m_2 \geq m_1$). As their work is the staring point for this study, we will revisit their main results in Chapter 1.\\
}

\an{
If we consider systems where the planets have inclined and/or eccentric orbits, the phase space is way more complex. New co-orbital configurations appear, such as quasi-satellites \citep{Namouni1999,MiIn2006} in the eccentric case and retrograde co-orbitals \citep{MoNa2013} in the inclined one. The full problem of the co-orbital resonance has many degrees of freedom and we can't pretend to know all the possible co-orbital configurations. In the Chapter 2 we develop a method to better understand the coplanar eccentric case, where we mainly focus on the changes occurring in the phase space as the eccentricities increases. Part of this method will be used in the inclined circular case as well in the Chapter 3.\\
}

\an{
From the results of the chapter 1 to 3, we know that, especially for moderates eccentricities and planetary masses, large areas of the phase space contain long-term stable orbits. However, no co-orbital exoplanets were found so far. This can be due to their absence for formation or evolution reasons, but also to the difficulty to detect these configurations with the traditional detection methods \citep[see the introduction in][for more details]{LeRoCo2015}. In Chapter 4 we discuss how to adapt the detection methods (radial velocity, astrometry, TTVs, ...) to the detection of co-orbital exoplanets. \\
}

\an{
In the last chapter, we discuss the effect of the orbital perturbations on the spin dynamics of an asymmetric rigid body. We will introduce a generic method that we apply in two cases: the co-orbital eccentric case, and the circumbinary case.  
}

  \section*{Notations}



\begin{table}[h]
 \rowcolors{0}{gray!25}{white}
\begin{tabular}{c c c}
$AL_k$ & & famille de points fixes \'emergent de $L_k$ (probl\`eme moyen) \\
& & \an{family of fixed point emerging from $L_k$ (averaged problem)}\\
$a_j$ & & demi-grand axe de la plan\`ete $j$\\
& & \an{semi-major axis of body $j$}\\
$\bar{a}$ & & demi-grand axe moyen (calcul\'e \`a partir de $\eta$)\\
& & \an{mean semi-major axis (computed from $\eta$)}\\
$\beta_j$ & $=\frac{m_0m_j'}{m_0+\eps m_j'}$ & \\
& & \\
$C$ & $=2J_1$ & moment cin\'etique total \\
& $=\sum_{j\in\{1,2\}} \Lambda_j-i x_j \xt_j$ & \an{total angular momentum}\\
$\Delta a$ & $=a_1-a_2$ & \\
& & \\
$\Delta \varpi$ & $=\varpi_1-\varpi_2$ & \\
& & \\
$\delta$ & $=\frac{m_2}{m_1+m_2}$ & \\
& & \\
$e_j$ & & excentricit\'e du corps $j$\\
 & & \an{eccentricity of body $j$}\\
$\eps$ & sec. \ref{sec:deb} & petit param\`etre  \\
& & \an{small mass parameter}\\
$\eta$ & & moyen mouvement moyen\\
& & \an{mean mean motion}\\
$\cF^j_k$ & sec. \ref{sec:gen} &r\'eunion d'orbites p\'eriodiques - prblm moyen\\
& &\an{reunion of periodic orbits - averaged problem}\\
$\ol \cF^j_k$ & sec. \ref{sec:gen} &r\'eunion d'orbites quasi-p\'eriodiques (2 freq.) - prblm moyen\\
& &\an{reunion of quasi-periodic orbits (2 freq.) - averaged problem}\\
$\ol \cF$ & sec. \ref{sec:rdp} &r\'eunion d'orbites p\'eriodiques - prblm moyen r\'eduit\\
& &\an{reunion of periodic orbits - reduced averaged problem}\\
$f_j$ & & longitude vraie of planet $j$\\
& & \an{true longitude of body $j$}\\
$\cG$ & & constante gravitationnelle \\
& & \an{gravitational constant} \\
$\gH \cS$ & & orbite en fer-\`a-cheval \\
& & \an{horseshoe orbit} \\
$J$ & & inclinaison mutuelle\\
& & \an{mutual inclination} \\
\end{tabular}
\end{table} 

\begin{table}[h!]
 \rowcolors{0}{gray!25}{white}
\begin{tabular}{c c c}
$L_j$ & & $j$\`eme \'equilibre de Lagrange\\
& & \an{$j$th Lagrangian equilibrium}\\
$\Lambda_j$ & $=\beta_j\sqrt{\mu_ja_j}$ & variable canonique de Poincar\'e (conj. $\lambda_j$)\\
& & \an{canonical variable of Poincar\'e (conj. $\lambda_j$)}\\
$\lambda$ &  & longitude moyenne du corps $j$\\
& & \an{mean longitude of $j$th body}\\
$m_j$ & & masse du corps $j$\\
& & \an{mass of body $j$}\\
$m'_j$ & $=m_j/\eps$ & masse r\'eduite du corps $j$ \\
& &  \an{reduced mass of body $j$}\\
$\mu$ & $= \frac{m_1+m_2}{m_0+m_1+m_2}$ & \\
& & \\
$\mu_j$ & $= \cG(m_0+\eps m_j')$ & \\
& & \\
$n$ & &  mouvement moyen\\
& & \an{mean motion}\\
$\nu$ &  & fr\'equence fondamentale associ\'ee \`a $\zeta$ \\
& & \an{fundamental frequency associated to $\zeta$}\\
$P_n$ & $=2\pi /n$ & p\'eriode orbitale\\
& & \an{orbital period}\\
$\varPi$ & $= i(x_2 \xt_2 - x_1 \xt_1)/2$ & conjugu\'e canoniquement \`a $\Delta \varpi$\\
& & \an{canonicaly conjugated to $\Delta \varpi$}\\
$\varpi_j$ & & longitude du p\'erih\'elie de la plan\`ete $j$\\
& & \an{longitude of periastron of the $j$th planet}\\
$\cQ \cS$ & & quasi-satellite\\
& & \an{quasi-satellite}\\
$r_j$ & & distance \`a l'\'etoile de la $j$\`eme plan\`ete\\
& & \an{distance to the star of the $j$th planet}\\
$\bold{r_j}$ & vecteur & position h\'eliocentrique de la plan\`ete $j$\\
& & \an{heliocentric position of the $j$th planet} \\
 $\tilde{\bold{r_j}}$ & vecteur  & impulsion barycentrique normalis\'ee de la plan\`ete $j$ \\
& & \an{normalized barycentric impulsion of the $j$th planet} \\
$\Sigma$ & sec \ref{sec:VaRe} & volume de l'espace des phases repr\'esent\'e par $\cV$\\
 &  & \an{volume of the phase space represented by $\cV$}\\
$\cV$ & sec \ref{sec:VaRe} & vari\'et\'e repr\'esentative\\
 &  & \an{representative manifold}\\
$x_j$ & $=\sqrt{\Lambda_j}\sqrt{1-\sqrt{1-e_j^2}}\operatorname{e}^{i\varpi_j}$ & variable canonique de Poincar\'e (conj. $\xt_j$)\\
& & \an{canonical variable of Poincar\'e (conj. $\xt_j$)}\\
$\xt_j$ & $=-\bar{x}_j$ & variable canonique de Poincar\'e (conj. $x_j$)\\
& & \an{canonical variable of Poincar\'e (conj. $x_j$)}\\
$y_j$ & $=\sqrt{\Lambda_j}\sqrt{\sqrt{1-e_j^2}(1-\cos{i})}\operatorname{e}^{i\Omega_j}$ & variable canonique de Poincar\'e (conj. $\yt_j$)\\
& & \an{canonical variable of Poincar\'e (conj. $\yt_j$)}\\
$\yt_j$ & $=-\bar{y}_j$ & variable canonique de Poincar\'e (conj. $y_j$)\\
& & \an{canonical variable of Poincar\'e (conj. $y_j$)}\\
$Z$ & $=\Lambda_1-\Lambda_1^0$ & variable conjugu\'ee canoniquement \`a $\zeta$\\
& & canonical conjugate variable to $\zeta$\\
$z$ & $=\zeta-\pi/3$ & \\
& & \\
$\zeta$ & $=\lambda_1-\lambda_2$ & angle r\'esonant\\
& & \an{resonant angle}\\
\end{tabular}
\end{table}

  \chapter{Coorbitaux quasi-circulaires coplanaires}

Sans ce chapitre, nous rappelons les principaux résultats sur la dynamique et la stabilité des coorbitaux circulaires coplanaires. Notre objet d'étude étant les systèmes planétaires, on se limitera au cas $m_0 \gg m_2  \geq m_1$.  \\

\an{In this chapter we revisit the results concerning coplanar circular co-orbitals. We focus on the planetary case, with $m_0 \gg m_2  \geq m_1$. These results come from \cite{Ed1977} for the restricted case, and \cite{RoPo2013,RoNi2015} for the planetary case.\\}

\section{Formulation hamiltonienne de la résonance 1:1 pour des orbites quasi-circulaires}

\textit{Les transformations du hamiltonien de la résonance 1:1 en moyen mouvement présenté ici sont issues de l'article de \cite{RoNi2015}.\\}  

\subsection{Formulation hamiltonienne du problème planétaire moyen général}
\label{sec:deb}


Le hamiltonien du problème à trois corps dans les coordonnées canoniques héliocentriques est le suivant \citep{LaRo1995,RoNi2015}:
\begin{equation}
\gH = \gH_K({\bf r}_j) + \eps \gH_P({\bf r}_j,{\bf \tilde{r}}_j)\, ,
\label{eq:Ht}
\end{equation}
où
\begin{equation}
\gH_K = \sum^2_{j=1} \left( \frac{||{\bf \tilde{r}}_j||^2}{2 \beta_j}-\frac{ \mu _j \beta _j}{||{\bf r}_j||} \right)\, , 
\end{equation}
est la partie keplerienne du hamiltonien où ${\bf r}_j$ est la position du corps $j$ par rapport au corps central $m_0$, ${\bf \tilde{r}}_j$ est l'impulsion barycentrique normalisée par le petit paramètre $\eps$ de manière à ce que les masses $m'_k=m_k/\eps=\gO(1)$, avec $m_k=\max{(m_1,m_2)}$. $\beta_j$ est le rapport de masse réduit $\beta_j= \frac{m_0 m'_j}{m_0+ \eps m'_j}$ et $\mu _j= \cG (m_0+\eps m'_j)$ où $\cG$ est la constante universelle de gravitation. La partie perturbée du Hamiltonien s'écrit:
\begin{equation}
\gH_P = \frac{{\bf \tilde{r}}_1 \cdot {\bf \tilde{r}}_2}{m_0} - \cG \frac{m'_1 m'_2}{||{\bf r}_1-{\bf r}_2||}\, .
\end{equation}
Afin de se rapprocher des éléments elliptiques, plus adaptés à l'étude d'orbites kepleriennes perturbées, on réécrit le hamiltonien \ref{eq:Ht} en fonction des variables canoniques de Poincaré:
\begin{equation}
\gH = \gH_K(\Lambda_j) + \eps \gH_p(\Lambda_j,\lambda_j,x_j,\tilde{x}_j,y_j,\tilde{y}_j)\, ,
\label{eq:Htpc}
\end{equation}
avec
\begin{equation}
\begin{aligned}
\Lambda_j &=\beta_j\sqrt{\mu_ja_j}\, , \vspace{2cm} &\lambda_j &=\lambda_j\, , \\ 
x_j & =\sqrt{\Lambda_j}\sqrt{1-\sqrt{1-e_j^2}}\operatorname{e}^{i\varpi_j}\, , \vspace{2cm} &\tilde{x}_j &=-i\bar{x}_j\, , \\
y_j & =\sqrt{\Lambda_j}\sqrt{\sqrt{1-e_j^2}(1-\cos{i})}\operatorname{e}^{i\Omega_j}\, , \vspace{2cm} &\tilde{y}_j &=-i\bar{y}_j\, ,
\end{aligned}
\label{eq:poincvar}
\end{equation}
et
\begin{equation}
\gH_K =- \sum^2_{j=1}  \left( \frac{\mu_j^2 \beta_j^3}{2 \Lambda_j^2} \right)\, . 
\end{equation}

Par définition, la résonance coorbitale exacte est atteinte quand $(\Lambda_1,\Lambda_2)=(\Lambda^0_1,\Lambda^0_2)$, de manière à ce que les moyen mouvements des deux planètes soient identiques (nous noterons cette fréquence $\eta$), donc, lorsque:
\begin{equation}
\frac{\partial \gH_K}{\partial \Lambda_1}(\Lambda_1^0,\Lambda_2^0) = \frac{\partial \gH_K}{\partial \Lambda_2}(\Lambda_1^0,\Lambda_2^0) = \eta\, ,
\end{equation}
ou
\begin{equation}
\frac{\mu^2_1 \beta^3_1}{(\Lambda_1^0)^3} = \frac{\mu^2_2 \beta^3_2}{(\Lambda_2^0)^3} = \eta\, .
\end{equation}
Plus généralement, quand deux corps sont en résonance coorbitale, ils librent autour des points de résonance exacte. Le moyen mouvement des deux corps étant `proche' à tout instant, la différence des longitudes moyennes évolue lentement par rapport à chacune des longitudes. Plus précisément, l'angle résonant $\zeta=\lambda_1-\lambda_2$ est régi par la fréquence fondamentale $\nu \propto \sqrt{\eps}n$. Nous effectuons donc le changement de variable canonique suivant:
\begin{equation}
 \begin{pmatrix}
  \zeta \\
  \zeta_2
   \end{pmatrix} =
   \begin{pmatrix}
  1 & -1 \\
  0 & 1
   \end{pmatrix} 
      \begin{pmatrix}
  \lambda_1 \\
  \lambda_2
   \end{pmatrix}
      , \hspace{1cm}
       \begin{pmatrix}
  Z \\
  Z_2
   \end{pmatrix} =
   \begin{pmatrix}
  1 & 0 \\
  1 & 1
   \end{pmatrix} 
      \begin{pmatrix}
  \Lambda_1-\Lambda_1^0 \\
  \Lambda_2-\Lambda_2^0
   \end{pmatrix}. 
   \label{eq:tl1}
   \end{equation}
qui met en évidence l'angle semi-rapide $\zeta$ et un angle rapide $\zeta_2$ ainsi que leur variable conjuguée $Z$ et $Z_2$. On obtient le hamiltonien suivant:
\begin{equation}
H = H_K(Z,Z_2) + \eps H_P(\zeta,\zeta_2,Z,Z_2,x_j,-i\bar{x}_j,y_j,-\bar{y}_j)+ \gO(\eps^2), 
\label{eq:Hts}
\end{equation}

avec

\begin{equation}
H_K(Z,Z_2) = -\frac{\beta_1^3\mu_1^2}{2(\Lambda_1^0+Z)^2} -\frac{\beta_2^3\mu_2^2}{2(\Lambda_2^0-Z+Z_2)^2}.
\end{equation}

Dans le hamiltonien (\ref{eq:Hts}), une troisième échelle de temps lente, dite séculaire, apparaît pour l'évolution des variables $x_j$ et $y_j$ ($\dot{x}_j=\eps \partial H_P / \partial (\tilde{x}_j) =  \gO({\eps})$). Les différentes échelles de temps qui apparaissent dans le hamiltonien nous permettent de le moyenner sur l'angle rapide $\zeta_2$. En pratique, cette moyenne se fait par l'application au temps $1$ du flot de l'hamiltonien $\ol{\cW}$:
\begin{equation}
 \begin{split}
\ol{\cW}(\zeta,Z,Z_2,x_j,\xt_j,y_j,\yt_j) = \eps \frac{1}{2 \pi} \int_0^{\zeta_2} \left[ \Hb_P- H_P \right] d\zeta_2\, ,
\end{split}
\end{equation}
où
\begin{equation}
 \begin{split}
\Hb_P(\zeta,Z,Z_2,x_j,\xt_j,y_j,\yt_j) & = \frac{1}{2 \pi} \int_0^{2 \pi}  H_P d\zeta_2 \, .
\end{split}
\end{equation}
On obtient alors le hamiltonien moyen:
\begin{equation}
\Hb = \cL_{\ol{\cW}} H. 
\label{eq:Hb}
\end{equation}
où $\cL_{\ol{\cW}}$ est la transformée de Lie:
\begin{equation}
\cL_{\cW}= Id + \{\cW,\cdot\} + \{\cW,\{\cW,\cdot\}\} + ...
\label{eq:Lietrans}
\end{equation}
avec $\{\cdot,\cdot\}$ le crochet de Lie. On note $\chi_m$ le changement de variables canonique proche de l'identité:
\begin{equation}
\chi_m = \cL_{-\ol{\cW}}\, ,
\label{eq:Hbtrans}
\end{equation}
où les anciennes variables s'écrivent en fonction des nouvelles sous la forme:
\begin{equation}
(\zeta,\zeta_2,Z,Z_2,x_j,\xt_j,y_j,\yt_j)=\chi_m(\zeta',\zeta'_2,Z',Z'_2,x'_j,\xt'_j,y'_j,\yt'_j)
\label{eq:Hbvar}
\end{equation}
Le hamiltonien $\ol{\cW}$ étant de taille $\eps$, les nouvelles variables ne diffèrent des anciennes que par des quantités de taille $\eps$. Dans la suite on écrira également les variables moyennes: ($\zeta,\zeta_2,Z,Z_2,x_j,\xt_j,y_j,\yt_j$). Nous obtenons donc:
\begin{equation}
\Hb = H_K(Z,Z_2) + \eps \Hb_P(\zeta,Z,Z_2,x_j,\xt_j,y_j,\yt_j) + \gO(\eps^2)\, , 
\label{eq:Hbp}
\end{equation}
car $\ol{\cW}$ est de taille $\eps$. Notons que $Z_2$ est une constante du mouvement de ce hamiltonien moyen. Quitte à renormer le temps $t$, on peut choisir $\eta$ de manière à ce que la variable $Z_2$ soit nulle:
\begin{equation}
\eta=\left( \frac{ \Lambda_1+\Lambda_2}{\mu_1^{2/3}\beta_1 + \mu_2^{2/3}\beta_2} \right)^3.
\label{eq:etaval}
\end{equation}
On obtient donc le hamiltonien moyen suivant:
\begin{equation}
\Hb = H_K(Z) + \eps \Hb_P(\zeta,Z,x_j,\xt_j,y_j,\yt_j) + \gO(\eps^2). 
\label{eq:Hb}
\end{equation}

\an{
We start from the 3-body problem Hamiltonian (\ref{eq:Ht}) in canonical heliocentric coordinates. We introduce the small mass parameter $\eps$, such that $m'_k=m_k/\eps =\gO(1)$, with $m_k=\max{(m_1,m_2)}$. We make the canonical changes of variables (\ref{eq:poincvar}) and (\ref{eq:tl1}) in order to bring out the resonant angle $\zeta=\lambda_1-\lambda_2$. We then average the Hamiltonian over the fast angle $\zeta_2$ to obtain an Hamiltonian of the form (\ref{eq:Hb}). The independence on the value of the integral of motion $Z_2$ is obtained by renormalising the time: we obtain the expression (\ref{eq:etaval}) for the mean mean-motion $\eta$ common to both co-orbitals.\\}

\subsection{Invariance des variétés circulaire et plane}
\label{sec:ivcp}
Dans un certain domaine \citep[voir][]{RoNi2015}, le hamiltonien $\Hb_P$ est analytique, et peut être développé en série de Taylor au voisinage de ($x_1$,$x_2$,$y_1$,$y_2$)=($0$,$0$,$0$,$0$):
\begin{equation}
\Hb_P(\zeta,Z,x_j,\xt_j,y_j\yt_j) = \sum_{(p,\tilde{p},q,\tilde{q}) \in \mathbb{N}^8} C_{p,\tilde{p},q,\tilde{q}}(\zeta,Z) x_1^{p_1} x_2^{p_2}\xt_1^{\tilde{p}_1} \xt_2^{\tilde{p}_2}y_1^{q_1} y_2^{q_2}\yt_1^{\tilde{q}_1} \yt_2^{\tilde{q}_2},
\label{eq:Hpt}
\end{equation}
où les coefficients $C_{p,\tilde{p},q,\tilde{q}}$ sont non nuls si et seulement si $(p,\tilde{p},q,\tilde{q}) \in \mathbb{N}^8$ respectent la règle de D'Alembert:
\begin{equation}
p_1 + p_2 + q_1 + q_2 = \tilde{p}_1 + \tilde{p}_2 + \tilde{q}_1 + \tilde{q}_2.
\label{eq:DAr}
\end{equation}
Ceci est équivalent au fait que la quantité $\sum_{j=1}^2 ( x_j\tilde{x}_j + y_j\tilde{y}_j)$ est une intégrale première du mouvement moyen. Le développement (\ref{eq:Hpt}) n'admet donc pas de termes linéaires en les variables $x_j$, $\tilde{x}_j$, $y_j$ et $\tilde{y}_j$. La variété 
\begin{equation}
\mathcal{C}_{0} = \{(\zeta,Z,x_j,\tilde{x}_j,y_j,\tilde{y}_j ) / x_j=\tilde{x}_j=y_j=\tilde{y}_j=0 \}
\label{eq:circplanvar}
\end{equation}
est donc une variété invariante par le flot du hamiltonien moyenné (\ref{eq:Hb}). D'autre part, le développement (\ref{eq:Hpt}) est pair en les puissances des ($y_j$,$\tilde{y}_j$). Ceci est dû à l'invariance par symétrie du hamiltonien par rapport au plan Oxy \citep[][page 26]{La89}. De la parité du développement (\ref{eq:Hpt}) en les puissances des ($y_j$,$\tilde{y}_j$), nous pouvons déduire que la variété:
\begin{equation}
\mathcal{C}_{y=0} = \{(\zeta,Z,x_j,\tilde{x}_j,y_j,\tilde{y}_j ) / y_j=\tilde{y}_j=0 \}
\label{eq:planvar}
\end{equation}
est également une variété invariante par le flot du hamiltonien moyenné (\ref{eq:Hb}). En effet, les termes de la série de Taylor (\ref{eq:Hpt}) commencent au degré deux en $y$, $\tilde{y}$. Pour finir, si le développement de (\ref{eq:Hpt}) est pair à la fois en les variables ($x_j$, $\tilde{x}_j$, $y_j$, $\tilde{y}_j$) et les variables ($y_j$, $\tilde{y}_j$), il est pair en les variables ($x_j$, $\tilde{x}_j$). La variété 
 \begin{equation}
\mathcal{C}_{x=0} = \{(\zeta,Z,x_j,\tilde{x}_j,y_j,\tilde{y}_j ) / x_j=\tilde{x}_j=0 \}\, ,
\label{eq:circvar}
\end{equation}
est donc également invariante par le flot du hamiltonien moyen (\ref{eq:Hb}). \\

\an{
We can expand $\Hb_P$ (\ref{eq:Hb}) in Taylor series in the neighbourhood of ($x_1$,$x_2$,$y_1$,$y_2$)=($0$,$0$,$0$,$0$), equation (\ref{eq:Hpt}). The monomials of this expansion follow the D'Alembert rule (\ref{eq:DAr}): there are no odd monomials. Additionally, the considered Hamiltonian is invariant by symmetry with respect to the plane Oxy: the monomials are even with respect to the variables ($y_j$,$\tilde{y}_j$). Consequently, the circular manifold (\ref{eq:circvar}), the coplanar manifold (\ref{eq:planvar}) and their intersection the coplanar circular manifold (\ref{eq:circplanvar}) are invariant by the flow of the averaged Hamiltonian (\ref{eq:Hb}).\\}

\subsection{Stabilité des variétés circulaire et plane}
\label{sec:stabC0}

Nous avons vu précédemment que la variété $\mathcal{C}_{x=0} $ était invariante par le flot du hamiltonien (\ref{eq:Hb}). Nous allons maintenant étudier sa stabilité linéaire dans la direction ($x_j$,$\bar{x}_j$). Il a été montré par \cite{RoPo2013} que l'équation variationnelle dans la direction ($x_j$,$\bar{x}_j$) autour d'une trajectoire arbitraire $\zeta(t)$ de $\mathcal{C}_{x=0} $ s'écrit:
\begin{equation}
\dot{X}= M(t) X
\label{eq:eqvar1}
\end{equation}
où
\begin{equation}
X=
   \begin{pmatrix}
 x_1\\
 x_2
   \end{pmatrix} 
   \hspace{0.3cm}
   \text{et}
   \hspace{0.3cm}
 M(t)= i\eps \eta \frac{m'_1 m'_2}{m_0}
   \begin{pmatrix}
\frac{A(\zeta(t))}{m'_1} & \frac{\bar{B}(\zeta(t))}{\sqrt{m'_1 m'_2}}\\
\frac{B(\zeta(t))}{\sqrt{m'_1 m'_2}} & \frac{A(\zeta(t))}{m'_2} 
   \end{pmatrix}  \, , 
\label{eq:eqvar2}
\end{equation}
avec:
\begin{equation}
\begin{aligned}
A(\zeta) & = \frac{1}{4\Delta(\zeta)^5}(5 \cos 2\zeta - 13 + 8 \cos \zeta)- \cos \zeta,\\
B(\zeta) & =\operatorname{e}^{-2i\zeta}-\frac{1}{8\Delta(\zeta)^5}(\operatorname{e}^{-3i\zeta}+16 \operatorname{e}^{-2i\zeta} - 26 \operatorname{e}^{-i\zeta} +9 \operatorname{e}^{i\zeta}),\\
\Delta(\zeta) & =\sqrt{2-2\cos \zeta}.
\end{aligned} 
\label{eq:eqvar3}
\end{equation}
Ces équations sont obtenues par le développement à l'ordre $2$ en excentricité du hamiltonien moyen (\ref{eq:Hb}). Suivant le théorème de Floquet-Lyapunov \citep{MeHa1992}, si la trajectoire $\zeta(t)$ est $2\pi/\nu$ périodique, alors les solutions de l'équation aux variations prennent la forme:
\begin{equation}
Y(t) = P(\nu t) \operatorname{e}^{Ut},
\label{eq:eqvar4}
\end{equation}
où $U$ est une matrice constante et $P(\psi)$ une matrice dont les coefficients sont des fonctions $2\pi$ périodiques de $\psi$. Nous avons donc:
\begin{equation}
Y(t+2 \pi\nu^{-1}) = Y(t) \operatorname{e}^{2\pi \nu^{-1} U}.
\label{eq:eqvar5}
\end{equation}
La stabilité des solutions de l'équation aux variations ne dépend donc que des valeurs propres de la matrice $U$. Nous avons vu précédemment que la quantité $x_1\tilde{x}_1 + x_2 \tilde{x}_2$ est une intégrale de l'équation variationnelle (\ref{eq:eqvar1}). Ceci implique que ses solutions sont bornées. U est alors diagonalisable, et les parties réelles de ses valeurs propres sont nulles.  $\mathcal{C}_{x=0} $ est donc linéairement stable dans la direction ($x_j$,$\tilde{x}_j$).\\

Nous ne pouvons pas effectuer la même démonstration dans la direction des ($y_j$,$\tilde{y}_j$) à cause d'une forte dégénérescence du système linéaire. Nous verrons cependant dans le Chapitre \ref{chap:coi} que dans le cas planétaire, si l'inclinaison mutuelle initiale est faible, elle ne subit que de très faibles oscillations au cours du temps. \\

\an{In this section we recall the demonstration of \cite{RoPo2013} regarding the linear stability of the circular manifold. A similar method cannot be applied to the coplanar case because of a strong degeneracy of the linear system associated with the variables ($y_j,\yt_j$). However, we will see in chapter $3$ that if a system as a low initial mutual inclination, it will remain low along the orbit.\\} 

\section{Dynamique sur la variété circulaire plane}
\label{sec:Dyncp}

\subsection{Modèle à un degré de liberté}

Sur cette variété, la dynamique est donnée par la restriction du hamiltonien (\ref{eq:Hb}) sur $\mathcal{C}_0$ \citep{RoNi2015}:
\begin{equation}
\Hb_0 = \frac{\beta_1 \mu_1}{a_1}+\frac{\beta_2 \mu_2}{a_2} + \eps \cG m'_1 m'_2 \left( \frac{\cos \zeta}{\sqrt{a_1 a_2}}-\frac{1}{\sqrt{a_1^2+a_2^2-2 a_1 a_2 \cos \zeta}} \right),
\label{eq:H0b}
\end{equation}
où:
\begin{equation}
a_j = \mu_j^{-1} \beta_j^{-2} ( \Lambda^0_j + (-1)^{j+1} Z)^2.
\label{eq:a1a2}
\end{equation}
Une approximation intégrable de $\Hb_0$ peut être obtenue en effectuant un développement limité de l'expression (\ref{eq:H0b}) en $Z$ et $\eps$ à l'ordre ($Z^2$,$\eps Z^0$):
\begin{equation}
\tilde{H}_0 = -\frac{3}{2} (\eta)^{4/3} ( \beta_1 \mu_1^{-2/3}+\beta_2 \mu_2^{-2/3} ) Z^2 + \eps \mu_0^{2/3} (\eta)^{2/3} \frac{m'_1 m'_2}{m_0} F (\zeta),
\label{eq:H0i}
\end{equation}
où
\begin{equation}
F (\zeta)= \cos \zeta - \frac{1}{\sqrt{2-2 \cos \zeta}}.
\label{eq:Fzet}
\end{equation}
Bien que le hamiltonien $\tilde{H}_0$ soit intégrable, ses trajectoires ne peuvent être données explicitement. Cependant, celles-ci vérifient les équations suivantes:
\begin{equation}
\left\{ 
\begin{aligned}
\dot{Z} & = - \frac{\partial \tilde{H}_0}{\partial \zeta} = & \eps \mu_0^{2/3} (\eta)^{2/3} \frac{m'_1 m'_2}{m_0} \left( 1- (2- 2\cos \zeta)^{-3/2} \right) \sin \zeta \\
\dot{\zeta} & = \phantom{-}  \frac{\partial \tilde{H}_0}{\partial Z} = & -3 \frac{m'_1 + m'_2}{m'_1 m'_2} (\eta)^{4/3} \mu_0^{-2/3} Z
\end{aligned}
\right.
\label{eq:syserdi}
\end{equation}
et peuvent donc être aisément intégrées numériquement. Ce système peut également être écrit sous la forme d'une équation différentielle d'ordre $2$, généralisant l'équation d'\cite{Ed1977} dans le cas planétaire:
\begin{equation}
\ddot{\zeta}=-3 \eps \eta^2 \frac{m'_1+m'_2}{m_0} \left( 1- (2- 2\cos \zeta)^{-3/2} \right) \sin \zeta\, .
\label{eq:eqerdi}
\end{equation}
 En utilisant la transformation linéaire (\ref{eq:tl1}), ainsi que l'application ${\cal C}$, on peut écrire l'évolution temporelle approchée des variables initiales $\lambda_j,\Lambda_j$ sur $\mathcal{C}_0 $, à partir des solutions du système (\ref{eq:syserdi}):
\begin{equation}
\begin{aligned}
\lambda_1 & = \eta t + \delta \zeta(t) + \lambda_0+\gO(\eps)\, ,\\
\lambda_2 & = \eta t - (1-\delta) \zeta(t) + \lambda_0+\gO(\eps),\\
\Lambda_1 & = \Lambda^0_1 + Z(t)+\gO(\eps),\\
\Lambda_2 & = \Lambda^0_2 - Z(t)+\gO(\eps),\\
\end{aligned}
\label{eq:varerdi}
\end{equation}
avec $\delta= \frac{m'_2}{m'_1+m'_2}$, et
\begin{equation}
\begin{aligned}
\lambda_0 & =	(1-\delta) \lambda_1(0) + \delta \lambda_2(0),\\
\zeta(0) & = \lambda_1(0)-\lambda_2(0),\\
Z(0) & = \delta (\Lambda_1(0) - \Lambda^0_1) - (1-\delta) (\Lambda_2(0) - \Lambda^0_2).
\end{aligned}
\label{eq:varerdici}
\end{equation}

\an{Starting from the restriction of the Hamiltonian (\ref{eq:Hb}) on the manifold representing the circluar coplanar case (\ref{eq:H0b}), we expand its expression in $Z$ and $\eps$ at the order ($Z^2$,$\eps Z^0$). We obtain the expression (\ref{eq:H0i}), whose trajectories verify the equations (\ref{eq:syserdi}) canonically associated to this Hamiltonian. These equations are equivalent to a second order differential equation (\ref{eq:eqerdi}), generalising the results of \cite{Ed1977} to the planetary case. From these equations, recalling the change of variable (\ref{eq:tl1}), we can compute the evolution of the variables ($\Lambda_j,\lambda_j$), see equations (\ref{eq:varerdi}) and (\ref{eq:varerdici}).}

\subsection{Portrait de Phase sur la variété circulaire plane}
\label{sec:pperdi}

\begin{figure}[h!]
\begin{center}
\includegraphics[width=0.6\linewidth]{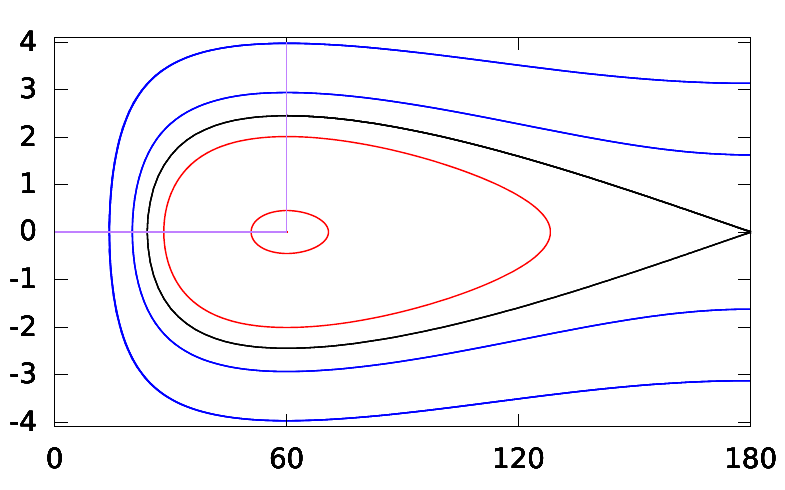}\\
  \setlength{\unitlength}{1cm}
\begin{picture}(.001,0.001)
\put(-5,3.5){{$\frac{\dot{\zeta}}{\sqrt{\mu}}$}}
\put(0,0.5){{$\zeta$}}
\end{picture}
\caption{\label{fig:ppH0b} Portrait de phase de l'équation (\ref{eq:eqerdi}) tracé pour des masses égales. Le point $L_4$ se situe en $(60,0)$. Autour de ce point, en rouge, librent les configurations troyennes. la courbe noire représente la séparatrice issue de l'équilibre instable $L_3$ $(180,0)$. Au delà de cette séparatrice, les configurations fers à cheval (en bleu) librent autour des points $L_3$, $L_4$ et $L_5$. Ce portrait de phase est symétrique par rapport à $\zeta=180^\circ$. Les droites violettes seront utilisées comme conditions initiales pour l'étude de stabilité menée en section suivante. Voir le texte pour plus de détails.
}
\end{center}
\end{figure}
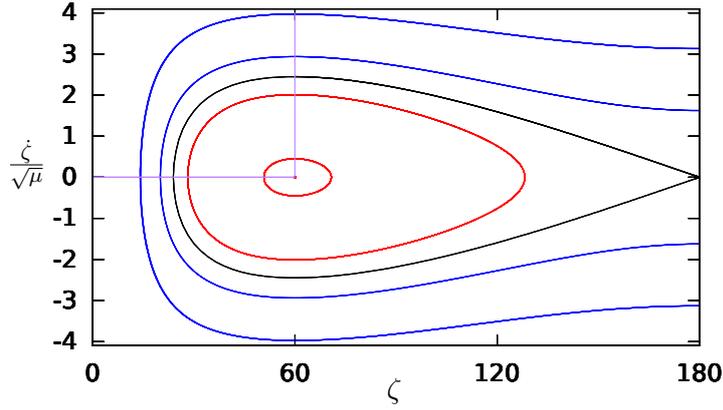

Le portrait de phase du hamiltonien à un degré de liberté $\Hb_0$ est représenté en Fig. \ref{fig:ppH0b} dans le plan ($\zeta,\dot{\zeta}/\sqrt{\mu}$), avec:
\begin{equation}
\mu = \eps \frac{m'_1+m'_2}{m_0+\eps(m'_1+m'_2)}\, .
\label{eq:mu}
\end{equation}
%

Cette figure a été obtenue pour des valeurs particulières des masses et des paramètres données dans la description de la figure, mais la topologie de l'espace des phases ne dépend pas de ces valeurs. 
%
%
Le portrait de phase de $\Hb_0$ est invariant par symétrie par rapport à $\zeta=180^\circ$: $(\zeta,\dot{\zeta}/\sqrt{\mu}) \rightarrow (360^\circ-\zeta,\dot{\zeta}/\sqrt{\mu})$. On concentrera donc notre description de l'espace des phases sur $\zeta \in [0,180^\circ]$. De plus, quand les coorbitaux sont de masse égale, le hamiltonien $\Hb_0$ et donc son portrait de phase, sont invariants par la symétrie $(\zeta,\dot{\zeta}/\sqrt{\mu}) \rightarrow (\zeta,-\dot{\zeta}/\sqrt{\mu})$. On rappelle que $\dot{\zeta}$ est proportionnel à la variable $Z$ (eq. \ref{eq:syserdi}).
%

Comme nous pouvons le voir Figure~\ref{fig:ppH0b}, toutes les trajectoires solutions de l'équation (\ref{eq:eqerdi}) peuvent être entièrement déterminées par des conditions initiales de la forme $(t_0,\zeta_0)$ tel que $\zeta(t_0)=\zeta_0$ et $\dot\zeta(t_0) =0$, où $\zeta_0$ est la valeur minimale de $\zeta$ sur sa trajectoire, et $t_0$ est le premier instant positif pour lequel $\zeta_0$ est atteint. 
Les valeurs possibles de $\zeta_0$, représentées par la ligne horizontale violette, sont incluses dans intervalle $]0^\circ,60^\circ]$; $\zeta_0 = 60^\circ$ correspond à l'équilibre $L_4$, situé en:
\be
\zeta =60^\circ \ \ \ \hspace{1cm} \text{et} \ \ \hspace{1cm} \  Z = \frac\eps6 \frac{m'_1-m'_2}{m'_1+m'_2}\frac{m'_1m'_2}{m_0} \mu_0^{2/3} \eta^{-1/3} + \gO(\eps^2)\, .
\label{eq:L4loc}
\ee
Les orbites troyennes sont situées en $\zeta_0 \in ]\zeta_s, 60^\circ[$, avec $\zeta_s \approx 23.9^\circ$, pour lesquelles quelques exemples ont été tracés en rouge. Cette région, qui a une épaisseur maximale de l'ordre de $\eps^{1/2}$ \citep{RoPo2013} est délimitée par la séparatrice émergeant du point fixe hyperbolique $L_3$ situé en:
\be
\zeta = 180^\circ \ \ \ \hspace{1cm} \text{et} \ \ \hspace{1cm} \  Z = \frac\eps2 \frac{m'_1-m'_2}{m'_1+m'_2}\frac{m'_1m'_2}{m_0} \mu_0^{2/3} \eta^{-1/3} + \gO(\eps^2)\, .
\label{eq:L3loc}
\ee
A l'extérieur de cette courbe se trouvent les configurations en fer-à-cheval, représentées ici en bleu. La largeur de cette région est de l'ordre de $\eps^{1/3}$ dans la direction $Z$, et $\zeta$ libre avec une grande amplitude tel que:
\be
\zeta \in \left[ \zeta_m , 2 \pi-\zeta_m \right] \ \text{avec}\ 0<\gO(\eps^{1/3})<\zeta_m<2 \arcsin((\sqrt{2}-1)/2)+ \gO(\eps).
\label{eq:HSlib}
\ee

Les formes des orbites du mouvement relatif sont entièrement déterminées par la quantité $\zeta_0$. En revanche, $t_0$ est nécessaire pour connaître la position des corps sur leur orbite à un instant donné, et $\eta\sqrt{\mu}$ est nécessaire pour connaître l'échelle dans le temps et dans la direction $Z$ du mouvement coorbital.\\ 

\an{
The phase portrait of the 1-degree of freedom Hamiltonian $\Hb_0$ is given figure~\ref{fig:ppH0b} in the ($\zeta,\dot{\zeta}/\sqrt{\mu}$) plane where $\mu$ is given eq. (\ref{eq:mu}). It was plotted for a given value of the masses, but the topology of the phase space does not depend on their value. The phase space is symmetric with respect to $\zeta=180^\circ$: $(\zeta,\dot{\zeta}/\sqrt{\mu}) \rightarrow (360^\circ-\zeta,\dot{\zeta}/\sqrt{\mu})$, we will hence focus our description on $\zeta < 180^\circ$. In addition, when $m_1=m_2$, the Hamiltonian $\Hb_0$ is invariant by the symmetry $(\zeta,\dot{\zeta}/\sqrt{\mu}) \rightarrow (\zeta,-\dot{\zeta}/\sqrt{\mu})$. We recall that $\dot \zeta$ is proportional to the variable $Z$ (see eq. \ref{eq:syserdi}). Note that any trajectory in this phase space can be identified by its initial conditions $(t_0,\zeta_0)$ such that $\zeta(t_0)=\zeta_0$ and $\dot\zeta(t_0) =0$, where $\zeta_0$ is the minimal value of $\zeta$ on its trajectory, and $t_0$ is the first positive instant when $\zeta_0$ is reached. The possible values for $\zeta_0$ are given by the purple horizontal line for the trojan ($\zeta < 180^\circ$) and horseshoe configurations.  \\ 
The position of the $L_4$ and $L_3$ equilibriums are given by the equations (\ref{eq:L4loc}) and (\ref{eq:L3loc}). The tadpole domain lie within the black separatrix emanating from $L_3$, hence for $\zeta_0 \in ]\zeta_s, 60^\circ[$ with $\zeta_s \approx 23.9^\circ$. This domain has a maximum width of the order of $\eps^{1/2}$ \citep{RoPo2013}.\\
The horseshoe domain lies outside the separatrix, with a maximum width of the order of $\eps^{1/3}$. In this domain, $\zeta$ librates with a large amplitude (eq. \ref{eq:HSlib}).\\
Note that the shape of the orbit is set by the value of $\zeta_0$. $t_0$ gives the position of the bodies at a given time, and $\eta\sqrt{\mu}$ gives the time scale of the resonant motion and the scale in the $Z$ direction. 
}

\subsection{Évolution des élément orbitaux dans le cas quasi-circulaire}

Le hamiltonien moyen (\ref{eq:Hb}) étant quadratique en les variables $x_j$, donc en excentricité, le système d'équation (\ref{eq:syserdi}) reste valable à l'ordre $1$ en excentricité. En revenant aux expressions des éléments orbitaux à l'ordre 1 en excentricité, nous pouvons écrire:
\be
a_j = \ab\left(1  + (-1)^j\frac23 \frac{m'_k}{m'_1+m'_2}\frac{\dot\zeta}{n}\right) +\gO(\eps,e\sqrt{\eps},e^2), \quad k\neq j.
\label{eq:a_sol}
\ee
où $\bar{a}$ est le demi grand axe associé au moyen mouvement $\eta$. Nous obtenons:
\begin{equation}
\begin{aligned}
&r_j = \ab\left(1 - e_j\cos(\lam_j -\varpi_j) + (-1)^j\frac23 \frac{m'_k}{m'_1+m'_2}\frac{\dot\zeta}{n} \right) +\gO(\eps,e\sqrt{\eps},e^2)\, , \\
&f_j = \lam_j + 2e_j\sin(\lam_j -\varpi_j)  +\gO(\eps,e\sqrt{\eps},e^2)\, .
\end{aligned}
\label{eq:polar}
\end{equation}
où $\lambda_j$ est donné par l'expression (\ref{eq:varerdi}).


 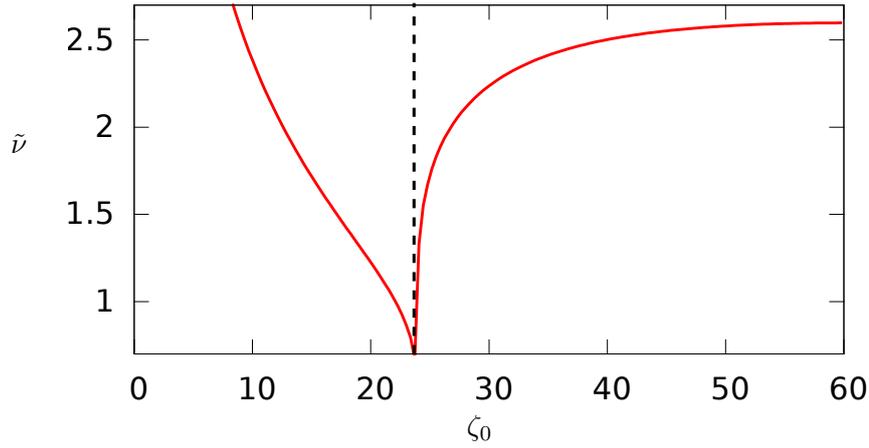
\begin{figure}[h!]
\centering
 \includegraphics[width=.7\linewidth]{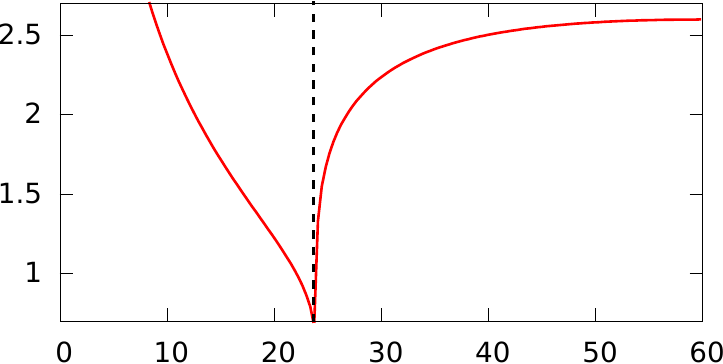}\\
   \setlength{\unitlength}{0.1\linewidth}
\begin{picture}(.001,0.001)
\put(-4,2.5){{$\tilde \nu$}}
\put(0,0){{$\zeta_0$}}
\end{picture}
\caption{Variation de la fréquence de libration normalisée $\tilde{\nu}=\nu/(\eta\sqrt{\mu})$ le long de la ligne violette horizontale de la figure~\ref{fig:ppH0b}. Dans la région troyenne, la fréquence normalisée décroît de $\sqrt{27/4}$ en $L_4$ jusqu'à $0$ à la séparatrice. Cette fréquence réaugmente dans le domaine des fers à cheval et tend vers l'infini quand $\zeta_0$ tend vers $0^\circ$, ce qui n'a pas de signification physique car les approximations nécessaires à l'obtention du modèle à un degré de liberté ne sont plus valables si proches de la collision.
%
}
\label{fig:zr}
\end{figure}

En développant l'équation (\ref{eq:eqerdi}) au voisinage de $\zeta=\pi/3$, on obtient \citep{Charlier1906}:
\be
\nu_0 = \eta \sqrt{\frac{27}{4}\mu}.
\label{eq:nu_L}
\ee
Plus généralement, à l'exception de la séparatrice, les solutions de l'équation (\ref{eq:eqerdi}) sont périodiques. La fréquence associée $\nu$ dépend de l'orbite. Les variations de la fréquence normalisée $\tilde{\nu}=\nu/(\eta\sqrt{\mu})$ le long de la ligne violette horizontale de la figure \ref{fig:ppH0b} sont représentées en figure \ref{fig:zr}. Cette courbe est obtenue en intégrant numériquement le modèle à un degré de liberté (eq. \ref{eq:eqerdi}). En configuration troyenne, cette fréquence sans dimension reste presque constante dans un voisinage de l'équilibre de Lagrange $L_4$ où $\nu \approx \nu_0$ (Eq.\,\ref{eq:nu_L}) et tend vers $0$ quand on atteint la séparatrice en $\zeta_0 = \zeta_s  \approx 23.9^\circ$. En configuration fer-à-cheval, $\nu$ varie grandement. Sur la figure~\ref{fig:zr}, on peut voir que tant que on reste suffisamment loin de la collision, $\tilde{\nu}$ est toujours proche de l'unité ou moins. Cela confirme que l'angle résonant $\zeta$, est lent par rapport à l'échelle de temps orbital, i.e. $\nu \ll n$ tant que $\sqrt{\eps} \ll 1$. \\
%
 
 \an{
 Up to order $1$ in eccentricity, the equations (\ref{eq:a_sol}) and (\ref{eq:polar}) give the semi-major axis $a_k$ the distance to the star $r_k$, and the true longitude $f_k$ in the co-orbital case, where $\bar a$ is the averaged semi-major axis associated to the mean mean-motion $\eta$.\\
In the vicinity of the Lagrangian equilibriums, the fundamental frequency $\nu$ is given by the equation (\ref{eq:nu_L}). The evolution of this frequency along the horizontal purple line on the figure~\ref{fig:ppH0b} is given in figure~\ref{fig:zr} (we show the normalised frequency $\tilde{\nu}=\nu/(\eta\sqrt{\mu})$). This graph is obtained by integrating the 1-degree of freedom model (this model is not valid in the vicinity of the collision $\zeta_0=0$). Note that as long as we are far from the collision, the normalised frequency is always of the order of the unity. We can hence consider that $\nu \ll \eta$ if $\sqrt{\mu} \ll 1$.  
 }

\section{Stabilité des orbites coorbitales quasi-circulaires}
\label{sec:stabcirc}

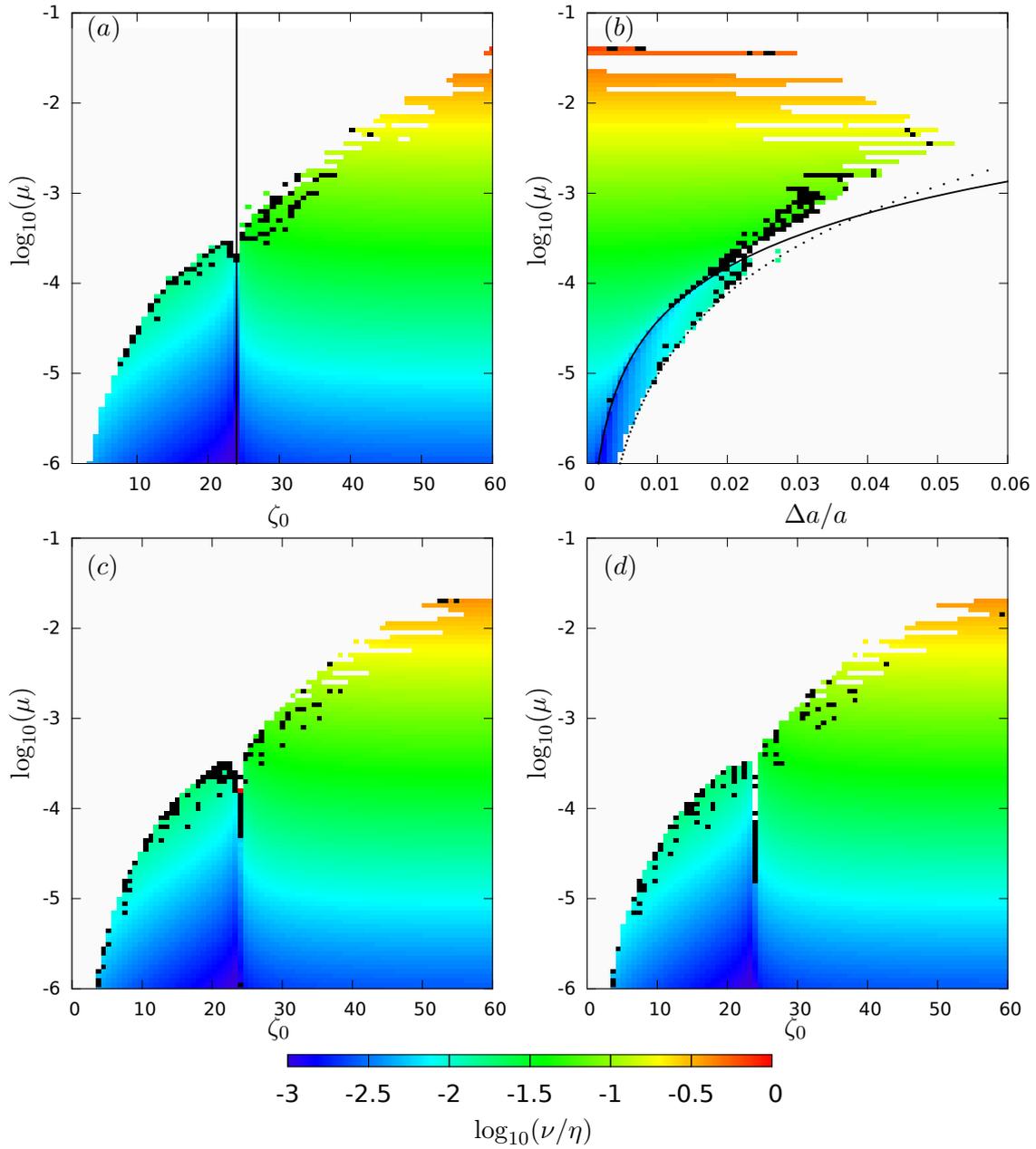
\begin{figure}[h!]
\begin{center}
\includegraphics[width=.49\linewidth]{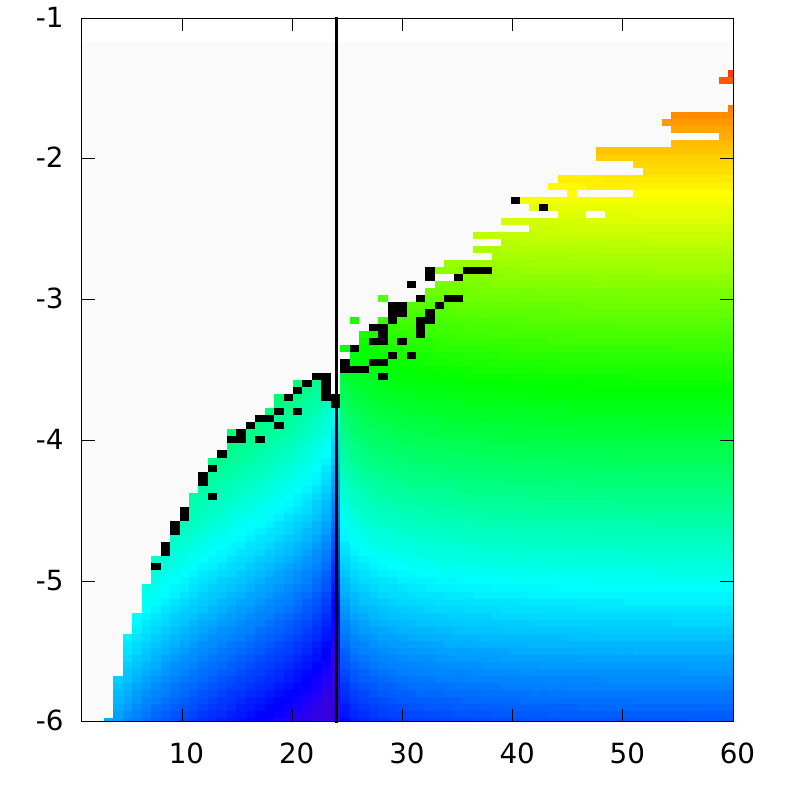}
\includegraphics[width=.49\linewidth]{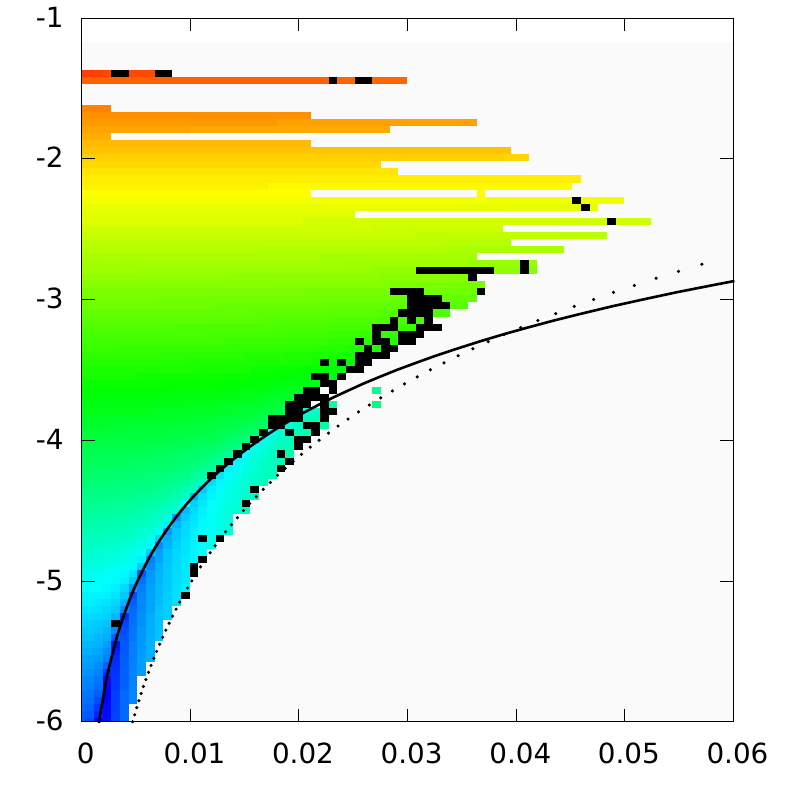}\\
\vspace{0.3cm}
\includegraphics[width=.49\linewidth]{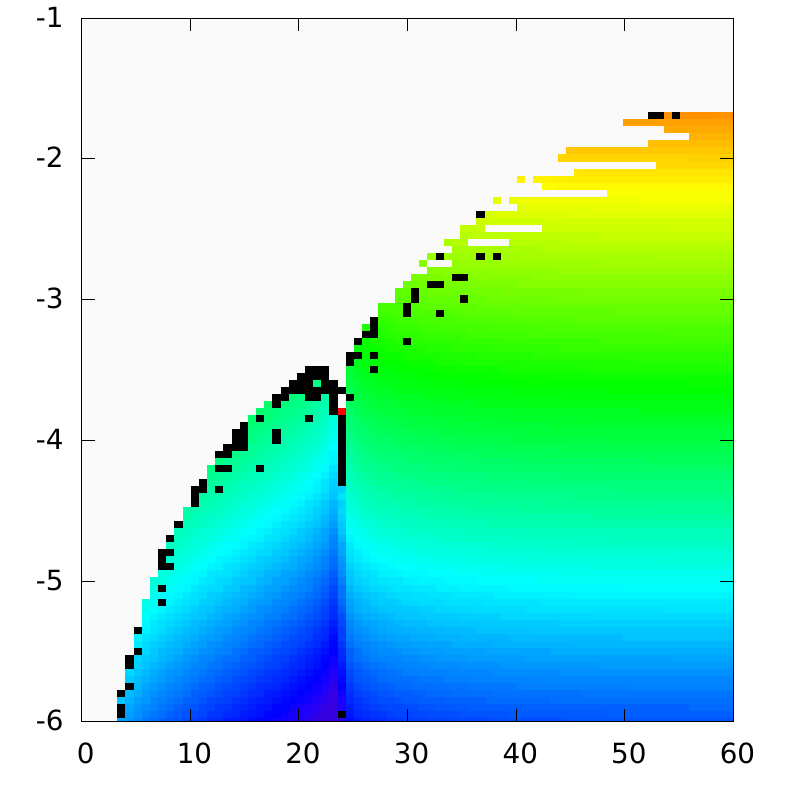}
\includegraphics[width=.49\linewidth]{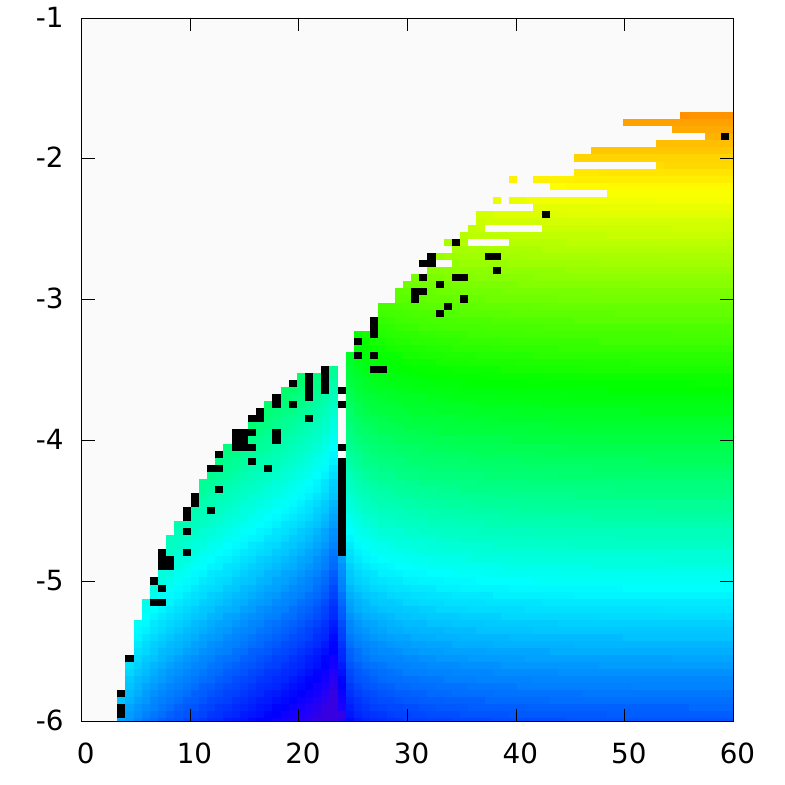}\\
\vspace{0.3cm}
\includegraphics[width=.49\linewidth]{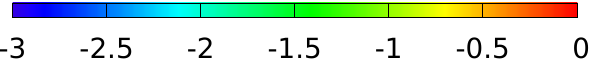}\\
  \setlength{\unitlength}{0.1\linewidth}
\begin{picture}(.001,0.001)
\put(-5,8.5){\rotatebox{90}{$\log_{10}(\mu)$}}
\put(0,8.5){\rotatebox{90}{$\log_{10}(\mu)$}}
\put(-5,3.5){\rotatebox{90}{$\log_{10}(\mu)$}}
\put(0,3.5){\rotatebox{90}{$\log_{10}(\mu)$}}
\put(-2.5,1){{$\zeta_0$}}
\put(2.5,1){{$\zeta_0$}}
\put(-2.5,6){{$\zeta_0$}}
\put(2.5,6){{$\Delta a/a$}}
\put(-4.25,10.75){{$(a)$}}
\put(0.75,10.75){{$(b)$}}
\put(-4.25,5.5){{$(c)$}}
\put(0.75,5.5){{$(d)$}}
\put(-0.5,0){{$\log_{10}(\nu/\eta)$}}
\end{picture}
\vspace{0.5cm}
\caption{\label{fig:stabzr} Stabilité des coorbitaux quasi-circulaires coplanaires en fonction de $\log_{10}(\mu)$ et $\zeta_0$ pour les graphes (a), (c) et (d), et $\Delta a/a$ pour le graphe (b). Pour les graphes (a) et (b) $m_2=m_1$, pour le graphe (c) $m_2=10 m_1$ et pour le graphe (d) $m_2=100 m_1$.  La ligne noire continue sur les graphes (a) et (b) indique la position de la séparatrice entre le domaine des troyens et des fer-à-cheval. Le code couleur indique la valeur de la fréquence de libration. Voir le texte pour plus de détails.
}
\end{center}
\end{figure}

Nous nous intéressons ici à l'influence qu'ont les masses des coorbitaux sur la stabilité des configurations coplanaires quasi-circulaires dans le cas planétaire. Pour cette étude, nous allons intégrer numériquement le problème à trois corps en balayant l'ensemble des configurations coorbitales possibles ($\zeta_0 \in ] 0, 60^\circ ] $) pour $\mu= \frac{m_1+m_2}{m_0+m_1+m_2} \in [10^{-6 },10^{-1}]$ dans trois cas: le cas $m_2=m_1$ (figure \ref{fig:stabzr} - (a)), le cas $m_2=10 m_1$ (c) et le cas $m_2 = 100 m_1$ (d). Dans chacune de ces figures, on considère deux planètes orbitant une étoile de masse $m_0=1\,M_\odot$, avec les éléments orbitaux initiaux suivants fixés: $a_1 = a_2 =1$~ua, $e_1 = e_2 =0.05$, and $\lambda_1=\varpi_1=0^\circ$. Nous faisons varier les éléments initiaux suivants $\lambda_2=\varpi_2=-\zeta_0$ et leur masse. Pour chaque jeu de conditions initiales, les systèmes sont intégrés pendant 5 millions d'années au moyen de l'intégrateur symplectique SABA4 \citep{LaRo2001} avec un pas d'intégration de $0.0101$~an. Pour la figure (b), on prend les conditions initiales sur la ligne violette verticale de la figure \ref{fig:ppH0b}, en gardant $\zeta_2=\varpi_2=-60^\circ$ pour chaque intégration avec $m_1=m_2$.

Les systèmes fortement chaotiques qui quittent la résonance coorbitale avant la fin de l'intégration, ou qui subissent des rencontres proches (variation relative d'énergie au dessus de $10^{-6}$), sont retirés de l'intégration. Dans ce cas, la couleur blanche est assignée à leurs conditions initiales $(\zeta_0,\mu)$ sur les figures (\ref{fig:stabzr}). Ces instabilités à court terme sont principalement dues au recouvrement de résonances secondaires \citep{RoGa2006,PaEf2015}. Après l'élimination de ces conditions initiales, la diffusion à long terme le long de ces résonances secondaires peut déstabiliser les configurations coorbitales sur des temps beaucoup plus longs. Une façon d'identifier cette diffusion est de mesurer la variation temporelle de la fréquence de libration \citep{Laskar1990,La99}. La couleur noire est assignée aux conditions initiales pour lesquelles la fréquence $\nu$ a une variation relative de plus de $10^{-6}$ entre la première moitié et la seconde moitié des 5 millions d'années d'intégration. Les trajectoires ainsi identifiées sont principalement situées le long de la séparatrice et proches des configurations éjectées. Dans les régions restantes, les faibles variations de $\nu$ garantissent, dans la plupart des cas, la stabilité sur des milliard d'années \citep{Laskar1990,RoGa2006}. Pour les systèmes stables sur le long terme, la couleur assignée représente la valeur de la fréquence de libration (voir le code couleur en bas de la figure~\ref{fig:stabzr}).
%
%
%

Nous observons que pour des masses proches de $\mu\approx 0.037$ \citep{Ga1843}, l'amplitude de libration des orbites stables est très faible et le domaine de stabilité est fortement perturbé par des résonances secondaires. 
Le chaos induit par ces résonances ($\nu = \eta/2$, $\nu = \eta/3$, et $\nu = \eta/4$) réduit la région de stabilité significativement, la confinant au voisinage des configurations équilatérales $L_4$ et $L_5$ \citep[voir][]{Robe2002,Nauenberg2002}. Quand $\mu$ diminue, la région troyenne stable s'élargit dans la direction des $\zeta_0$, augmentant l'amplitude de libration maximale que peuvent atteindre les configurations troyennes. L'influence déstabilisatrice des résonances secondaires ne reste dominante qu'au bord des régions stables \citep[voir][pour le problème restreint]{PaEf2015,RoGa2006,ErNa2007}.

Quand $\mu \approx 3\times 10^{-4} \approx 2 M_{Saturne}/M_{\odot}$, l'ensemble du domaine troyen devient stable à l'exception d'une petite région au voisinage de la séparatrice ($\zeta_0 = \zeta_s \approx 23.9^{\circ}$). De l'autre coté de la séparatrice, pour $\zeta_0<\zeta_s$, des fers à cheval stables commencent à apparaître \citep[voir][]{LauCha2002}. Pour des masses plus faibles, La taille du domaine des fers à cheval augmente quand $\mu$ diminue jusqu'à atteindre la limite extérieure de la sphère de Hill à une distance de la collision de l'ordre de $\mu^{1/3}$ \citep[voir][]{RoPo2013}.

La comparaison des figures (a), (c) et (d) montre que la répartition de masse entre les deux coorbitaux n'a que peu d'influence sur la stabilité. On constate cependant une légère diminution des domaines de stabilité, notamment dans le voisinage immédiat de la séparatrice.

La figure \ref{fig:stabzr} (b) représente les mêmes configurations que la figure (a) avec des conditions initiales le long de la ligne violette verticale (fig~\ref{fig:ppH0b}). Sur cette figure, le domaine troyen se situe au-dessus de la ligne noire continue d'équation $ \Delta a = 2\sqrt{2}/\sqrt{3} \sqrt{\mu}$ \citep{RoPo2013}. Dans ce cas, contrairement à la direction $\zeta_0$ où la largeur de la région troyenne stable est grossièrement monotone en $\mu$, la largeur de la région stable atteint ici un maximum en $\Delta a/a \approx 0.052$ pour $\mu = 3.5\times 10^{-3}$ puis tend vers $0$ quand $\mu$ décroit. La courbe en pointillé est proportionnelle à $\mu^{1/3}$ (d'équation $ \Delta a = 0.47 \mu^{1/3}$) et a été ajustée à la frontière extérieure du domaine de stabilité des fers à cheval \citep[voir][]{RoPo2013}. La largeur de ce domaine tend aussi vers $0$ quand $\mu$ décroit, mais moins vite que celle du domaine des troyens. Par conséquence, le domaine des fers à cheval stables devient plus important que celui des configuration troyennes quand la masse des coorbitaux tend vers $0$ \citep{DeMu1981a}.\\

\an{In this section we study numerically the stability of quasi-circular coplanar co-orbitals. We integrate the 3-body problem for a grid of initial conditions. The results are displayed in figure \ref{fig:stabzr}. As we saw in section \ref{sec:pperdi}, taking initial conditions in the $\zeta_0$ direction while studying $t_0=0$ allows to study all the possible co-orbital configurations of the coplanar circular case. We hence take ($\zeta_0 \in [ 0, 60^\circ ] $) and $\mu= \frac{m_1+m_2}{m_0+m_1+m_2} \in [10^{-6 },10^{-1}]$ for our grid of initial conditions for the graphs (a), (c) and (d). On the graph (b), we check the width of the stability domain in the direction $Z$. We set $m_0=1\,M_\odot$, $a_1 = a_2 =1$~AU, $e_1 = e_2 =0.05$, $\varpi_2=\lambda_2$, and $\lambda_1=\varpi_1=0^\circ$. One can check \citep{LeRoCo2015} (in annex), or the section \ref{sec:Coo2meg}, for more details regarding the integration. \\
The mass of each planet is given by the $y$ coordinate (the value of $\mu$) and the relation between $m_1$ and $m_2$. For the graphs (a) and (b) $m_2=m_1$, for (c) we have $m_2=10 m_1$, and for (d) $m_2=100 m_1$.
We integrate each initial condition over $5 \times 10^{6}$~orbital periods. Trajectories ejected from the resonance before the end of the integration are identified by white pixels. These short term instabilities are generally due to the overlap of secondary resonances \citep{RoGa2006,PaEf2015}. The black pixels identifies the initial conditions for which the diffusion of the libration frequency $\nu$ between the first and second half of the integration is higher than  $10^{-6}$. Most of the black pixels are close from the stability boundary or the separatrix. Other trajectories are expected to be stable for a duration long with respect to $10^7$ orbital periods \citep{Laskar1990,RoGa2006}. For these trajectories, the color code gives the value of $log_{10}(\nu/\eta)$.
\\}

\an{For $\mu$ close to the Gascheau's criterion value ($\mu\approx 0.037$), stable orbits stays in the vicinity of the Lagrangian equilibrium, confined by the chaos induced by the resonances $\nu = \eta/2$, $\nu = \eta/3$, and $\nu = \eta/4$. As $\mu$ decreases, orbits with larger amplitude of libration become stable, until stable horseshoe configurations appear for $\mu \approx 3\times 10^{-4}$ or less. For these lower value of $\mu$, the instability induced by the resonances is significant only near the stability border \citep[see][in the restricted case]{PaEf2015,RoGa2006,ErNa2007}. The stability domain of the horseshoe configuration in the $\zeta_0$ direction is bound by the hill sphere around the collision, of width $\mu^{1/3}$ \citep[see][]{RoPo2013}. \\     
The graphs (a), (c) and (d), show that the mass repartition between co-orbitals does not impact much the stability, excepted in the vicinity of the separatrix. \\
Finally, the graph (b) represents another section of the same phase space than the graph (a): the initial conditions are taken along the purple vertical line on figure~\ref{fig:ppH0b}. The black curves delimit the trojan and horseshoe domains. The equation of the continuous line is $ \Delta a = 2\sqrt{2}/\sqrt{3} \sqrt{\mu}$, and the dashed line is fitted proportionally to $\mu^{1/3}$ (its equation is $ \Delta a = 0.47 \mu^{1/3}$). Combining the information of the graphs (a) and (b), we find that the co-orbital domain is at its largest for $10^{-3} < \mu < 10^{-2}$, and that the horseshoe domain ($\propto \mu^{1/3}$) becomes larger than the tadpole one ($\propto \mu^{1/2}$) as $\mu$ tends to $0$ \citep{DeMu1981a}.       
}

  \chapter{Dynamique des co-orbitaux excentriques coplanaires}

Nous avons vu dans le Chapitre 1 que la dynamique sur la variété circulaire coplanaire se réduit à celle d'un système à un degré de liberté dont le portrait de phase est bien connu (voir section \ref{sec:pperdi}): le problème comporte 2 points d'équilibre stables, les points $L_4$ et $L_5$ qui sont des points fixes du problème moyen. Au voisinage de ces positions d'équilibre se trouve le domaine des configurations troyennes qui s'étend jusqu'à la séparatrice émergeant du point d'équilibre instable $L_3$. Au delà de la séparatrice se trouve le domaine des fers-à-cheval. 

Les équilibres $L_3$, $L_4$ et $L_5$ existent également dans le cas excentrique pour $e_1=e_2$, avec $\zeta=\Dv=180^\circ$ pour $L_3$ ($\Dv=\varpi_1-\varpi_2$), $\zeta=\Dv=60^\circ$ pour $L_4$ et $\zeta=\Dv=-60^\circ$ pour $L_5$. Dans le problème moyen, en plus de ces familles de points fixes, des familles d'orbites périodiques $AL_3$, $AL_4$ et $AL_5$ émergent des équilibres $L_k$ circulaires. Pour de faibles excentricités, la dynamique du degré de liberté associé à ($Z,\zeta$) est inchangée (l'équation \ref{eq:eqerdi} reste valable à l'ordre $1$ en excentricité): les orbites troyennes librent autour de $L_4$, $L_5$, $AL_4$ ou $AL_5$ et les orbites en fer-à-cheval gardent une grande amplitude de libration de l'angle $\zeta$.

On s'intéresse dans ce chapitre à des excentricités modérées à grandes (jusqu'à $0.7$). Nous avons vu dans le Chapitre 1 que la variété des coorbitaux coplanaires ($\mathcal{C}_{y=0}$) est invariante, nous pouvons donc considérer le cas excentrique plan. Comme nous allons le voir dans ce chapitre, les familles de points fixes du problème moyen ainsi que les familles d'orbites quasi-périodiques de dimension non maximale entraînent des changements de topologie importants quand l'énergie cinétique totale du système augmente (les excentricités augmentent) et sont au centre des différentes régions stables et instables de l'espace des phases. Ce chapitre porte donc à la fois sur l'évolution de ces familles, et sur la dynamique des différentes configurations co-orbitales qui librent autour de celles-ci: les configurations troyennes, fers-à-cheval et d'autres qui apparaîtront à plus haute excentricité.  

\vspace{1cm}

\an{The dynamics of coplanar circular co-orbitals is well known (see chapter 1, \cite{Ed1977,RoPo2013}). There are two stable equilibrium points $L_4$ and $L_5$ which are fixed points of the averaged problem. Tadpole orbits librate around these equilibriums. The horseshoe domain lies outside the separatrix emanating from the unstable equilibrium $L_3$ (see figure~\ref{fig:ppH0b}).\\
These $3$ equilibriums exist also in the eccentric case for $e_1=e_2$, with $\zeta=\Dv=180^\circ$ for $L_3$ ($\Dv=\varpi_1-\varpi_2$), $\zeta=\Dv=60^\circ$ for $L_4$ and $\zeta=\Dv=-60^\circ$ for $L_5$. In addition, the new families of periodic (in the averaged problem) orbits $AL_3$, $AL_4$ and $AL_5$ emerge from the circular equilibriums $L_k$. For low eccentricities, the dynamics of the degree of freedom ($Z$,$\zeta$) is similar to the circular case (the equation \ref{eq:eqerdi} is valid up to the first order in eccentricities): tadpole orbits librate around $L_4$, $L_5$, $AL_4$ or $AL_5$ and the horseshoe orbits librate with large amplitude. \\
In this chapter, we consider moderate to large eccentricities (up to $0.7$). We saw in chapter 1 that the manifold of coplanar co-orbitals $C_{y=0}$ is invariant by the flow of the averaged Hamiltonian, we can thus consider coplanar eccentric orbits. We will see in section \ref{sec:Coo2meg} that the families of quasi-periodic orbits of non-maximal dimension will undergo strong modification when the total angular momentum of the system increases for a fixed values of the semi-major axis (hence when the eccentricities increase). These modifications will cause significant topological changes in the phase space. This chapter is hence focused on the evolution of these families and the dynamics of the various orbital configurations that librate around them: tadpole orbits, horseshoe orbits, and others which will appear for higher eccentricities.}

\section{Approche du problème}
\label{sec:adp}

 L'espace des phases complet pour le problème excentrique plan est de dimension $8$. La moyenne sur l'angle rapide $\zeta_2=\lambda_2$ peut réduire le nombre de dimensions à $6$ (voir Chapitre 1). Une réduction à $4$ est possible en utilisant la conservation du moment cinétique total. Pour finir, une seconde moyenne sur l'angle résonant $\zeta$ est envisageable pour obtenir un espace des phases de dimension $2$. Nous voulons travailler autant que possible sur un espace des phases de dimension minimale pour des raisons évidentes de facilité d'interprétation des résultats.  \\

Pour des excentricités modérées, la moyenne sur l'angle rapide peut être effectuée analytiquement comme nous l'avons fait au Chapitre 1. Si l'excentricité n'est pas petite, les moyennes doivent être effectuées numériquement, mais l'intégration d'une trajectoire avec moyenne numérique est longue, et la 2ème moyenne (sur l'angle semi-rapide $\zeta$) nécessite en plus le calcul de coordonnées action angle pour chaque orbite \citep{BeaugeR01}. Cependant, si on considère des coorbitaux de masse suffisamment faible, la dynamique du problème moyenné sera très proche de la dynamique des variables équivalentes dans le problème complet. En effet, si $\sqrt{\eps}<<1$ (et donc a fortiori $\eps <<1$), l'influence des courtes périodes sur les variables du problème moyen est négligeable (perturbations de taille $\eps$), et l'échelle de temps semi-rapide affecte peu le problème séculaire \citep[perturbations de taille $\sqrt{\eps}$, voir][]{Morbidelli02}.

Dans cette étude, nous considérons le cas planétaire. Si la masse des planètes est de l'ordre de celle de Jupiter ($\eps>10^{-3}$), nous ne pourrons que constater le résultat d'intégrations numériques du problème complet. Pour de telles masses, les domaines de stabilités sont relativement restreints dans le cas circulaire (voir  figure \ref{fig:stabzr}) et dans le cas excentrique \citep[voir][]{GiuBeMiFe2010}. De plus, ils sont perturbés par des résonances entre le moyen mouvement et la fréquence de libration: $\eta/\nu \approx 10$ dans le cas de Jupiter, \cite[voir][]{RoGa2006}.
En revanche, si la masse des planètes est de l'ordre de celle de saturne ou moins ($\eps\leq 10^{-4}$), ces résonances sont moins importantes. De plus, nous pourrons considérer que $\sqrt{\eps}\ll 1$. Dans ce cas, l'évolution des variables semi-rapides (resp. séculaire) lors de l'intégration du problème complet sera proche de leur évolution dans le problème moyen (resp. doublement moyenné). Nous pouvons donc comparer le résultats de calculs analytiques et semi-analytiques obtenus dans le problème moyen, aux résultats d'intégrations numériques du problème complet.\\

Dans un premier temps, nous étudierons le voisinage des équilibres de Lagrange circulaires analytiquement à l'aide d'une approximation quadratique du hamiltonien moyen (\ref{eq:Hb}), puis nous étendrons nos résultats au reste de l'espace des phases à l'aide de méthodes numériques. Enfin, nous vérifierons analytiquement que la position des objets mis en évidence de manière numérique et semi-analytique correspond à celle attendue.

\vspace{1cm}

\an{The phase space of the coplanar excentric co-orbital problem has 8 dimensions. The averaging on the fast angle $\lambda_j$ can reduce it to 6. The conservation of the total angular momentum can bring it down to $4$ dimensions. Finally, if we average the system over the semi-fast resonant angle $\zeta$ we are left with 2 dimensions. We obviously want to study a phase space of minimal dimension.\\
For moderate eccentricities, the average over the fast angle $\zeta_2=\lambda_2$ can be done analytically. However, for higher eccentricities, the averaging must be done numerically. It requires time, and the second averaging over the semi-fast angle $\zeta$ presents technical difficulties: it requires the computation of action-angle variables for each orbit \citep{BeaugeR01}. However if we consider co-orbitals with small masses with respect to the star, the dynamics of the averaged problem is very close to the dynamics of the associated variables in the full 3-body problem. Indeed, if $\sqrt{\eps} \ll 1$ (we recall that $\eps=\max{(\frac{m_1}{m_0},\frac{m_2}{m_0})}$), the impact of the short period on the variables of the average problem is negligible (of order $\eps$), and the semi-rapid time-scale does not impact much the secular problem \citep[perturbation of size $\sqrt{\eps}$, see][]{Morbidelli02}.  
Thus, we can integrate the full 3-body problem with the masses of the co-orbital below $10^{-4} m_0$ and consider on the one hand the evolution of the variables ($Z,\zeta$) as their evolution in the averaged problem, and on the other hand the evolution of the variables $e_j$ and $\varpi_j$ as their evolution in the secular problem. The studies of co-orbital of masses of the order of $10^{-3} m_0$ is of limited interest because the stability domain shrinks very fast in the eccentric case \citep{GiuBeMiFe2010}.\\
}

\an{
At first, we study the neighbourhood of the circular Lagrangian equilibriums in the eccentric direction using a quadratic approximation of the averaged Hamiltonian (\ref{eq:Hb}). Then, we will observe how the global phase space fares when the total angular momentum increases, using semi-analytical and numerical methods. Finally, we will head back to analytical methods to confirm the identification of the families of quasi-periodic orbits that we observed numerically.
}

\section{Dynamique au voisinage des points fixes circulaires $L_k$}
\label{sec:H0pf}

On s'intéresse ici à la dynamique au voisinage des équilibres de Lagrange circulaires $L_3$, $L_4$ et $L_5$, notamment au calcul de la tangente aux familles $L_k$ et $AL_k$ excentrique. Pour ce faire, on utilisera la partie quadratique du hamiltonien moyen (\ref{eq:Hb}) développé au voisinage de l'équilibre de Lagrange $L_k$ circulaire en fonction des variables $x_j$, $\tilde{x}_j$, $Z$ et $z=\zeta-\zeta_{L_k}$. La dynamique au voisinage de $L_4$ et $L_5$ étant équivalente pour des raisons de symétrie, les calculs ne seront développés que pour les points $L_4$ et $L_3$. Le hamiltonien moyen au voisinage de l'équilibre $L_k$ circulaire peut se décomposer de la manière suivante:
\begin{equation}
\ol H_{L_k}(\zeta,Z,x_j,\xt_j) = \sum_{(p,\tilde{p}) \in \mathbb{N}^6} C_{p,\tilde{p}}z^{p_0} Z^{\tilde{p}_0} x_1^{p_1} x_2^{p_2}\xt_1^{\tilde{p}_1} \tilde{x}_2^{\tilde{p}_2},
\label{eq:HbLk}
\end{equation}
 Où la partie quadratique de $\ol H_{L_4}$ s'écrit:
 \begin{equation}
 \begin{aligned}
\ol H^{(2)}_{L_4} = & h_{4,0} + h_{4,Z} Z + h_{4,Z^2} Z^2 + h_{4,z^2} z^2 \\
 & + h_{4,x_1 \xt_1} x_1 \xt_1  +  h_{4,x_2 \xt_2}x_2\xt_2+ h_{4,x_2 \xt_1} x_2 \xt_1 +  h_{4,x_1 \xt_2} x_1 \xt_2\, , 
\end{aligned}
\label{eq:Hquad1L4}
\end{equation}
avec
 \begin{equation}
 \begin{aligned}
&h_{4,Z}&=& + \eps \frac{\eta}{2}(m'_2-m'_1)  \\
& h_{4,Z^2}&= & + \frac{3}{2} \frac{\eta^{4/3}}{\cG^{2/3}m_0^{2/3}} [-(\frac{1}{m'_1}+\frac{1}{m'_2}) + \eps \frac{1}{m_0} (\frac{m'_1}{m'_2}+\frac{m'_2}{m'_1}-\frac{7}{3})]  \\
& h_{4,z^2}&=         & - \eps \frac{9}{8} \cG^{2/3} \frac{\eta^{2/3}}{m_0^{1/3}} m'_1 m'_2 \\
& h_{4,x_1 \xt_1}&=        & +i \eps \eta \frac{27}{8} \frac{m'_2 }{m_0} \\
 & h_{4,x_2 \xt_2}&=        & +i \eps \eta \frac{27}{8} \frac{m'_1}{m_0}  \\
 & h_{4,x_2 \xt_1}&=        & -i \eps \eta \frac{27}{8} \frac{\sqrt{m'_1m'_2}}{m_0}   \e^{i\pi/3} \\
 &   h_{4,x_1 \xt_2}&=        &+ \bar{h}_{x_2 \xt_1}\, .
\end{aligned}
\label{eq:Hquad1L4coef}
\end{equation}
  Pour $\ol H_{L_3}$ nous avons:
 \begin{equation}
 \begin{aligned}
\ol H^{(2)}_{L_3} = & h_{3,0} + h_{3,Z} Z + h_{3,Z^2} Z^2 + h_{3,z^2} z^2 \\
 & + h_{3,x_1 \xt_1} x_1 \tilde{x}_1  +  h_{3,x_2 \tilde{x}_2}x_2\xt_2+ h_{3,x_2 \tilde{x}_1} x_2 \tilde{x}_1 +  h_{3,x_1 \tilde{x}_2} x_1 \tilde{x}_2\, , 
\end{aligned}
\label{eq:Hquad1L3}
\end{equation}
avec
 \begin{equation}
 \begin{aligned}
&h_{3,Z}&=& + \eps \frac{3}{2} \eta(m'_2-m'_1)  \\
& h_{3,Z^2}&= & - \frac{3}{2} \frac{\eta^{4/3}}{\cG^{2/3}m_0^{2/3}} [(\frac{1}{m'_1}+\frac{1}{m'_2}) + \eps \frac{1}{m_0} (\frac{m'_1}{m'_2}+\frac{m'_2}{m'_1}-\frac{4}{3})]  \\
& h_{3,z^2}&=         &  \eps \frac{7}{16} \cG^{2/3} \frac{\eta^{2/3}}{m_0^{1/3}} m'_1 m'_2 \\
& h_{3,x_1 \tilde{x}_1}&=        & + i \eps \eta \frac{7}{8} \frac{m'_2 }{m_0} \\
 & h_{3,x_2 \tilde{x}_2}&=        & + i \eps \eta \frac{7}{8} \frac{m'_1}{m_0}  \\
 & h_{3,x_2 \tilde{x}_1}&=        & + i \eps \eta \frac{7}{8} \frac{\sqrt{m'_1m'_2}}{m_0}  \\
 &   h_{3,x_1 \tilde{x}_2}&=        &+ h_{3,x_2 \tilde{x}_1}\, .
\end{aligned}
\label{eq:Hquad1L3coef}
\end{equation}

Dans un premier temps, on se débarrasse des termes linéaires avec le changement de variable $\chi^{(1)}$ (\ref{eq:cv1f}). Ensuite, nous pouvons diagonaliser indépendamment la partie dépendant des variables $(z,Z)$ et la partie dépendant des variables $(x_j,\tilde{x}_j)$ par les transformations suivantes:\\
Pour $L_4$:\\
\begin{equation}
  \chi_{L_4}(z_0,\tilde{z}_0,z_1,\tilde{z}_1,z_2,\tilde{z}_2)= (z,Z,x_1,\xt_1,x_2,\xt_2)
\label{eq:cv1f}
\end{equation}
Pour $L_3$:\\
\begin{equation}
  \chi_{L_3}(z_0,\tilde{z}_0,z_1,\tilde{z}_1,z_2,\tilde{z}_2)= (z,Z,x_1,\xt_1,x_2,\xt_2)
\label{eq:cv1f}
\end{equation}

 Les matrices de passage de cette transformation peuvent être trouvées en annexe \ref{an:MP}. Nous rappellerons juste ici que la diagonalisation des variables $(x_j,\tilde{x}_j)$ s'effectue par la matrice des vecteurs propres de $M$ (eq. \ref{eq:eqvar2}) calculée en l'équilibre considéré ($L_4$ ou $L_3$). Nous appelons également $\ol H_{L_4}$ le hamiltonien du problème moyen dont la partie quadratique est diagonale et égale à:
 \begin{equation}
 \begin{aligned}
\ol H_{L_4}^{(2)} = &  \omega_0 z_0\tilde{z}_0 + \omega_1 z_1\tilde{z}_1 + \omega_2 z_2\tilde{z}_2 \, , 
\end{aligned}
\label{eq:HquadL4}
\end{equation}
où les variables $\tilde{z}_j$ vérifient la relation $\tilde{z}_j=-i \bar{z}_j$, et:
 \begin{equation}
 \begin{aligned}
&\omega_0 = - i \sqrt{\eps}  \eta \sqrt{ \frac{27}{4}  \frac{m'_1+m'_2}{m_0} } \, , \\
&\omega_1 = - i \eps  \eta \frac{27}{8}  \frac{m'_1+m'_2}{m_0} \, ,   \\
&\omega_2 = \phantom{-} 0 \, . 
\end{aligned}
\label{eq:HquadL4c}
\end{equation}
Pour le voisinage de $L_3$:
 \begin{equation}
 \begin{aligned}
\ol H_{L_3}^{(2)} = &  \omega_0 z_0\tilde{z}_0 + \omega_1 z_1\tilde{z}_1 + \omega_2 z_2\tilde{z}_2 \, ,  
\end{aligned}
\label{eq:HquadL3}
\end{equation}
avec:
 \begin{equation}
 \begin{aligned}
&\omega_0 = -\sqrt{\eps}\eta \sqrt{\frac{21}{8}\frac{m'_1+m'_2}{m_0}} \, ,  \\
&\omega_1  = +i \eps \eta  \frac{7}{8}  \frac{(m'_1+m'_2)}{m_0}   \, , \\
&\omega_2  = \phantom{-} 0 \, .  
\end{aligned}
\label{eq:HquadL3c}
\end{equation}
notons que les variables ($z_j$,$\tilde{z}_j$) sont canoniquement conjuguées mais ne vérifient pas de relations particulières.\\
 
\begin{rem}
Les variables $z_j$ et $\tilde{z}_j$ introduites ici dépendent de l'équilibre considéré. Les variables ($z_j$,$\tilde{z}_j$) qui diagonalisent le hamiltonien dans le voisinage du point $L_4$ n'ont aucun lien avec les variables ($z_j$,$\tilde{z}_j$) qui le diagonalisent au voisinage de $L_3$. Pour ne pas introduire de notation supplémentaire, on parlera du voisinage de $L_4$ si rien n'est précisé. 
\end{rem} 
 
Nous pouvons d'ores et déjà retrouver que les équilibres de Lagrange $L_4$ et $L_5$ sont des points d'équilibres stables du problème moyen, avec deux directions elliptiques et une direction dégénérée; alors que le point $L_3$ est hyperbolique dans la direction $z_0$.
Nous allons exprimer ces différentes directions au voisinage des équilibres de Lagrange en fonction des éléments elliptiques. Cela est effectué en inversant les matrices de passages (\ref{eq:P2B2L4}) et (\ref{eq:P2B2L3}). Pour les familles émergeant de $L_4$, nous avons:
\begin{equation}
 \begin{aligned}
z_0 & = \frac{\alpha z-iZ/\alpha}{\sqrt{2}} &\ \text{où}\ & \alpha= \left(\eps \frac{3}{4}\cG^{4/3} m_0^{1/3} \eta^{-2/3} \frac{(m'_1m'_2)^2}{m'_1+m'_2} \right)^{1/4} \, , \\ 
z_1 & = \frac{\sqrt{m'_2} \e^{-i\pi/3}x_1 - \sqrt{m'_1}x_2}{\sqrt{m'_1+m'_2}} \, ,            &  &\\
z_2 & = \frac{\sqrt{m'_1} \e^{-i\pi/3}x_1 + \sqrt{m'_2}x_2}{\sqrt{m'_1+m'_2}} \, .          &      &
\label{eq:L4exz}
\end{aligned}
\end{equation}
Au voisinage de $L_4$, la famille des anti-Lagrange $AL_4$ est tangente à $z_1= c \e^{h_{z_1\tilde{z}_1}}$ où $c$ est un réel positif quelconque. Cette tangente est donnée par le système suivant:
\begin{equation}
 \begin{aligned}
z_0 & = z_2 = 0 \, , \\ 
z_1 & = c \e^{\omega_1 t}\, .
\label{eq:ML3}
\end{aligned}
\end{equation}
 Nous obtenons donc:
 \begin{equation}
 \begin{aligned}
z&=0\, , \hspace{1cm}   & Z &=0\, , \\
x_1&= \frac{\sqrt{m'_1+m'_2}}{\sqrt{2}(1+m'_1/m'_2)} c \e^{i (\pi/3 +gt) } \, , \hspace{1cm} & x_2 &= \frac{\sqrt{m'_1}}{\sqrt{m'_2}} \e^{i(2\pi/3)} x_1 \, .
\label{eq:ML3}
\end{aligned}
\end{equation}
Ce qui signifie que au voisinage de $L_4$, la famille $AL_4$ est tangente à la direction $a_1=a_2$, $\zeta=\pi/3$, $e_2= \frac{m_1}{m_2} e_1$ et $\Delta \varpi = \zeta + \pi$ dans les coordonnées elliptiques. De plus, les deux ellipses précessent à la fréquence $g_{1,L_4}$. De la même manière on retrouve que, au moins au voisinage de $L_4$, la famille des $L_4$ elliptiques est donnée par $a_1=a_2$, $\zeta=\Delta \varpi =\pi/3$, $e_2= e_1$, et que la fréquence de précession, $g_{2,L_4}$, est nulle. Cette expression est en fait valable pour toute valeur du paramètre $e_1$.\\

De manière analogue, nous avons pour $L_3$:
 \begin{equation}
 \begin{aligned}
z_0 & =   \frac{i \alpha z+ Z/\alpha}{\sqrt{2}} &\ \text{où}\ & \alpha= \sqrt{i} \left(\eps \frac{1}{4}\cG^{4/3} m_0^{1/3} \eta^{-2/3} \frac{(m'_1m'_2)^2}{m'_1+m'_2} \right)^{1/4} \, , \\  
\tilde{z}_0 & =  \frac{- \alpha z -i Z/\alpha}{\sqrt{2}}\, ,            &  &\\
z_1 & =  \frac{\sqrt{m'_2} x_1 + \sqrt{m'_1}x_2}{\sqrt{m'_1+m'_2}}\, ,            &  &\\
z_2 & =\frac{\sqrt{m'_1} x_1 - \sqrt{m'_2}x_2}{\sqrt{m'_1+m'_2}} \, .          &      &
\label{eq:L3exz}
\end{aligned}
\end{equation}
On rappelle que nous n'obtenons pas la relation $\tilde{z}_0=-i\bar{z}_0$ par la matrice de passage (\ref{eq:P2A2}) car $\alpha$ n'est pas un réel. De l'expression (\ref{eq:L3exz}) nous pouvons déduire que au voisinage de $L_3$, la famille $AL_{3}$ est tangente à la direction $a_1=a_2$,  $\zeta=\pi$, $e_2= \frac{m_1}{m_2} e_1$ et $\Delta \varpi = 0$. Les deux ellipses précessent à la fréquence $g_{L_3}$. De la même manière, on peut montrer que, au moins au voisinage de $L_3$, la famille $L_3$ des Eulers elliptiques est donnée par $a_1=a_2$, $\zeta=\Delta \varpi =\pi$, $e_2= e_1$, et que la fréquence de précession, $g_{2,L_3}$, est nulle. Cette expression est en fait valable pour toute valeur du paramètre $e_1$. \\

Dans cette section nous avons retrouvé les résultats de \cite{RoPo2013}. Nous connaissons maintenant la dynamique coorbitale au voisinage des équilibres de Lagrange circulaires et la tangente aux familles $L_k$ et $AL_k$ excentriques. Avant de poursuivre notre étude en dehors du voisinage des équilibres circulaires, nous allons introduire plus formellement les différents objets qui émergent de ces équilibres.

\vspace{1cm}

\an{
In this section we recall the results of \cite{RoPo2013}. We diagonalise the quadratic part of the averaged Hamiltonian (\ref{eq:Hb}) in the neighbourhood of the Lagrangian equilibrium points $L_3$, $L_4$ and $L_5$. The expression of the diagonalised quadratic Hamiltonian can be found in equations (\ref{eq:HquadL4}) and (\ref{eq:HquadL3}) (in this chapter we develop only the cases of the $L_3$ and $L_4$ equilibriums, as the $L_5$ case can be deduced from the $L_4$ case by symmetry). the change of variables used to diagonalise the Hamiltonian is detailed in Appendix \ref{an:MP}. We can see from the expression of the three eigenvalues (eq. \ref{eq:Hquad1L4coef} and \ref{eq:Hquad1L3coef}) that the $L_4$ equilibrium has $2$ elliptic directions and one neutral direction, and $L_3$ as one elliptic direction, one hyperbolic direction and one neutral direction.\\
These directions are tangent to the families of fixed points and periodic orbits emanating from the circular equilibriums $L_k$: in the case of $L_4$ and $L_5$, the direction $z_0$ ($z_1=z_2=0$) is tangent to the family of circular co-orbital orbits which are periodic orbits of the averaged problem. In each case, the direction $z_1$ is tangent to the families $AL_k$ which are also a reunion of periodic orbits of the averaged problem (the expression of the tangent to these families will be recalled in section \ref{sec:DFFbLc}). The direction $z_2$ is tangent to the families of fixed point of the averaged problem: the $L_k$ eccentric equilibriums.
}

\section{Dynamique au voisinage d'un point fixe}
\label{sec:Dvpf}



Considérons un hamiltonien $H'$ décrivant un système à trois degrés de liberté au voisinage d'un point fixe $L_k$ situé à l'origine des coordonnées.

\vspace{1cm}

\an{
In this section we give a precise (local) definition of the objects that we study in this chapter: the families of quasi-periodic orbits of non-maximal dimension. We consider a Hamiltonian $H'$ describing a system with 3 degrees of freedom in the neighbourhood of a fixed point $L_k$ located at $z_0=z_1=z_2=0$. }

\subsection{Approximation quadratique}
\label{sec:genquad}

Soit $\cQ'$ la partie quadratique du hamiltonien $H'$. On effectue un changement de variables canonique visant à diagonaliser $\cQ'$. On appellera $H$ le hamiltonien dans ces nouvelles variables et on écrit sa partie quadratique:
\begin{equation}
\cQ=\omega_0 z_0\zt_0 + \omega_1 z_1\zt_1 + \omega_2 z_2\zt_2\, ,
\label{eq:expQ}
\end{equation}
où les couples de variables $(z_j,\zt_j)$ sont canoniquement conjugués, et les $\omega_j$ sont les valeurs propres du système d'équation associé à $\cQ'$. Nous définirons deux types de variétés:\\
\noindent - Les variétés $z_j=\zt_j=0$. Qui sont des objets de dimension $4$.\\
\noindent - Les variétés $z_m=\zt_m=z_l=\zt_l=0$ ($j$, $l$ et $m$ deux à deux distincts) qui sont l'intersection des variétés $z_l=\zt_l=0$ et $z_m=\zt_m=0$, donc des objets de dimension $2$.\\
Notons que toutes ces variétés sont invariantes par le flot du hamiltonien $\cQ$.
Le théorème du centre de Lyapunov \citep{MeHa1992} stipule que si $\omega_j$ est imaginaire pure, non nulle, et n'est pas résonante avec les deux autres valeurs propres, alors il existe une famille d'orbites périodiques à un paramètre issu de $L_k$, orbites dont la période tend vers $2\pi/i\omega_j$ quand on s'approche du point d'équilibre. Cette famille, que nous appellerons $\cF^j_k$, est tangente à l'origine au sous-espace propre associé à $\omega_j$: la variété $z_m=\zt_m=z_l=\zt_l=0$.                

\vspace{1cm}

\an{
Considering the diagonalised quadratic expression (\ref{eq:expQ}), we can identify two kind of manifolds:\\
\noindent - The manifolds $z_j=\zt_j=0$ which are $4$-dimensional.\\
\noindent - The manifolds $z_m=\zt_m=z_l=\zt_l=0$ ($j$, $l$ and $m$ pairwise distinct) which are the intersection of the manifolds $z_l=\zt_l=0$ and $z_m=\zt_m=0$, hence $2$-dimensional objects.\\
} 

\an{
Note that these manifolds are invariant by the flow of the Hamiltonian $\cQ$.\\
The Lyapunov's center theorem \citep{MeHa1992} implies that if $\omega_j$ is an imaginary number, non-zero, and is not resonant with the other two eigenvalues, then there is a one-parameter family of periodic orbits emanating from $L_k$. The period of these orbits tends to $2\pi/i\omega_j$ as they get close to the equilibrium point. This family, that we call $\cF^j_k$, is tangent in $L_k$ to the eigenspace associated with $\omega_j$: the manifold $z_m=\zt_m=z_l=\zt_l=0$.                
} 


\subsection{Au-delà de l'approximation quadratique}
\label{sec:gen}

Supposons ici que les valeurs propres $\omega_j$ sont soit imaginaires pures, soit nulles. On peut obtenir une approximation intégrable de $H'$ dans le voisinage de $L_k$ à un ordre supérieur à $2$ en calculant la forme normale de Birkhoff. On rappelle qu'une forme normale est une expression invariante par le flot du système hamiltonien engendré par sa partie quadratique. 

Si les $\omega_j$ sont non résonants et non nuls, la forme normale $\cN$, dans les variables canoniques ($Z_j,\Zt_j$), ne contient que des monômes de la forme $\prod_{j=0}^2 (Z_j\Zt_j)^{p_j}$ avec $p_j \in \mathbb{Z}$. 

Dans le cas résonant, si par exemple $\omega_m$ et $\omega_p$ vérifient $k_1\omega_m=k_2\omega_l$, $k_j \in \mathbb{Z}$, alors la forme normale comprendra tous les monômes de la forme $\prod_{j=0}^2 (Z_j\Zt_j)^{p_j}(Z_m^{k_1}\Zt_l^{k_2})^{q_1}(Z_l^{k_2}\Zt_m^{k_1})^{q_2}$, $q_j \in \mathbb{Z}$.

 Pour finir, si $\omega_l=0$, la forme normale comprendra l'ensemble des monômes de la forme $\prod_{j=0}^2 (Z_j\Zt_j)^{p_j}Z_l^{q_1}\Zt_l^{q_2}$.

Effectuons maintenant le changement de variable suivant:
\begin{equation}
\begin{aligned}
 (Z_j, \tilde{Z}_j) & = \chi_{\cN}(I_j,\theta_j)\\
                  & = (\sqrt{2I_j} \e^{i \theta_j},\sqrt{2I_j} \e^{-i \theta_j}).
\end{aligned}
\label{eq:cv5}
\end{equation}
Si les $\omega_j$ sont tous non nuls et non résonants, nous obtenons la forme normale à l'ordre $p$:
\begin{equation}
\cN' = \sum_{d=2}^p \left( \sum_{k_0+k_1+k_2=d} C_{k_0,k_1,k_2} I_0^{k_0} I_1^{k_1} I_2^{k_2} \right)\, .
  \label{eq:H6}
\end{equation}
Une orbite générique du voisinage de $L_k$ vit sur un tore invariant de dimension $3$ paramétré par les actions $I_0$, $I_1$ et $I_2$. Si $\omega_2$ est nulle, nous avons:
\begin{equation}
\cN' = \sum_{d=2}^p \left( \sum_{k_0+k_1+k_2+\bar{k}_2=d}   C_{k_0,k_1,k_2} I_0^{k_0} I_1^{k_1} \sqrt{I_2}^{k_2+\bar{k}_2} \e^{i(k_2-\bar{k}_2) \theta_2} \right)\, .
  \label{eq:H6o2n}
\end{equation}
Notons qu'ici $I_2$ n'est pas une intégrale problème du mouvement ($\dot I_2 =\partial\cN'/\partial \theta_2 \neq 0$).\\

Ainsi, les coordonnées qui réduisent le hamiltonien à sa forme normale fournissent une paramétrisation naturelle des familles d'orbites de Lyapunov. En effet, dans un voisinage du point fixe, $\cF^j_k$ est paramétrée par l'action $I_j$ pour $J_k=J_l=0$ ($j$, $k$, $l$ deux à deux distincts). On notera que cette paramétrisation est possible même si la fréquence associée à l'action est nulle. De manière analogue, on définit la variété $\ol \cF^j_k$ par $I_j=0$. Cette variété est tangente à l'origine à la variété $z_j=\zt_j=0$. Nous avons les relations suivantes: 
\begin{equation}
\begin{aligned}
&\cF^j_k \subset \ol \cF^l_k\, \hspace{2cm} & \forall j \neq j \\
 &\ol \cF^l_k\ \cap  \ol \cF^m_k = \cF^j_k\,  & \text{$j$, $l$, $m$ deux à deux distincts} \\
  \label{eq:Flkrel}
  \end{aligned}
\end{equation}

Les orbites des familles $\cF^j_k$ évoluent sur des tores de dimension $1$ paramétrés par l'action $I_j$. Ces familles, réunions d'orbites périodiques, sont donc des objets de dimension $2$, sauf si il s'agit de familles de points fixes (quand $\omega_j=0$), auquel cas ce sont des objets de dimension $1$. 

Les variétés $\ol \cF^j_k$ sont la réunion d'orbites quasi-périodiques vivant sur des tores de dimension $2$ paramétrés par les actions $I_l$ et $I_m$, ces variétés sont donc des objets de dimension $4$.

\vspace{1cm}

\an{
If we consider that the $\omega_j$ are either imaginary or null, we can obtain an integrable approximation of the Hamiltonian $H'$ in the neighbourhood of $L_k$ at a higher order by computing the Birkhoff's normal form. Depending on the occurrence of resonances between the $\omega_j$ (including the case where one of them is null), the normal form can contain various monomials in addition to the classic $\prod_{j=0}^2 (Z_j\Zt_j)^{p_j}$ with $p_j \in \mathbb{Z}$.\\
Let us consider two cases:\\
- The case where the $\omega_j$ are all non null and non resonant, then we can write the normal form as the expression (\ref{eq:H6}) (using the change of variables eq. (\ref{eq:cv5})). In this case, a generic orbit in the neighbourhood of $L_k$ evolves on an 3-dimensional invariant torus parametrised by the action $I_0$, $I_1$ and $I_2$.\\
- The case $\omega_2=0$, where the normal form can be written as eq. (\ref{eq:H6o2n}) (using the change of variables eq. \ref{eq:cv5}). Here $I_2$ is not an integral of the motion ($\dot I_2 =\partial\cN'/\partial \theta_2 \neq 0$).\\
  }

\an{
The change of coordinates which turns the Hamiltonian $H'$ into its normal form $\cN'$ gives a natural parametrisation of the Lyapunov orbits: in the neighbourhood of $L_k$, the family $\cF^j_k$ is parametrised by the action $I_j$ for $J_m=J_l=0$ ($j$, $m$ and $l$ pairwise distinct). This parametrisation holds when $\omega_j=0$. Similarly, we define $\ol \cF^j_k$ by $I_j=0$. This manifold is tangent to the manifold $z_j=\tilde{z}_j=0$, and we have the relations (\ref{eq:Flkrel}).\\ 
  } 
  
 \an{
The orbits of the family $\cF^j_k$ evolve on a 1-dimensional torus parametrized by the action $I_j$. The family $\cF^j_k$, as reunion of these orbits, is hence a 2-dimensional object (except when it is a family of fixed points ($\omega_j=0$) in which case it is an 1-dimensional object).\\
Since the manifold $\ol \cF^j_k$ is a reunion of quasi-periodic orbits evolving on a 2-dimensional torus parametrised by the actions $I_l$ and $I_m$, it is a 4-dimensional object.
  }  
  
\subsection{Dynamique le long des familles $\cF^2_k$}
\label{sec:genF2}
On s'intéresse maintenant à la dynamique du système dans le voisinage d'une des orbites de la famille $\cF^2_k$. Considérons dans un premier temps que les $\omega_j$ soient non nulles et non résonantes. On peut écrire $\cN_{\cF^2_k}$, la forme normale du hamiltonien au voisinage de ce point, en translatant $\cN$ le long de la famille $\cF^2_k$. Cela s'effectue naturellement au moyen du changement de variable canonique suivant:
\begin{equation}
 \begin{aligned}
I_2 &=I^0_2+ I'_2 \\
I_0 &= I_0\\
I_1 &= I_1\, .
\end{aligned}
\label{eq:transIj}
\end{equation}
où $I_j^0$ est une constante qui paramètre la translation. Nous obtenons le nouveau hamiltonien:
\begin{equation}
\cN_{\cF^2_k} = \sum_{d=2}^p \left( \sum_{k_0+k_1+k_2=d}   C_{k_0,k_1,k_2}(I^0_2) I_0^{k_0} I_1^{k_1} I_2^{'k_2}  \right)\, ,
  \label{eq:H7}
\end{equation}
qui représente la dynamique dans le voisinage du tore invariant ($I_0=0,I_1=0,I_2=I^0_2$). 

Considérons maintenant que la valeur propre $\omega_2$ soit nulle, et que les deux autres valeurs propres soient non nulles et non résonantes. Comme nous l'avons vu en section précédente, la transformation effectuée ne permet à priori pas d'obtenir des variables actions-angles dans le voisinage de $L_k$ ($I_2$ n'est pas une intégrale première du mouvement). Cependant, si on suppose que la famille $\cF^2_k$ est une famille de points fixes du problème moyen, alors la décomposition (\ref{eq:H6o2n}) ne comprend pas de termes tels que $k_0=k_1=0$, $k_2\neq 0$, $\bar{k}_2 \neq 0$ \citep{RoPo2013}. Les tores de la forme ($I_0=0,I_1=0,I_2=I^0_2$) sont donc invariants par le flot de (\ref{eq:H6o2n}). Nous pouvons donc translater le hamiltonien le long de la famille $\cF^2_k$ afin de décrire les orbites dans le voisinage de cette famille.

Pour le moment, on se contente de l'approximation quadratique au voisinage des équilibres circulaire dont les résultats sont résumés dans la section suivante (\ref{sec:DFFbLc}). On étudiera la dynamique le long des familles $\cF^2_k$ en fin de chapitre, dans la section \ref{sec:DynFl}.

\begin{rem}
Dans ce cas la translation le long de cette famille peut s'effectuer de manière équivalente dans n'importe quelles variables. En effet les membres de $\cF^2_k$ sont des points fixes, et ne nécessitent donc pas de passer aux variables action-angles pour faire apparaitre des intégrales premières du mouvement. 
\end{rem}

\an{
When the orbits of the family $\cF^2_k$ are periodic, we can study the dynamics in the neighbourhood of a member of this family by doing the translation (\ref{eq:transIj}), as long as the frequencies $\omega_0$ and $\omega_1$ are non-null and non-resonant. The result of this translation is given by the Hamiltonian (\ref{eq:H7}), which represents the dynamics in the neighbourhood of the invariant torus ($I_0=0,I_1=0,I_2=I^0_2$).\\
In the case $\omega_j=0$, \cite{RoPo2013} showed that the expression (\ref{eq:H6o2n}) does not contain terms such that $k_0=k_1=0$, $k_2\neq 0$ and $\bar{k}_2 \neq 0$. The torus ($I_0=0,I_1=0,I_2=I^0_2$) is hence invariant by the flow of (\ref{eq:H6o2n}). We can thus translate the Hamiltonian in this case as well, in order to study the dynamics in the neighbourhood of one of the fixed points of the family $\cF^2_k$.\\ 
This translation will be done later in the chapter, in section \ref{sec:DynFl}.\\
}

\an{
Remark: in the case where $\cF^2_k$ is a family of fixed points, the translation along the family $\cF^2_k$ can be done in any set of variables, since the variable of the normal form are not required to bring out the integrals of motion. 
}
 


\subsection{Approximation quadratique des familles $\cF$ et $\ol \cF$ au voisinage des équilibres de Lagrange circulaires}
\label{sec:DFFbLc}


%
%

L'ensemble de ces familles de Lyapunov à un paramètre qui émergent des points $L_3$, $L_4$ et $L_5$ est maintenant bien connu: nous avons calculé leur expression quadratique au voisinage des $L_k$ circulaires dans la section \ref{sec:H0pf}. Pour résumer, nous avons: \\

\noindent - Les familles $\cF_{4}^0$ et $\cF_{5}^0$ tangentes au sous-espace engendré par ($z_0$,$\zt_0$) en $L_4$ et $L_5$, respectivement. Ces deux familles sont des représentations locales de la famille des co-orbitaux circulaires que nous avons étudiée dans le Chapitre 1 et que nous noterons $\cF^0$. Elles sont tangentes à la variété $e_1=e_2=0$.\\
- La famille $\cF^1_{4}$ (resp $\cF^1_{5}$) émergeant de $L_4$ (resp. $L_5$) dans la direction $z_1$ est la famille des anti-Lagrange $AL_4$ (resp. $AL_5$), mise en évidence numériquement par \cite{GiuBeMiFe2010}. Elle est tangente à la variété définie par $\zeta=\pi/3$, $e_2= \frac{m_1}{m_2} e_1$ et $\Delta \varpi = \zeta + \pi$.\\
- La famille $\cF^2_{4}$ (resp $\cF^2_{5}$) émergeant de $L_4$ (resp. $L_5$), portée par la direction $z_2$ est la famille des configurations équilatérales excentriques. Elle est tangente à la variété $\zeta=\Dv=\pi/3$, $e_2= e_1$. \\
- La famille $\cF^1_{3}$ émergeant de $L_3$, portée par la direction $z_1$ est la famille que nous appellerons anti-Euler ($AL_3$), calculée numériquement par \cite{HaPsyVo2009}. Elle est tangente à la variété $\zeta=\pi$, $e_2=\frac{m_1}{m_2}e_1$ et $\Dv=0$.\\
- La famille $\cF^2_{3}$ émergeant de $L_3$, portée par la direction $z_2$ est la famille des Euler excentriques. Elle est tangente à la variété $\zeta=\Dv=\pi$ et $e_2=e_1$. \\

Nous choisirons $\zeta_0$ comme paramètre pour la famille $\cF^0$ et l'excentricité du corps $1$ pour les autres. Il est important de rappeler que les familles $\cF^2_{3}$, $\cF^2_{4}$ et $\cF^2_{5}$ sont des familles de points fixes du problème moyen ($\omega_2=0$, voir section \ref{sec:Dvpf}). 

Les variétés d'expression quadratique tangentes aux familles $\overline{\cF}^l_{k}$ en les équilibres de Lagrange circulaires $L_4$ et $L_3$ apparaissent également de manière évidente dans les variables $zj$. 
Les équations (\ref{eq:L4exz}) et (\ref{eq:L3exz}) nous donnent leur expression en les variables elliptiques. Pour le voisinage de $L_4$:\\
\noindent - La famille $\ol \cF^0_{4}$ est tangente à la variété $z=Z=0$\\
- La famille $\ol \cF^1_{4}$ est tangente à la variété $e_1=e_2$ et $\Dv=\pi/3$.\\
- La famille $\ol \cF^2_{4}$ est tangente à la variété  $m_1e_1=m_2e_2$ et $\Dv=-2\pi/3$. \\

Des intégrations numériques d'orbites dans le voisinage de ces familles à plus haute excentricité sont représentées sur la figure~\ref{fig:FbL4orb} et section \ref{sec:DFFN}.

\noindent Pour le voisinage de $L_3$:\\
- La famille $\ol \cF^0_{3}$ est tangente à la variété $z=Z=0$.\\
- La famille $\ol \cF^1_{3}$ est tangente à la variété $e_1=e_2$ et $\Dv=\pi$.\\
- La famille $\overline{\cF}^2_{3}$ est tangente à la variété $m_1 e_1=m_2 e_2$ et $\Dv=0$ .\\

 Notons que dans le cadre de nos approximations (voisinage du cas circulaire à l'ordre 2 en les variables $z_j$) ces directions ne dépendent pas de $\eps$.\\
 
 \an{ 
 All the one-parameter Lyapunov families emanating from $L_3$, $L_4$ and $L_5$ are well known. We computed their expression in the neighbourhood of the $L_k$ circular equilibriums in section \ref{sec:H0pf}. We have: \\
\noindent - The families $\cF_{4}^0$ and $\cF_{5}^0$ are tangent to the linear span of ($z_0$,$\zt_0$) in $L_4$ and $L_5$, respectively. They are local representations of the circular co-orbitals family that we studied in chapter 1 (we call this family $\cF^0$). These families are tangent to the manifold $e_1=e_2=0$.\\
- The family $\cF^1_{4}$ (resp. $\cF^1_{5}$) emanating from $L_4$ (resp. $L_5$) in the $z_1$ direction is the anti-Lagrange family $AL_4$ (resp. $AL_5$), found numerically by \cite{GiuBeMiFe2010}. This family is tangent to the direction $\zeta=\pi/3$, $e_2= \frac{m_1}{m_2} e_1$ and $\Delta \varpi = \zeta + \pi$.\\
- The family $\cF^2_{4}$ (resp. $\cF^2_{5}$) emanating from $L_4$ (resp. $L_5$), in the $z_2$ direction is the family of the eccentric equilateral configurations. It is tangent to the manifold $\zeta=\Dv=\pi/3$, $e_2= e_1$. \\
- The family $\cF^1_{3}$ emanating from $L_3$, in the direction $z_1$. It is the anti-Euler family ($AL_3$), computed numerically by \cite{HaPsyVo2009}. It is tangent to the manifold $\zeta=\pi$, $e_2=\frac{m_1}{m_2}e_1$ and $\Dv=0$.\\
- The family $\cF^2_{3}$ emanating from $L_3$, in the direction $z_2$. It is the eccentric Euler family. It is tangent to the manifold $\zeta=\Dv=\pi$ and $e_2=e_1$. \\}

\an{
We chose $\zeta_0$ as parameter for the family $\cF^0$ and $e_1$ for the other families. Note that $\cF^2_{3}$, $\cF^2_{4}$ et $\cF^2_{5}$ are families of fixed points of the averaged problem ($\omega_2=0$, see section \ref{sec:Dvpf}). 
}

\an{The expression of the quadratic approximation of the families $\overline{\cF}^l_{k}$ in the neighbourhood of $L_4$ and $L_3$ appears trivially in the variable $z_j$. The equations (\ref{eq:L4exz}) and (\ref{eq:L3exz}) give their expression in the elliptical variables. In the neighbourhood of $L_4$:\\
\noindent - $\ol \cF^0_{4}$ is tangent to the manifold $z=Z=0$\\
-  $\ol \cF^1_{4}$ is tangent to the manifold $e_1=e_2$ and $\Dv=\pi/3$.\\
-  $\ol \cF^2_{4}$ is tangent to the manifold  $m_1e_1=m_2e_2$ and $\Dv=-2\pi/3$. \\
Numerical integration of trajectories in the neighbourhood of these manifolds for higher eccentricities are plotted on figure~\ref{fig:FbL4orb} (section \ref{sec:DFFN}).\\}

\an{
\noindent In the neighbourhood of $L_3$:\\
- $\ol \cF^0_{3}$ is tangent to the manifold $z=Z=0$.\\
- $\ol \cF^1_{3}$ is tangent to the manifold $e_1=e_2$ and $\Dv=\pi$.\\
- $\overline{\cF}^2_{3}$ is tangent to the manifold $m_1 e_1=m_2 e_2$ and $\Dv=0$ .\\}

 \an{
 Note that within our approximation (quadratic approximation near the circular case) all these directions do not depend neither on $\eps$. }

\section{Réduction du problème}
\label{sec:rdp}

Avant de poursuivre sur l'étude globale de l'espace des phases, il est intéressant de remarquer que, comme proposé par \cite{GiuBeMiFe2010}, le système peut être réduit de 1 degré de liberté supplémentaire en prenant le moment cinétique total comme paramètre du problème. Repartant du hamiltonien (\ref{eq:Ht}) dans les variables de Poincaré, nous effectuons le changement de variable canonique suivant\footnote{Ce changement de variable comporte une singularité pour des excentricités nulles}


\begin{equation}
\begin{aligned}
\zeta & =\lambda_1 - \lambda_2; & Z &=(\Lambda_1-\Lambda_2)/2\\
\Delta \varpi & =\varpi_1 - \varpi_2; & \varPi &= i(x_2 \xt_2 - x_1 \xt_1)/2\\
q &= \varpi_1+\varpi_2; & J_1 &= (\Lambda_1+\Lambda_2-i(x_1 \xt_1+x_2 \xt_2))/2\\
 Q &= \lambda_1+ \lambda_2 - q; & J_2 &=(\Lambda_1+\Lambda_2)/2\, .
\end{aligned}
\label{eq:CIred}
\end{equation}
En notant ce changement de variable $\chi_{\cR}$, nous obtenons le hamiltonien réduit suivant:
\begin{equation}
\gH_\cR = \gH \circ \chi^{-1}_{\cR} (\zeta,Z,\Delta\varpi,\varPi,J_1,J_2,q).
\label{eq:HR}
\end{equation}
Notons que $J_1$ est la moitié du moment cinétique total des deux coorbitaux. Il s'agit d'une intégrale première du mouvement. Comme précédemment, nous pouvons moyenner le hamiltonien sur l'angle rapide $Q$, et la quantité $J_2$ devient également une intégrale première du mouvement. Son interprétation physique est celle d'un facteur d'échelle: le demi grand-axe moyen des coorbitaux n'influe pas sur la dynamique, excepté pour la valeur du moyen mouvement moyen. En notant le changement de variable associé à cette moyenne $\chi_{\cM}$, nous obtenons le hamiltonien suivant:
\begin{equation}
\gH_{\cR\cM} = \gH_{\cR} \circ \chi^{-1}_{\cM} (\zeta,Z,\Delta\varpi,\varPi,J_1,J_2).
\label{eq:HRM}
\end{equation}
où les grandeurs $J_1$ et $J_2$ sont des intégrales premières du mouvement. Suite à ces deux transformations, le hamiltonien devient:
\begin{equation}
\gH_{\cR\cM}= \gH^K_{\cR\cM}(Z) + \eps \gH^P_{\cR\cM}(Z,\zeta,\Delta\varpi,\varPi)\, , 
\label{eq:expHRM}
\end{equation}
Nous obtenons donc pour le problème moyen réduit un espace des phases de dimension 4, avec pour variables $\zeta$, $\Delta \varpi$, $Z$ et $\varPi$. Par analogie au problème moyen non réduit, on appellera $\nu$ la fréquence fondamentale de l'angle de libration semi-rapide $\zeta$ et $g$ la fréquence fondamentale séculaire d'évolution des excentricités et de la différence des périhélies. \\

Dans la section \ref{sec:gen} nous avons introduit les familles $\cF^j_k$ et $\ol \cF^j_k$ à partir d'un équilibre circulaire $L_k$ donné. Tout d'abord, notons que les familles $\cF^0_{k}$ (paramétrant les co-orbitaux circulaires) ne sont plus décrites par le nouvel hamiltonien car nous avons une singularité pour les excentricités nulles. Après réduction du problème, les familles $\cF^l_{k},\ l\in\{1,2\}$ sont réduites à leur représentant pour la valeur du moment cinétique choisi, donc à des points fixes du problème moyen réduit car $Z$, $\zeta$, $\varPi$ et $\Dv$ sont constants pour chacune de ces orbites. Leur valeur dépend de l'élément de la famille $\cF^l_{k}$. Nous continuerons à appeler $L_k$ le représentant de la famille $\cF^2_{k}$ et $AL_k$ le représentant de la famille $\cF^1_{k}$. Toute mention d'un équilibre $L_k$ à partir de maintenant se réfèrera donc à l'équilibre de Lagrange excentrique pour la valeur du moment cinétique considérée, sauf si le contraire est précisé.

Comme nous nous intéressons désormais à l'ensemble de l'espace des phases, nous introduisons les réunions de familles d'orbites périodiques du problème moyen réduit $\ol \cF^\nu$ et $\ol \cF^g$, avec $\ol \cF^\nu$ toute orbite ne subissant pas de variation à la fréquence semi-rapide $\nu$ (donc les orbites $2\pi/g$ périodiques du problème moyen réduit) et $\ol \cF^g$ toute orbite ne subissant pas de variations à la fréquence $g$ (donc les orbites $2\pi/\nu$ périodiques du problème moyen réduit). La famille $\ol \cF^\nu$ est représentée localement au voisinage d'un point de $L_k$ par $\ol \cF^0_k$ et la famille $\ol \cF^g$ est paramétrée localement par $\ol \cF^1_k$ au voisinage de $L_k$ et par $\ol \cF^2_k$ dans le voisinage de $AL_k$. On pourra faire référence à une famille donnée des réunions $\ol \cF^g$ et $\ol \cF^\nu$ en appelant par exemple $\ol \cF^g_k$ la famille $\ol \cF^g$ passant par l'équilibre $L_k$.


\begin{rem} Dans la suite, l'espace des phases de dimension 4 sera exploré à la fois par intégration numérique du problème à trois corps et par des méthodes semi-analytiques. 
La moyennisation sur l'angle $\cQ$ est identique à celle décrite dans le Chapitre 1. Dans le cas de l'étude semi-analytique, cette moyennisation sera effectuée numériquement. Dans le cas d'intégrations du problème à trois corps, on étudiera la dynamique et la stabilité en fonction des variables $\zeta$, $Z$, $\Delta\varpi$ et $\varPi$, variables qui seront $\eps$ proches de leur équivalent dans le problème moyen réduit (voir la section \ref{sec:adp} pour plus d'explication).
\end{rem}



Représenter l'ensemble de la dynamique d'un espace des phases de dimension $4$ n'est pas facile. Nous allons voir que dans certains cas il est possible d'avoir une bonne représentation de l'ensemble des orbites d'une configuration donnée en ne considérant qu'une variété de dimension $2$ de conditions initiales.

\an{
In the eccentric case, the conservation of the total angular momentum allows for the reduction of the problem to 2 degrees of freedom. Departing from the Hamiltonian of the 3-body problem (\ref{eq:Ht}), we use the canonical change of variables (\ref{eq:CIred}) \citep{GiuBeMiFe2010} in order to obtain the reduced Hamiltonian $\gH_\cR$ (\ref{eq:HR}). By averaging over the fast angle $Q$, we obtain the 2-degrees of freedom Hamiltonian $\gH_{\cR\cM}$, where the variables $J_1$ and $J_2$ are invariant integrals of the motion. $J_1$ is half the total angular momentum of the system, and $J_2$ is a scaling factor: the value of the mean semi-major axis does not impact the dynamics, it is only a scaling factor for the mean mean-motion (thus for all the fundamental frequencies). By analogy with the non-reduced averaged problem, we call $\nu$ the fundamental frequency of the degree of freedom ($Z,\zeta$), and $g$ the fundamental frequency of the secular motion ($\varPi,\Dv$).\\
In section \ref{sec:Dvpf}, we introduced the families $\cF^j_k$ and $\ol \cF^j_k$ that we defined in the vicinity of a given circular Lagrangian equilibrium $L_k$. First, note that the families $\cF^0_k$ (parametrizing the circular co-orbitals) are not described by the new Hamiltonian (\ref{eq:HRM}) since the change of variables (\ref{eq:CIred}) is not defined if $e_1$ or $e_2$ are equal to zero. \\
 In the reduced averaged phase space, the families $\cF^j_k$ ($l\in \{1,2\}$) are reduced to an unique orbit for the chosen value of the total angular momentum $J_1$, which is a fixed point of the reduced averaged problem. We call $L_k$ the member of the $\cF^2_k$ family and $AL_k$ the member of the $\cF^1_k$ family. From now on, the equilibrium point $L_k$ refer to the given eccentric equilibrium point except if `circular' is mentioned.\\
As we now study the whole phase space, we introduce the reunion of all the periodic orbits of the reduced averaged phase space: the manifolds $\ol \cF^\nu$ and  $\ol \cF^g$, where $\ol \cF^\nu$ contains the orbits that are $2\pi/g$ periodic and $\ol \cF^g$ the orbits that are $2\pi/\nu$ periodic. The family $\ol \cF^0_l$ are local representations of $\ol \cF^\nu$ and the families $\ol \cF^1_k$ (resp. $\ol \cF^2_k$) are local representations of $\ol \cF^g$ near the $L_k$ (resp. $AL_k$) equilibrium. $\ol \cF^\nu$ and  $\ol \cF^g$ are 2-dimensional objects in the reduced averaged phase space.\\
In the following sections, we can refer to a given family of the reunions $\ol \cF^\nu$ and $\ol \cF^g$ by noting for example $\ol \cF^\nu_k$ the family of $\ol \cF^\nu$ going through the $L_k$ equilibrium.
}

\subsection{Variété de référence}
\label{sec:VaRe}

Dans le cas circulaire plan, nous avons vu que le problème moyen possédait un seul degré de liberté, et donc qu'une trajectoire pouvait être déterminée de manière unique par un jeu de conditions initiales ($\zeta_i$,$Z_i$). Nous avons vu section \ref{sec:pperdi} que ces conditions initiales étaient équivalentes à $\zeta_0$ et $t_0$, où $t_0$ ne représentait que la position initiale de la trajectoire sur une orbite donnée, qui elle est définie de manière unique par $\zeta_0$. Nous avons choisi de définir $\zeta_0$ à l'instant où $\dot{\zeta}=0$ (ce qui est vérifié une infinité de fois dans la résonance coorbitale). Étudier uniquement l'ensemble des trajectoires partant de $t_0=0$ permet donc d'avoir un aperçu exhaustif de l'espace des phases en n'explorant qu'une seule dimension de conditions initiales. 

Dans le cas elliptique, les 4 dimensions du problème moyen réduit nécessitent 4 conditions initiales ($\zeta_i$, $\Delta \varpi_i$, $Z_i$ et $\varPi_i$) pour définir une trajectoire. 
A l'instar du cas circulaire nous voulons construire une variété de conditions initiales $\cV'$ de dimension $2$ (la moitié de la dimension de l'espace des phases), de manière à ce que les trajectoires prenant leurs conditions initiales sur cette variété soient représentatives de l'ensemble des orbites comprises dans un volume $\Sigma'$ de l'espace des phases. Ici, "représentatif" signifie que toutes les orbites de $\Sigma'$ passent aussi proches que l'on veut de la variété $\cV'$ en un temps fini, on fait donc l'hypothèse que les trajectoires émergentes de $\cV'$ balayent l'ensemble des comportements des orbites de $\Sigma$. Nous appellerons $\cV'$ la variété représentative de $\Sigma'$ \citep[similaire au plan représentatif introduit par][]{MiFeBe2006}.

Le couplage entre les degrés de liberté ($\zeta$,$Z$) et ($\Delta \varpi$,$\varPi$) empêche une détermination triviale d'un jeu de conditions initiales du type ($\zeta_0,t_0$), mais une construction heuristique reste néanmoins possible si on considère que les orbites étudiées sont quasi-périodiques. On travaille ici sur des temps finis, on peut donc considérer qu'une trajectoire est quasi-périodique si ses fréquences fondamentales n'évoluent pas de manière significative sur les échelles de temps étudiées \citep[on peut étudier de manière équivalente la diffusion du demi-grand axe moyen d'un des corps,][]{RoLa2001}. En vérifiant cela sur une grille de conditions initiales dans un domaine donné, on admettra que la mesure de l'ensemble des conditions initiales conduisant à des trajectoires quasi-périodiques est grande dans le domaine de l'espace des phases étudié.\\

\an{
In the circular coplanar case, we saw in chapter 1 that the co-orbital dynamics was reduced to a 1-degree of freedom problem. In that case, we saw in section \ref{sec:Dyncp} that the initial conditions of the system were equivalent to a couple ($\zeta_0,t_0$) where $\zeta_0$ is the minimal value of $\zeta$ on the orbit and $t_0$ is the first instant when this minimum is reached. Therefore, all trajectories with the same $\zeta_0$ are on the same orbit. We can hence explore the characteristics of all the trajectories of the phase space by studying only the trajectories having for initial condition ($\zeta_0$, $t_0=0$). This reduces the relevant space of initial conditions to a 1-dimensional space.\\
In the eccentric case, the $4$ dimensions of the reduced restricted phase space require 4 initial conditions ($\zeta_i$,$\Dv_i$,$Z_i$ and $\varPi_i$) to define a given trajectory. Following the circular case, we want to determine a 2-dimensional manifold $\cV'$ of initial conditions which would be representative of a volume $\Sigma'$ of the 4-dimensional phase space. Here, `representative' means that any trajectory taking its initial conditions in the volume $\Sigma'$ will, in a finite time, get as close as we want to a point of the manifold $\cV'$. If that is so, we consider that the trajectories emanating from the manifold $\cV'$ are dense in the volume $\Sigma'$. They can thus span all behaviours of the orbits included in $\Sigma'$. We call $\cV'$ the representative manifold of the volume $\Sigma'$ (similar to the representative plan of initial conditions introduced by \cite{MiFeBe2006}).\\
The coupling between the semi-fast degree of freedom ($Z$,$\zeta$) and the secular degree of freedom ($\varPi$,$\Dv$) prevents a trivial determination of the manifold $\cV'$, but a heuristic construction is possible if we consider that the studied orbits are quasi-periodic. We are working on finite durations, relevant to planetary systems. We hence consider that a trajectory is quasi-periodic if its fundamental frequencies do not evolve significantly over the studied time-scale \citep[we can equivalently study the diffusion of the mean semi-major axis][]{RoLa2001}. By checking this on a grid of initial conditions in a given domain, we will admit that the measure of the set of initial conditions leading to quasi-periodic trajectories is large in the studied domain of the phase space.
}

\subsubsection{Construction de $\cV'$}

La construction du jeu de conditions initiales s'effectue en deux étapes. La première est de déterminer une variété de dimension $3$: $\cV_Z=\{(\zeta,\Dv,Z,\varPi_i) / Z=Z_\cV \}$ (resp. $\cV_\varPi=\{(\zeta,\Dv,Z,\varPi) / \varPi=\varPi_\cV \}$) de manière à ce que l'adhérence des trajectoires issues de $\cV_Z$ (resp $\cV_\varPi$) forme un volume que nous appellerons $\Sigma_{Z}$ (resp. $\Sigma_\varPi$). Les trajectoires étudiées étant supposées quasi-périodiques, il existe pour toute trajectoire de $\Sigma_{Z_\cV}$ un nombre infini d'instants $t^k_{Z_\cV}$ (resp. $t^k_{\varPi_\cV}$), avec $k\in \mathbb{N}$, tels que:
\begin{equation}
 (\zeta,\Delta \varpi,Z,\varPi)(t^k_{Z_\cV}) =(\zeta ,\Delta \varpi,Z=Z_\cV,\varPi)(t^k_{Z_0}) \ .
\end{equation}
Respectivement, il existe pour toute trajectoire comprise dans $\Sigma_{\varPi_\cV}$ un nombre infini d'instants $t^k_{\varPi_\cV}$ tels que:
\begin{equation}
 (\zeta,\Delta \varpi,Z,\varPi) (t^k_{\varPi_\cV}) =(\zeta ,\Delta \varpi,Z,\varPi=\varPi_\cV)(t^k_{\varPi_\cV}).
\end{equation}

\begin{rem}
On exclut ici le cas particulier où l'adhérence de l'orbite est tangente à la variété $\cV_Z$ (resp. $\cV_\varPi$) sans y être confinée.
\end{rem}

 
Considérons désormais que l'intersection $\tilde{\Sigma}=\Sigma_{Z_\cV} \cap \Sigma_{\varPi_\cV}$ est non vide. 
La seconde étape est de vérifier que pour tout $\epsilon_t > 0$, il existe un couple d'entiers $k$ et $n$ tels que
\begin{equation}
 |t^k_{\varPi_\cV}- t^n_{Z_\cV}| < \epsilon_t.
 \label{eq:ingetVaRe}
\end{equation}
Nous appellerons $\Sigma'$ l'adhérence des trajectoires de $\tilde{\Sigma}$ vérifiant cette condition. Comme les trajectoires de $\Sigma'$ sont continues, on peut choisir $\epsilon_t$ de manière à ce qu'elles passent $\epsilon_\Sigma/2$ proches d'un point de la variété $\cV'$:
%
%
\begin{equation}
\begin{aligned}
\cV' =\cV_Z \cap \cV_\varPi.
\end{aligned}
\label{eq:VaRe}
\end{equation}
%
%
Si ces deux étapes sont validées, puisque les orbites de $\Sigma$ sont quasi-périodiques, le théorème de récurrence de Poincaré nous indique qu'une trajectoire de $\Sigma$ repasse un nombre infini de fois $\epsilon_\Sigma/2$ proche d'un point qui est lui même $\epsilon_\Sigma/2$ proche d'un des points de $\cV'$. Le flot du hamiltonien (\ref{eq:Ht}) étant continu, Le théorème de Cauchy-Lipschitz implique que la trajectoire étudiée reste $\epsilon$ proche de l'orbite issue de $\cV'$ sur un temps $\pm t$, où $t$ dépend de la valeur de $\epsilon_\Sigma$. 

Notons que si $\epsilon_t$ est aussi petit que l'on veut, $\epsilon_\Sigma$ l'est également, et par continuité du flot du hamiltonien et le théorème de Cauchy Lipschitz, toute orbite de $\Sigma$ est identique à une orbite passant par un point de $\cV'$. En pratique, on vérifiera qu'une trajectoire appartient à $\Sigma$ en intégrant le problème à trois corps sur une durée longue devant la plus longue période du système, et on en vérifiant qu'il existe un temps $t$ tel que:
\begin{equation}
(Z-Z_0)^2+(\varPi-\varPi_0)^2 < \epsilon_\Sigma\, , 
\label{eq:condVaRe}
\end{equation}
%
%
%
%
Nous avons donc construit $\cV'$, une variété représentative de la région $\Sigma'$ de l'espace des phases, au sens où l'étude du comportement des trajectoires émergeant de cette variété nous informe sur l'ensemble des trajectoires comprises dans $\Sigma'$.  

Par la suite, notre but est que $\Sigma'$ englobe la totalité des trajectoires stables d'une ou plusieurs configurations coorbitales pour une valeur des masses et du moment cinétique donné, si elle existe, on notera ce volume $\Sigma$ et $\cV$ la variété représentative associée. Notons qu'une condition nécessaire est donc que $\Sigma$ comprenne l'ensemble des points fixes et des variétés $\ol \cF_k$ existant dans la ou les configurations coorbitales que nous voulons représenter.\\

\an{
The first step to construct $\cV'$ is to determine the 3-dimensional manifolds $\cV_Z = \{(\zeta,\Dv,Z,\varPi) / Z=Z_\cV \}$ and $\cV_\varPi=\{(\zeta,\Dv,Z,\varPi) / \varPi=\varPi_\cV \}$, such that the adherence of the trajectories taking as initial condition a point of $\cV_Z$ (resp. $\cV_\varPi$) forms a volume in the 4-dimensional phase space that we call $\Sigma_Z$ (resp. $\Sigma_\varPi$). Since we only consider quasi-periodic orbits, each of them goes through $\cV_Z$ (resp. $\cV_\varPi$) an infinite number of time $t^k_{Z_\cV}$ (resp. $t^k_{Z_\varPi}$). We do not consider here the special case where the trajectory is tangent to $\cV_Z$ or $\cV_\varPi$ without staying on that manifold.\\
We now consider that the intersection $\tilde \Sigma= \Sigma_{Z_\cV} \cap \Sigma_{\varPi_\cV}$ is not empty. The second step is to check that for all $\epsilon_t>0$, there is a couple of integer $k$ and $n$ such that the equation (\ref{eq:ingetVaRe}) is fulfilled. We call $\Sigma'$ the adherence of the trajectories of $\tilde \Sigma$ that fulfil this condition. The trajectories of $\Sigma'$ are continuous, we can thus choose $\epsilon_t$ such that they pass $\epsilon_\Sigma/2$ close to a point of the manifold $\cV'=\cV_Z \cap \cV_\varPi$ (eq. \ref{eq:VaRe}).\\
Since the trajectories of $\Sigma'$ are quasi-periodic, the Poincaré recurrence theorem implies that a trajectory of $\Sigma'$ goes an infinite number of time $\epsilon_\Sigma/2$ close to a point which is itself  $\epsilon_\Sigma/2$ close from $\cV'$. The flow of the Hamiltonian (\ref{eq:Ht}) being continuous, the Cauchy-Lipschitz theorem implies that the considered trajectories stay $\epsilon$ close to an orbit taking its initial condition on the manifold $\cV'$. In our case, we will check that a given trajectory belongs to $\Sigma'$ by integrating this trajectory numerically and checking the equation (\ref{eq:condVaRe}).\\
In the following sections, our goal is that $\Sigma'$ contains all the stable trajectories of one or several co-orbital configurations for a given value of the masses of the co-orbitals and of the total angular momentum. If such manifold exists, we call it $\cV$. Notate that a necessary condition for $\cV$ to represent all the orbits of a given configuration is that its associated volume $\Sigma$ contains all the families $\ol \cF$ included in that configuration. 
}

\subsubsection{Expression de $\cV$}


Nous développerons en section \ref{sec:VaRemdif} un algorithme visant à déterminer une variété de référence pour une configuration coorbitale donnée à masses et valeur du moment cinétique fixés. Dans certains cas, nous pouvons déterminer une expression approchée de $\cV$. Nous le ferons pour le voisinage de l'équilibre de Lagrange circulaire $L_4$ en section \ref{sec:VaReL4}, et dans le cas d'excentricité modérée (ordre $1$) pour des coorbitaux à masses égales en section \ref{sec:VaRemeq}.\\    

\an{
In section \ref{sec:VaRemdif} we will describe an algorithm to determine the position of $\cV$ for a given configuration and a fixed value of the masses and total angular momentum (we will take as example the neighbourhood of $L_4$ for $e_1=e_2=0.4$). In some cases, we can give an analytical approximation for the position of $\cV$: the neighbourhood of the circular $L_4$ equilibrium (section \ref{sec:VaReL4}), and in the case of equal mass co-orbitals, see section \ref{sec:VaRemeq}.      
}
  
\subsection{Séparation des échelles de temps et invariance adiabatique}  
\label{sec:set}

Rappelons qu'il existe deux échelles de temps pour le problème moyen réduit: une échelle de temps semi-rapide de fréquence fondamentale $\nu =\gO(\sqrt{\eps})$ qui est associée à l'évolution de l'angle résonant $\zeta$, et une échelle de temps dite séculaire, associée à l'évolution des excentricités et des périhélies. Dans le problème moyen réduit, la fréquence fondamentale associée aux variables $\Delta \varpi$ et $\varPi$ est notée $g=\gO(\eps)$. La séparation de ces échelles de temps dans le cas de résonances en moyen mouvement est une approche classique \citep[][par exemple]{HeCa1989,Morbidelli02,BaMo2013,Delisle2012,Delisle2014}. En théorie, cette séparation nous permettrait de moyenner le hamiltonien sur l'angle semi-rapide, et donc d'obtenir le hamiltonien du problème séculaire. En pratique cette opération est assez délicate car ($Z$,$\zeta$) ne sont pas proches de variables action-angle \citep[][dans le cas restreint]{Morais1999,Morais2001,BeaugeR01,PaEf2015}. Supposons cependant que nous effectuons cette seconde moyenne. Nous obtiendrions un hamiltonien à 1 degré de liberté avec un paramètre supplémentaire $J_0$ qui serait la variable d'action associée au degré de liberté ($Z$,$\zeta$). La transformation canonique associée à cette moyennisation pour les variables $\Delta \varpi$ et $\varPi$ ne diffère de l'identité que par des coefficients de taille $\gO(\sqrt{\eps})$ \citep[voir][]{Morbidelli02}.  

La possibilité d'effectuer cette moyenne nous apporte des informations importantes sur la dynamique. Premièrement, dans le problème moyen réduit (à deux degrés de liberté), l'évolution des variables $\Delta \varpi$ et $\varPi$ sur des temps de l'ordre de $1/\nu$ se limite à des perturbations de taille $\sqrt{\eps}$. Par conséquent, les variables $\Delta \varpi$ et $\varPi$ peuvent être considérées constantes sur des temps courts devant $1/g$. Pour des coorbitaux de masses suffisamment faibles, nous pouvons donc faire l'hypothèse que $\Delta \varpi$ et $\varPi$ sont des invariants adiabatiques.

Cette séparation des échelles de temps induit une propriété intéressante des variétés de référence définies par l'équation (\ref{eq:VaRe}). Considérons à nouveau que nous ayons effectué la seconde moyenne sur l'angle $\zeta$. Le système à 1 degré de liberté ($\Delta\varpi,\varPi$) est donc intégrable et ses trajectoires sont périodiques. Distinguons trois cas: si l'intersection entre la famille $\ol \cF^g_{k}$ et la variété $\cV_\varPi$ est non vide, les trajectoires qui ont pour condition initiale un des points de cette intersection sont des points fixes du problème séculaire. Leur orbite est donc représentée par un unique point de la variété $\cV_\varPi$. Le cas où $\dot{\varPi}=0$ sur la variété $\cV_\varPi$ sans que la trajectoire soit membre de $\ol \cF^g_{k}$ est un cas particulier à priori exclu par construction de $\cV$. Dans tous les autres cas, les trajectoires périodiques passent par au moins deux points de la variété $\cV_\varPi$, un pour $\dot{\varPi}>0$ et un pour $\dot{\varPi}<0$.

 Rappelons que les points de la variété $\cV$ sont par définition à la fois dans les variétés $\cV_\varPi$ et $\cV_Z$. Considérons les orbites passant par un ou plusieurs points de $\cV_\varPi$. Comme évoqué précédemment, on peut considérer que sur une échelle de temps courte devant $1/g$, $\varPi=\varPi_\cV$ et $\Delta \varpi$ sont des constantes. Sur cette échelle de temps, la dynamique est réduite à un degré de liberté, celui des variables ($Z,\zeta$). Pour ces orbites périodiques, nous pouvons avoir une discussion similaire à celle conduite pour les variables ($\Delta\varpi,\varPi$) dans le cas du problème doublement moyenné. Les trajectoires génériques issues de la variété de référence passent dans le voisinage de plusieurs points distincts de la variété. Le nombre de points par lequel passe la trajectoire est réduit si la condition initiale appartient à une des familles $\overline{\cF}_{k}$. Au cas où la condition initiale appartient à la fois aux familles $\ol \cF^\nu_{k}$ et $\ol \cF^g_{k}$, alors l'orbite est un point fixe du problème réduit moyen, et la trajectoire est réduite à sa condition initiale.\\

 \an{
 The reduced averaged problem possesses two time scales: the semi-fast time scale of fundamental frequency $\nu=\gO(\sqrt{\eps})$, associated with the evolution of the resonant angle $\zeta$, and the secular time scale of fundamental frequency $g=\gO(\eps)$, associated with the evolution of the eccentricities and the arguments of perihelia. The separation of these two time scales is a classical approach for the study of mean motion resonances \citep{HeCa1989,Morbidelli02,BaMo2013,Delisle2012,Delisle2014}. In theory, this separation allows for the averaging of the Hamiltonian over the semi-fast angle $\zeta$ to obtain the secular Hamiltonian: a 1-degree of freedom Hamiltonian with an additional parameter $J_0$ (the action variable associated with the degree of freedom $Z$,$\zeta$). The canonical transformation associated with this averaging for the variables $\varPi$ and $\Dv$ differs from the identity only with coefficients of the order of $\gO(\sqrt \eps)$ \citep{Morbidelli02}.\\
 This averaging is rather difficult to execute because the variables $Z$ and $\zeta$ are not close to action-angle variables \citep{Morais1999,Morais2001,BeaugeR01,PaEf2015}, but the possibility to do it gives us important information on the dynamics of the system: in the averaged reduced problem (2 degree of freedom), the evolution of the variables $\varPi$ and $\Dv$ is of the order of $\gO(\sqrt{\eps})$ over durations of the order of $1/\nu$. This yields two interesting approaches: \\
First, the $\varPi$ and $\Dv$ can be considered as constant on a time scale short with respect to $1/g$. For sufficiently low-mass co-orbitals, we can hence consider that the variables $\varPi$ and $\Dv$ are adiabatic invariants. \\
Second, we can do numerical integrations of the full 3-body problem and study the evolution of these variables as if we were observing their evolution in the double-averaged problem, as their evolution in the full 3-body problem only differs from their evolution in the secular problem by coefficients of the order of $\gO(\sqrt \eps)$.\\}

\an{
Combining these two approaches, we can deduce an interesting propriety of the reference manifold that we introduced in the previous section: let us consider a generic trajectory (which does not belong to neither $\ol \cF^\nu$ nor $\ol \cF^g$). This trajectory goes twice through the 3-dimensional $\cV_Z$ manifold over a period of $\nu$ and twice through the 3-dimensional $\cV_\varPi$ manifold over a period of $g$. Let us consider one instant when the trajectory goes through the manifold $\cV_\varPi$. On a time scale short with respect to $1/g$, the system undergo a full period associated with the frequency $\nu$ and thus cross the manifold $\cV_Z$ in two different places, while $\Dv$ and $\varPi$ can be considered as constant. Then, the system will evolve away from these manifolds, until in crosses once again the manifold $\cV_\varPi$ in another point. Following the same approach, we can deduce that the system will once again cross the manifold $\cV_Z$ in two points while it stays in the neighbourhood of the same point of $\cV_\varPi$. These generic trajectories will hence quasi-periodically get close to $4$ points of the manifold $\cV=\cV_Z \cap \cV_\varPi$. Moreover, these 4 points are two pairs of points having the same value of $\Dv$ and $\varPi$.\\
 If the considered trajectory belongs to one of the family $\ol \cF^\nu$ or $\ol \cF^g$, then this trajectory is periodic and get close to two points of the manifold $\cV$. If it is a fixed point of the reduced averaged problem, then the whole trajectory is an unique point of $\cV$.}

\subsection{Variété de référence au voisinage de l'équilibre de Lagrange circulaire}
\label{sec:VaReL4}

Le voisinage de l'équilibre de Lagrange circulaire est un des cas où nous pouvons avoir une approximation analytique de la variété $\cV$. L'équation (\ref{eq:syserdi}) reste valable pour de petites excentricités. Donc, indépendamment de $\Delta \varpi$, nous pouvons voir sur le portrait de phase figure \ref{fig:ppH0b} que l'ensemble des trajectoires passe deux fois par la variété $\dot{\zeta}=0$. On sait que $L_4$ est en $Z_{L_4}= \frac\eps6 \frac{m'_1-m'_2}{m'_1+m'_2}\frac{m'_1m'_2}{m_0} \mu_0^{2/3} \eta^{-1/3} + \gO(\eps^2)$ (équation \ref{eq:L4loc}). Nous définissons donc $\cV_Z$ par $Z_\cV=Z_{L_4}$.
   
Nous rappelons qu'au voisinage de la variété $\mathcal{C}_0 $, l'évolution des variables $x_1$ et $x_2$ est donnée par le système différentiel (\ref{eq:eqvar1}). Au voisinage de l'équilibre de Lagrange $L_4$, les solutions de ce système se décomposent sur la base des vecteurs propres $V^1_{\pi/3}$ et $V^2_{\pi/3}$. Nous avons donc:
\begin{equation}
\begin{aligned}
x_1 & =\alpha \sqrt{m_2} \e^{i (\frac{\pi}{3}+g t)} + \beta \sqrt{m_1} \e^{i\frac{\pi}{3}}\, , \\
x_2 & =- \alpha \sqrt{m_1} \e^{i g t} + \beta \sqrt{m_2} .
\end{aligned}
\label{eq:x1x2L}
\end{equation}
avec $\alpha$ et $\beta$ complexes, dépendants des valeurs initiales de $x_1$ et $x_2$. En injectant ces expressions dans celle de $\varPi$ eq.~(\ref{eq:CIred}), et en notant $\alpha \bar{\beta}= C \e^{i c}$, nous obtenons:
\begin{equation}
\begin{aligned}
\varPi & = (\alpha\bar{\alpha}-\beta\bar{\beta})(m_1-m_2)-2C\sqrt{m_1 m_2} \cos (g t + c)\, ,
\end{aligned}
\label{eq:PiL4}
\end{equation}
d'autre part,
\begin{equation}
\begin{aligned}
\Dv=\arg(x_1\bar{x}_2) \ \text{avec}\ 
x_1\bar{x}_2 &=[\sqrt{m_1m_2}(\beta\bar{\beta}-\alpha\bar{\alpha})+Cm_2\e^{i(gt+c)}-Cm_1\e^{-i(gt+c)}]\e^{i\pi/3}\, ,
\end{aligned}
\label{eq:DpiL4}
\end{equation}
%
L'ensemble des orbites dans le voisinage de Lagrange passe donc par la variété $\cV_\varPi$, définie par $\varPi_\cV=(\alpha\bar{\alpha}-\beta\bar{\beta})(m_1-m_2)$ quand $t_k=(\pi/2+k\pi-c)/g$. A ces mêmes instants, $\Dv$ atteint ses extrema, donc $\dot{\Dv}=0$. Notons que la variété $\cV_\varPi$ est le plan d'équation $\varPi=0$ sur le voisinage de $L_4$ dans le cas où $m_1=m_2$. Par ailleurs, les orbites génériques passent deux fois par $\cV_Z$ avec une fréquence $\nu$ et deux fois par $\cV_\varPi$ avec une fréquence $g$. Toute orbite générique du voisinage de l'équilibre de Lagrange passe donc au voisinage de 4 points de $\cV$. Si les fréquences $g$ et $\nu$ ne sont pas résonantes, alors la condition (\ref{eq:ingetVaRe}) est vérifiée. La variété $\cV=\cV_Z \cap \cV_\varPi$ est donc une variété représentative de conditions initiales pour l'ensemble du voisinage de $L_4$.\\

\an{
Near the circular $L_4$ equilibrium, we can obtain an analytical approximation for the position of $\cV$. Indeed, the equation (\ref{eq:syserdi}) is relevant at order $1$ in $e_j$. Hence, independently from the value of $\varPi$ and $\Dv$, we can see in the phase space represented in figure~\ref{fig:ppH0b} that each trajectory passes twice through the manifold $Z=Z_{L_4}=\frac\eps6 \frac{m'_1-m'_2}{m'_1+m'_2}\frac{m'_1m'_2}{m_0} \mu_0^{2/3} \eta^{-1/3} + \gO(\eps^2)$ (equation \ref{eq:L4loc}) during each period. We hence take $Z_\cV=Z_{L_4}$. \\
Near the circular Lagrangian equilibrium, the evolution of the variables $x_1$ and $x_2$ is given by the differential system (\ref{eq:eqvar1}). The solutions of this system are given by the equation (\ref{eq:x1x2L}) we can hence deduce the expression of $\varPi$ and $\Dv$ (equations \ref{eq:PiL4} and \ref{eq:DpiL4}).\\
All the orbits in the neighbourhood of the circular Lagrangian equilibrium thus pass through the manifold $\cV_\varPi$, defined by $\varPi_\cV=(\alpha\bar{\alpha}-\beta\bar{\beta})(m_1-m_2)$, when $t_k=(\pi/2+k\pi-c)/g$. the instants $t_k$ are also the instants where $\Dv$ reaches its extremal values, hence $\Delta \dot \varpi = 0$. Note that $\varPi_\cV=0$ when $m_1=m_2$.\\
A generic trajectory in the neighbourhood of $L_4$ goes twice through $\cV_Z$ each period $2\pi/\nu$ and twice through $\cV_\varPi$ with a period $2\pi/g$. If these frequencies are not resonant, all the trajectories verify the condition (\ref{eq:ingetVaRe}). The manifold $\cV=\cV_Z \cap \cV_\varPi$ is hence a representative manifold of the neighbourhood of $L_4$.   
}

\subsection{Variété de référence pour des coorbitaux à masses égales}
\label{sec:VaRemeq}

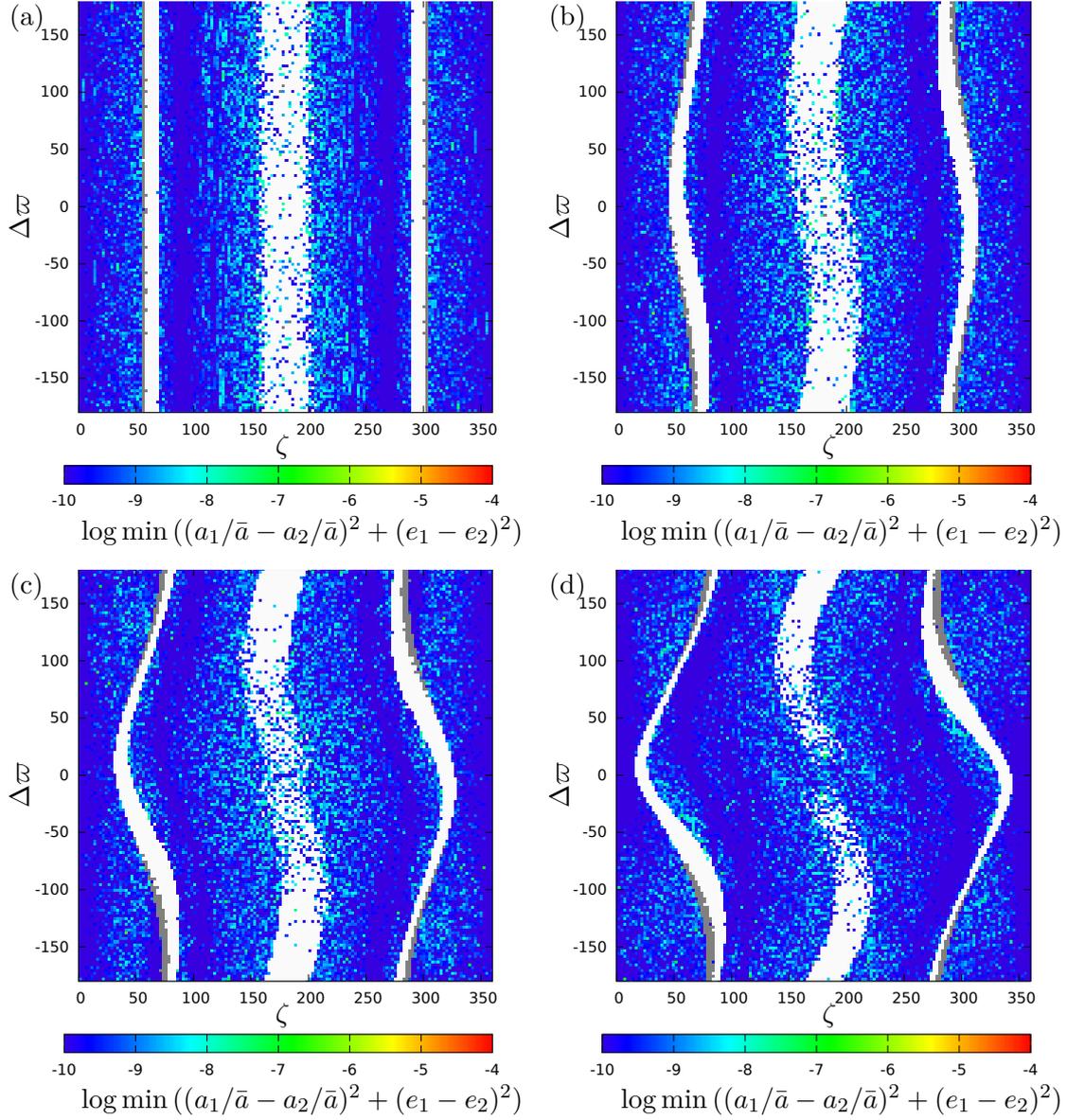
\begin{figure}[h!]
\begin{center}
 \includegraphics[width=0.49\linewidth]{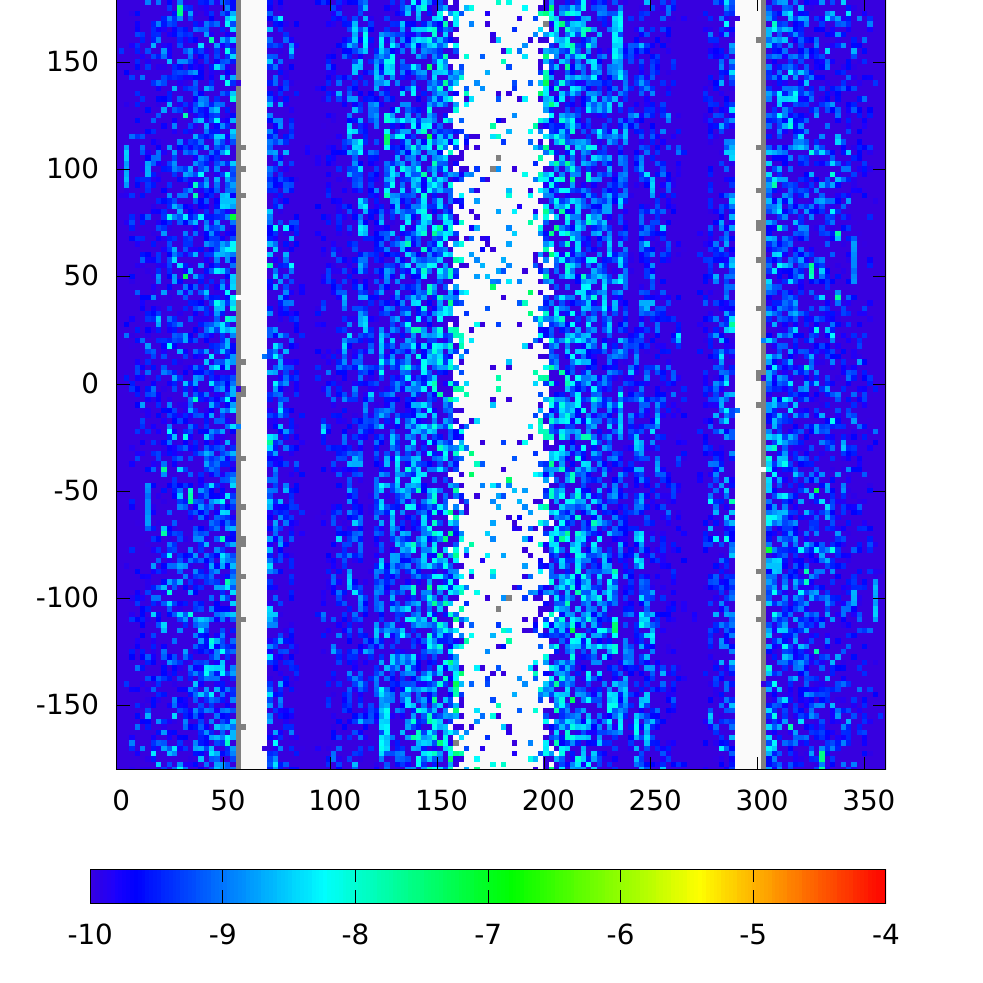}  \includegraphics[width=0.49\linewidth]{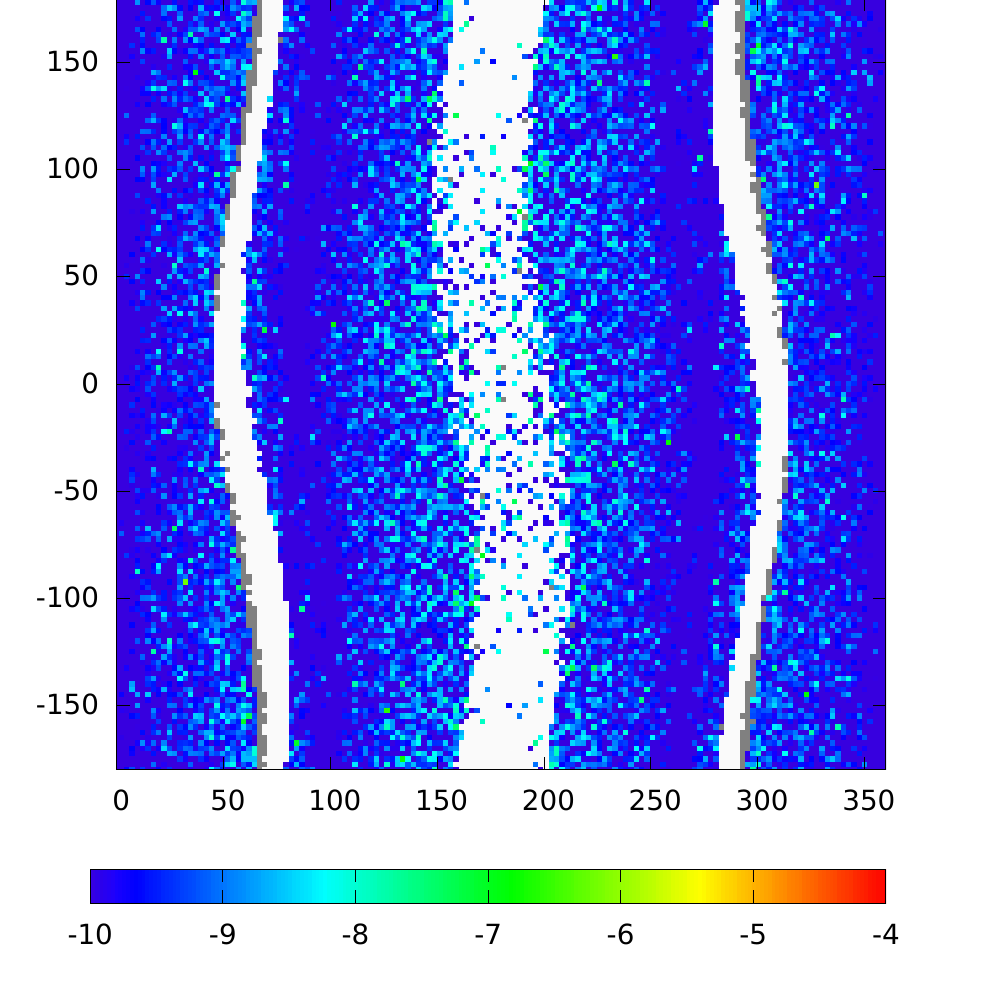}\\
   \setlength{\unitlength}{1cm}
\begin{picture}(.001,0.001)
\put(-7.5,4.5){\rotatebox{90}{$\Dv$}}
\put(0,4.5){\rotatebox{90}{$\Dv$}}
\put(-7.5,7.5){(a)}
\put(0,7.5){(b)}
\put(-3.8,1.5){{$\zeta$}}
\put(3.8,1.5){{$\zeta$}}
\put(-6.5,0.3){{$\log\min{} ((a_1/\bar{a}-a_2/\bar{a})^2+(e_1-e_2)^2)$}}
\put(1,0.3){{$\log\min{}((a_1/\bar{a}-a_2/\bar{a})^2+(e_1-e_2)^2)$}}
\end{picture}

 \vspace{0.1cm}
 
  \includegraphics[width=0.49\linewidth]{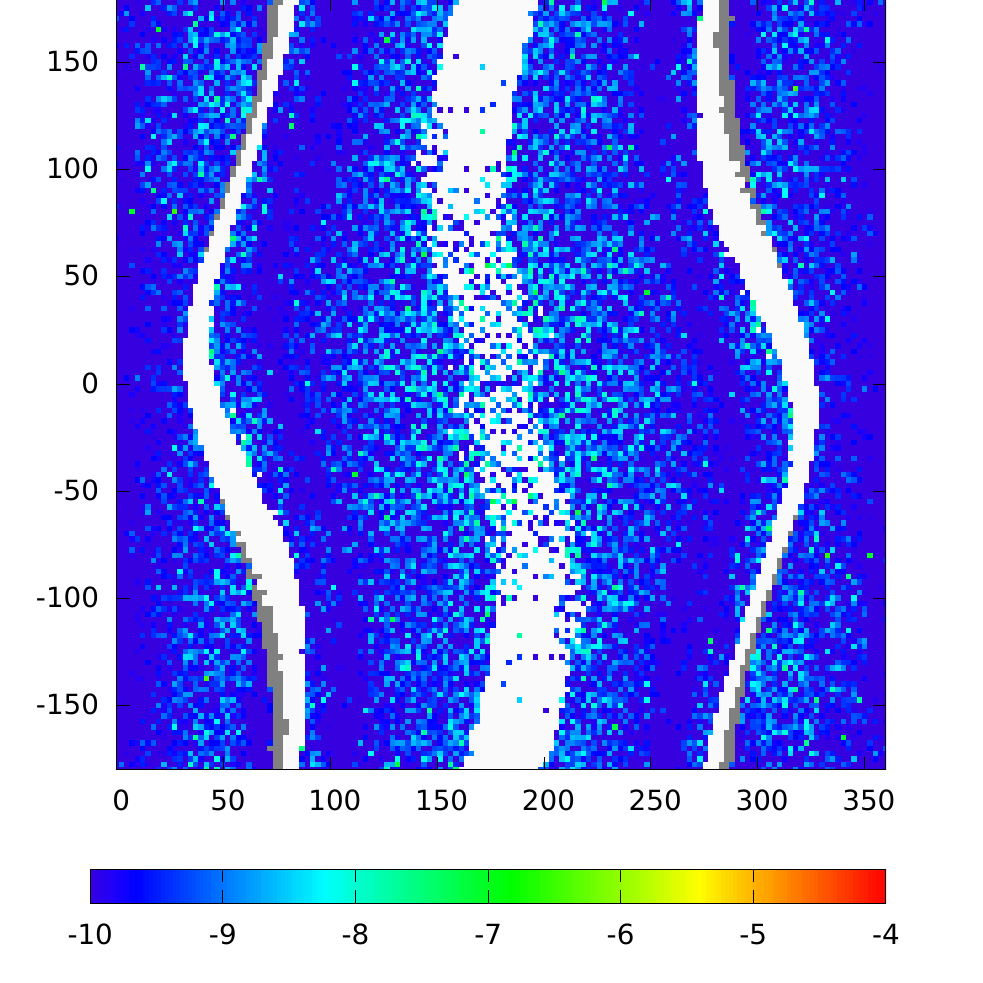}
   \includegraphics[width=0.49\linewidth]{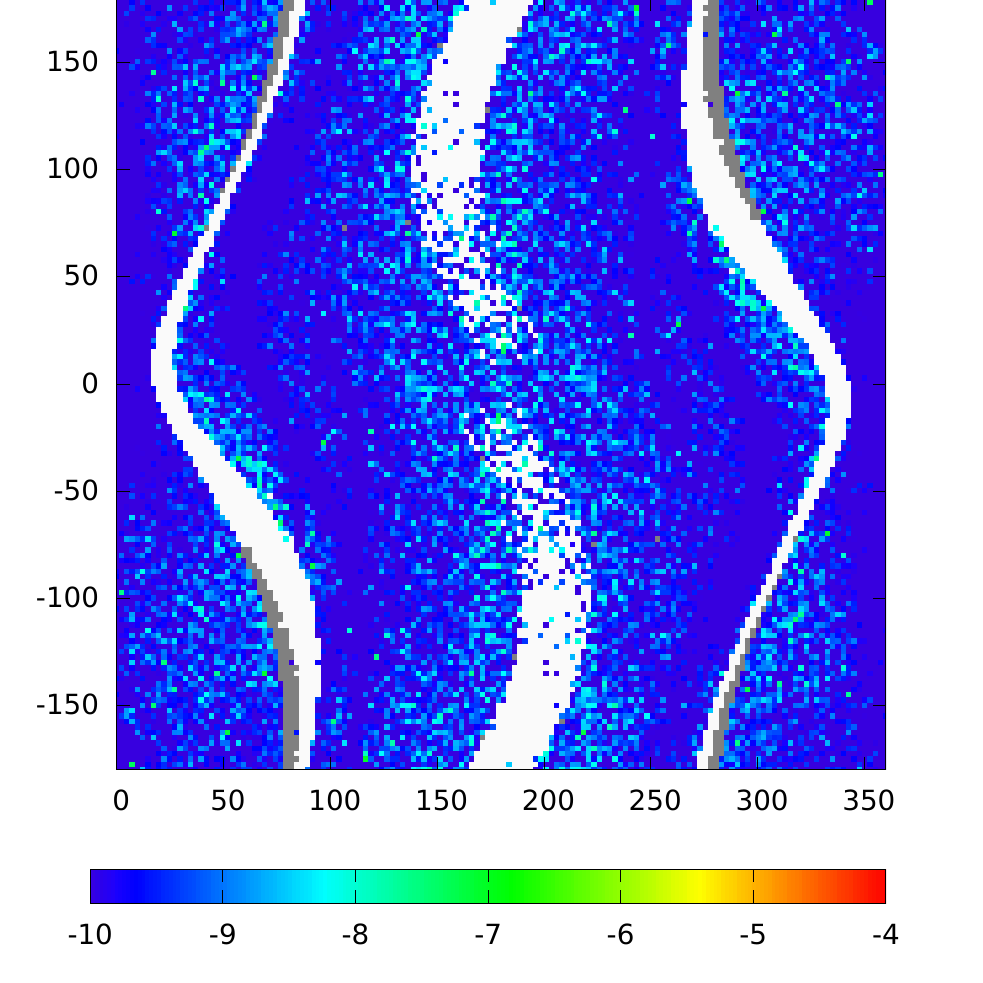}\\
  \setlength{\unitlength}{1cm}
\begin{picture}(.001,0.001)
\put(-7.5,4.5){\rotatebox{90}{$\Dv$}}
\put(0,4.5){\rotatebox{90}{$\Dv$}}
\put(-7.5,7.5){(c)}
\put(0,7.5){(d)}
\put(-3.8,1.5){{$\zeta$}}
\put(3.8,1.5){{$\zeta$}}
\put(-6.5,0.3){{$\log\min{}((a_1/\bar{a}-a_2/\bar{a})^2+(e_1-e_2)^2)$}}
\put(1,0.3){{$\log\min{}((a_1/\bar{a}-a_2/\bar{a})^2+(e_1-e_2)^2)$}}
\end{picture}
\caption{\label{fig:varefmeqe04} Valeur minimale de $\log \left( (a_1/\bar a-a_2/\bar a)^2+(e_1-e_2)^2 \right)$ sur $10\times 10^{5}$~an avec un pas de $0.01$~an pour une valeur fixée du moment cinétique $J_1(e_1=e_2=0.4)$ avec les conditions initiales suivantes $a_1=a_2=1$, $m_1=m_2=10^{-5} m_0$. Sur la figure (a) $e_1=0.00$ et $e_2 \approx 0.55$; (b) $e_1=0.10$ et $e_2 \approx 0.54$; (c) $e_1=0.20$ et $e_2 \approx 0.52$; (d) $e_1=0.30$ et $e_2 \approx 0.47$. Les trajectoires éjectées avant la fin de l'intégration sont identifiées par des pixels blancs. Voir la section \ref{sec:Coo2meg} pour plus de détails concernant les intégrations. La couleur grise indique les trajectoires dont la diffusion du demi-grand axe moyen est supérieure à $\epsilon_a=10^{-5.5}$.\\
} 
\end{center}
\end{figure}

\begin{figure}[h!]
\begin{center}
 \includegraphics[width=0.49\linewidth]{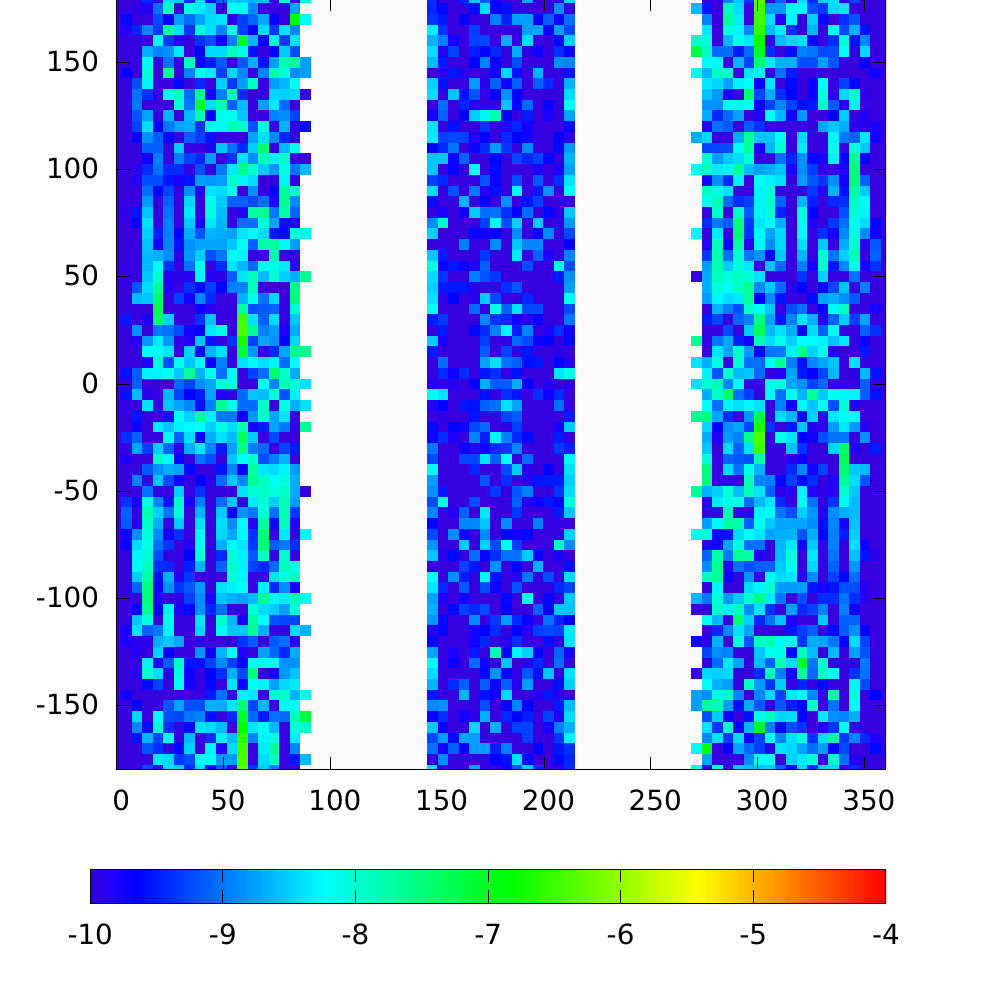}  \includegraphics[width=0.49\linewidth]{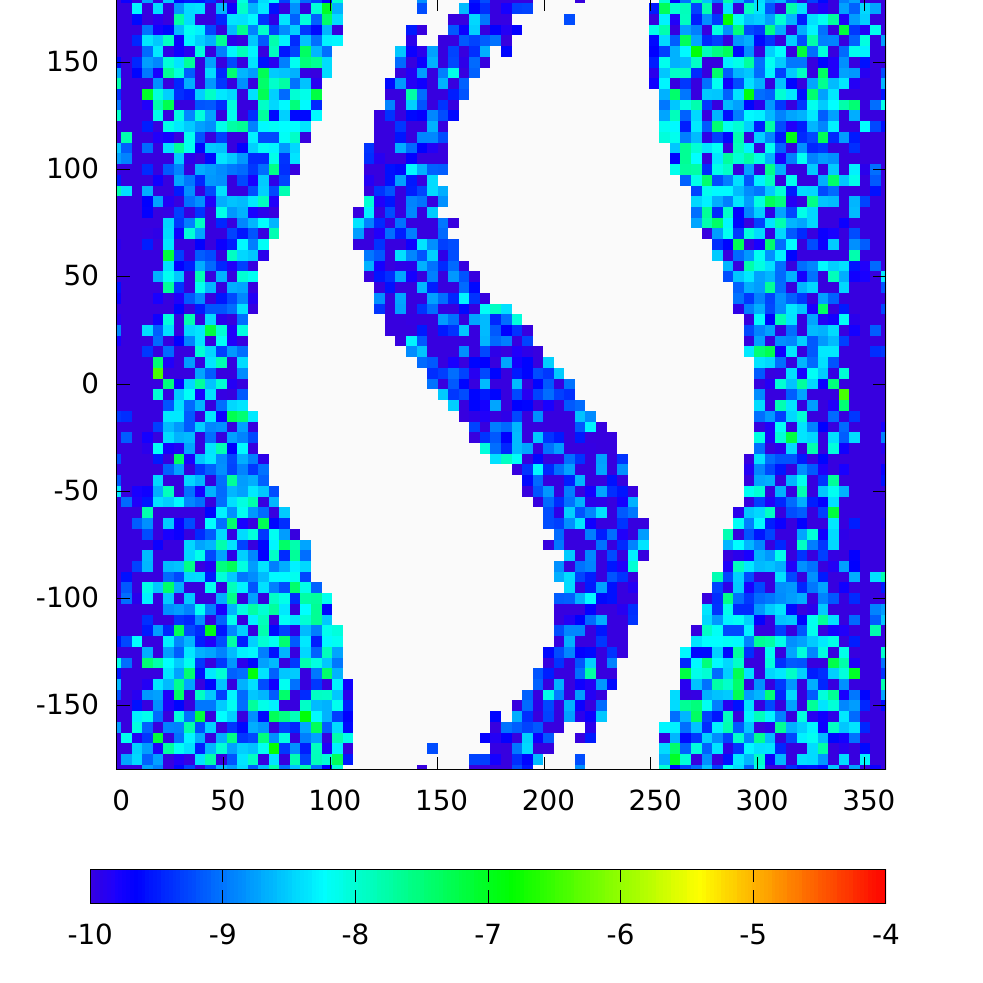}\\
   \setlength{\unitlength}{1cm}
\begin{picture}(.001,0.001)
\put(-7.5,4.5){\rotatebox{90}{$\Dv$}}
\put(0,4.5){\rotatebox{90}{$\Dv$}}
\put(-7.5,7.5){(a)}
\put(0,7.5){(b)}
\put(-3.8,1.5){{$\zeta$}}
\put(3.8,1.5){{$\zeta$}}
\put(-6.5,0.3){{$\log\min{}((a_1/\bar{a}-a_2/\bar{a})^2+(e_1-e_2)^2)$}}
\put(1,0.3){{$\log\min{}((a_1/\bar{a}-a_2/\bar{a})^2+(e_1-e_2)^2)$}}
\end{picture}

 \vspace{0.1cm}
 
  \includegraphics[width=0.49\linewidth]{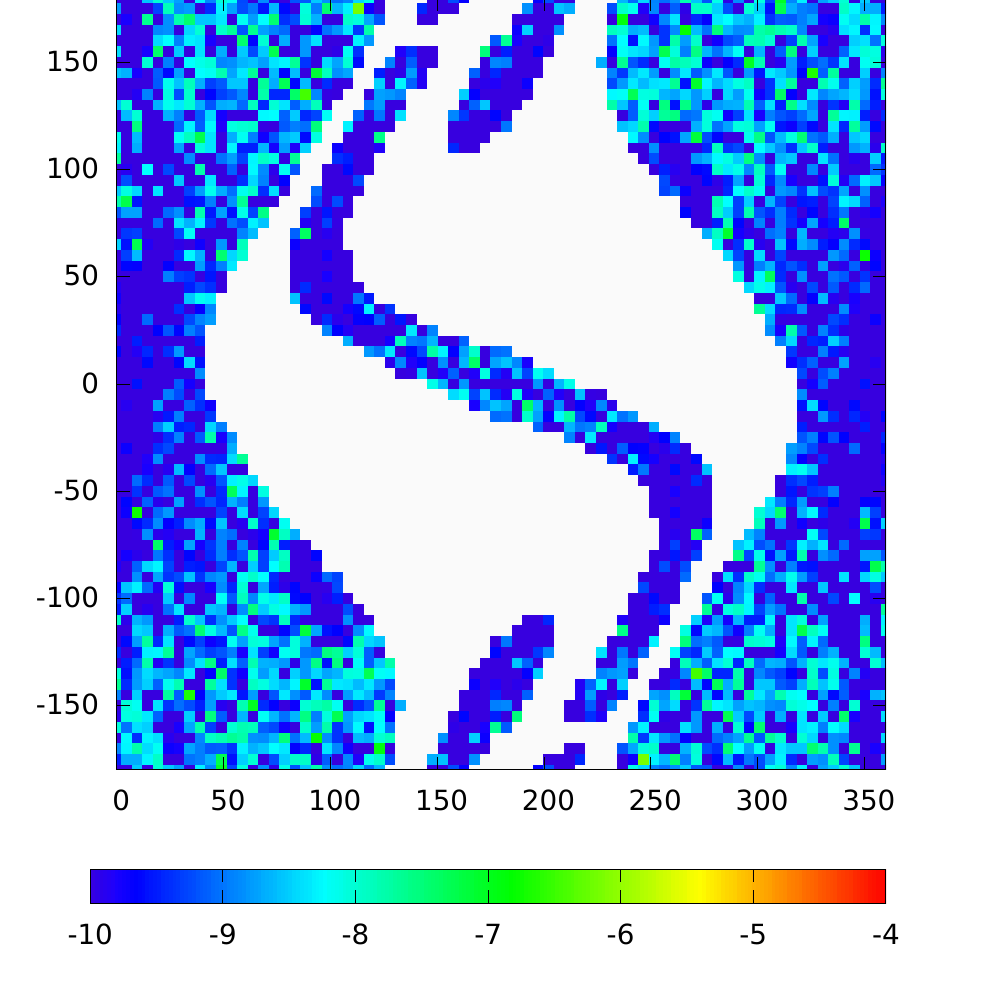}
   \includegraphics[width=0.49\linewidth]{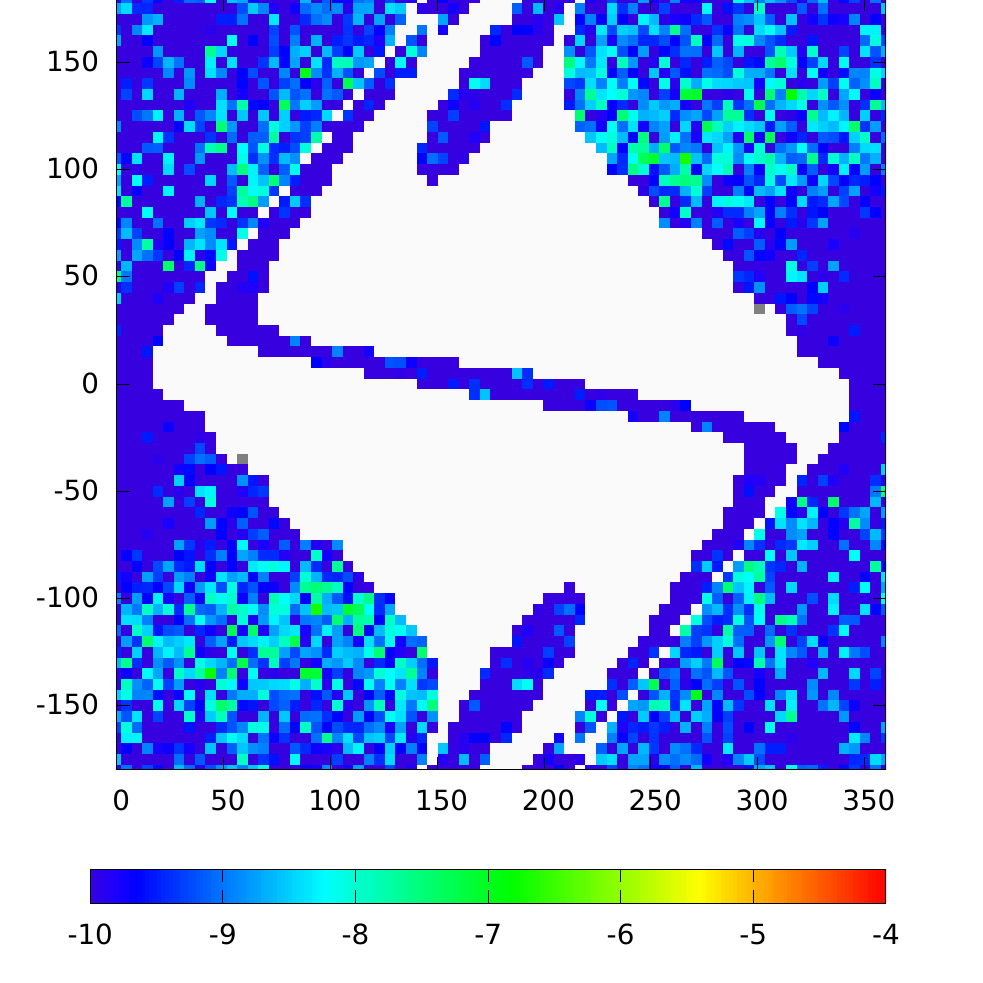}\\
  \setlength{\unitlength}{1cm}
\begin{picture}(.001,0.001)
\put(-7.5,4.5){\rotatebox{90}{$\Dv$}}
\put(0,4.5){\rotatebox{90}{$\Dv$}}
\put(-7.5,7.5){(c)}
\put(0,7.5){(d)}
\put(-3.8,1.5){{$\zeta$}}
\put(3.8,1.5){{$\zeta$}}
\put(-6.5,0.3){{$\log\min{}((a_1/\bar{a}-a_2/\bar{a})^2+(e_1-e_2)^2)$}}
\put(1,0.3){{$\log\min{}((a_1/\bar{a}-a_2/\bar{a})^2+(e_1-e_2)^2)$}}
\end{picture}
\caption{\label{fig:varefmeqe07} Valeur minimale de $\log \left( (a_1/\bar a-a_2/\bar a)^2+(e_1-e_2)^2 \right)$ sur $10\times 10^{5}$~an avec un pas de $0.01$~an pour une valeur fixée du moment cinétique $J_1(e_1=e_2=0.7)$ avec les conditions initiales suivantes $a_1=a_2=1$, $m_1=m_2=10^{-5} m_0$. Sur la figure (a) $e_1=0.00$ et $e_2 \approx 0.90$; (b) $e_1=0.20$ et $e_2 \approx 0.89$; (c) $e_1=0.40$ et $e_2 \approx 0.86$; (d) $e_1=0.60$ et $e_2 \approx 0.78$. Les trajectoires éjectées avant la fin de l'intégration sont identifiées par des pixels blancs. Voir la section \ref{sec:Coo2meg} pour plus de détails concernant les intégrations. La couleur grise indique les trajectoires dont la diffusion du demi-grand axe moyen est supérieure à $\epsilon_a=10^{-5.5}$.\\
} 
\end{center}
\end{figure}

\begin{figure}[h!]
\begin{center}
 \includegraphics[width=0.49\linewidth]{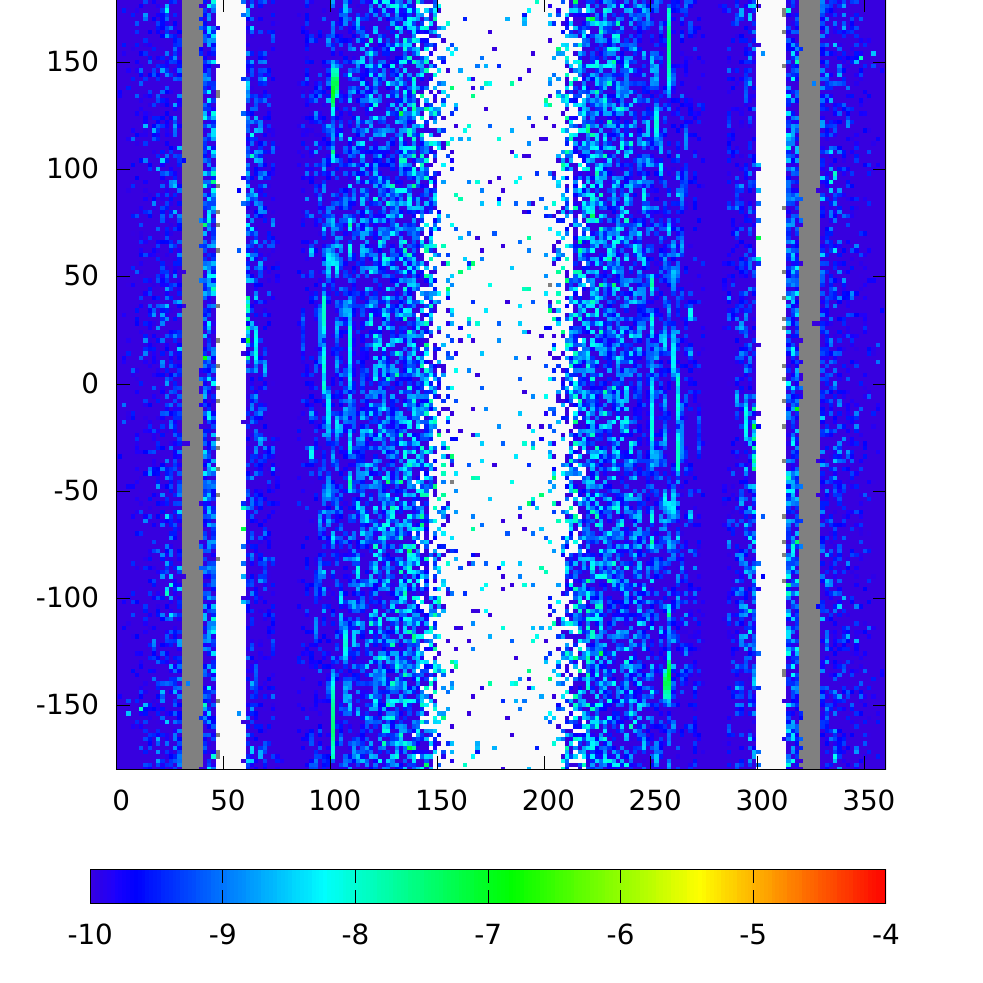}  \includegraphics[width=0.49\linewidth]{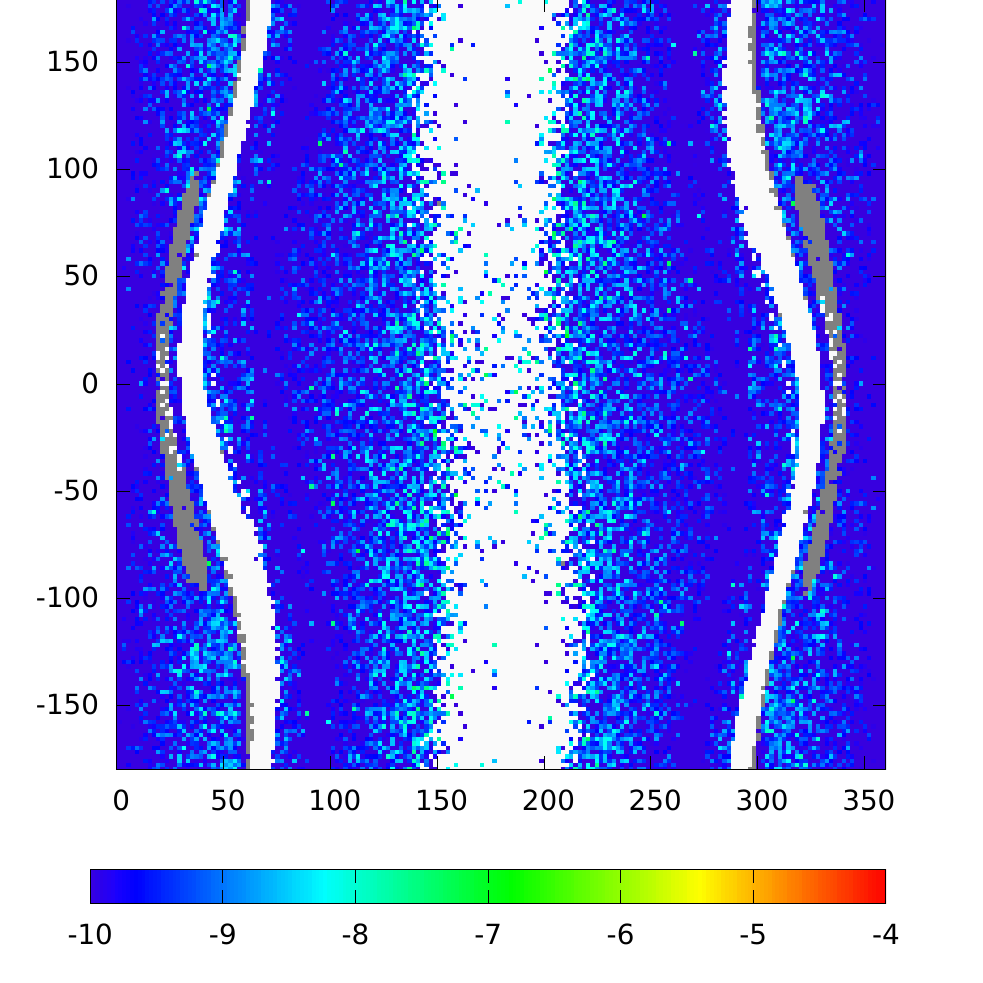}\\
   \setlength{\unitlength}{1cm}
\begin{picture}(.001,0.001)
\put(-7.5,4.5){\rotatebox{90}{$\Dv$}}
\put(0,4.5){\rotatebox{90}{$\Dv$}}
\put(-7.5,7.5){(a)}
\put(0,7.5){(b)}
\put(-3.8,1.5){{$\zeta$}}
\put(3.8,1.5){{$\zeta$}}
\put(-6.5,0.3){{$\log\min{}((a_1/\bar{a}-a_2/\bar{a})^2+(e_1-e_2)^2)$}}
\put(1,0.3){{$\log\min{}((a_1/\bar{a}-a_2/\bar{a})^2+(e_1-e_2)^2)$}}
\end{picture}

 \vspace{0.1cm}
 
  \includegraphics[width=0.49\linewidth]{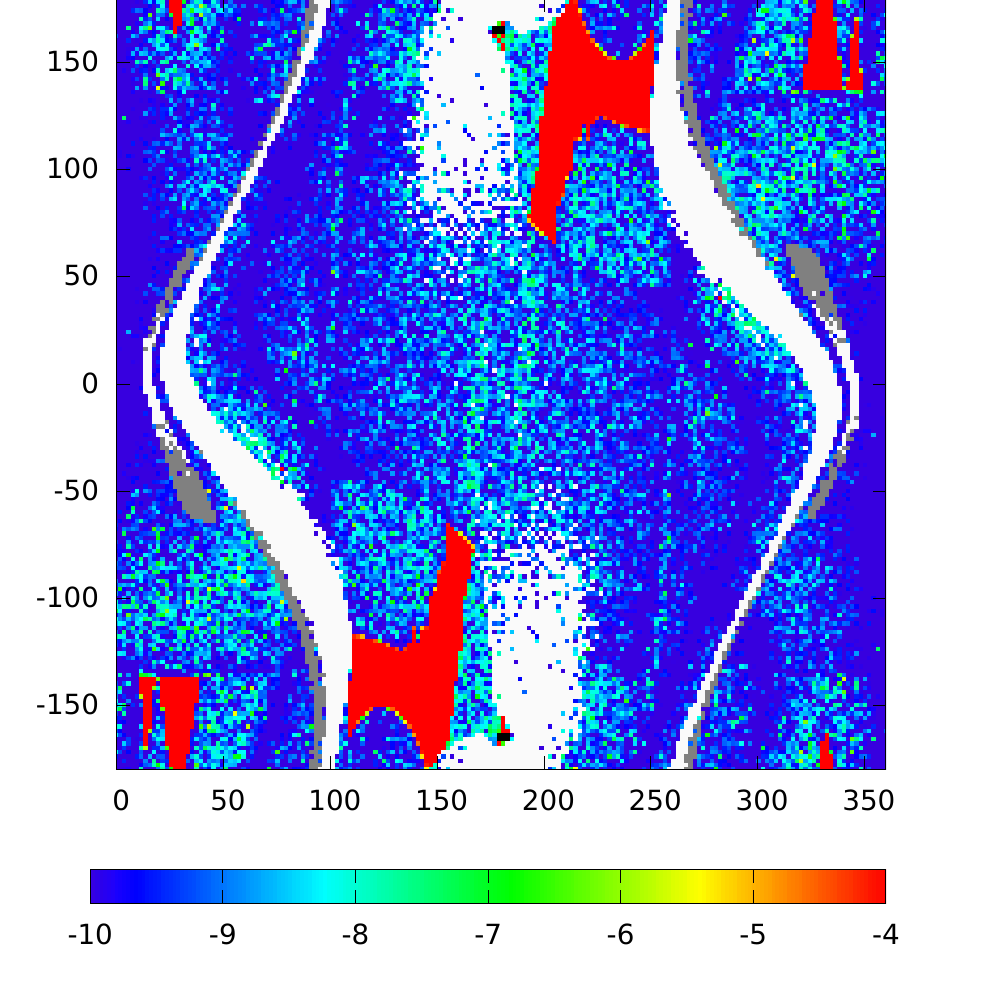}
   \includegraphics[width=0.49\linewidth]{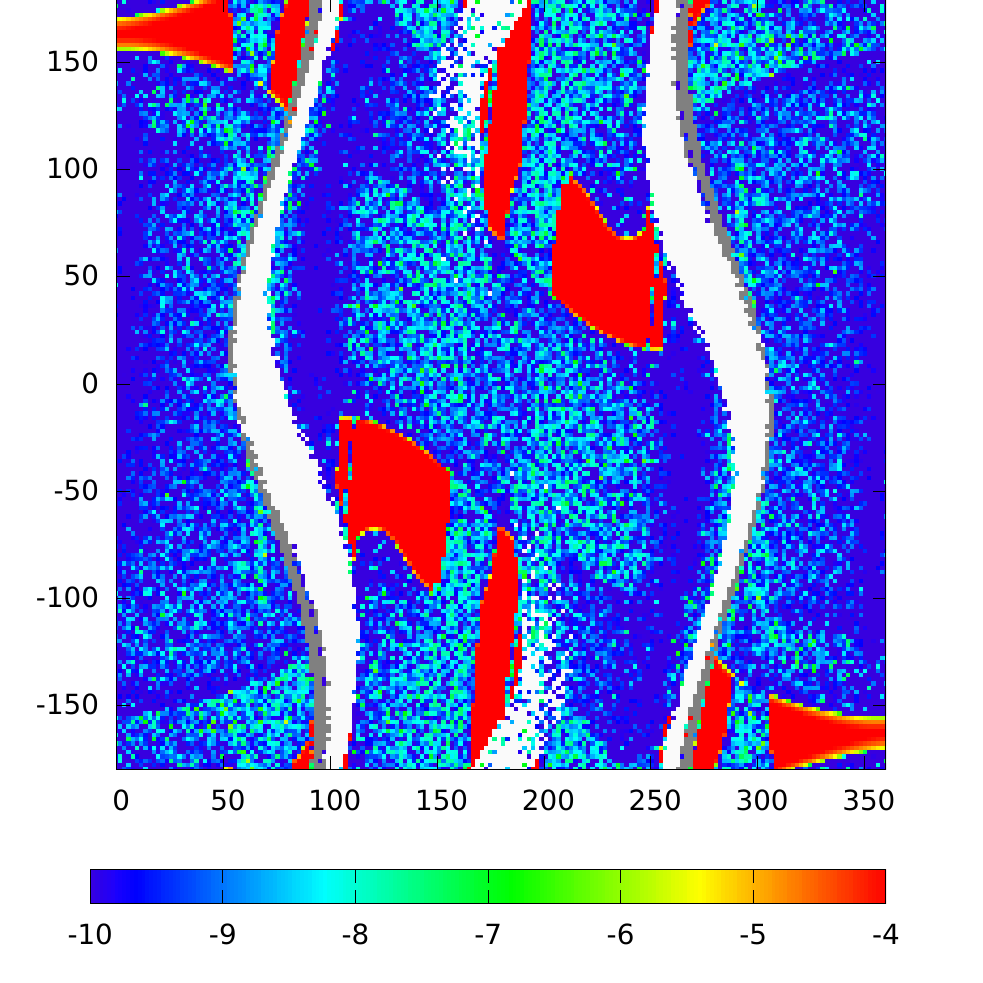}\\
  \setlength{\unitlength}{1cm}
\begin{picture}(.001,0.001)
\put(-7.5,4.5){\rotatebox{90}{$\Dv$}}
\put(0,4.5){\rotatebox{90}{$\Dv$}}
\put(-7.5,7.5){(c)}
\put(0,7.5){(d)}
\put(-3.8,1.5){{$\zeta$}}
\put(3.8,1.5){{$\zeta$}}
\put(-6.5,0.3){{$\log\min{}((a_1/\bar{a}-a_2/\bar{a})^2+(e_1-e_2)^2)$}}
\put(1,0.3){{$\log\min{}((a_1/\bar{a}-a_2/\bar{a})^2+(e_1-e_2)^2)$}}
\end{picture}
 \caption{\label{fig:VaRetestdiff} Valeur minimale de $\log \left( (a_1/\bar a-a_2/\bar a)^2+(e_1-e_2)^2 \right)$ sur $10\times 10^{5}$~an avec un pas de $0.01$~an pour une valeur fixée du moment cinétique $J_1(e_1=e_2=0.4)$ avec les conditions initiales suivantes $a_1=a_2=1$, $m_2=3 m_1= 1.5 \times 10^{-5} m_0$. Sur la figure (a) $e_1=0.00$ et $e_2 \approx 0.55$; (b) $e_1=0.10$ et $e_2 \approx 0.54$; (c) $e_1=0.20$ et $e_2 \approx 0.52$; (d) $e_1=0.30$ et $e_2 \approx 0.47$. Les trajectoires éjectées avant la fin de l'intégration sont identifiées par des pixels blancs. Voir la section \ref{sec:Coo2meg} pour plus de détails concernant les intégrations. La couleur grise indique les trajectoires dont la diffusion du demi-grand axe moyen est supérieure à $\epsilon_a=10^{-5.5}$.\\
 }
 \end{center}
 \end{figure}


Dans le cas de coorbitaux à masses égales nous pouvons avoir une expression simple de $\cV$ dans l'ensemble de l'espace des phases à l'ordre $1$ en excentricité. Nous avons vu au Chapitre 1, section \ref{sec:pperdi} que le hamiltonien moyen du cas circulaire plan est symétrique par rapport à la variété $Z=0$. La quantité $\dot{\zeta}$ s'annule donc sur cette variété. On fait ici l'hypothèse que cette propriété reste vraie pour des coorbitaux excentriques et on définit $\cV_Z$ par $Z=0$.


On rappelle que l'équation variationnelle dans la direction ($x_j$,$\xt_j$) autour d'une trajectoire arbitraire $\zeta(t)$ du cas circulaire plan est donnée par les équations (\ref{eq:eqvar1}) à (\ref{eq:eqvar3}). Les coefficients de la matrice (\ref{eq:eqvar2}) dépendent donc du temps et des conditions initiales de la trajectoire $\zeta(t)$ choisie. Cependant, comme l'évolution des $x_j$ est lente par rapport à celle des variables ($Z,\zeta$), nous pouvons obtenir une approximation de l'évolution séculaire des $x_j$ en moyennant temporellement les coefficients de cette matrice sur une période $2\pi/\nu$. De plus en prenant $m_1=m_2$, cette équation variationnelle moyennée s'écrit sous la forme: 
\begin{equation}
\begin{aligned}
\dot{x_1} & = i(a x_1 + \bar{c} x_2)\ ,\ \dot{\tilde{x}}_1 & =-i(a \tilde{x}_1 + c \tilde{x}_2)\, ,\\
\dot{x_2} & = i(c x_1 + a x_2)\ ,\ \dot{\tilde{x}}_2 & =-i(\bar{c} \tilde{x}_1 + a \tilde{x}_2)\, ,\\
\end{aligned}
\label{eq:x1x2meq}
\end{equation}
avec $a$ réel et $c$ complexe. De ces relations, nous pouvons déduire que:
\begin{equation}
\frac{d^2}{dt^2} |x_1|^2 = -\frac{d^2}{dt^2} |x_2|^2=-2|c|^2(|x_1|^2-|x_2|^2)\, .
\label{eq:osci}
\end{equation}
d'où
\begin{equation}
\frac{d^2}{dt^2} (|x_1|^2- |x_2|^2)=-4|c|^2(|x_1|^2-|x_2|^2)\, .
\label{eq:osci2}
\end{equation}

On reconnaît ici l'équation de l'oscillateur harmonique. $|x_1|=|x_2|$ étant une solution admissible du système, on en déduit que les solutions du système (\ref{eq:eqvar2}) oscillent autour de $|x_1|=|x_2|$ à la fréquence $2|c|$. $|x_1|=|x_2|$ est équivalent à $\varPi =0$ quand $m_1=m_2$ (et donc $e_1=e_2$). On définit donc $\cV_\varPi$ par $\varPi=0$. La variété $\cV =\cV_Z \cap \cV_\varPi$ est représentative des coorbitaux excentriques tant que l'approximation quadratique du hamiltonien (\ref{eq:Hpt}) est valable.

Nous allons verifier numériquement si cela reste vrai pour de plus grandes excentricités. Comme énoncé en section \ref{sec:VaRe}, nous allons chercher le volume de l'espace des phases dans lequel l'ensemble des trajectoires vérifie la relation (\ref{eq:condVaRe}). Ici, comme $m_1=m_2$, ce critère est équivalent à
\begin{equation}
\frac{(a_1-a_2)^2}{\bar{a}^2}+(e_1-e_2)^2 < \epsilon'_\Sigma
\label{eq:critVaRemeq}
\end{equation}

%
%
%
%
Comme $\Sigma \subset \Sigma_{Z_0}$, nous prenons nos conditions initiales sur la variété $Z=0$ et nous explorons l'espace des phases en faisant varier $\varPi$, $\zeta$ et $\Delta\varpi$. En pratique nous regardons des cartes pour $\zeta \in [ 0^\circ : 360^\circ ] $ et $\Dv \in [-180^\circ:180^\circ]$ et différentes valeurs de $\varPi$ pour $J_1$ fixé. Pour les excentricités, cela implique:
\begin{equation}
\begin{aligned}
e_2=\sqrt{1-\frac{2J_1}{\Lambda^0_1}-\sqrt{1-e_1^2}}
\end{aligned}
\label{eq:e2J1cst}
\end{equation}
La figure \ref{fig:varefmeqe04} représente le cas de deux coorbitaux de masses égales $m_1=m_2=10^{-5}$ pour diverses valeurs de $\varPi$ à $J_1(e_1=e_2=0.4)$ fixé (pour des raisons de clarté on donnera les valeurs des $e_j$ équivalentes). Les intégrations sont effectuées sur des temps de $4/\eps$ périodes orbitales, correspondant à quelques périodes de $g$ seulement. Pour chacune des valeurs initiales de $\varPi$ choisies, la condition (\ref{eq:critVaRemeq}) est vérifiée par l'ensemble des trajectoires non éjectées pour $\epsilon'_\Sigma \approx 10^{-8}$. Bien que non exhaustive, cette vérification suggère que la variété de référence choisie représente une part conséquente de l'espace des phases du problème moyen réduit. Il n'est cependant pas exclu que, même avec $m_1=m_2$, des domaines stables apparaissent à haute excentricité avec $e_1 \neq e_2$ sur toute l'orbite. Cependant aucun d'entre eux n'a été découvert pendant cet étude. La figure~\ref{fig:varefmeqe07} montre une vérification similaire pour $J_1(e_1=e_2=0.7)$.

Pour comparaison, la figure~\ref{fig:VaRetestdiff} montre que, pour $m_1 \neq m_2$, le critère (\ref{eq:critVaRemeq}) n'est pas vérifié partout pour $\epsilon'_\Sigma=10^{-4}$. Il y a donc des orbites de l'espace des phases qui ne sont pas représentées par des conditions initiales prises sur le plan $e_1=e_2$, $a_1=a_2$.\\

\an{
In the case of equal mass co-orbitals, we can have a simple expression for $\cV$ in the whole phase space at order $1$ in eccentricity. We saw in chapter 1, section \ref{sec:pperdi} that the average Hamiltonian of the planar circular case is symmetrical with respect to $Z=0$. We hence have $\dot \zeta=0$ when $Z=0$. From now we make the hypothesis that this is also the case for the eccentric co-orbitals and we define $\cV_Z$ such that $Z_\cV=0$.\\
Near a solution of the circular coplanar case $\zeta(t)$, the equation of variation in the direction ($x_j,\xt_j$) is given by the matrix (\ref{eq:eqvar2}). For a given trajectory $\zeta(t)$, since $\nu \gg g$, we can obtain an approximation of the secular dynamics in the direction ($x_j,\xt_j$) by averaging the expression of the matrix (\ref{eq:eqvar2}) over a period $2\pi/\nu$ with respect to the time $t$.\\
For equal mass co-orbitals, the symmetries of this matrix give relations of the form (\ref{eq:x1x2meq}) with $a$ real and $c$ complex. Therefore, we can deduce the equation (\ref{eq:osci2}): when $m_1=m_2$, the quantity $e_1^2-e_2^2$ behaves like an harmonic oscillator, librating around $0$ with the frequency $2|c|$. We hence define $\cV_\varPi$ such that $\varPi=0$. The manifold $\cV =\cV_Z \cap \cV_\varPi$ represents the phase space as long as the quadratic approximation of the Hamiltonian (\ref{eq:Hpt}) is relevant.\\}

\an{
We want to check numerically if this definition of $\cV$ holds for any value of the eccentricities, and which volume of the phase space is represented by $\cV$. As explained in section \ref{sec:VaRe}, we check which volume of the phase space verifies the condition (\ref{eq:condVaRe}). Since $m_1=m_2$, this criterion is equivalent to equation (\ref{eq:critVaRemeq}).\\
Since $\Sigma \subset \Sigma_{Z_0}$, we take our initial conditions on the plane $Z=0$ and we explore the phase space for different values of $\varPi$, $\zeta$ et $\Delta\varpi$. We check the grid of initial conditions for $\zeta \in [ 0^\circ : 360^\circ ] $ and $\Dv \in [-180^\circ:180^\circ]$ and several values of $\varPi$ for a fixed value of $J_1$. The corresponding values of the eccentricities are given by equation (\ref{eq:e2J1cst}).\\}

\an{
We checked the criterion (\ref{eq:critVaRemeq}) for several value of $\varPi$ for both $J_1(e_1=e_2=0.4)$ and $J_1(e_1=e_2=0.7)$ when $m_1=m_2$, figures \ref{fig:varefmeqe04} and \ref{fig:varefmeqe07}. The integrations are conducted over $10/\eps$ orbital periods, hence only a few times $2\pi/g$ at best. For all initial conditions in both graphs, the criterion (\ref{eq:critVaRemeq}) is verified for $\epsilon'_\Sigma \approx 10^{-8}$. Although this verification is not exhaustive, it suggests that the chosen reference manifold represents a significant part of the phase space of the averaged reduced problem. However it is possible that, especially at high eccentricity, stable domains appear for which the orbits never reach $e_1=e_2$ even in the case $m_1=m_2$, but none was discovered during this study.\\}

\an{
In order to compare with the case $m_1 \neq m_2$, the figure ~\ref{fig:VaRetestdiff} shows that there are areas of the phase space where the criterion (\ref{eq:critVaRemeq}) is not verified for even for $\epsilon'_\Sigma=10^{-4}$. There are hence orbits in the phase space that are not represented by the trajectories taking their initial conditions on the plane $e_1=e_2$, $a_1=a_2$.   
}

\subsection{Identification des familles $\overline{\cF}$}

\label{sec:SAIF}

A la section \ref{sec:DFFbLc} nous avons donné une approximation quadratique de la position des familles $\overline{\cF}_{k}$ au voisinage des points fixes du problème circulaire. Il est possible de calculer leur position à n'importe quel ordre dans le voisinage des équilibres de Lagrange elliptique, c'est ce que nous montrerons en section \ref{sec:DynFl}. La séparation des échelles de temps exposée en section \ref{sec:set} va nous permettre d'identifier les familles $\overline{\cF}$ d'un point de vue global à l'aide de méthodes semi-analytiques et numériques.\\

  Commençons par faire l'hypothèse d'invariance adiabatique pour les variables $\Delta \varpi$ et $\varPi$, comme exposé en section \ref{sec:set}. Ce faisant, nous pouvons étudier le système à 1 degré de liberté ($\zeta$,$Z$) sur des temps courts devant $2\pi/g$. En partant du hamiltonien dans les variables du problème réduit (\ref{eq:HR}), nous pouvons calculer une estimation en tout point de l'espace des phases du hamiltonien réduit moyenné $\gH_{\cR\cM}$ (\ref{eq:HRM}) en effectuant une moyenne numérique sur l'angle rapide $Q$. On rappelle que les familles $\ol \cF^\nu$ sont composées d'orbites dont les termes de fréquence $\nu$ sont d'amplitude nulle. Sur des temps courts devant $2\pi/g$, nous pouvons considérer qu'il s'agit d'une famille de points fixes. Les $\ol \cF^\nu$ sont donc l'ensemble des points de l'espace des phases pour lesquels nous avons:
\begin{equation}
\frac{\partial }{\partial Z}\gH_{\cR\cM} = \frac{\partial }{\partial \zeta} \gH_{\cR\cM}= 0\, . 
\label{eq:condFb0}
\end{equation}
On se place sur la variété $\dot{\zeta}=0$ ($Z=0$ quand $m_1=m_2$). Sur cette variété, les $\ol \cF^\nu$ se trouvent donc pour $\dot{Z}= \frac{\partial }{\partial \zeta} \gH_{\cR\cM}= 0$. Cette dérivée numérique, effectuée en tout point $\zeta_k$ d'une discrétisation de l'espace des phases dans la direction $\zeta$ est calculée de la manière suivante: 
\begin{equation}
\frac{\partial }{\partial \zeta} \gH_{\cR\cM}|_{\zeta=\zeta_k} = \frac{\gH_{\cR\cM}(Z,\zeta_{k+1},\Delta\varpi,\varPi) -\gH_{\cR\cM}(Z,\zeta_{k-1},\Delta\varpi,\varPi) }{|\zeta_{k+1}-\zeta_{k-1}|}\, . 
\label{eq:expZp}
\end{equation}
On considèrera que la condition (\ref{eq:condFb0}) est remplie au point ($\zeta_k,\Delta \varpi$) si on a:
\begin{equation}
\frac{\partial }{\partial \zeta} \gH_{\cR\cM}|_{\zeta=\zeta_k} \times \frac{\partial }{\partial \zeta} \gH_{\cR\cM}|_{\zeta=\zeta_{k+1}} < 0\, .
\label{eq:condFb0n}
\end{equation}
Dans les variables utilisées, l'expression de (\ref{eq:expHRM}) $\gH_{\cR\cM}$ montre de manière évidente que la dérivée $\frac{\partial }{\partial \zeta} \gH_{\cR\cM}$ ne dépend pas de $\eps$ dans le cas de notre approximation tant que nous nous trouvons sur des variétés à $Z$ constant. Cette méthode permet de trouver des points de l'espace des phases vérifiant l'équation (\ref{eq:condFb0}) pour le problème moyen, mais certains de ces points peuvent être situés dans des zones où les orbites ne sont pas quasi-périodiques dans le problème complet (voir les figures de la section \ref{sec:Coo2meg}).

Cette méthode semi-analytique n'est pas utilisable pour les familles $\ol \cF^g$ en l'état car elle nécessiterait d'effectuer la seconde moyenne sur l'angle semi-lent $\zeta$. Cependant, comme expliqué en sections \ref{sec:rdp} et \ref{sec:set}, l'intégration du problème à trois corps représente correctement la dynamique du problème moyen réduit tant que la masse des coorbitaux est suffisamment faible. Nous avons également vu dans ces sections que les variations de $\Dv$ dues à la fréquence $\nu$ sont de taille $\sqrt{\eps}$. Pour les orbites membres des familles $\ol \cF^g_{k}$, les variations de $\Delta \varpi$ sont de taille $\sqrt{\eps}$ sur toute échelle de temps. Nous allons donc utiliser la quantité $(\max{(\Delta \varpi)}-\min{(\Delta \varpi)} )$ comme un traceur des orbites membres du voisinage des familles $\ol \cF^g_{k}$. Nous ferons l'hypothèse suivante: on considèrera donc membre du voisinage des familles $\ol \cF^g_{k}$ toute orbite située dans une zone régulière de l'espace des phases (loin des séparatrices et des zones instables) vérifiant la condition suivante:
\begin{equation}
(\max{(\Delta \varpi)}-\min{(\Delta \varpi)} ) < \epsilon_g\, , 
\label{eq:condFb1}
\end{equation}
avec $\epsilon_g \propto \sqrt{\eps}$. On pourra comparer les membres de $\ol \cF^g$ ainsi déterminés avec ceux calculés analytiquement au voisinage des familles de points fixes du problème moyen dans la section \ref{sec:DynFl}. \\

Dans le cadre de ces intégrations du problème à trois corps, une constatation sur l'ensemble des simulations qui seront exposées par la suite nous amène à la détermination d'un traceur équivalent pour les orbites proches des familles $\ol \cF^\nu$. En effet, la quantité $(\max{(Z)}-\min{(Z)} )$ semble peu affectée par la fréquence $g$. En complément du critère (\ref{eq:condFb0}), nous ferons l'hypothèse suivante: on considèrera donc membre du voisinage des familles $\ol \cF^\nu$ toute orbite située dans une zone régulière de l'espace des phases (loin des séparatrices et des zones instables) vérifiant la condition suivante:
\begin{equation}
(\max{(Z)}-\min{(Z)} ) < \epsilon_\nu\, , 
\label{eq:condFb02}
\end{equation}
avec $\epsilon_\nu \propto \sqrt{\eps}$.\\

\an{
In this section we give a method to identify the position of the families $\ol \cF$ in the phase space. In section \ref{sec:DFFbLc} we gave a quadratic approximation of the position of these families near the Lagrangian circular equilibrium. It is possible to have an approximation at any order in the neighbourhood of the eccentric Lagrangian equilibriums, as we will show in section \ref{sec:DynFl}.\\
The separation of the time scales that we explained in section \ref{sec:SAIF} allows us to identify the position of the $\ol \cF$ anywhere in the phase space by using semi-analytical and numerical techniques.\\}

\an{
First, we look for the position of $\ol \cF^\nu$. Let us use the hypothesis of adiabatic invariant for the variables $\varPi$ and $\Dv$. By doing so, we can study the degree of freedom ($Z$,$\zeta$) on a short time scale with respect to $1/g$. $\ol \cF^\nu$ is hence made of orbits that behave as fixed points of the reduced averaged problem on a time scale short with respect to $1/g$. Starting from the reduced Hamiltonian (\ref{eq:HR}), we can estimate the value of the averaged Hamiltonian in any point of the phase space by doing a numerical averaging over the fast angle $Q$. The orbits belonging to $\ol \cF^\nu$ are hence all the orbits verifying the equation (\ref{eq:condFb0}).\\
Let us consider the manifold $\dot{\zeta}=0$ ($Z=0$ when $m_1=m_2=0$). On this manifold, we can identify the orbits belonging to $\ol \cF^\nu$ by finding the orbits for which $\dot Z = 0$. For a given value of the constants $\varPi$ and $\Dv$, we take a grid of value for $\zeta$ and estimate the averaged Hamiltonian at each point. We can then have the approximate position of the points where $\dot Z = 0$ by finding the positions of the grid where equation (\ref{eq:condFb0n}) is satisfied (the expression of the numerical derivative is given by eq. \ref{eq:expZp}).\\}

\an{
First, this method clearly shows that the position of $\ol \cF^\nu$ does not depends on $\eps$ for a fixed value of $Z$. It allows to find all the points of the phase space that verify equation (\ref{eq:condFb0}) for a fixed value of $\varPi$ and $\Dv$, in the averaged problem. However, it is not guaranteed that the associated trajectory in the full 3-body problem is quasi-periodic. If not, the averaged problem does not have any physical meaning in that area of the phase space. We hence need to check (by numerical integration) that the orbit can be considered quasi-periodic before asserting that a given orbit belongs to $\ol \cF^\nu$.\\}

\an{This method cannot be used directly to find the manifold $\ol \cF^g$ because it requires to numerically average the Hamiltonian over the semi-fast angle $\zeta$. However, we saw in sections \ref{sec:rdp} and \ref{sec:set} that the evolution of the variables $\Dv$ and $\varPi$ during numerical integrations of the 3-body problem are $\gO(\sqrt \eps)$ close to their evolution in the secular problem. Since orbits belonging to $\ol \cF^g$ are fixed points in the secular problem (1 degree of freedom), we can make the following hypothesis:} 
\an{all orbits in a regular area of the phase space (far from the separatrix, the chaotic and the unstable areas) that verify the equation (\ref{eq:condFb1}) with $\epsilon_g \propto \sqrt{\eps}$ are in the neighbourhood of the manifold $\ol \cF^g$.}
\an{We can check that such orbits (represented by black pixel in the figures of the sections \ref{sec:Coo2meg} and \ref{sec:mdif}) are indeed in the neighbourhood of the analytical approximation of the positions of $\ol \cF^1_4$ and $\ol \cF^2_4$,}
\an{ see section \ref{sec:DynFl}.\\}

\an{
Numerical integrations allow us to determine a similar criterion for a numerical determination of the position of $\ol \cF^\nu$. In the various integrations that we computed through the following sections, we noted that the amplitude of variation of $Z$ (hence $a_1-a_2$) seems not to be impacted much by the frequency $g$. We hence make the hypothesis that if an orbit in a regular area of the phase space verifies the condition (\ref{eq:condFb02}) with $\epsilon_\nu \propto \sqrt{\eps}$, this orbit is in the neighbourhood of the manifold $\ol \cF^\nu$. One can check in figures \ref{fig:glob_e01} to \ref{fig:glob_e7}, where the quasi-periodic orbits that verify equation (\ref{eq:condFb02}) are identified by brown pixels, and the result of the semi-analytical method (equation \ref{eq:condFb0}) are identified by purple or red lines, that both methods yield very similar results in the regular areas of the phase space.
}

\subsubsection{Identification des familles $\ol \cF^\nu$ dans le cas de deux masses égales}

\begin{figure}[h!]
\begin{center} 
\includegraphics[width=0.4\linewidth]{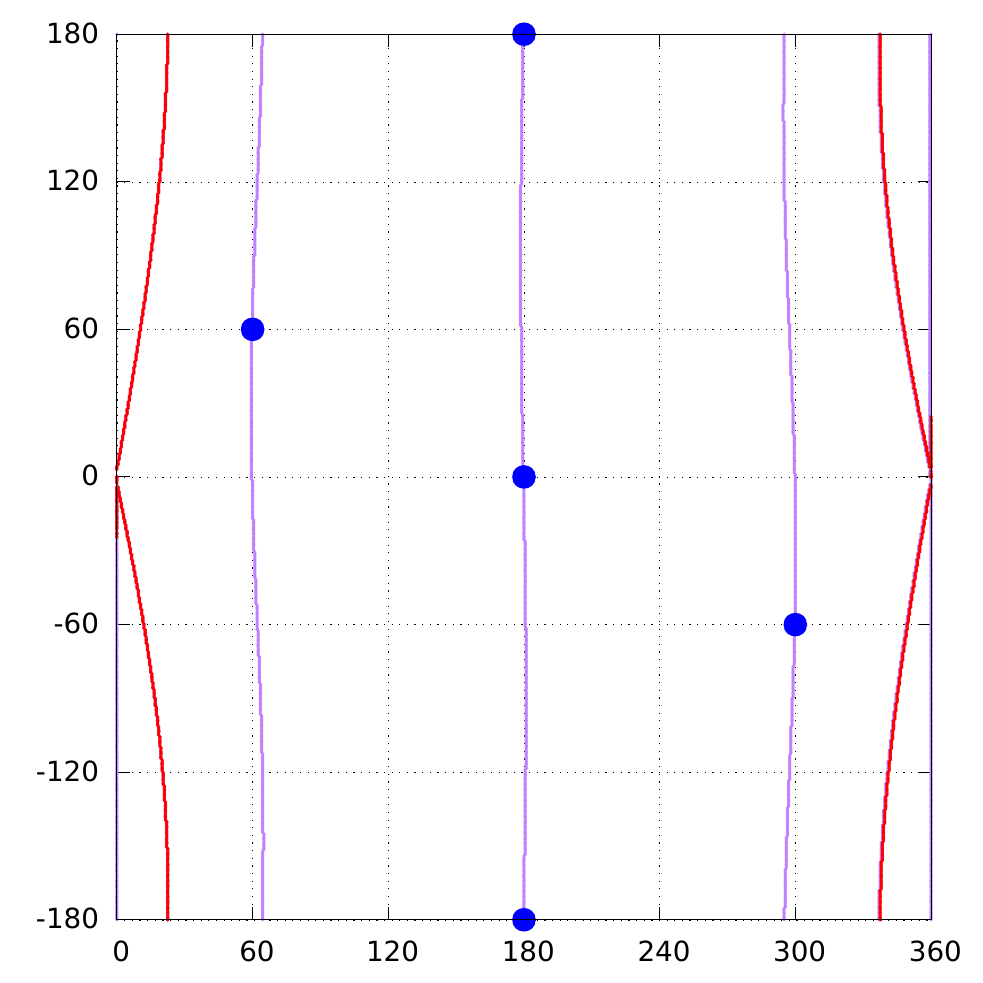}
\includegraphics[width=0.4\linewidth]{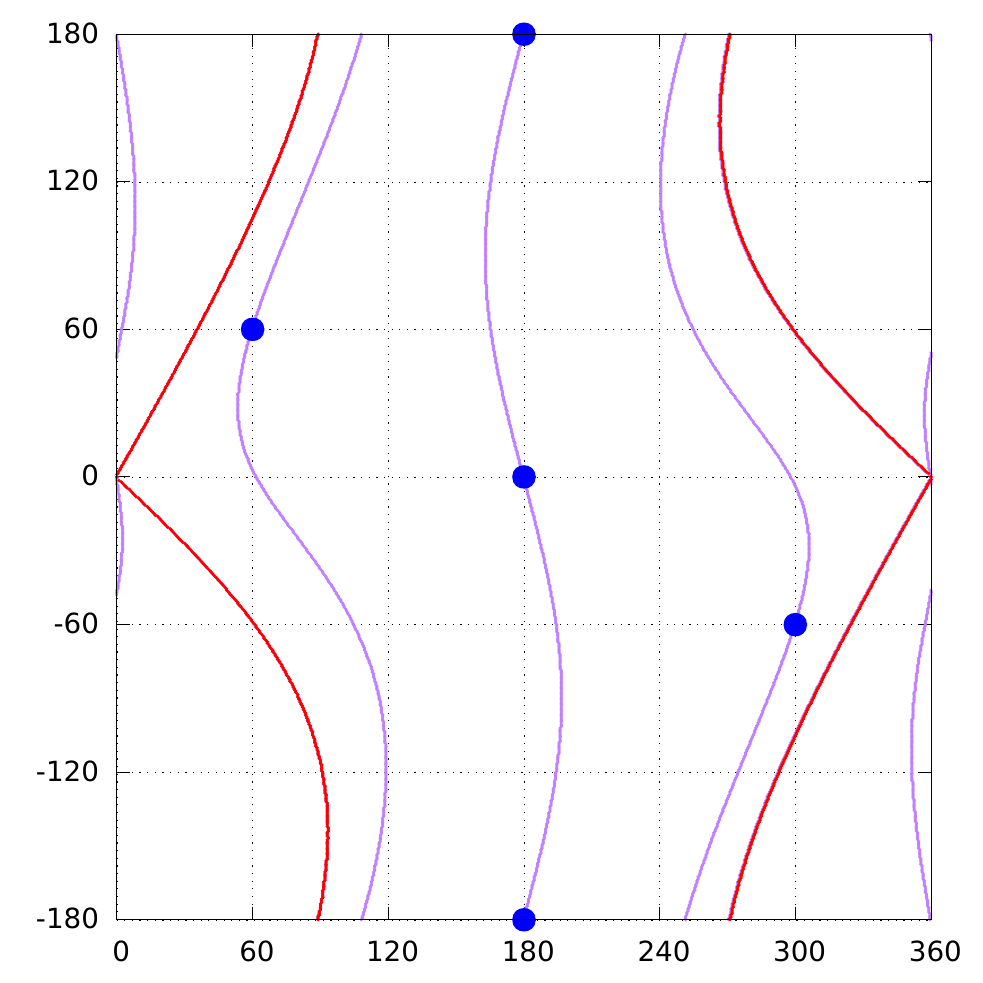}\\
\includegraphics[width=0.4\linewidth]{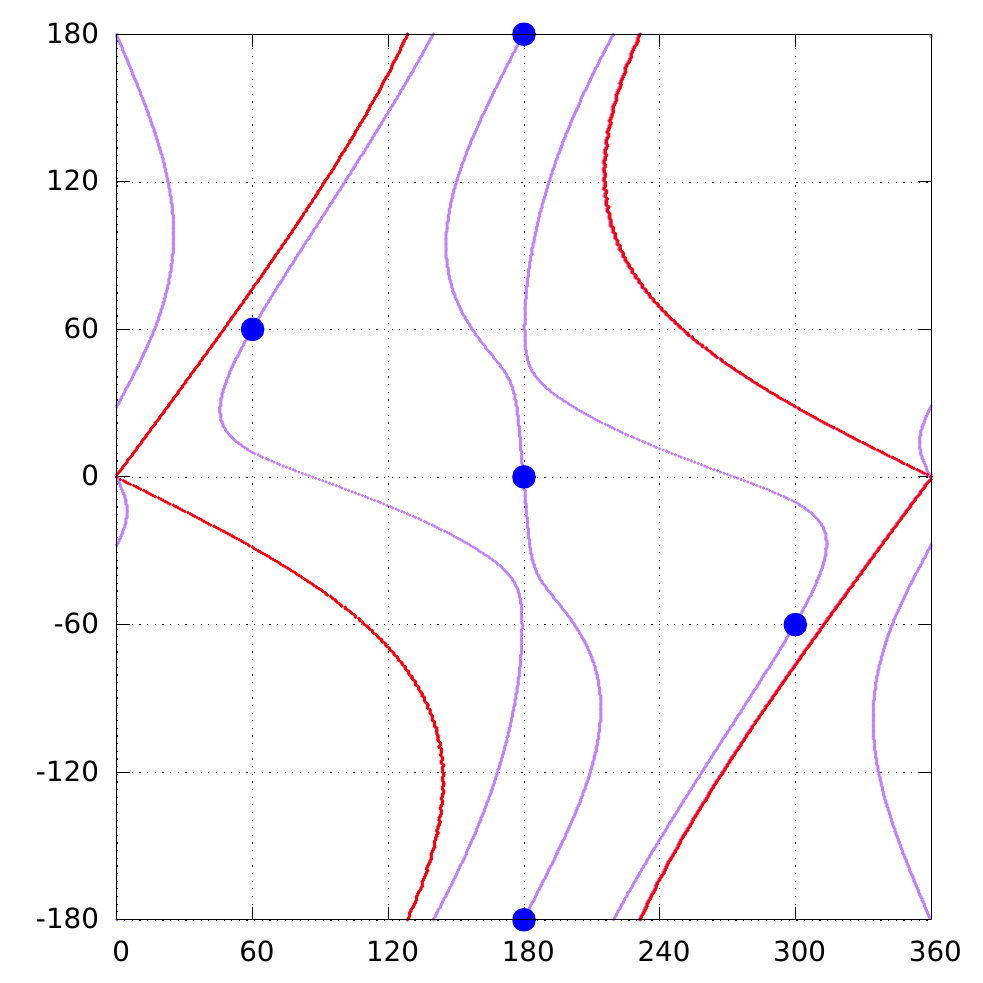}
\includegraphics[width=0.4\linewidth]{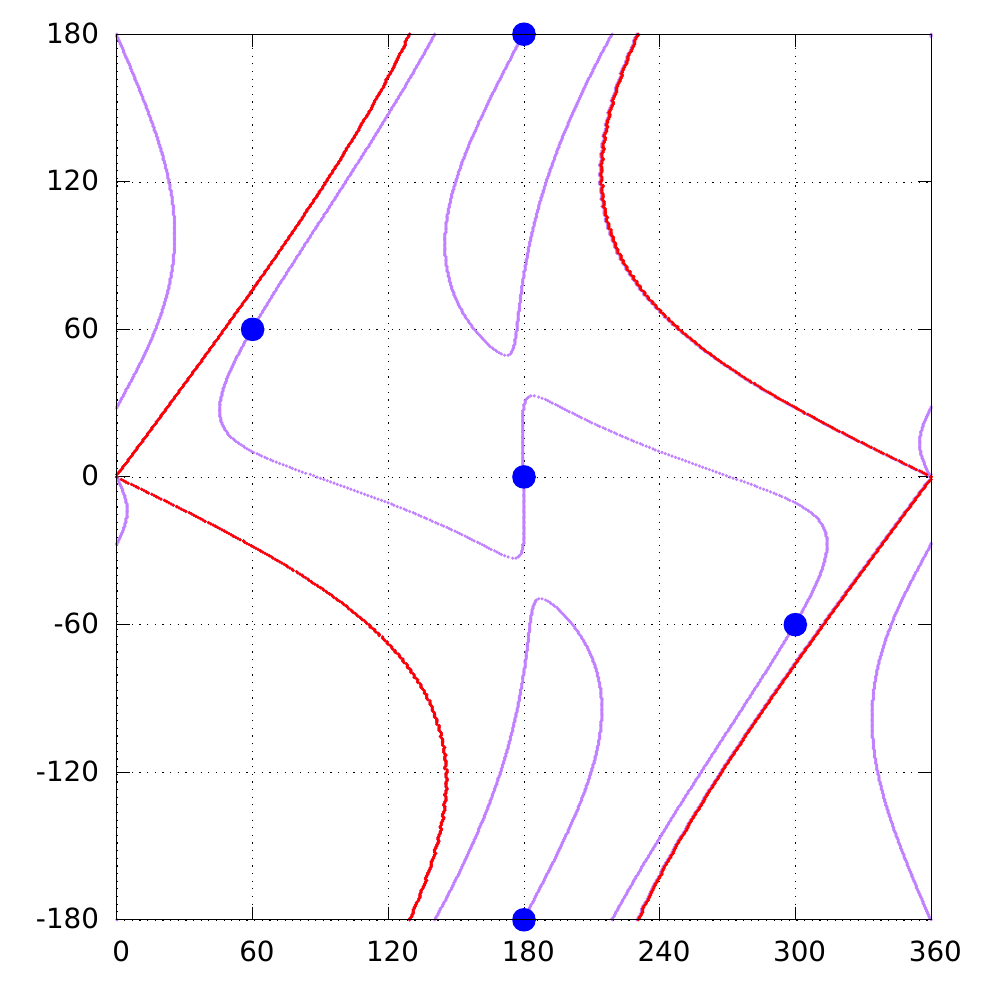}\\
\includegraphics[width=0.4\linewidth]{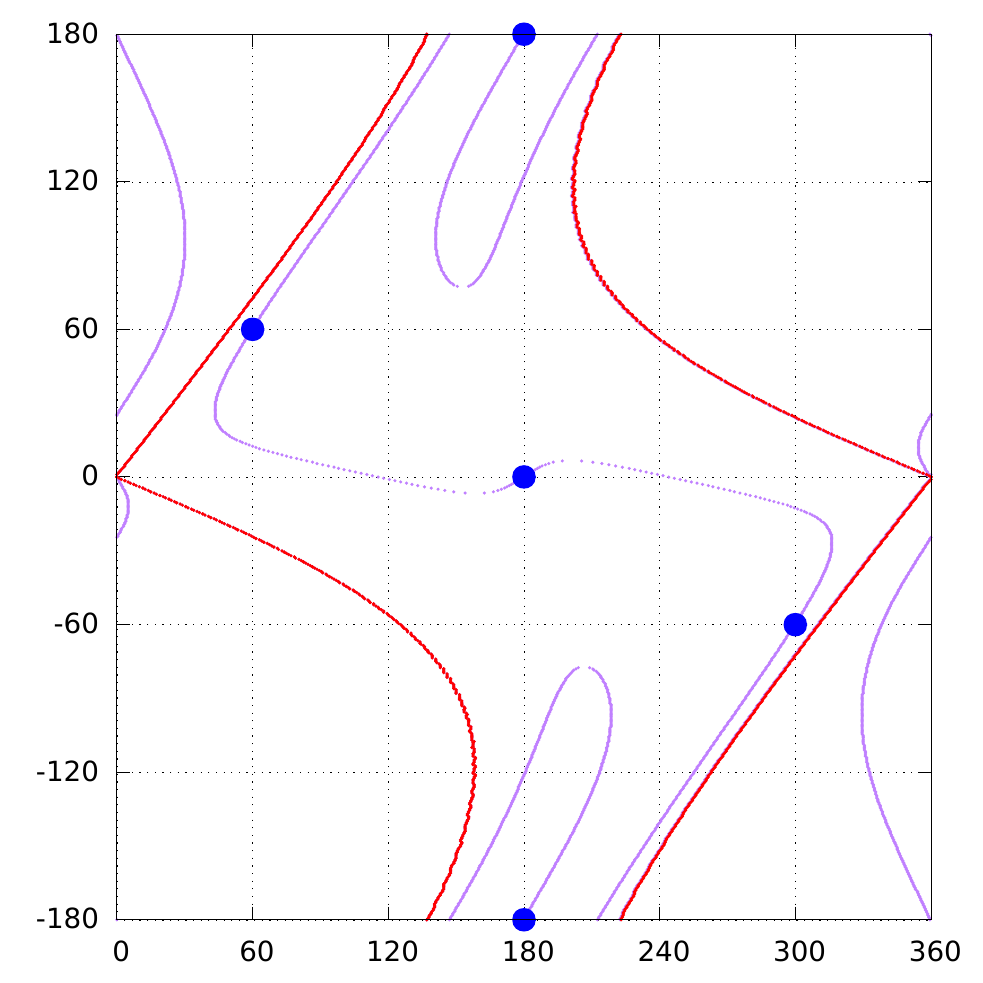}
\includegraphics[width=0.4\linewidth]{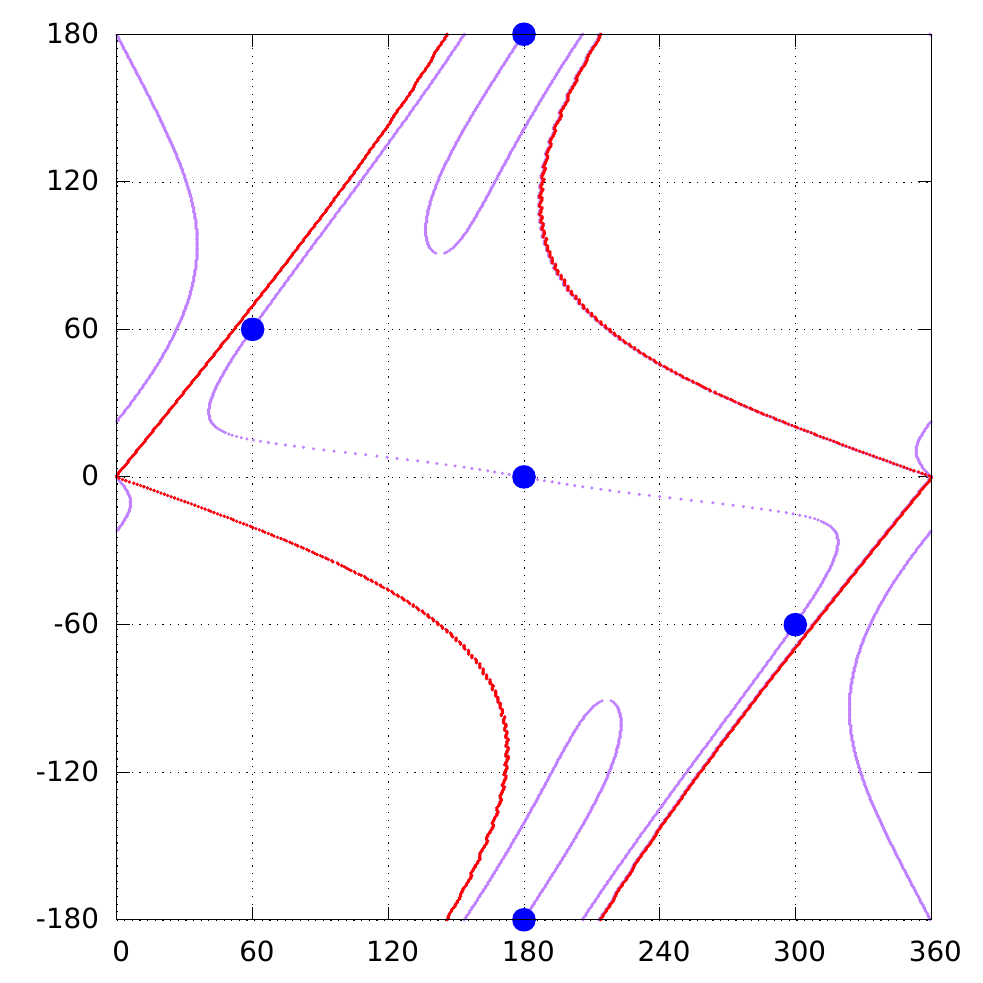}\\
 \setlength{\unitlength}{0.1\linewidth}
\begin{picture}(.001,0.001)
\put(-4.1,2.1){\rotatebox{90}{$\Dv$}}
\put(-4.1,6.1){\rotatebox{90}{$\Dv$}}
\put(-4.1,10.2){\rotatebox{90}{$\Dv$}}
\put(-2,0){{$\zeta$}}
\put(2.1,0){{$\zeta$}}
\end{picture}
\caption{\label{fig:Fb0meq}  Trace de $\ol \cF^\nu$ sur la variété représentative $a_1=a_2$ et $e_1=e_2$. Les familles $\ol \cF^\nu$ sont tracées en violet et les courbes dues à la variété de collision sont représentées en rouge. (a) $e_j=0.1$; (b) $e_j=0.4$; (c) $e_j=0.6$; (d) $e_j=0.605$; (e) $e_j=0.65$ et (f) $e_j=0.7$. } 
\end{center}
\end{figure}
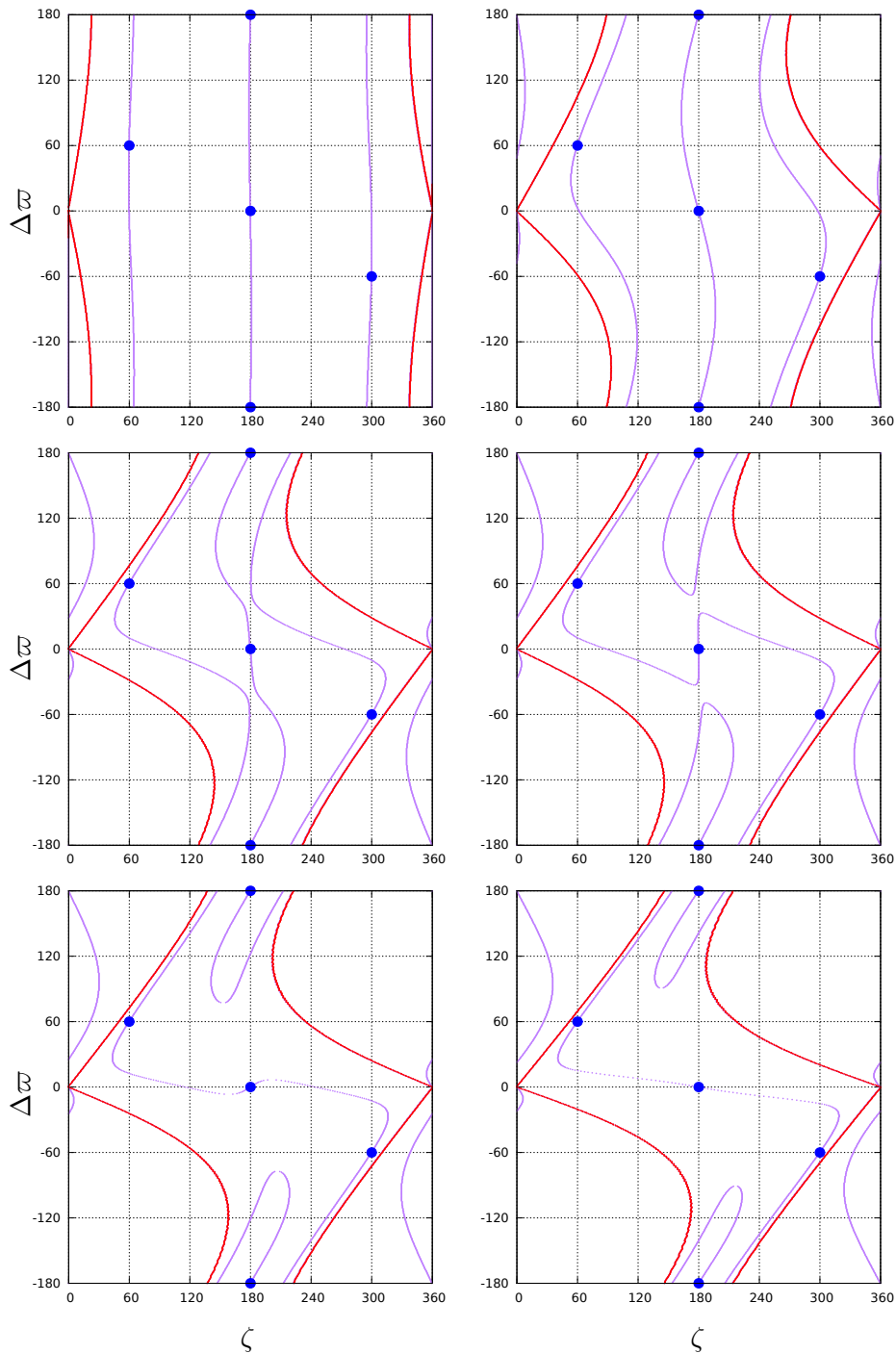

Nous savons déjà que la famille $\ol \cF^\nu_{k}$ se trouve sur la variété $Z=0$ dans le cas de masses égales. Les figures~\ref{fig:Fb0meq} montrent l'ensemble des points de la variété ($Z=0$, $\varPi=0$) pour lesquels la condition (\ref{eq:condFb0n}) est vérifiée. Ces cartes sont tracées grâce à une estimation de $\gH_{\cR\cM}$ sur une grille de conditions initiales ($\zeta$, $Z=0$, $\Delta\varpi$, $\varPi=0$) avec $\zeta$ et $\Delta\varpi$ sur des intervalles de taille $360^\circ$ et un pas de $0.5^\circ$.  

Nous savons que la variété de collision passe par le point ($0,0$), mais nous ne connaissons pas son expression en dehors de ce point. Cependant, des courbes issues de ($0,0$) et vérifiant la condition (\ref{eq:condFb0n}) semblent correspondre à la position de la variété de collision dans les intégrations numériques du problème à trois corps complet (voir les figures de la section \ref{sec:Coo2meg}). Une explication possible est que $\frac{\partial }{\partial \zeta} \gH_{\cR\cM}$ tend vers $+\infty$ en se rapprochant d'un côté de la collision et $-\infty$ de l'autre, ce qui expliquerait pourquoi la collision vérifie la condition (\ref{eq:condFb0n}). A partir de maintenant on fera cette hypothèse et on identifiera ainsi la position de la variété de collision.

Pour des excentricités faibles ($\leq 0.1$) nous sommes au voisinage de la variété circulaire et la direction $\Delta \varpi$ influe peu sur la position des familles $\ol\cF^\nu$. Nous avons $\ol\cF^\nu_{4}$ en $\zeta=60^\circ$, $\ol\cF^\nu_{3}$ en $\zeta=180^\circ$ et $\ol\cF^\nu_{5}$ en $\zeta=300^\circ$, ce qui correspond à l'approximation quadratique effectuée en section \ref{sec:DFFbLc}. Pour $e_1=e_2=0.1$, on observe une courbe supplémentaire sur laquelle la condition (\ref{eq:condFb0n}) est vérifiée en $\zeta \approx 0^\circ$ qui n'est pas la collision (cette courbe est confondue avec l'axe $\zeta=0$ sur la figure~\ref{fig:Fb0meq} (a)). Comme on le verra par la suite, cette variété coupe le domaine des quasi-satellites. Nous l'appellerons donc $\ol\cF^\nu_{\cQ \cS}$. 

Une augmentation de l'excentricité entraîne une dépendance croissante de la position des $\ol\cF^\nu$ en fonction de $\Delta \varpi$. Jusqu'à $e_j=0.6$, l'excentricité n'a pour effet que de tordre les $\ol\cF^\nu$ existantes. On vérifiera analytiquement la position de ces famille en section \ref{sec:DynFl}. Cependant, entre $e_j=0.6$ et $e_j=0.605$, survient un changement de topologie important: les $\ol\cF^\nu$ se reconnectent de manière à former une seule famille continue passant par l'ensemble des $L_k$ et des $AL_k$ pour $k \in \{3,4,5\}$. Comme nous le verrons dans les sections suivantes, cette reconnexion semble entraîner une modification globale de l'espace des phases de la résonance coorbitale excentrique.

\begin{rem} \label{rq:6} Dans ce plan, la variété de collision représentée en rouge sur les figures~\ref{fig:Fb0meq} se rapproche des équilibres de Lagrange et anti-Lagrange excentriques $L_4$ et $L_5$ quand $e_j$ augmente. Cela diminue la taille possible du domaine de stabilité dans cette direction pour les régions troyennes et augmente la taille du domaine des quasi-satellites. Dans ce plan, la variété de collision semble se rapprocher plus rapidement de $L_4$ (resp. $L_5$) que de $AL_4$ (resp. $AL_5$). Ce qui est cohérent avec l'observation de \cite{GiuBeMiFe2010} selon laquelle la taille de la région stable au voisinage de $L_4$ diminue plus vite que celle au voisinage de $AL_4$ à mesure que l'excentricité des coorbitaux augmente.
\end{rem}

\begin{rem} \label{rq:7} \cite{RoPo2013} conjecturent que les familles $L_3$, $AL_4$ et $AL_5$ se confondent quand l'excentricité des deux corps tend vers $1$. Cette reconnexion est consistante avec cette conjecture.
\end{rem}

\begin{rem} \label{rq:8} On rappelle que la méthode semi-analytique utilisée ici est effectuée dans le cadre du problème moyen. La vérification de la condition (\ref{eq:condFb0}) est donc une condition nécessaire pour que l'orbite associée du problème à trois corps excentrique plan (donc à 4 degrés de liberté) soit une orbite quasi-périodique à 3 fréquences. Il reste cependant à vérifier que cette orbite est effectivement quasi-périodique (et non chaotique/instable).
\end{rem}

\an{
We apply here the semi analytical method to identify the position of $\ol \cF^\nu$ in the case $m_1=m_2$. This case is easier than $m_2 \neq m_1$ because we know that the manifold $\dot{\zeta}=0$ is located at $Z=0$, and the equilibriums $L_k$ and $AL_k$ are all located in the plane $\varPi =0$ ($e_1=e_2$). We can hence explore the manifold $\cV =\{Z,\zeta , \varPi ,\Dv / Z=\varPi =0 \}$ that we know to be a representative manifold (see section \ref{sec:VaRemeq}). To explore this manifold, we chose a grid of initial conditions for $\zeta$ and $\Dv$ with a step of $0.5^\circ$, and we compute numerically the averaged Hamiltonian in each point of the grid. In the figures \ref{fig:Fb0meq} We show all the points of $\cV$ that verify the condition (\ref{eq:condFb0n}). Each figure corresponds to a different value of the total angular momentum (thus a different value of $e_1=e_2$).\\} 

\an{
We know that the collision manifold contains the point $(0,0)$ of these figures, but we do not know its position besides that. However, there are curves satisfying the relation (\ref{eq:condFb0n}) that emanate from $(0,0)$ and seem to be located at a relevant position for the collision manifold (see the red curves in the unstable areas in the figures \ref{fig:glob_e4} to \ref{fig:glob_e7}). A possible explanation is that $\frac{\partial}{\partial \zeta} \gH_{\cR \cM}$ tends to $- \infty$ when we get close to the collision from one side and $+ \infty$ from the other side. That would explain why the collision manifold would satisfy the equation (\ref{eq:condFb0n}). From now on, we consider that this is the case, and we also use this method to identify the collision manifold.\\}

\an{
We now describe the figures \ref{fig:Fb0meq}. For low eccentricities ($\leq 0.1$), we are in the neighbourhood of the circular case and the direction $\Dv$ does not impact much the position of the $\ol \cF_k^\nu$: $\ol\cF^\nu_{4}$ is located in $\zeta=60^\circ$, $\ol\cF^\nu_{3}$ in $\zeta=180^\circ$ and $\ol\cF^\nu_{5}$ in $\zeta=300^\circ$, which agrees with the quadratic approximation (section \ref{sec:DFFbLc}). For $e_1=e_2=0.1$, a new curve appears for $\zeta \approx 0^\circ$ (the curve is mingled with the axis $\zeta=0$ in the figure (a)). This branch of $\ol \cF^\nu$ intersects the domain of the quasi-satellite configuration, we hence call it $\ol \cF^\nu_{\cQ \cS}$.\\
When we increase the eccentricity, there is an increasing dependence on the direction $\Dv$ for the position of the $\ol \cF^\nu_k$. Until $e_1=e_2\approx 0.6$, the sole effect of the increasing eccentricity is to twist the existing $\ol \cF^\nu_k$. We check analytically the position of these family in section \ref{sec:DynFl}. However, between $e_j=0.6$ and $e_j=0.605$, an important topological change occurs: in the averaged problem, the $\ol \cF^\nu_k$ reconnect in order to create a single family of periodic orbits that goes through all the $L_k$ and $AL_k$ for $k \in \{1,2,3\}$. As we will see in the coming sections, this reconnection seems to lead to a modification of the whole phase space of the eccentric co-orbital resonance.\\}

\an{ Remark \ref{rq:6}: In the considered plane, the collision manifold plotted in red in the figures~\ref{fig:Fb0meq} gets closer to the $L_k$ and $AL_k$ equilibriums as $e_j$ increases. It reduces the size of the maximum possible stable areas in that direction for the trojan configuration and increases the quasi-satellite area. In this plane, the collision manifold gets close to the $L_4$ equilibrium faster than to the $AL_4$ equilibrium. This is in agreement with the observation of \cite{GiuBeMiFe2010}, in which the stability domain around $L_4$ shrinks faster that the stability domain around $AL_4$ when the eccentricity increases.\\}

\an{Remark \ref{rq:7}: \cite{RoPo2013} conjecture that $L_3$, $AL_4$ and $AL_5$ tend to the same position in the phase space when the eccentricities of the two bodies tend to $1$. This reconnection is consistent with this conjecture.\\}

\an{Remark \ref{rq:8}: Note that the semi-analytical method used here is relevant only when the averaged reduced Hamiltonian is. To verify the equation (\ref{eq:condFb0}) is hence a necessary condition for the associated orbit of the full planar 3-body problem to be a quasi-periodic orbit with $3$ fundamental frequencies, but we still need to check that this orbit is indeed quasi-periodic (not unstable/chaotic).}

\section{Étude des coorbitaux excentriques à deux masses égales}
\label{sec:Coo2meg}

A l'aide des outils et études précédemment développés, nous allons maintenant étudier la dynamique et la stabilité de deux coorbitaux excentriques. Comme nous l'avons vu en section \ref{sec:VaRemeq}, l'expression de la variété de référence est triviale dans le cas de co-orbitaux de masses égales, c'est pourquoi nous nous limitons à ce cas pour l'instant. Le moment cinétique total $J_1$ étant un paramètre du problème moyen réduit, lié à la valeur de l'excentricité des corps, nous allons étudier son influence sur l'espace des phases.

Nous verrons que la valeur de $m_1/m_0=m_2/m_0=\eps$ ne semble pas influer significativement la dynamique co-orbitale pour les masses choisies dans cette étude (entre $\eps =10^{-6}$ et $10^{-4}$). On verra cependant en section \ref{sec:meqstab} que ce paramètre influe grandement sur la taille des domaines de stabilité.\\ 

Nous prendrons pour l'ensemble des intégrations $a_1=a_2=1\ ua$ ($J_2$ jouant le rôle d'un facteur d'échelle), $m_0$ égal à une masse solaire. L'éventuelle fréquence de précession des périhélies étant très faible devant la fréquence orbitale, nous pouvons prendre en condition initiale $\lambda_1=\varpi_1=0$ sans perte de généralité. Le reste des conditions initiales est déterminé par les coordonnées du point considéré sur la variété de référence ainsi que par la valeur de $J_1$ considérée. En pratique on donnera la valeur de $e_1=e_2$ pour des raisons de clarté.

Pour chaque condition initiale, le système est intégré pendant $10/\eps$ ans en utilisant l'intégrateur symplectique SABA4 \citep{LaRo2001} avec un pas de temps de $0.01001$ ans. Ce pas de temps est suffisament faible pour des excentricités allant jusqu'à $\approx 0.6$. Au delà, il est préférable de prendre $0.01001$ ans pour éviter les éjections dues aux erreurs numériques. Les conditions initiales d'orbites fortement chaotiques, ou qui quittent la résonance coorbitale avant la fin de l'intégration sont identifiées dans tous les graphes par des pixels blancs. De plus, afin d'identifier les orbites qui sont instables sur des temps longs devant $1/\eps$, on calcule la diffusion de la valeur moyenne du demi-grand axe de la planète $m_1$ entre la première et la seconde moitié de l'intégration. La couleur grise sera attribuée aux conditions initiales pour lesquelles cette diffusion dépasse la valeur $\epsilon_a$.

\begin{rem} \label{rq:9} L'espace des phases est symétrique par rapport au point ($\zeta=0,\Dv=0$). On n'intègrera et on ne décrira donc que la partie de l'espace des phases pour laquelle $\zeta \in [0:180]$. La seconde moitié sera affichée pour faciliter l'appréhension de la totalité de l'espace des phases et pour pouvoir représenter les points de la variété de référence au voisinage desquels passent quelques orbites qu'on décidera de représenter.
\end{rem}

\an{
Using the method developed in the previous sections, we now study the dynamics and stability of two eccentric co-orbitals. As we saw in section \ref{sec:VaRemeq}, the expression of the reference manifold $\cV$ is trivial in the case $m_1=m_2$, that is why we will study this case at first. We will study the impact of the total angular momentum $J_1$ (thus the value of $e_1=e_2$) on the dynamics and the stability of the co-orbitals.\\
The value of the parameter $\eps=m_1/m_0=m_2/m_0$ seems to not impact significantly the dynamics of the co-orbitals. However, as we will see in section \ref{sec:meqstab}, it impacts the size of the stability domain.\\}

\an{
For all integrations in this section, we take as initial conditions $a_1=a_2$ ($=1$~au, but the actual value of the semi-major axis is only a scale factor), $m_0$ equal to one solar mass. The precession rate of the periastron is very low with respect to the mean mean-motion, so the initial values of $\lambda_1$ and $\varpi_1$ do not change anything. We can thus chose $\lambda_1=\varpi_1=0^\circ$. The other initial conditions are given by the coordinate of the point on the manifold of reference and by the chosen value of $J_1$ (we give the equivalent value of $e_1=e_2$ for clarity).\\
For each set of initial condition, the system is integrated over $10/\eps$ orbital periods using the symplectic integrator SABA4 \citep{LaRo2001} with a time step of $0.01001$ orbital periods (eccentricities higher that $0.6$ might require to take $0.001$ in order to avoid to eject stable orbits for numerical reasons). The initial conditions that lead to highly chaotic orbits, or that quit the resonance before the end of the integration are identified by a white pixel in the figure. Moreover, in order to identify the orbits that are not stable on a time scale that is long with respect to $10/\eps$, we compute the diffusion of the average value of the semi-major axis of the planet $m_1$ between the first and the second half of the integration. The grey pixels identify the initial conditions for which this diffusion is higher than a given small parameter $\epsilon_a$.\\}

\an{Remark \ref{rq:9}: The phase space is symmetric with respect to the point ($\zeta=0,\Dv=0$), we hence compute and describe only half of the phase space ($\zeta \in [0,180]$). The other half is also displayed for a better understanding of the whole phase space.}

\subsection{Dynamique des coorbitaux excentriques à masses égales}

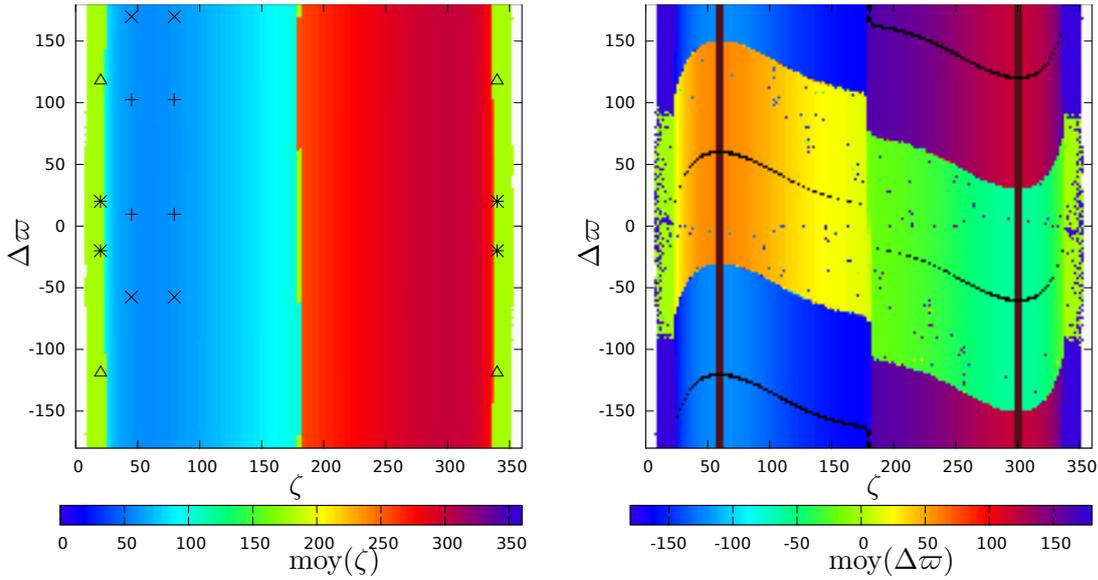
\begin{figure}[h!]
\begin{center}
\includegraphics[width=0.5\linewidth]{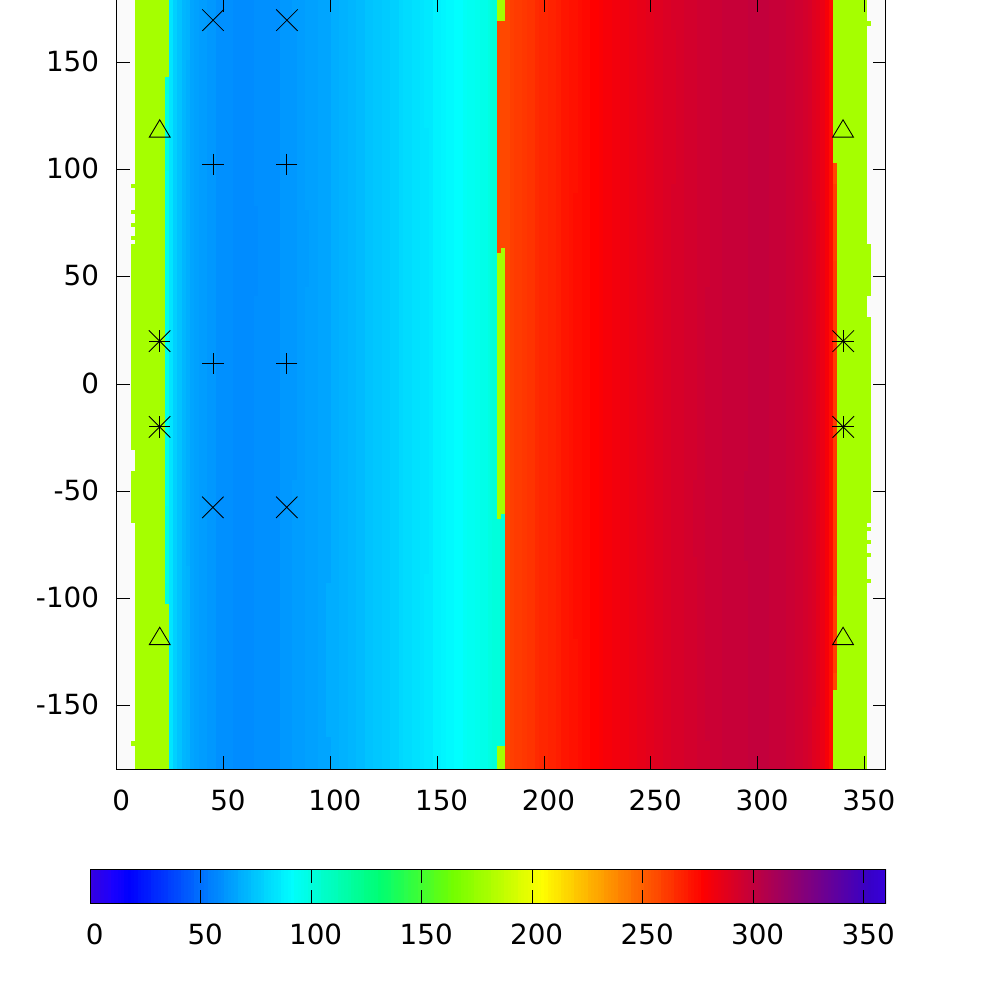}\includegraphics[width=0.5\linewidth]{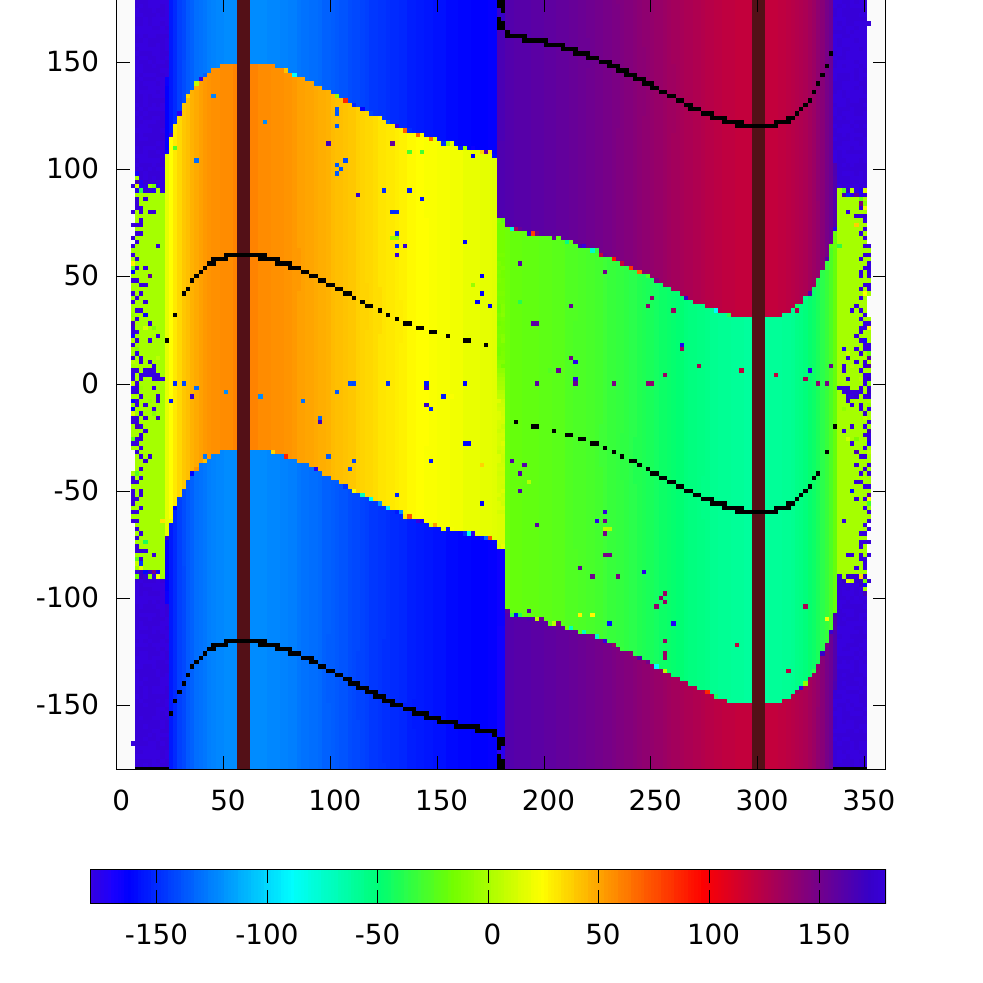}\\
  \setlength{\unitlength}{1cm}
\begin{picture}(.001,0.001)
\put(-7.5,4.5){\rotatebox{90}{$\Dv$}}
\put(0,4.5){\rotatebox{90}{$\Dv$}}
\put(-3.8,1.5){{$\zeta$}}
\put(3.8,1.5){{$\zeta$}}
\put(-3.8,0.5){{moy($\zeta$)}}
\put(3.3,0.5){{moy($\Dv$)}}
\end{picture}
\caption{\label{fig:glob_e01} Plan de référence pour $\eps=m_1/m_0=m_2/m_0=10^{-5}$ et $e_1=e_2=0.01$. Le plan est défini par $a_1=a_2=1\,ua$, et $e_1=e_2$. Chaque point de ce plan est une condition initiale d'orbite. Sur la figure de gauche, le code couleur correspond à la valeur moyenne de $\zeta$ sur cette orbite. Les divers marqueurs représentent les points près desquels passent les orbites de la figure~\ref{fig:orbe01}. Sur la figure de droite le code couleur représente la valeur moyenne de $\Dv$ sur chaque orbite. Les excentricités sont faibles et susceptibles de s'annuler, la détermination de $\Dv$ est donc difficile. Les orbites proches de $\ol \cF^\nu$, donc celles vérifiant la relation (\ref{eq:condFb02}) avec $\epsilon_\nu=10^{-3.5}$ sont représentées en marron. Les orbites proches de $\ol \cF^g$, donc celles vérifiant la relation (\ref{eq:condFb1}) avec $\epsilon_g= 3^\circ$ sont représentées en noir. $\epsilon_a=10^{-5.5}$.\\
}
\end{center}
\end{figure}

\an{
In the figures \ref{fig:glob_e01} to \ref{fig:glob_e7} we show the integration of the grid of initial condition of $\cV$ for $e_j=0.01$, $0.4$, $0.6$ and $0.7$. In each case, $m_1=m_2=10^{-5} m_0$. The left pictures represent the mean value of $\zeta$ over the integration and the right ones represent the mean value of $\Dv$. When some markers such as $\times$ or $+$ are displayed on the left pictures, they indicate the point of the manifold of reference in the neighbourhood of which the orbit passes quasi-periodically. On the right plot, the numerical criteria (\ref{eq:condFb02}) and (\ref{eq:condFb1}) are used to identify the position of the intersection between $\cV$ and $\ol \cF^\nu$ in brown and $\ol \cF^g$ in black. We also plot on these graphs the result of the semi-analytical method to identify the position of $\ol \cF^\nu$ (in red or purple in the figures \ref{fig:glob_e4} to \ref{fig:glob_e7}). Note that the numerical criterion and the semi analytical method yield very similar results, as the red curves are on top of the brown pixels in each stable area of the phase space. To have an idea on the long-term stability of a given orbit, we displayed in grey the initial conditions that lead to a diffusion of the average semi-major axis of the planet $1$ higher than $\epsilon_a= 10^{-5.5}$ between the first half and the second half of the $10/\eps=10^6$ orbital periods integration. We recall that orbits that quit the resonance before the end of the integration are identified by a white pixel.
}

\subsubsection{quasi-circular case, $e_1=e_2=0.01$}

Prenons des masses relativement faibles mais consistantes avec un cas planétaire $\eps=1\, 10^{-5}$. Partons du cas quasi-circulaire et augmentons les excentricités progressivement. La figure~\ref{fig:glob_e01} représente les valeurs moyennes des angles $\zeta$ et $\Dv$ quand les conditions initiales parcourent la variété $\cV$ pour $e_1=e_2=0.01$. Les orbites voisines des variétés $\ol \cF_{k}$ sont mises en évidence grâce aux critères développés en section \ref{sec:SAIF}: en noir pour $\ol \cF^g_{k}$ et en marron pour et $\ol \cF^\nu_{k}$. A l'intersection de ces variétés se trouvent les $L_k$ et $AL_k$ excentriques qui sont des points fixes du problème réduit moyen. Les familles $\ol \cF_{3}$, ainsi que les points $L_3$ et $AL_3$ ne sont pas mis en évidence par cette méthode car ils sont situés le long de la séparatrice.

 On peut vérifier que chacun de ces points fixes est situé à l'endroit prévu par le développement du hamiltonien à faible excentricité, section \ref{sec:H0pf}. Nous pouvons vérifier également que la direction des familles $\overline{\cF}_{k}$ à partir des $L_k$ est bien celle prévue par l'étude menée en section \ref{sec:DFFbLc}, c'est à dire une tangente en $\Delta\varpi$ constant pour les familles $\ol \cF^g_{k}$ au voisinage de $L_4(0.01)$ et $L_{5}(0.01)$, ainsi que $\zeta$ constant pour la direction des $\ol \cF^\nu_{k}$ au voisinage de ces deux mêmes points.

D'un point de vue global, la dynamique mise en évidence sur cette figure est en accord avec l'étude analytique: le degré de liberté ($Z,\zeta$) évolue presque indépendamment du degré de liberté ($\varPi,\Delta\varpi$) et est semblable au cas circulaire plan rappelé au Chapitre 1. On retrouve en particulier les régions appelées troyennes au voisinage des $L_4$ et $L_5$ circulaires (donc les bandes autour de $\zeta=60^\circ$ et $300^\circ$), la séparatrice entre régime troyen et fer à cheval pour $\zeta \approx 24^\circ$ et $ \approx 336^\circ$ (passant également au voisinage de cette variété pour $\zeta=180^\circ$), le régime fer à cheval, et la zone instable centrée sur la collision en $\zeta=0^\circ$. On peut voir cependant par la dépendance de la position des familles $\ol \cF^g_{k}$ en $\zeta$ que la dynamique du degré de liberté ($\varPi,\Delta\varpi$) est influencée par celle de ($Z,\zeta$)\footnote{cf. la dépendance en ($Z,\zeta$) du système d'équation aux variations (\ref{eq:eqvar1})}. Notons par ailleurs que dans la région des fers-à-cheval, deux groupements d'orbites vérifient la condition (\ref{eq:condFb1}), un centré en $\Delta \varpi=0^\circ$ (davantage visible sur la figure~\ref{fig:glob_e4}) et l'autre en $\Delta \varpi=180^\circ$. Par analogie avec les familles émergeant des équilibres de Lagrange, on suppose que ces orbites sont dans le voisinage d'une variété d'orbites périodiques du problème moyen réduit de fréquence $\nu$. Ces variétés seront nommées $\overline{\cF}^g_{\gH \cS}$ (pour l'anglais horseshoe) pour celle en $\Delta \varpi=0^\circ$ et $\overline{\cF}^g_{\gH \cS '}$ pour celle en $\Delta \varpi=180^\circ$.\\

Pour une condition initiale dans ce plan, la trajectoire qui en est issue passe une infinité de fois dans le voisinage de $4$ points du plan pour une trajectoire générique, $2$ points pour les membres des familles $\overline{\cF}_{k}$ et $1$ point pour les points fixes du problème moyen réduit comme nous l'avons vu en section \ref{sec:VaRemeq}. Les projections sur les plans ($Z,\zeta$) et ($\varPi,\Dv$) de quelques orbites génériques ayant pour conditions initiales des points de $\cV$ sont représentées en annexe sur la figure~\ref{fig:orbe01}. Les points auprès desquels repassent chaque trajectoire sont représentés par le même symbole sur la figure \ref{fig:glob_e01}. Dans la région troyenne, les $4$ points représentant la même trajectoire se répartissent dans les 4 quadrants délimités par les $\ol \cF_{4}$, soit autour de $L_4$, soit autour de $AL_4$. Dans la région des fers-à-cheval, il y a deux points représentant la même trajectoire dans la zone $\zeta<180^\circ$ et deux points dans la zone $\zeta>180^\circ$, chaque couple de points étant de part et d'autre de $\overline{\cF}^g_{\gH \cS}$ ou de $\overline{\cF}^g_{\gH \cS '}$ selon sa condition initiale. Il y a donc, à la fois pour les orbites troyennes et les orbites en fer-à-cheval, une transition dans l'espace des phases entre les orbites qui librent autour de $L_4$ ou $AL_4$ d'une part et $\overline{\cF}^g_{\gH \cS }$ ou $\overline{\cF}^g_{\gH \cS '}$ d'autre part. Cette transition peut être mise en évidence de manière simple en traçant la moyenne de l'angle $\Delta \varpi$ pour chaque point de la variété, figure~\ref{fig:glob_e01} à droite\footnote{on remarque que pour de nombreuses conditions initiales la valeur moyenne de $\Dv$ semble discontinue sur $\cV$ dans des régions qui semblent cependant régulières sur le graphe de gauche. Cela est dû aux difficultés rencontrées lors de la détermination de $\Dv$ qui n'est pas défini quand une des excentricités est nulle.}. En effet, la discontinuité de la valeur moyenne de $\Delta \varpi$ sépare les orbites librant autour de points fixes ou de variétés différentes. Cette frontière n'est cependant pas une séparatrice, au sens où on peut passer continûment d'une orbite du voisinage de $L_4$ (resp. $\overline{\cF}^1_{\gH \cS }$) à une orbite au voisinage de $AL_4$ (resp. $\overline{\cF}^1_{\gH \cS '}$), la discontinuité apparente dans le comportement de $\Delta \varpi$ étant due à l'annulation d'une des excentricités le long de la frontière (on rappelle que la variable $\Delta \varpi$ n'est pas définie quand une des excentricités est nulle). \\

\an{
In the quasi-circular case (figure~\ref{fig:glob_e01}), the dynamics of the degree of freedom ($Z,\zeta$) is very close to the circular case (the equation \ref{eq:eqerdi} is relevant at the order one in eccentricities): we still have a tadpole and a horseshoe domain, with the separatrix located in $\zeta \approx 24^\circ$ and $\approx 336^\circ$, and the initial value of $\Dv$ does not impact much the average value of $\zeta$ on the orbit (left graph). On the right graph, we represented the positions of the $\ol \cF$. The positions of the union of orbits $\ol \cF^\nu$ and $\ol \cF^g$ (and their intersections, the $L_k$ and $AL_k$) are consistent with their quadratic approximation (section \ref{sec:DFFbLc}). However, the families emanating from the $L_3$ circular equilibrium cannot be identified by the criterion we developed in section \ref{sec:SAIF} because they are near the separatrix located at $\zeta=180^\circ$. In addition to the families that we defined in the neighbourhood of a circular equilibrium, there are branches of the reunion $\ol \cF^g$ in the horseshoe domain. We name $\ol \cF^g_{\gH \cS}$ the family located at $\Dv=0^\circ$, around which librate the orbits of the green area (right graph), and an other is located at $\Dv=180^\circ$ around which librate the orbits of the blue area (right graph). The domains of trojan orbits and horseshoe orbits are hence split in two: for the trojan orbit, $\Dv$ oscillates either around its value in $\ol \cF^g_{L_k}$, or near its value in $\ol \cF^g_{AL_k}$. In the horseshoe domain, it oscillates either near $\Dv=0^\circ$ or $\Dv=180^\circ$. Note that this "split" occurring within the tadpole and horseshoe domains results from our choice of variables: there is no separatrix between these domains. We can, for example, pass continuously from the neighbourhood of $L_4$ to the neighbourhood of $AL_4$ by cancelling one of the eccentricities, which lead to a discontinuity in the value of $\Dv$.  \\}

\an{
The markers on the left graph indicate the points of $\cV$ near which passes the $4$ orbits whose projection on the ($Z,\zeta$) and ($e_1-e_2,\Dv$) plane are represented in figure~\ref{fig:orbe01} (in annex). We represented orbits in the neighbourhood of $L_4$, $AL_4$, and the two types of horseshoe orbits. As we explained in section \ref{sec:set}, each of these generic orbits passes near $4$ different points of $\cV$, and these points can be divided in 2 pairs which have the same value of $\Dv$. Note that the 4 points representing a given orbit are each time positioned in a different quadrant (quadrants that are delimited by the $\ol \cF^\nu$ and $\ol \cF^g$ families).} 

\subsubsection{$e_1=e_2=0.4$}

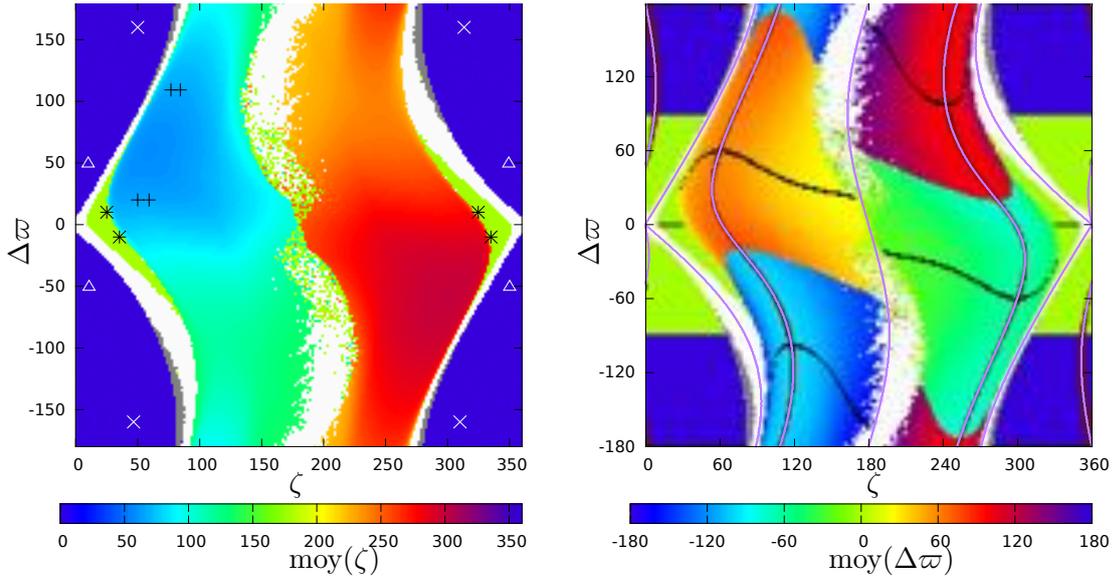
\begin{figure}[h!]
\begin{center}
\includegraphics[width=0.5\linewidth]{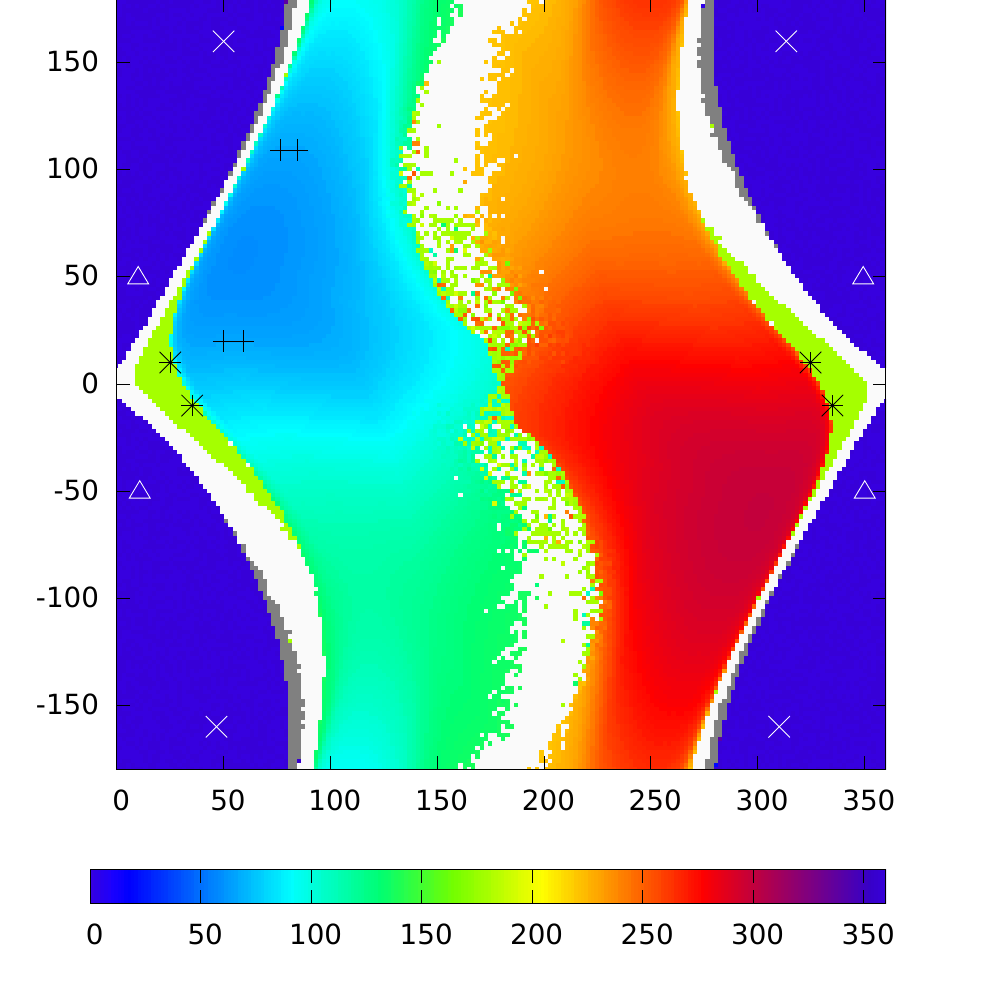}\includegraphics[width=0.5\linewidth]{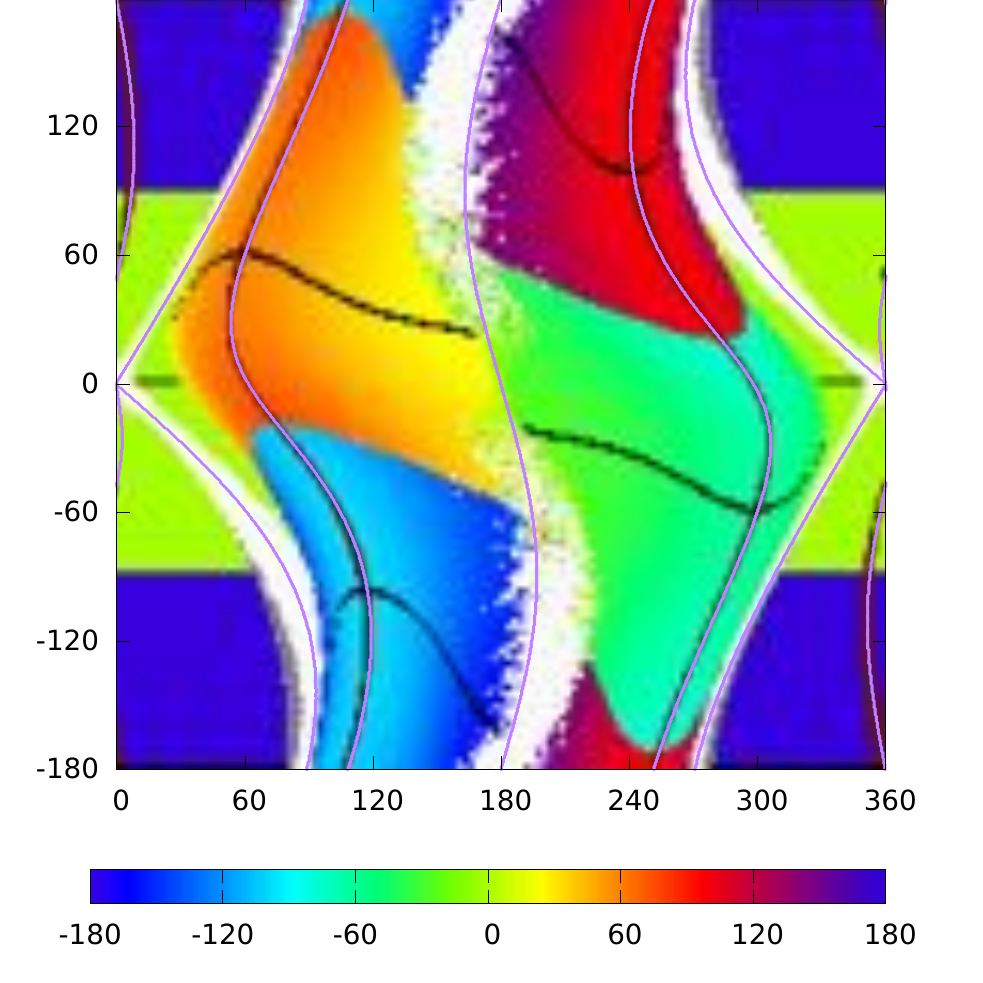}\\
  \setlength{\unitlength}{1cm}
\begin{picture}(.001,0.001)
\put(-7.5,4.5){\rotatebox{90}{$\Dv$}}
\put(0,4.5){\rotatebox{90}{$\Dv$}}
\put(-3.8,1.5){{$\zeta$}}
\put(3.8,1.5){{$\zeta$}}
\put(-3.8,0.5){{moy($\zeta$)}}
\put(3.3,0.5){{moy($\Dv$)}}
\end{picture}
\caption{\label{fig:glob_e4} Plan de référence pour $\eps=10^{-5}$ et $e_1=e_2=0.4$. Le plan est défini par $a_1=a_2=1\,ua$, et $e_1=e_2$. Chaque point de ce plan est une condition initiale d'orbite. Sur la figure de gauche, le code couleur correspond à la valeur moyenne de $\zeta$ sur cette orbite. Les divers marqueurs représentent les points près desquels passent les orbites de la figure~\ref{fig:orbe4}. Sur la figure de droite le code couleur représente la valeur moyenne de $\Dv$ sur chaque orbite. Les orbites proches de $\ol \cF^\nu$, donc celles vérifiant la relation (\ref{eq:condFb02}) avec $\epsilon_\nu=10^{-3.5}$ sont représentées en marron. Les résultats de la méthode semi-analytique sont surimprimés en rouge (relation \ref{eq:condFb0}). Les orbites proches de $\ol \cF^g$, donc celles vérifiant la relation (\ref{eq:condFb1}) avec $\epsilon_g= 3^\circ$ sont représentées en noir. $\epsilon_a=10^{-5.5}$.\\
}
\end{center}
\end{figure}

\begin{figure}[h!]
\begin{center}
\includegraphics[width=0.5\linewidth]{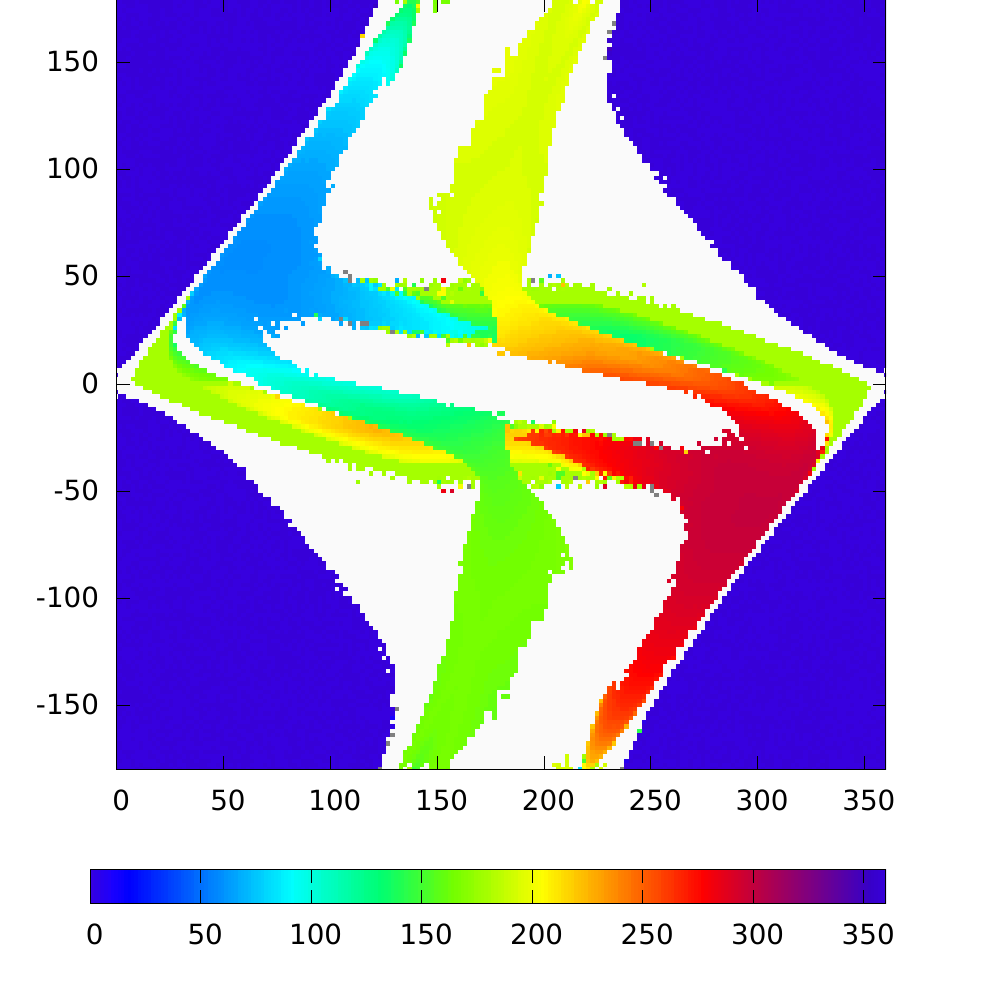}\includegraphics[width=0.5\linewidth]{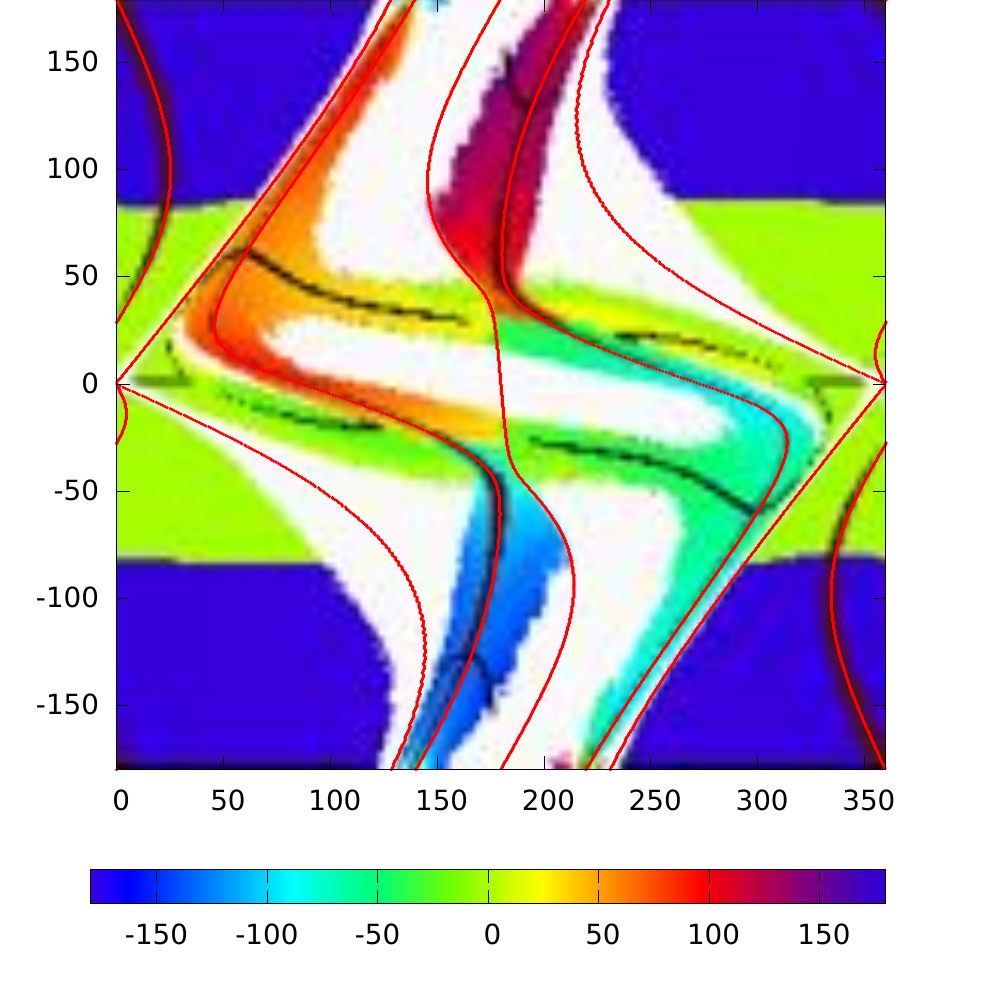}\\
  \setlength{\unitlength}{1cm}
\begin{picture}(.001,0.001)
\put(-7.5,4.5){\rotatebox{90}{$\Dv$}}
\put(0,4.5){\rotatebox{90}{$\Dv$}}
\put(-3.8,1.5){{$\zeta$}}
\put(3.8,1.5){{$\zeta$}}
\put(-3.8,0.5){{moy($\zeta$)}}
\put(3.3,0.5){{moy($\Dv$)}}
\end{picture}
\caption{\label{fig:glob_e6} Plan de référence pour $\eps=10^{-5}$ et $e_1=e_2=0.6$. Le plan est défini par $a_1=a_2=1\,ua$, et $e_1=e_2$. Chaque point de ce plan est une condition initiale d'orbite. Sur la figure de gauche, le code couleur correspond à la valeur moyenne de $\zeta$ sur cette orbite. Sur la figure de droite le code couleur représente la valeur moyenne de $\Dv$ sur chaque orbite. Les orbites proches de $\ol \cF^\nu$, donc celles vérifiant la relation (\ref{eq:condFb02}) avec $\epsilon_\nu=10^{-3.5}$ sont représentées en marron. Les résultats de la méthode semi-analytique sont surimprimés en rouge (relation \ref{eq:condFb0}). Les orbites proches de $\ol \cF^g$, donc celles vérifiant la relation (\ref{eq:condFb1}) avec $\epsilon_g= 3^\circ$ sont représentées en noir. $\epsilon_a=10^{-5.5}$.\\
}
\end{center}
\end{figure}

\begin{figure}[h!]
\begin{center}
\includegraphics[width=0.5\linewidth]{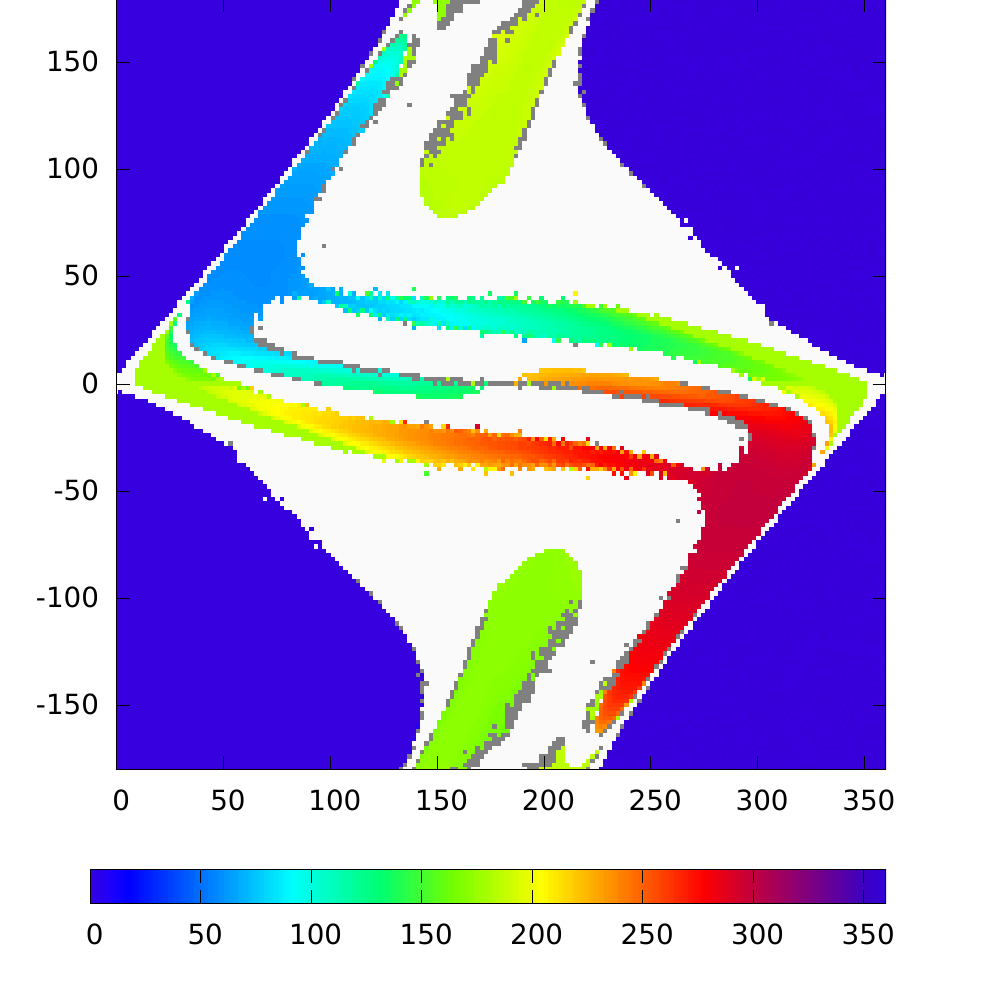}\includegraphics[width=0.5\linewidth]{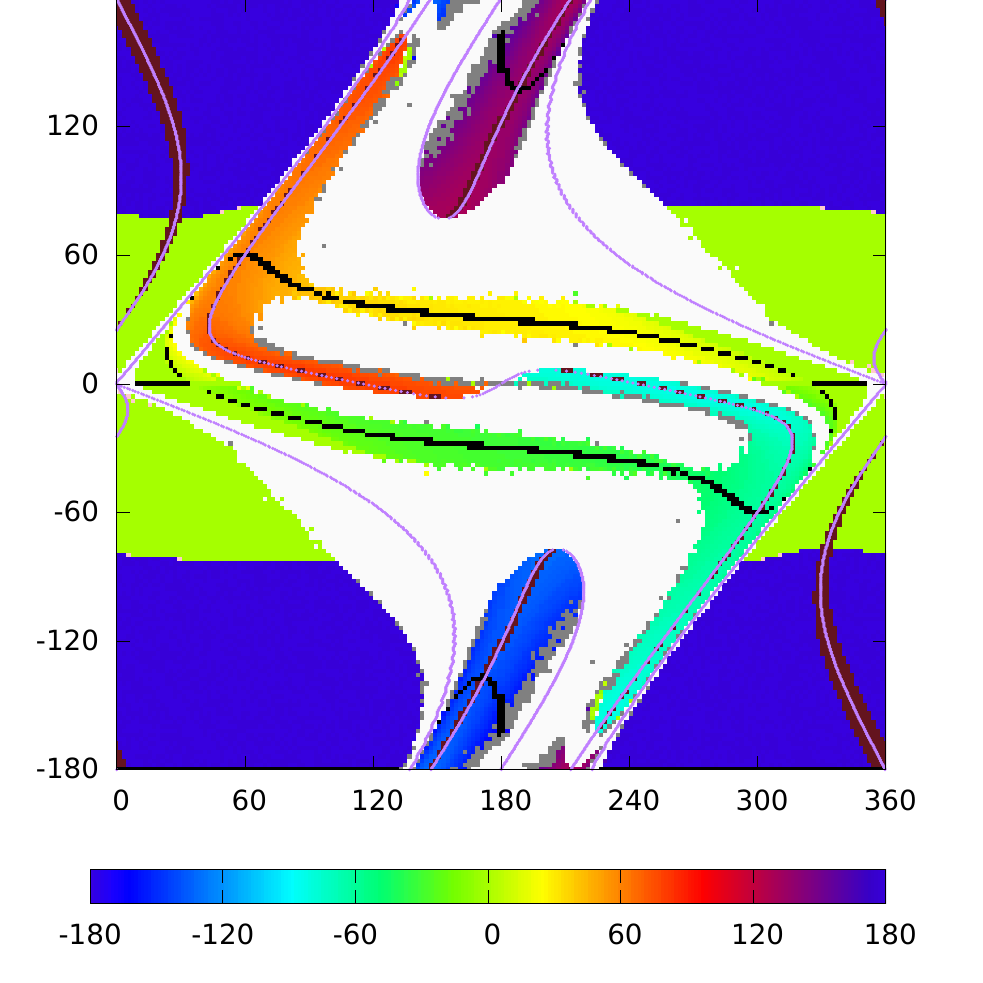}\\
  \setlength{\unitlength}{1cm}
\begin{picture}(.001,0.001)
\put(-7.5,4.5){\rotatebox{90}{$\Dv$}}
\put(0,4.5){\rotatebox{90}{$\Dv$}}
\put(-3.8,1.5){{$\zeta$}}
\put(3.8,1.5){{$\zeta$}}
\put(-3.8,0.5){{moy($\zeta$)}}
\put(3.3,0.5){{moy($\Dv$)}}
\end{picture}
\caption{\label{fig:glob_e65}Plan de référence pour $\eps=10^{-5}$ et $e_1=e_2=0.65$. Le plan est défini par $a_1=a_2=1\,ua$, et $e_1=e_2$. Chaque point de ce plan est une condition initiale d'orbite. Sur la figure de gauche, le code couleur correspond à la valeur moyenne de $\zeta$ sur cette orbite. Sur la figure de droite le code couleur représente la valeur moyenne de $\Dv$ sur chaque orbite. Les orbites proches de $\ol \cF^\nu$, donc celles vérifiant la relation (\ref{eq:condFb02}) avec $\epsilon_\nu=10^{-3.5}$ sont représentées en marron. Les résultats de la méthode semi-analytique sont surimprimés en rouge (relation \ref{eq:condFb0}). Les orbites proches de $\ol \cF^g$, donc celles vérifiant la relation (\ref{eq:condFb1}) avec $\epsilon_g= 3^\circ$ sont représentées en noir. $\epsilon_a=10^{-5.5}$. \\
}
\end{center}
\end{figure}

\begin{figure}[h!]
\begin{center}
\includegraphics[width=0.5\linewidth]{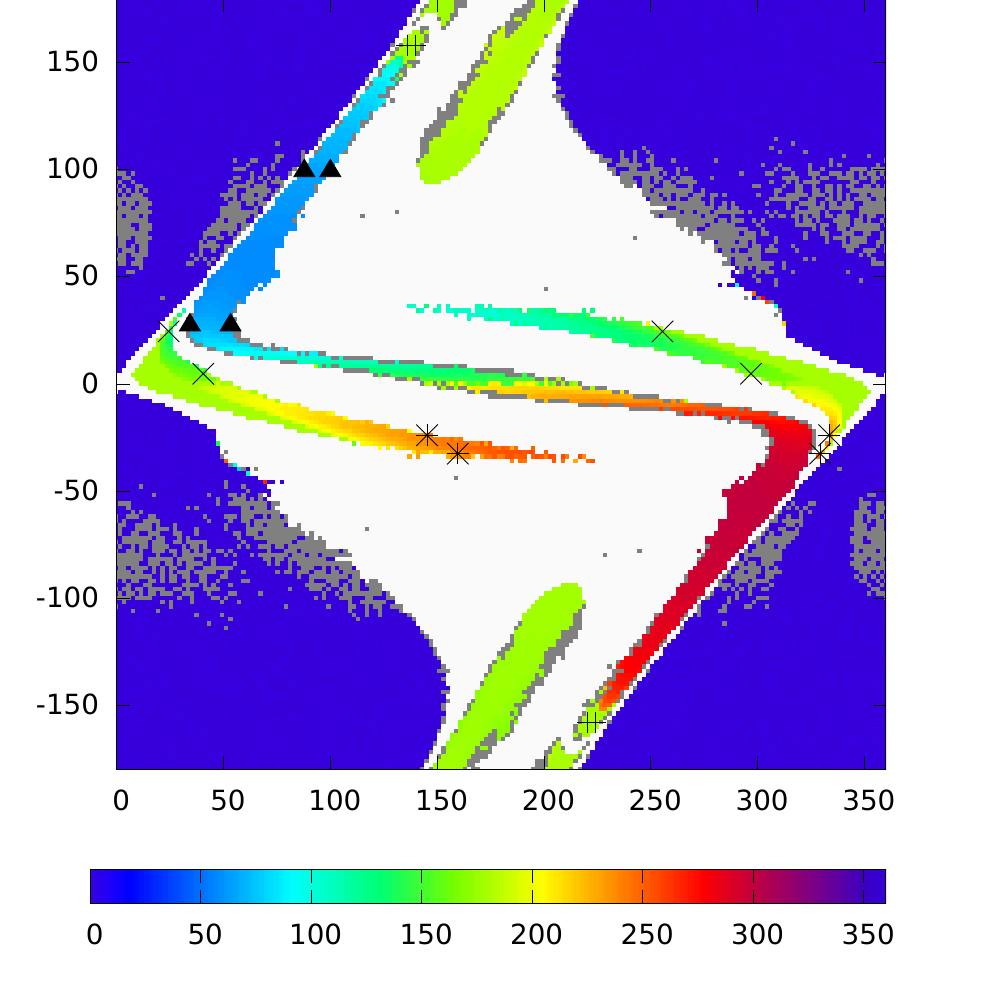}\includegraphics[width=0.5\linewidth]{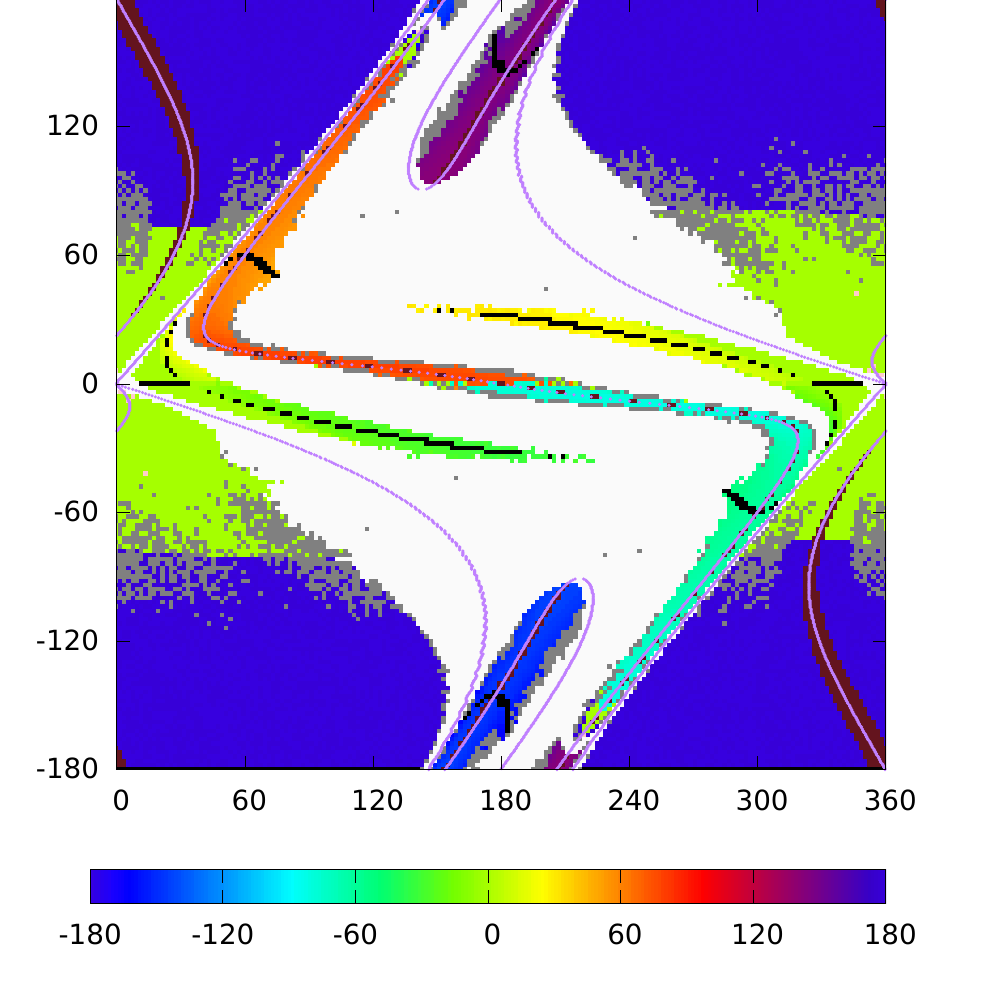}\\
  \setlength{\unitlength}{1cm}
\begin{picture}(.001,0.001)
\put(-7.5,4.5){\rotatebox{90}{$\Dv$}}
\put(0,4.5){\rotatebox{90}{$\Dv$}}
\put(-3.8,1.5){{$\zeta$}}
\put(3.8,1.5){{$\zeta$}}
\put(-3.8,0.5){{moy($\zeta$)}}
\put(3.3,0.5){{moy($\Dv$)}}
\end{picture}
\caption{\label{fig:glob_e7} Plan de référence pour $\eps=10^{-5}$ et $e_1=e_2=0.7$. Le plan est défini par $a_1=a_2=1\,ua$, et $e_1=e_2$. Chaque point de ce plan est une condition initiale d'orbite. Sur la figure de gauche, le code couleur correspond à la valeur moyenne de $\zeta$ sur cette orbite. Sur la figure de droite le code couleur représente la valeur moyenne de $\Dv$ sur chaque orbite. Les orbites proches de $\ol \cF^\nu$, donc celles vérifiant la relation (\ref{eq:condFb02}) avec $\epsilon_\nu=10^{-3.5}$ sont représentées en marron. Les résultats de la méthode semi-analytique sont surimprimés en rouge (relation \ref{eq:condFb0}). Les orbites proches de $\ol \cF^g$, donc celles vérifiant la relation (\ref{eq:condFb1}) avec $\epsilon_g= 3^\circ$ sont représentées en noir. $\epsilon_a=10^{-5.5}$. Les projections des trajectoires représentées par les différents symboles sur la figure de gauche sont tracées en annexe (figure~\ref{fig:orbe7})\\
}
\end{center}
\end{figure}

\begin{figure}[h!]
\begin{center}
\includegraphics[width=0.7\linewidth]{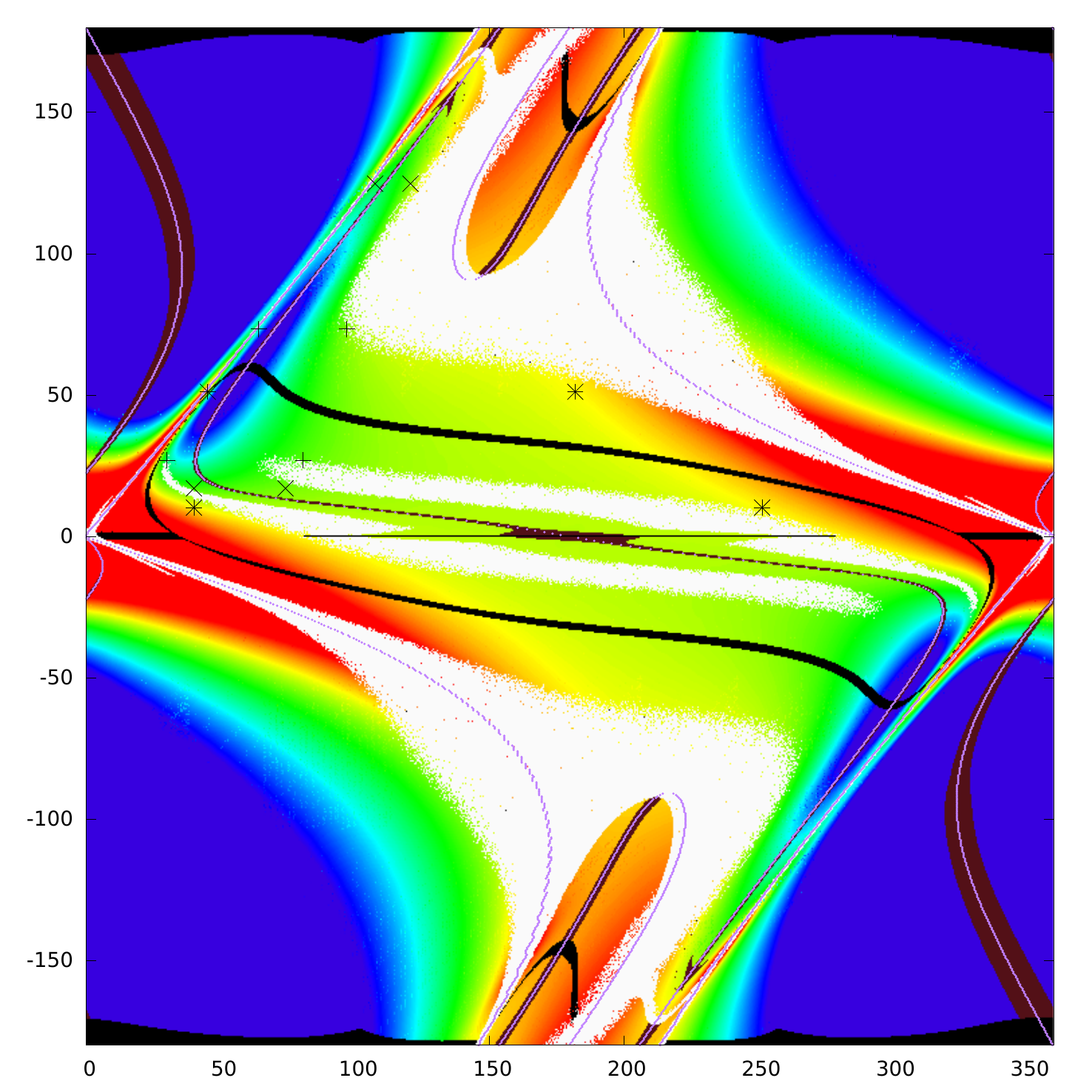}\\
  \setlength{\unitlength}{0.07\linewidth}
\begin{picture}(.001,0.001)
\put(-5,5){\rotatebox{90}{$\Dv$}}
\put(0,0.3){$\zeta$}
\end{picture}
\caption{\label{fig:glob_e7_m6} Variations de l'énergie totale du problème moyen sur le plan de référence pour $\eps=10^{-6}$ et $e_1=e_2=0.7$. Le plan est défini par $a_1=a_2=1\,ua$, et $e_1=e_2$. Chaque point de ce plan est une condition initiale d'orbite, seules les orbites qui ne sont pas sorties de la résonance en $1/\eps$~an sont représentées. la couleur rouge représente un maximum d'énergie et la couleur bleue un minimum (échelle non linéaire). Les orbites proches de $\ol \cF^\nu$, donc celles vérifiant la relation (\ref{eq:condFb02}) avec $\epsilon_\nu=10^{-4}$ sont représentées en marron. Les résultats de la méthode semi-analytique sont surimprimés en violet (relation \ref{eq:condFb0}). Les orbites proches de $\ol \cF^g$, donc celles vérifiant la relation (\ref{eq:condFb1}) avec $\epsilon_g= 3^\circ$ sont représentées en noir.\\
}
\end{center}
\end{figure}

Quand le moment cinétique total augmente, on s'éloigne du cas circulaire et l'ensemble de l'espace des phases se modifie. Nous pouvons voir sur la figure \ref{fig:glob_e4} le cas de deux masses égales pour $e_1=e_2=0.4$. Par rapport au cas précédent, nous pouvons constater l'apparition du domaine des quasi-satellites \citep{GiuBeMiFe2010,Namouni1999}, centré sur le point que nous appellerons $\cQ \cS$: $\zeta=0^\circ$ et $\Dv =180^\circ$. Notons que ce point est à l'intersection d'un groupement d'orbites vérifiant la condition (\ref{eq:condFb0}), que nous supposerons être au voisinage d'une variété d'orbites périodiques $\overline{\cF}^\nu_{\cQ\cS}$, et d'un groupement d'orbites vérifiant la condition (\ref{eq:condFb1}), que nous supposerons être au voisinage d'une variété d'orbites périodiques $\overline{\cF}^g_{\cQ\cS}$. La variété de collision, ainsi que l'ensemble des $\overline{\cF}_{k}$, sont également déformés par l'augmentation du moment cinétique. Notons notamment que dans ce plan, l'aire des configurations troyennes et des fers-à-cheval diminue au profit de celle des quasi-satellites. De plus, la zone d'instabilité au voisinage de la variété $\overline{\cF}^\nu_{3}$ s'épaissit. La dynamique dans les régions troyennes et fer-à-cheval reste sensiblement la même, excepté que le domaine des fers-à-cheval semble rétrécir au point de disparaitre dans le voisinage de $\overline{\cF}^g_{\gH\cS'}$ ($\Dv=180^\circ$).

\begin{rem} \label{rq:10}
Le plan de référence est censé représenter l'ensemble des configurations co-orbitales pour une valeur du moment cinétique donné. Cependant la taille relative de la coupe de deux domaines de stabilité par la variété de référence n'est généralement pas représentative du volume relatif de ces configurations. Les figures~\ref{fig:stabzr} du Chapitre 1 montrent par exemple que selon le plan choisi, l'intersection du domaine des fers-à-cheval avec ce plan peut être plus ou moins importante que celle avec le domaine troyen (pour des $\mu \approx 10^{-6}$).
\end{rem}

\begin{rem} \label{rq:11}
Nous avons deux méthodes pour identifier les familles $\overline{\cF}^\nu$, basées sur les critères (\ref{eq:condFb0n}) et (\ref{eq:condFb0}). Comme nous pouvons le voir sur les figures \ref{fig:glob_e01} à \ref{fig:glob_e7}, et plus facilement sur les figures \ref{fig:glob_e65} et \ref{fig:glob_e7_m6} (en annexe) ces deux méthodes donnent des résultats similaires. La méthode semi-analytique est basée sur le calcul du hamiltonien moyen et donne donc une position approchée de $\ol \cF^\nu$ sans prendre en compte les instabilités du problème réel, alors que la méthode numérique identifie les orbites stables qui sont susceptibles d'être au voisinage des $\overline{\cF}^\nu$.
\end{rem}

Dans le domaine des quasi-satellites, nous pouvons constater une frontière similaire à celle présente dans le domaine troyen: selon la position sur la variété de référence, les orbites oscillent soit autour du point $\cQ \cS$ en passant dans le voisinage d'un point dans chaque quadrant délimité par les familles $\overline{\cF}^0_{\cQ\cS}$ et $\overline{\cF}^1_{\cQ\cS}$; soit les orbites oscillent autour de la collision ($\Delta \varpi = 0^\circ$) avec également un point représentatif dans chaque quadrant. Les orbites situées sur cette frontière représentent les "exchange orbits" introduites par \cite{Nauenberg2002}. De manière analogue aux cas troyen et fer-à-cheval expliqués précédemment, nous pouvons passer continûment d'un type de quasi-satellite à l'autre.    \\

\begin{figure}[h!]
\begin{center}
\includegraphics[width=0.5\linewidth]{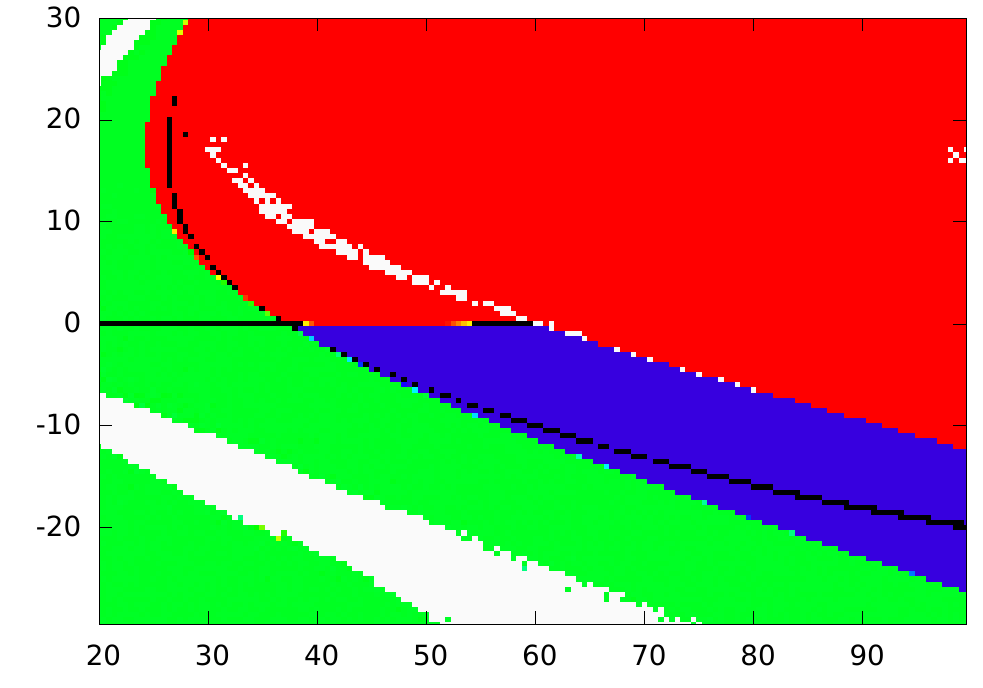}\includegraphics[width=0.5\linewidth]{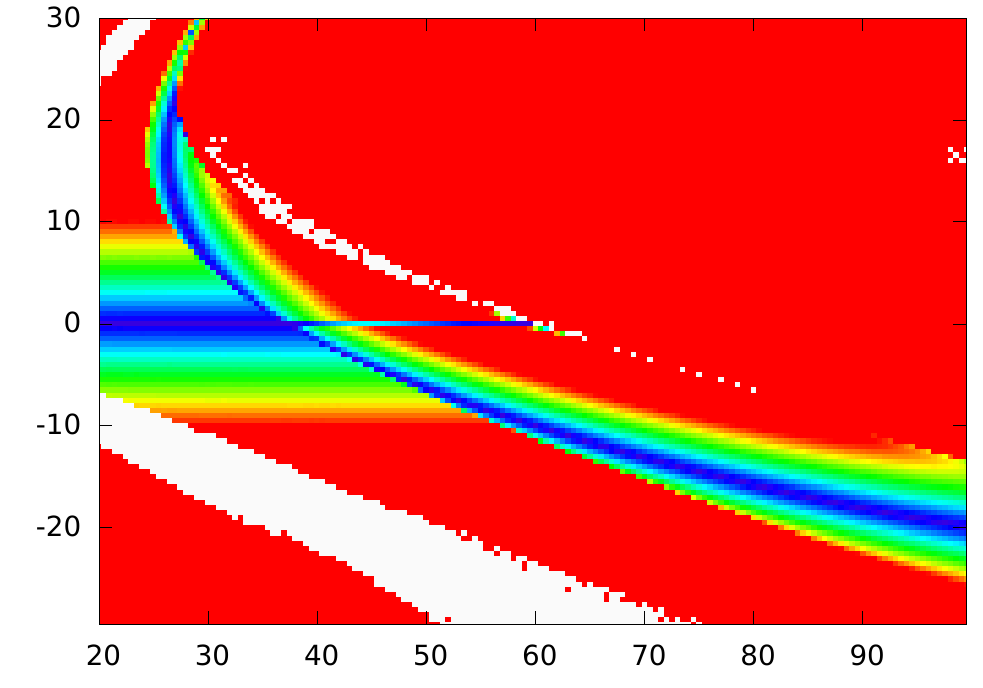}\\
  \setlength{\unitlength}{1cm}
\begin{picture}(.001,0.001)
\put(-7.5,3){\rotatebox{90}{$\Dv$}}
\put(0,3){\rotatebox{90}{$\Dv$}}
\put(-3.8,0.2){{$\zeta$}}
\put(3.8,0.2){{$\zeta$}}
\end{picture}
\caption{\label{fig:zoomAHS} Détail de la région fer-à-cheval du plan de référence pour $\eps=10^{-6}$ et $e_1=e_2=0.6$, $m_1=m_2$. Chaque point de ce plan est une condition initiale d'orbite. Sur la figure de gauche, la couleur verte est assignée quand $\Dv$ libre autour de $0$, rouge quand il libre autour d'une valeur supérieure à $+1^\circ$ et bleue pour une valeur inférieure à $-1^\circ$. Les orbites proches de $\ol \cF^g$, vérifiant la relation (\ref{eq:condFb1}) avec $\epsilon_g= 1^\circ$, sont représentées en noir. Sur la figure de droite le code couleur représente l'amplitude de variation $\max(\Dv)-\min(\Dv)$ sur chaque orbite. avec $\max(\Dv)-\min(\Dv)=0^\circ$ en bleu et $\max(\Dv)-\min(\Dv) \geq 20^\circ$ en rouge.\\
}
\end{center}
\end{figure}

\an{
We now increase the total angular momentum of the system: $e_1=e_2= 0.4$ in the figure~\ref{fig:glob_e4}. As we depart from the circular equilibrium, the phase space evolves. The quasi-satellite domain appears \citep{GiuBeMiFe2010,Namouni1999}, centred on the fixed point $\cQ\cS$ of the averaged reduced problem: $\zeta=0^\circ$, $\Dv=180^\circ$, located at the intersection of the families $\ol \cF^g_{\cQ \cS}$ and $\ol \cF^\nu_{\cQ \cS}$. We can observe on the right-hand graph that the quasi-satellite domain can also be splited in two kind of quasi-satellites: those for which $\Dv$ librate around $180^\circ$ and those for which it librates around $0^\circ$. The exchange orbits introduced by \cite{Nauenberg2002} are located at the frontier between these two kind of quasi-satellite orbits. 
\\}

\an{
 The dynamics in the trojan and horseshoe domain remains similar to the quasi-circular case, but the domain where the horseshoe orbits librate around $\Dv=180^\circ$ shrinks on this plane. This is due to the increase of the unstable area near the $\ol \cF^\nu_3$ family and the position of the collision manifold. Indeed, the collision manifold, as well as all the $\ol \cF$, is twisted as the total angular momentum increases (see figure \ref{fig:Fb0meq}). On this plan of initial conditions, this leads to the reduction of the stability domain for trojan and horseshoe configurations and the increase of the stability domain of quasi-satellites.\\} 
 
\an{ Remark \ref{rq:10}:
When $m_1=m_2$ we suppose that the reference manifold represents all the co-orbital configurations reaching $a_1=a_2$ on their orbit, for a given value of the total angular momentum. However, the relative size of the section of two stability domains by the reference manifold is generally not representative of the relative volume of these two configurations in the phase space. For example, the figures~\ref{fig:stabzr} (chapter 1) show that depending on the chosen section, the horseshoe domain might appear larger or smaller than the tadpole one (for $\mu \approx 10^{-6}$).
\\}

\an{ Remark \ref{rq:11}: We have 2 methods to identify $\overline{\cF}^\nu$, based on the criteria (\ref{eq:condFb0n}) and (\ref{eq:condFb0}). As we can see in the figures \ref{fig:glob_e01} to \ref{fig:glob_e7}, and more easily in the figure \ref{fig:glob_e7_m6}, both methods give very similar results. However the semi-analytical method is based on the average Hamiltonian and hence gives an approximation of the position of $\ol \cF^\nu$ without considering the instabilities of the full 3-body problem.
}

\subsubsection{Apparition des fers-à-cheval asymétriques} 

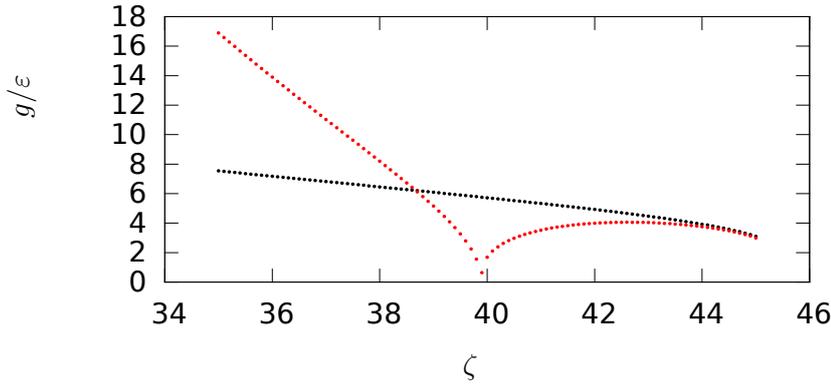
\begin{figure}[h!]
\begin{center}
\includegraphics[width=0.7\linewidth]{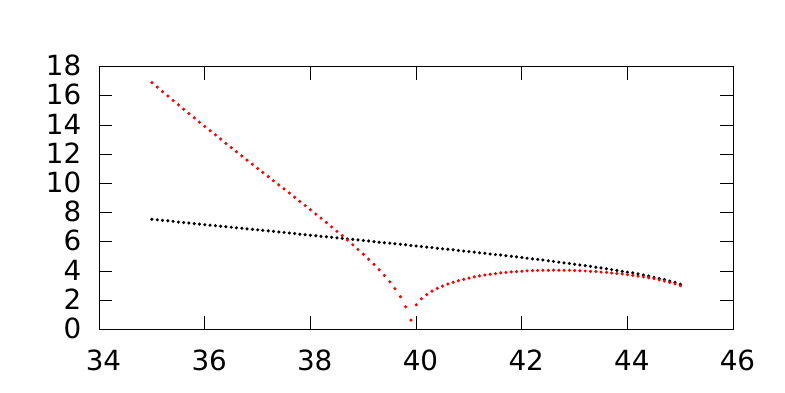}\\
 \setlength{\unitlength}{1cm}
\begin{picture}(.001,0.001)
\put(-6,3.5){\rotatebox{90}{$g/\eps$}}
\put(0,0.1){{$\zeta$}}
\end{picture}
\caption{\label{fig:freqge5} Évolution des fréquences $g/(n\eps)$ en rouge et $\nu/(n\sqrt \eps )$ en noir pour $\Dv=1^\circ$ constant et $e_1=e_2=0.55$.\\
}
\end{center}
\end{figure}

Pour $e_j\leq 0.5$, la variété $\nu=0$ (séparatrice) délimite le domaine troyen du domaine fer-à-cheval à l'instar du cas circulaire (voir figure~\ref{fig:zr}), alors que $g$ reste non nul dans l'ensemble de ces domaines.
En augmentant le moment cinétique total ($e_1=e_2=0.6$), nous voyons apparaître la première modification notable de l'espace des phases dans ces régions. A partir de $e_1=e_2\approx 0.5$, $g$ s'annule dans le domaine des fers-à-cheval sur une variété $g=0$ avant que la séparatrice $\nu=0$ soit rencontrée (figure \ref{fig:freqge5}). L'annulation de $g$ entraîne une bifurcation de la famille d'orbites périodiques du problème moyen réduit $\overline{\cF}^g_{\gH\cS}$ en deux familles d'orbites stables: $\overline{\cF}^g_{\gH\cS 1}$ (pour $\Delta \varpi >0$) et $\overline{\cF}^g_{\gH\cS 2}$ (pour $\Delta \varpi <0$); et une famille instable $\overline{\cF}^g_{\gH\cS i}$ pour $\Delta \varpi = 0$. Les orbites dans le voisinage des familles stables $\overline{\cF}^g_{\gH\cS 1}$ et $\overline{\cF}^g_{\gH\cS 2}$ librent autour de celles-ci, pour des valeurs moyennes de $\Dv$ différentes de $0^\circ$, ce qui diffère de la région des fers-à-cheval avant la bifurcation. Cette différence permet d'identifier les différents domaines des fers-à-cheval symétriques ($\gH\cS$) en vert sur la figure~\ref{fig:zoomAHS} (gauche), des fers-à-cheval asymétriques  $A\gH\cS_1$ en rouge et $A\gH\cS_2$ en bleu sur cette même figure. Cette dénomination "asymétrique" provient de la forme de la projection de ces orbites sur les variables ($Z,\zeta$), voir les exemples indiqués par les $\times$ et les $\ast$ sur la figure~\ref{fig:glob_e7} et tracés en annexe sur la figure~\ref{fig:orbe7} (b) et (c), à comparer avec le cas symétrique figure~\ref{fig:orbe4} (b). Notons que la transition des couleurs vert à bleu et vert à rouge permet de visualiser l'intersection de la variété $g=0$ avec le plan de référence. La variété instable $\overline{\cF}^g_{\gH\cS i}$ est quant à elle mise en évidence par la transition du rouge au bleu et semble se trouver pour $\Delta \varpi =0^\circ$ sur le plan de référence.\\

\an{
As long as $e_j \leq 0.5$, the manifold $\nu=0$ (separatrix emanating from the unstable family $\ol \cF^\nu_3$) splits the trojan domain from the horseshoe domain as in the circular case, and the frequency $g$ is always different from $0$ in these domains. By further increasing the total angular momentum, we can see the first notable modification of the phase space: when $e_1=e_2$ gets higher than $\approx 0.5$, the fundamental frequency $g$ vanishes in the horseshoe domain. As a result, the previously stable family of periodic orbits $\ol \cF^g_{\gH \cS}$ bifurcates into two stable families of periodic orbits $\ol \cF^g_{\gH \cS 1}$ for $\Dv >0^\circ$ and $\ol \cF^g_{\gH \cS 2}$ for $\Dv <0^\circ$, and one unstable family of periodic orbits $\cF^g_{\gH \cS i}$ located at $\Dv =0^\circ$. }
\an{Here, `stability' is in the sense of the eigenvalues: a family $\ol \cF^g$ is either stable (elliptic) in the direction associated with the frequency $g$ or unstable (hyperbolic) in the direction associated with the eigenvalue. The illustration of this transition can be found in the figures \ref{fig:zoomAHS} and \ref{fig:freqge5}. The figure \ref{fig:freqge5} shows the evolution of the frequencies $\nu$ and $g$ in the horseshoe domain for $e_1=e_2=0.55$ and $\Dv=1^\circ$.} 
\an{The other initial conditions are the same than for the other integrations of this section. 
We integrated the 3-body problem and made a frequency analysis of the angle $\zeta$ (resp. $\Dv$) to obtain $\nu$ (resp. $g$).} 
\an{We can see on this figure that $g$ vanishes before $\nu$.\\}
\an{The figure \ref{fig:glob_e6} represents a zoom in the reference manifold for $e_1=e_2=0.6$ (here $\eps=10^{-6}$). 
On the left hand graph, the green area centred on $\Dv=0^\circ$ is the horseshoe domain we had for lower eccentricities, and the black line in the middle of this domain is the family $\ol \cF^g_{\gH \cS}$. }
\an{The bifurcation occurs for $\zeta \approx 40^\circ$. After the bifurcation, we can see the stable families $\cF^g_{\gH \cS 1}$ and $\cF^g_{\gH \cS 2}$ identified by black pixels (some orbits in the close neighbourhood of $\cF^g_{\gH \cS i}$ can also verify the condition (\ref{eq:condFb1}) when integrated over a duration of the order of $1/\eps$ because $g$ tends to $0$ in this family). On the left of the bifurcation, the orbits librating around $\cF^g_{\gH \cS 1}$ are represented in red and those librating around $\cF^g_{\gH \cS 2}$ are represented in blue (the tadpole orbits beyond the separatrix $\nu=0$ are also represented in red).} 
\an{In these two domains, the projection of a given orbit on the ($Z,\zeta$) plane is not symmetric with respect to $\zeta=180^\circ$, see the examples figure \ref{fig:orbe7}. We hence call these domains asymmetric horseshoe $\cA\gH\cS_1$ and $\cA\gH\cS_2$. The red/green interface and blue/green interface mark the separatrix $g=0$. The position of $\cF^g_{\gH \cS i}$ can be identified by the transition from red to blue.
}

\subsubsection{$e_1=e_2=0.6$} 

La figure~\ref{fig:glob_e6} représente les valeurs moyennes des angles $\zeta$ et $\Dv$ quand les conditions initiales parcourent la variété $\cV$ pour $e_1=e_2=0.6$ et $\eps=10^{-5}$. On peut y voir la même bifurcation que sur la figure~\ref{fig:zoomAHS}, ce qui laisse penser que la valeur de $\eps$ a peu d'influence sur l'apparition de cette bifurcation. D'autre part on peut voir que les régions stables, même à court terme, se réduisent de plus en plus au voisinage des familles $\ol \cF$ pour les domaines troyen et fer-à-cheval, alors que le domaine de stabilité des quasi-satellites continue de s'étendre. Les familles $\ol \cF^\nu$ sont à la limite de la reconnexion mise en évidence en section \ref{sec:SAIF}.\\

\an{
The figure \ref{fig:glob_e6} gives the mean value of the angles $\zeta$ and $\Dv$ when the initial conditions are taken across $\cV$ for $e_1=e_2=0.6$ for $m_1=m_2=10^{-5} m_0$. The bifurcation of the family $\ol \cF^g_{\gH \cS}$ occurs for the same coordinates than the case  $m_1=m_2=10^{-6} m_0$ (figure~\ref{fig:zoomAHS}). The intersection between the manifold $g=0$ and $\cV$ may thus not depend on the value of $\eps$ (as it is the case for the intersection of $\nu=0$ and $\cV$). In the figure~\ref{fig:glob_e6}, the stable area shrinks into the neighbourhood of the families $\ol\cF$ in the tadpole and horseshoe area, while the quasi-satellite domain keeps increasing. On the right hand graph, for this value of the eccentricities, the families $\ol \cF^\nu$ are on the verge of the reconnection that we explained in section \ref{sec:SAIF}.} 

\subsubsection{$e_1=e_2=0.65$}

La figure \ref{fig:glob_e65} représente les valeurs moyennes des angles $\zeta$ et $\Dv$ quand les conditions initiales parcourent la variété $\cV$ pour $\eps=10^{-5}$ et $e_1=e_2=0.65$, donc après la reconnexion. Les domaines des anti-lagranges au voisinage des points $AL_4$ et $AL_5$ sont séparés des autres configurations par de larges régions instables, et les familles $\ol \cF^g_{\gH \cS k}$ semblent se reconnecter continûment avec les familles $\ol \cF^g_{k}$ issues des équilibres de lagrange elliptiques $L_4$ et $L_5$, en plus de la reconnexion des familles $\ol \cF^\nu$. Cette reconnexion des familles $\ol \cF^g$ permettrait de passer continûment du voisinage des points $L_4$ et $L_5$ aux orbites en fers-à-cheval asymétriques. Cette reconnexion est stable sur des temps long (diffusion de $\eta$ inférieure à $10^{-7}$ sur une durée de $100/\eps$) pour $\eps=10^{6}$ et $e_1=e_2=0.7$, voir section \ref{sec:meqstab} pour plus de détails.  \\

\an{
The figure \ref{fig:glob_e65} shows the manifold of reference after the reconnection, for $e_1=e_2=0.65$. The reconnection splits the domain of the trojan orbits librating around $AL_4$ from those librating around $L_4$. Moreover, in addition to the reconnection of the $\ol \cF^\nu$ families, the $\ol \cF^g$ manifold reconnects as well: they link the tadpole domain near $L_4$ to the asymmetric horseshoe domain. Eventually, We can pass continuously from a tadpole orbit to an asymmetric horseshoe orbit, for values of the eccentricities and masses for which this `path' is not disturbed by unstable areas, see the section \ref{sec:meqstab} for more details.}

\begin{figure}[h!]
\begin{center}
\includegraphics[width=0.7\linewidth]{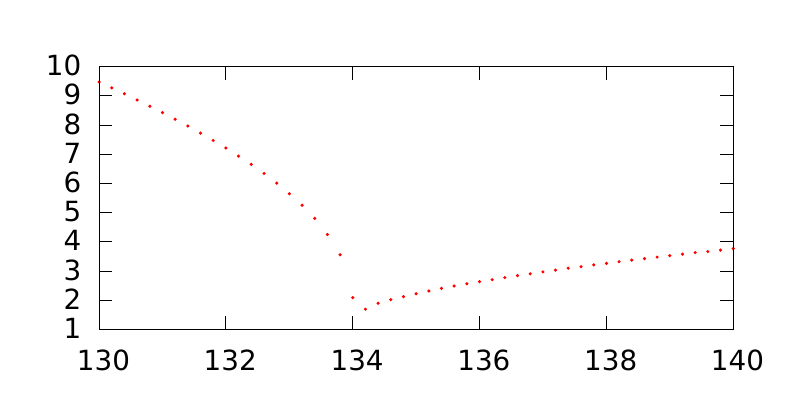}\\
 \setlength{\unitlength}{1cm}
\begin{picture}(.001,0.001)
\put(-6,3.5){\rotatebox{90}{$g/\eps$}}
\put(0,0.1){{$\zeta$}}
\end{picture}
\caption{\label{fig:freqg0} Évolution de la fréquence $g/(n\eps)$ le long de la famille $\ol \cF^\nu_{4}$ pour $e_1=e_2=0.7$ (figure~\ref{fig:glob_e7}).\\
}
\end{center}
\end{figure}

\subsubsection{$e_1=e_2=0.7$} 

Enfin, la figure \ref{fig:glob_e7} représente les valeurs moyennes des angles $\zeta$ et $\Dv$ quand les conditions initiales parcourent la variété $\cV$ pour $\eps=10^{-5}$ et $e_1=e_2=0.7$. Les domaines stables du voisinage des $L_k$ et $AL_k$ continuent à rétrécir. On note cependant l'apparition d'orbites quasi-périodiques (sur le temps de l'étude) au voisinage du point d'équilibre instable (au sens hyperbolique) $AL_3$. Les nombreux pixels gris dans le domaine des quasi-satellites sont susceptibles d'être dus à des problème numériques: à l'interface des domaines bleu et vert sur la figure de gauche, l'excentricité de chaque corps s'annule périodiquement, pendant que l'autre prend des valeurs proches de $0.99$. Ces orbites ne sont donc pas intégrées correctement avec le pas d'intégration choisi ($0.01$~an). Par ailleurs, en suivant la famille $\ol \cF^\nu_{4}$ partant de $L_4$ pour des $\Dv$ croissant, on rencontre une nouvelle variété $g=0$ (voir figure~\ref{fig:freqg0}) avant de rencontrer la bande d'orbites instables qui sépare la région troyenne de la région des anti-Lagrange. Au-delà de cette séparatrice, une nouvelle configuration stable apparait. Nous l'appellerons la configuration $G$. L'exemple de trajectoire indiqué par les symboles $+$ sur la figure~\ref{fig:glob_e7} est donné sur la figure \ref{fig:orbe7} (a). Ces orbites, comme nous pouvons le voir par la position des quatre $+$ sur la figure~\ref{fig:glob_e7}, passent au voisinage de la variété de référence à la fois proche des configurations troyennes librant autour de $L_4$ et celles librant autour de $L_5$. Ces trajectoires librent donc autour des familles $\ol \cF^g_{k}$ avec $k\in\{3,4,5\}$, où $\ol \cF^g_{4}$ et $\ol \cF^g_{5}$ sont stables et $\ol \cF^g_{5}$ est instable, à l'extérieur de la variété $g=0$ à la manière des trajectoires bleues sur la figure~\ref{fig:ppH0b} pour le degré de liberté ($Z,\zeta$). Il est important de noter que toute orbite dans le domaine $G$ voit l'excentricité des deux co-orbitaux osciller grandement (voir figure~\ref{fig:orbe7} (a)), dans le cas $m_1 \ll m_2$, l'oscillation est principalement subie par l'excentricité de la petite planète, ces configurations risquent donc d'être instables pour des coorbitaux de masses trop différentes.   

Pour comparaison, on trace la figure \ref{fig:glob_e7_m6}, pour $\eps=10^{-6}$. Notons que pour cette figure les trajectoires sont intégrées sur seulement $1/\eps=10^6$~an\footnote{Ce qui est de l'ordre d'une unique période $2\pi/g$. Le fait qu'une orbite ne soit pas éjecté ne donne donc pas d'information sur la stabilité à long terme de la dite orbite.}. La position de l'intersection des familles $\ol \cF$ avec la variété de référence semble ne pas dépendre fortement de la valeur de $\eps=m_2/m_0=m_1/m_0$\footnote{Nous le savions déjà pour $\ol \cF^\nu$, voir section \ref{sec:SAIF} et pour les position des points fixes, voir section \ref{sec:H0pf}.}. L'augmentation brusque du nombre d'orbites satisfaisant la  relation (\ref{eq:condFb0}) (représentées en marron) au voisinage de la famille $\ol \cF^g_{AL_3}$ et de la frontière $g=0$ est due à la mauvaise détermination des variations de $Z$ car $g$ tend vers $0$ le long de cette séparatrice, le temps d'intégration est donc court devant $1/g$. L'intersection des variétés $g=0$ et $\nu=0$ avec la variété de référence semble également être indépendante de $\eps$ (comparer par exemple les figures~\ref{fig:glob_e6} et \ref{fig:zoomAHS}). Sur la figure~\ref{fig:glob_e7_m6} nous avons marqué trois conditions initiales '$+$', '$\times$' et '$\ast$' dans le quadrant supérieur gauche de l'équilibre elliptique $L_4$ ($\zeta=\Dv=60^\circ$) délimité par $\ol \cF^\nu_{4}$ et $\ol \cF^g_{4}$. Ces conditions initiales sont prises dans le voisinage immédiat des orbites éjectées au voisinage de la variété de collision. En intégrant ces orbites sur quelques périodes $2\pi/g$, on identifie les $3$ autres points de $\cV$ au voisinage desquels passent ces orbites. Chaque passage s'effectue dans un quadrant différent, au voisinage de la limite de stabilité dans chaque direction. Ces frontières de stabilité semblent donc avoir la même origine: elles sont voisines de la variété de collision. De la même manière, on peut identifier deux à deux les différentes frontières qui délimitent le domaine de stabilité des configurations voisines de $AL_4$ et $AL_5$.\\

\an{
Finally, the figure~\ref{fig:glob_e7} shows the results of the integrations of $\cV$ for $e_1=e_2=0.7$. The trends we previously observed are still presents, as the stability domain for the tadpole and horseshoe configurations continue to shrink. However, the neighbourhood of the hyperbolic equilibrium $AL_3$ seems to harbour stable orbits (on the time scale of the integration). There are many grey pixels in the quasi-satellite domain, but it can be due to a numerical problem: between the blue and the green domain on the right hand graph, each eccentricity vanishes periodically, while the other gets close to $0.99$, so our integration step of $0.01$ orbital period cannot represent these orbits correctly.   
A new domain of stable orbits appears in this figure: following the $\ol \cF^\nu_4$ family emanating from $L_4$ as $\Dv$ increases, we meet a new separatrix $g=0$ (see figure~\ref{fig:freqg0}) before we meet the unstable domain. Beyond this separatrix lies a new stable domain that we call the $G$ configuration. An example of the $G$ trajectories is identified by the $4$ markers $+$ in the figure~\ref{fig:glob_e7} and is plotted in figure~\ref{fig:orbe7}. Each of the $G$ trajectories passes near both the trojan configuration librating around $L_4$ and $L_5$. These trajectories hence librate around the families $\ol \cF^g_k$ with $k \in \{3,4,5\}$ (where $\ol \cF^g_4$ and $\ol \cF^g_5$ are stable-elliptic, and $\ol \cF^g_3$ is unstable-hyperbolic), outside the separatrix $g=0$ in a similar way to the blue trajectories of the figure~\ref{fig:ppH0b}. The domain of $G$ seems to be split as well between orbits that librate around $\Dv=0^\circ$ and those librating around $\Dv=180^\circ$. Note that the eccentricities of the orbits in this domain have a huge amplitude of variation, therefore these orbits may not exist when the mass of one of the co-orbitals is significantly smaller than the mass of the other.
\\}

\an{
To compare with the figure~\ref{fig:glob_e7}, we plot the figure~\ref{fig:glob_e7_m6}, for $\eps=10^{-6}$. In this case the orbits are integrated only over $1/\eps=10^6$ orbital periods. By comparing these two figures we can see that the intersection of the families $\ol \cF$ (and the manifolds $g=0$ and $\nu=0$) with the manifold of reference seems not to depend on the value of $\eps$ ($=m_1/m_0=m_2/m_0$). The large amount of orbits satisfying the relation (\ref{eq:condFb02}) (in brown) when we get close to the separatrix is due to a bad determination of the amplitude of libration of $Z$ (the duration of the integration is small with respect to $2\pi/g$ when we approach the separatrix). In the figure~\ref{fig:glob_e7_m6}, we showed 3 initial conditions ($+$, $\times$ and $\*$) in the top left quadrant of $L_4$ (the quadrant are delimited by the families $\ol \cF^\nu$ and $\ol \cF^g$). These initial conditions are taken very close to the collision manifold. Integrating these trajectories over few periods of $g$, we identify for each of these orbits the $3$ other points of $\cV$ near which they pass. For a given trajectory, each point is in a different quadrant. The three trajectories pass near the instability border in each quadrant, these borders hence seem to have the same origin: they emanate from the collision manifold. Using a similar method, we can identify the borders of the stability domains of the $AL_4$ and $AL_5$ configurations.
}



\subsection{Stabilité}
\label{sec:meqstab}

\begin{figure}[h!]
\begin{center}
\includegraphics[width=0.5\linewidth]{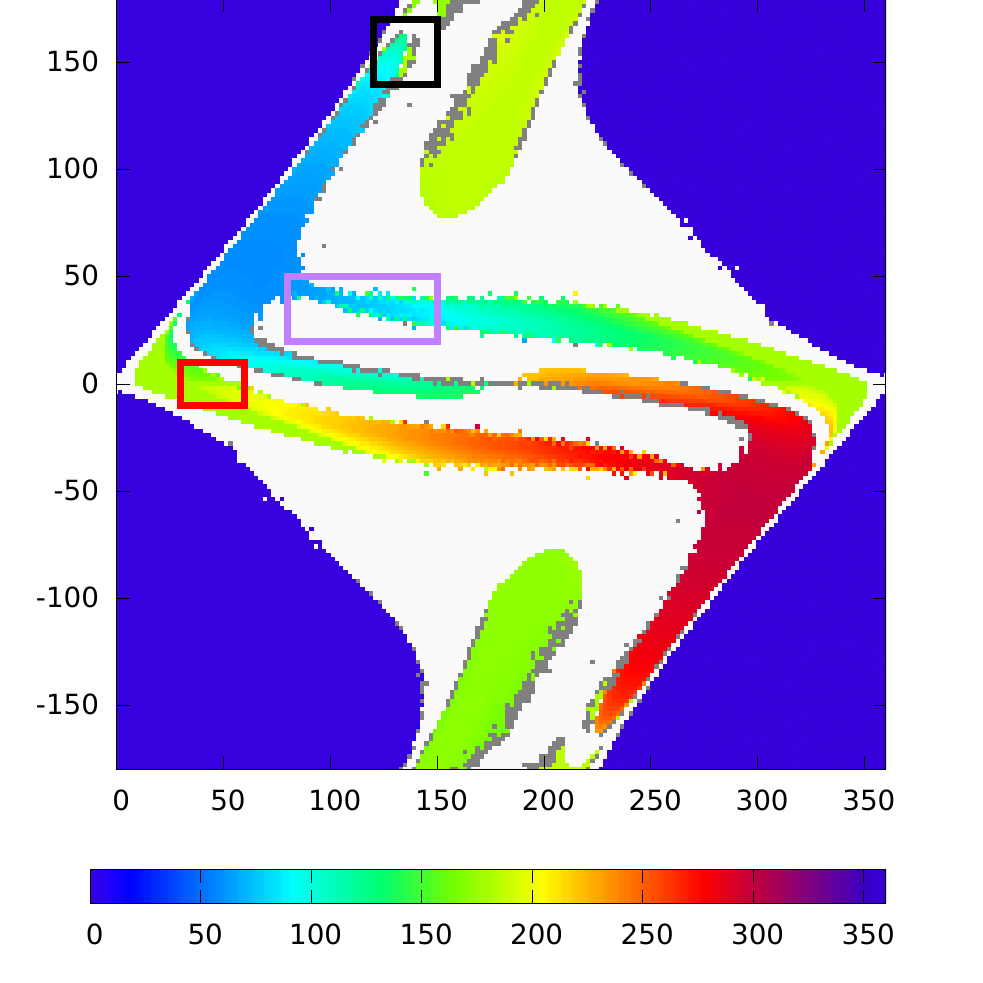}\\
  \setlength{\unitlength}{1cm}
\begin{picture}(.001,0.001)
\put(-3.7,4.6){\rotatebox{90}{$\Dv$}}
\put(0,1.5){{$\zeta$}}
\put(-0.5,0.5){{moy($\zeta$)}}
\end{picture}
\caption{\label{fig:glob_box} Identification des régions où la stabilité à long terme est étudiée: la configuration $G$ (rectangle noir), l'éventuelle reconnexion des familles $\ol \cF^g$ (rectangle violet) et le domaines de fers à cheval asymétriques (rectangle rouge).}
\end{center}
\end{figure}
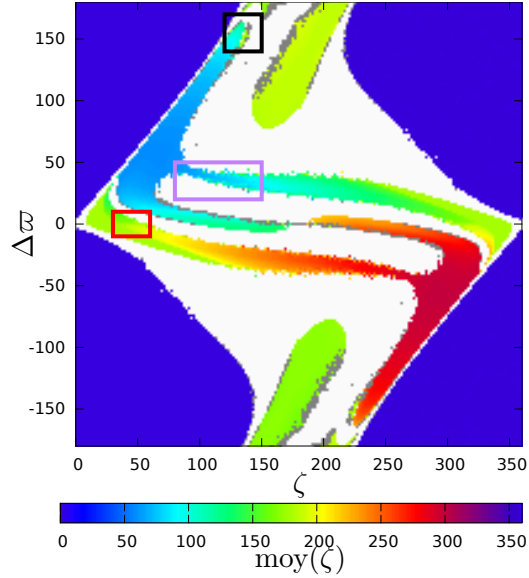

Dans cette section nous étudions la stabilité sur le long terme de certaines régions de l'espace des phases. On se concentrera sur les régions de la variété de référence qui sont identifiées par des rectangles sur la figure~\ref{fig:glob_box}: une région où apparaissent les fer-à-cheval asymétriques ($\cA\gH\cS$), identifiée par le rectangle rouge, une région où apparaissent les orbites $G$, identifiée par le rectangle noir, et la région où se produit l'éventuelle reconnexion entre les familles $\ol \cF^g$ (rectangle violet). Les rectangles de conditions initiales sont susceptibles d'être légèrement déplacés car la position sur la variété représentative de certaines configurations (par exemple les $\cA\gH\cS$) évolue avec l'excentricité. Notons que la région encadrée par le rectangle noir a un intérêt double: en plus de contrôler la stabilité des configurations de type $G$, elle contrôle également la stabilité des configurations du voisinage du point $\zeta=180^\circ$, $\Dv=0^\circ$ (qui est l'équilibre $AL_3$ car ici $m_1=m_2$ et $e_1=e_2$), car les orbites en bas à gauche de la variété $g=0$ sont celles qui passent également au voisinage de ce point.\\

Pour étudier la stabilité de ces configurations, évaluons la diffusion du moyen mouvement moyen $\eta$ \citep{RoLa2001}:
\begin{equation}
\Deta' = \log_{10} \left| \frac{\eta_2 -\eta_1}{\eta_1} \right| \, ,
\label{eq:Detap}
\end{equation}
où $\eta_1$ est calculé sur la première moitié de l'intégration et $\eta_2$ sur la deuxième moitié. Cette quantité est calculée pour chaque trajectoire prenant pour condition initiale $a_1=a_2=1$~ua, $e_1=e_2$, $\lambda_1=\varpi_1=0^\circ$, et $\lambda_2$ et $\varpi_2$ sont donnés par les coordonnées du point sur chaque figure.  

\an{
In this section we study the long term stability of some specific areas of the phase space. These areas are represented by coloured rectangles in figure~\ref{fig:glob_box}: the red rectangle for the asymmetric horseshoe configurations, the purple rectangle for the reconnection of the $\ol \cF^g$ families, and the black rectangle for the $G$ configuration. Note that in addition to indicate the stability of orbits in $G$ configuration, the latter rectangle allows to study the stability of the orbit crossing the representative manifold in the vicinity of the point $AL_3$ ($\zeta=180^\circ$, $\Dv=0^\circ$) as well, since the same orbits pass close to the bottom-left part of the black rectangle.  \\
To study the stability of these orbits, we integrate trajectories and we evaluate the diffusion of the mean mean-motion  \citep{RoLa2001} given by the equation (\ref{eq:Detap}), where $\eta_1$ is computed on the fist half of the integration and $\eta_2$ is computed on the second half. Since the studied trajectories are taken in the associated representative manifold, we have $a_1=a_2=1$~ua, $e_1=e_2$, $\lambda_1=\varpi_1=0^\circ$ as initial conditions, and $\lambda_2$ and $\varpi_2$ are given by the coordinates in the graph.
} 
 
 \subsubsection{Choix de l'intégrateur et méthode}
 \label{sec:choixint}

 \begin{figure}[h!]
\begin{center}
\includegraphics[width=0.5\linewidth]{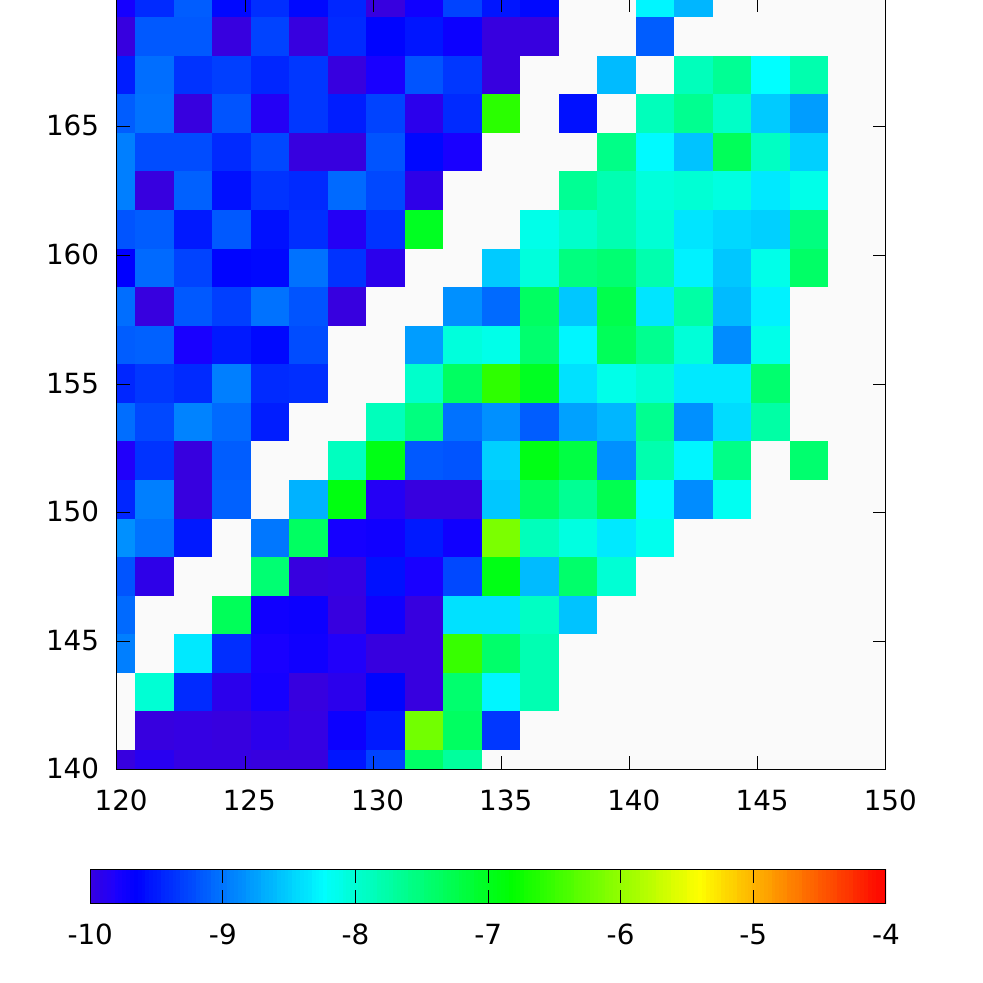}\includegraphics[width=0.5\linewidth]{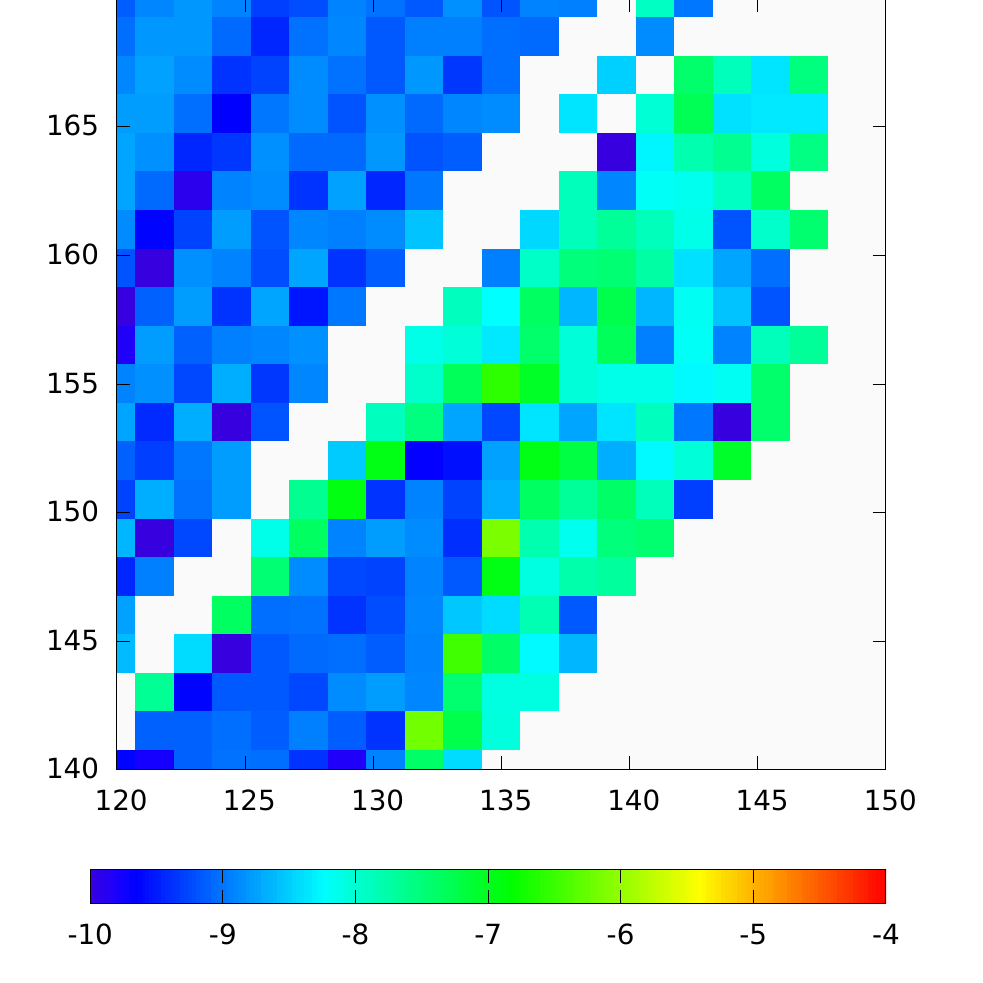}\\
  \setlength{\unitlength}{1cm}
\begin{picture}(.001,0.001)
\put(-7.5,4.5){\rotatebox{90}{$\Dv$}}
\put(0,4.5){\rotatebox{90}{$\Dv$}}
\put(-3.8,1.5){{$\zeta$}}
\put(3.8,1.5){{$\zeta$}}
\put(-3.8,0.5){{$\Deta'$}}
\put(3.3,0.5){{$\Deta'$}}
\end{picture}

\vspace{-0.05cm}

\includegraphics[width=0.5\linewidth]{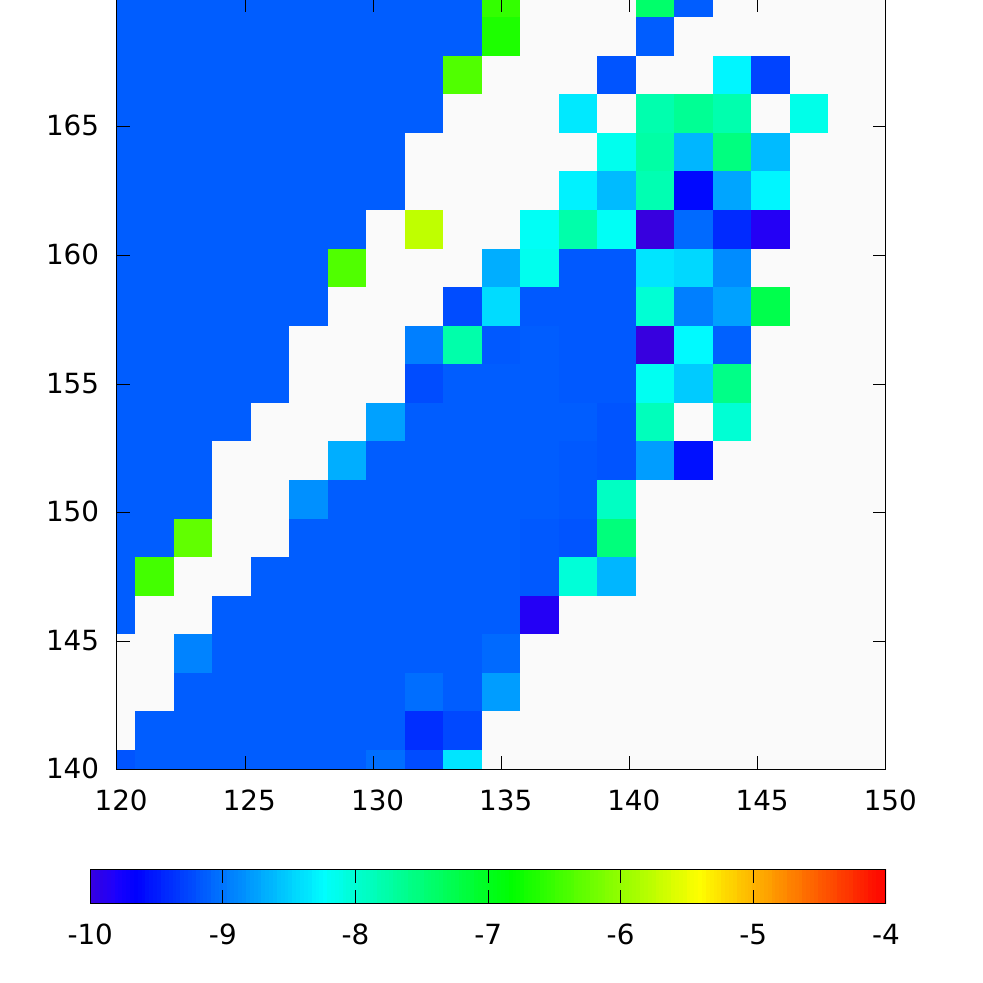}\includegraphics[width=0.5\linewidth]{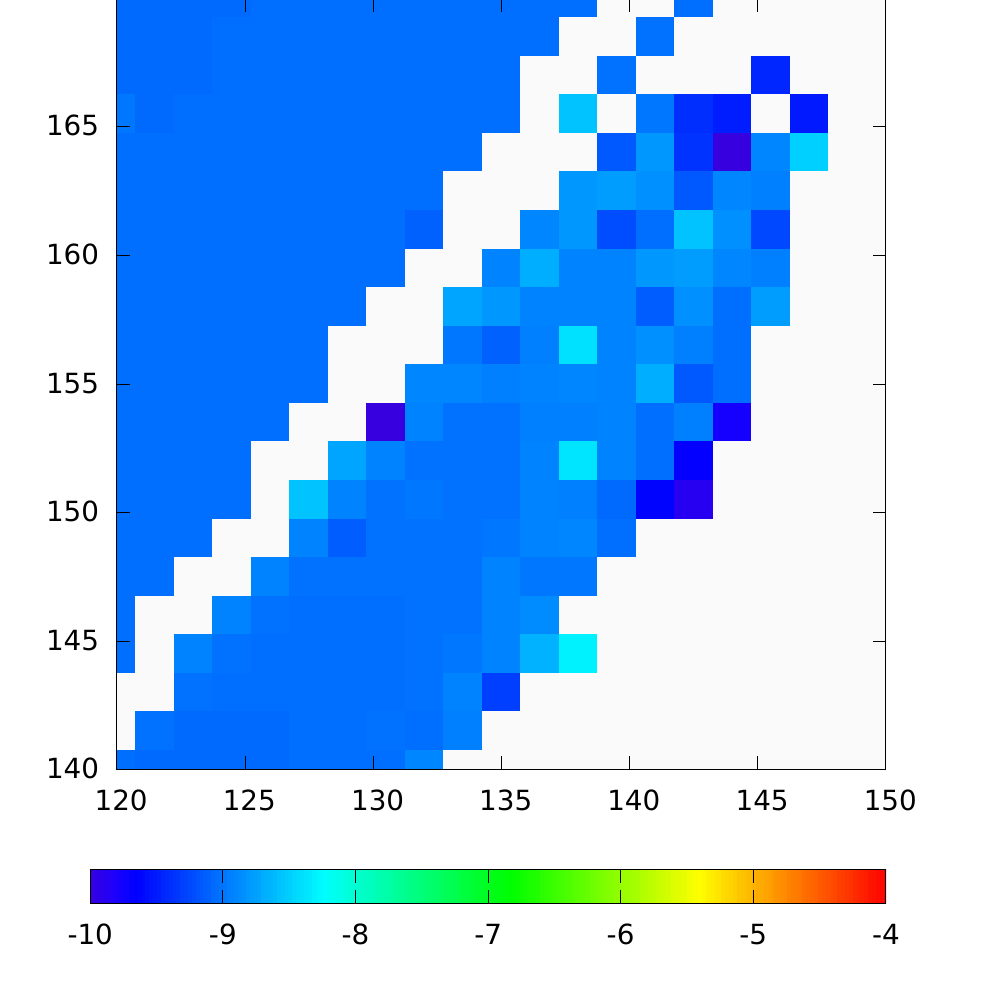}\\
  \setlength{\unitlength}{1cm}
\begin{picture}(.001,0.001)
\put(-7.5,4.5){\rotatebox{90}{$\Dv$}}
\put(0,4.5){\rotatebox{90}{$\Dv$}}
\put(-3.8,1.5){{$\zeta$}}
\put(3.8,1.5){{$\zeta$}}
\put(-3.8,0.5){{$\Deta'$}}
\put(3.3,0.5){{$\Deta'$}}
\end{picture}

\vspace{-0.5cm}

\caption{\label{fig:compint} diffusion de $\eta$ pour $e_1=e_2=0.7$, $m_1=m_2=10^{-5}$, avec à gauche l'intégrateur à pas constant SABA4 ($dt=0.001$~an) et à droite l'intégrateur DOPRI avec un pas variable (pas minimal $dt=0.001$) avec une précision de $10^{-14}$. Intégration sur $10/\eps$~an pour la ligne du haut et $100/\eps$~an sur la ligne du bas.}
\end{center}
\end{figure}
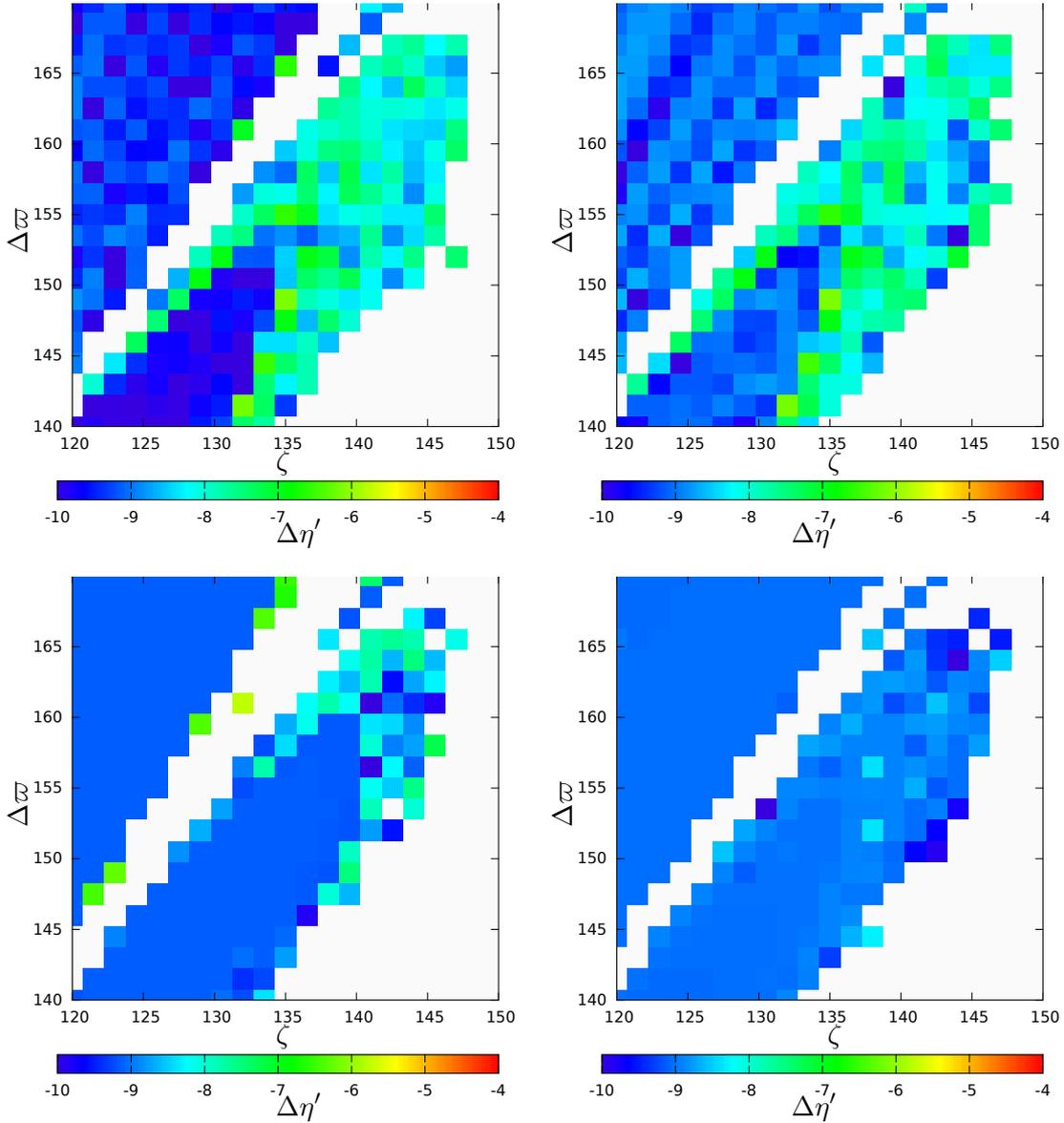

 L'intégrateur SABA4 \citep{LaRo2001} que nous avons utilisé jusqu'à présent a l'avantage de conserver les intégrales premières du mouvement. Il s'agit cependant d'un intégrateur à pas constant. Or, pour de grandes excentricités, ce pas doit être adapté au moment où les planètes sont proches de leur périastres respectifs ($dt=0.001$ semble être un bon compromis pour des conditions initiales $e_j=0.7$). Pour vérifier la stabilité de systèmes avec des masses de $\eps=10^{-6}$ sur des temps long devant la fréquence séculaire $g = \gO(\eps)$, nous verrons dans la prochaine section que nous devons intégrer les trajectoires sur des temps de l'ordre de $100/\eps$. Il est donc beaucoup plus rapide d'utiliser un intégrateur à pas variable pour ces cas excentriques (de plus nous n'intégrons que le problème plan). On utilise l'intégrateur DOPRI (Runge-Kutta (7)8) avec un pas minimal de $0.001$ et une précision locale de $10^{-14}$ (une tolérance plus grande dégrade significativement les résultats, une tolérance inférieure ralenti l'intégration). Néanmoins cet intégrateur ne conserve pas les intégrales premières du mouvement, et on constate une dérive linéaire de $\eta$ dans les zones régulières quand on augmente le temps d'intégration pour une trajectoire donnée \citep{RoLa2001}. \\
 
Sur la figure~\ref{fig:compint}, nous comparons les résultats des deux intégrateurs pour les conditions initiales de la région encadrée en noir sur la figure~\ref{fig:glob_box}. Les trajectoires des graphes de gauche ont été intégré par SABA4 et ceux de droite par DOPRI. En haut nous avons des intégrations sur $10/\eps$ périodes orbitales et en bas sur $100/\eps$. 

Les deux intégrateurs donnent des résultats similaires sur $10/\eps$ périodes orbitales. La diffusion de $\eta$ est plus importante dans le domaine des configurations $G$ (centré en $\zeta=140^\circ$, $\Dv=160^\circ$) que dans le domaine troyen (en bas à gauche) et le domaine quasi-satellite (en haut à gauche), mais la diffusion s'homogénéise pour des intégration sur $100/\eps$. Cette diffusion plus importante sur des temps de $10/\eps$ peut donc s'expliquer par une mauvaise estimation de $\eta$ sur la première et la seconde moitié de l'intégration. En effet, chaque tranche a une durée de $5/\eps$ période orbitale, et $g$ est de l'ordre de $\eps$. Si la valeur du moyen mouvement instantané $n$ est sensible au variations séculaires dans ce domaine, une intégration plus longue permet d'obtenir une meilleur estimation de sa valeur moyenne.

Sur $100/\eps$ périodes orbitales, la taille et la forme du domaine de stabilité est similaire dans les deux cas, mais la diffusion de $\eta$ dans les zones régulières de l'espace des phases a augmenté d'un facteur $10$ pour l'intégrateur DOPRI. La stabilité des différentes configurations est discutée dans la section suivante.

On constate que la plupart des trajectoires ayant une diffusion similaire à celle des zones régulières pour une durée d'intégration donnée ($10/\eps$ sur les graphes du haut) ne sont pas éjecté pendant la même simulation sur des temps $10$ fois plus long (graphes du bas).\\

A partir de maintenant nous utilisons l'intégrateur à pas variable pour vérifier la stabilité les différentes configurations identifiées sur la figure~\ref{fig:glob_box}. Avec cet intégrateur et une précision de $10^{-14}$, la diffusion dans les domaines régulier (voir par exemple les points aux alentour de $\zeta=120^\circ$ et $\Dv=170^\circ$ dans le domaine quasi-satellite) est de l'ordre de $\Deta' \approx 10^{-9}$ pour les intégrations sur $10^6$ périodes orbitales, $\Deta' \approx 10^{-8}$ pour $10^7$ périodes orbitales et $ \Deta' \approx 10^{-7}$ pour $10^8$ périodes orbitales, ce qui est cohérent avec la dérive linéaire observée par \cite{RoLa2001}. Si une trajectoire diffuse d'au plus $ \Deta' \approx 10^{k-15}$ pour une intégration sur $10^{k}$ périodes orbitales (avec l'intégrateur DOPRI), on considèrera qu'elle est stable sur des temps long devant $10^{k}$ périodes orbitales (au moins sur $10^{k+1}$ périodes orbitales). Nous définissons, à partir de l'équation (\ref{eq:Deta}), la diffusion $\Deta$ qui prend en compte la dérive due à l'erreur de méthode \citep{RoLa2001}:
\begin{equation}
\Deta = \Deta' - (k-6)\, ,
\label{eq:Deta}
\end{equation}

\an{
In this section we explain our choice regarding the integrator. The integrator SABA4 \citep{LaRo2001} that we used so far conserves the integrals of motion, which generally makes it ideal for long-term integrations. However, since its time step is fixed, it takes time to integrate highly eccentric orbits as it requires to chose a time step small enough to correctly integrate the trajectories when the planets are close from their periastron ($dt=0.001$ seems to be a good compromise for $e_j=0.7$ as initial condition). To check the stability of systems with earth masses ($\eps = 10^{-6}$) on a time scale long with respect to the secular frequency $g = \gO(\eps)$ requires to integrate over time of the order of $100/\eps$ (see next section).}
\an{ It is hence faster to use a variable-step integrator for the highly eccentric cases. We use the DOPRI integrator (Runge-Kutta (7)8) with a minimal step of $0.001$ and a local tolerance of $10^{-14}$ (a higher tolerance degrades the results, a lower tolerance slows down the integration). However this integrator does not conserve the integrals of motion, we hence observe a linear trend for the value of $\eta$ in the regular areas of the phase space as we increase the duration of the integrations for a given trajectory \citep{RoLa2001}.\\}
\an{In figure~\ref{fig:compint} we compare the results of both integrator (SABA4 and DOPRI), taking the black rectangle (figure~\ref{fig:glob_box}) as initial conditions. Left-hand graphs were obtained from SABA4, right-hand graph from DOPRI. Integrations over $10/\eps$~orbital periods in top graphs and $100/\eps$ in bottom graphs.\\}
\an{Both integrators give similar results over $10/\eps$~orbital periods: $\Deta'$ is larger in the $G$ domain (centred at $\zeta=140^\circ$, $\Dv=160^\circ$) than in the trojan (bottom left) and quasi-satellite (top left) domains. But the diffusion becomes more homogeneous when we integrate over $100/\eps$~orbital periods. The higher diffusion in the $10/\eps$~orbital periods case might be due to a bad determination of $\eta$: each half of the integration is $5/\eps$ orbital periods long, which might be not long enough to average the effect of the frequency $g=\gO(\eps)$.\\}
\an{We find that most of the trajectories having a diffusion similar to the orbits of the regular areas of the phase space over $10/\eps$ orbital periods are not ejected when we integrate them $10$ times longer.\\
From now on we use the variable-step integrator to check the stability of the areas identified in figure~\ref{fig:glob_box}.} 
\an{With this integrator and a local tolerance of $10^{-14}$, the diffusion in the vicinity of the regular areas (for example the vicinity of $\zeta=120^\circ$,$\Dv=170^\circ$ in the quasi-satellite domain) is of the order of $\Deta \approx 10^{-9}$ for integrations over $10^6$~orbital periods,}
\an{$\Deta \approx 10^{-8}$ for $10^7$~orbital periods and $ \Deta \approx 10^{-7}$ for $10^8$~orbital periods, 
which is consistent with the linear trend observed by \cite{RoLa2001}. }
\an{If for a given trajectory $ \Deta' \approx 10^{k-15}$ or less for an integration over $10^{k}$~orbital periods, we consider that this trajectory is stable for a time long with respect to $10^{k}$~orbital periods (at least $10^{k+1}$~orbital periods). From equation (\ref{eq:Deta}), we define the diffusion $\Deta$ (eq. \ref{eq:Deta}) that takes into account the diffusion due to the error of the method \citep{RoLa2001}. }

 \subsubsection{Configuration $G$ et voisinage de $AL_3$}

  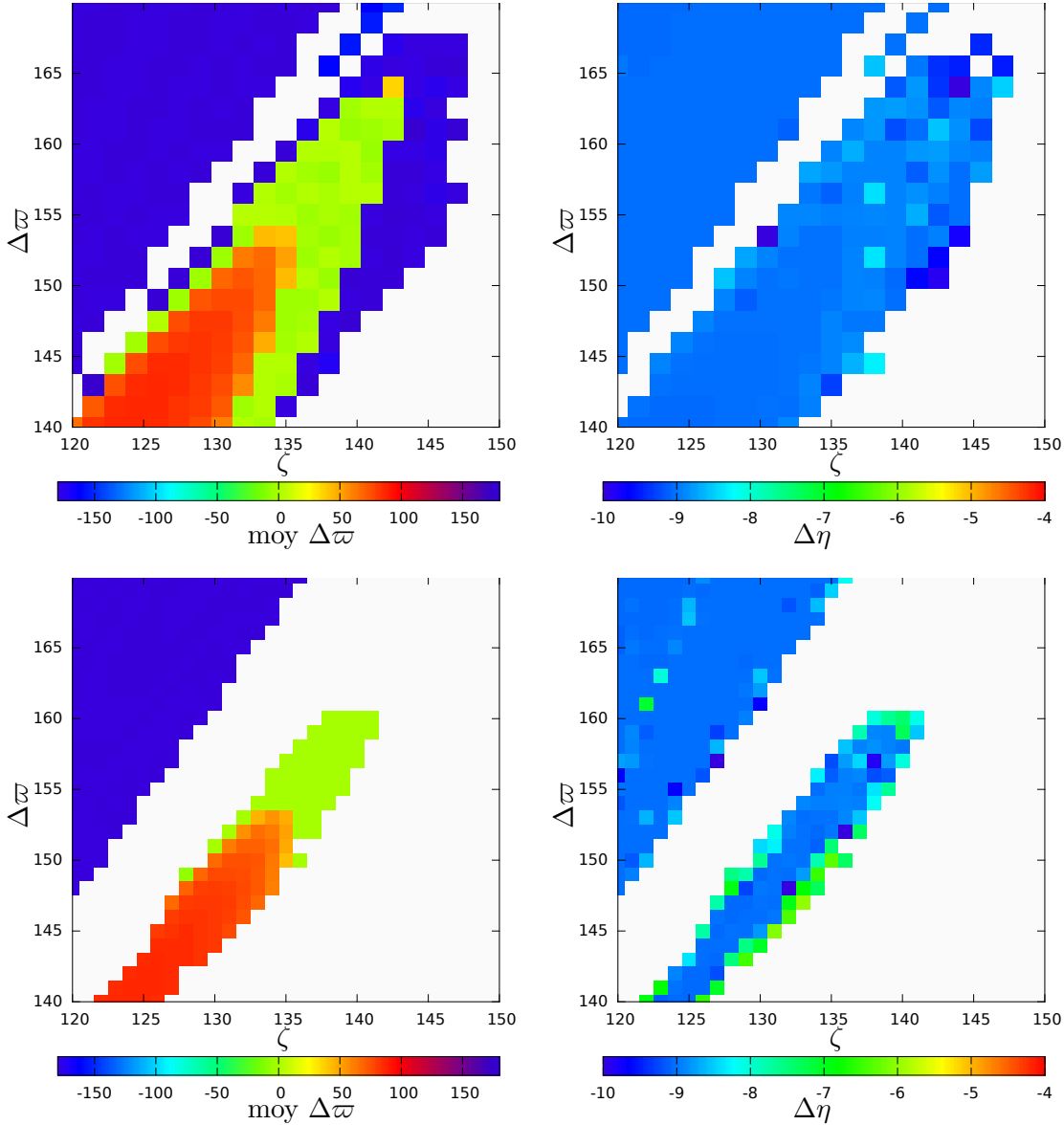
\begin{figure}[h!]
\begin{center}
\includegraphics[width=0.5\linewidth]{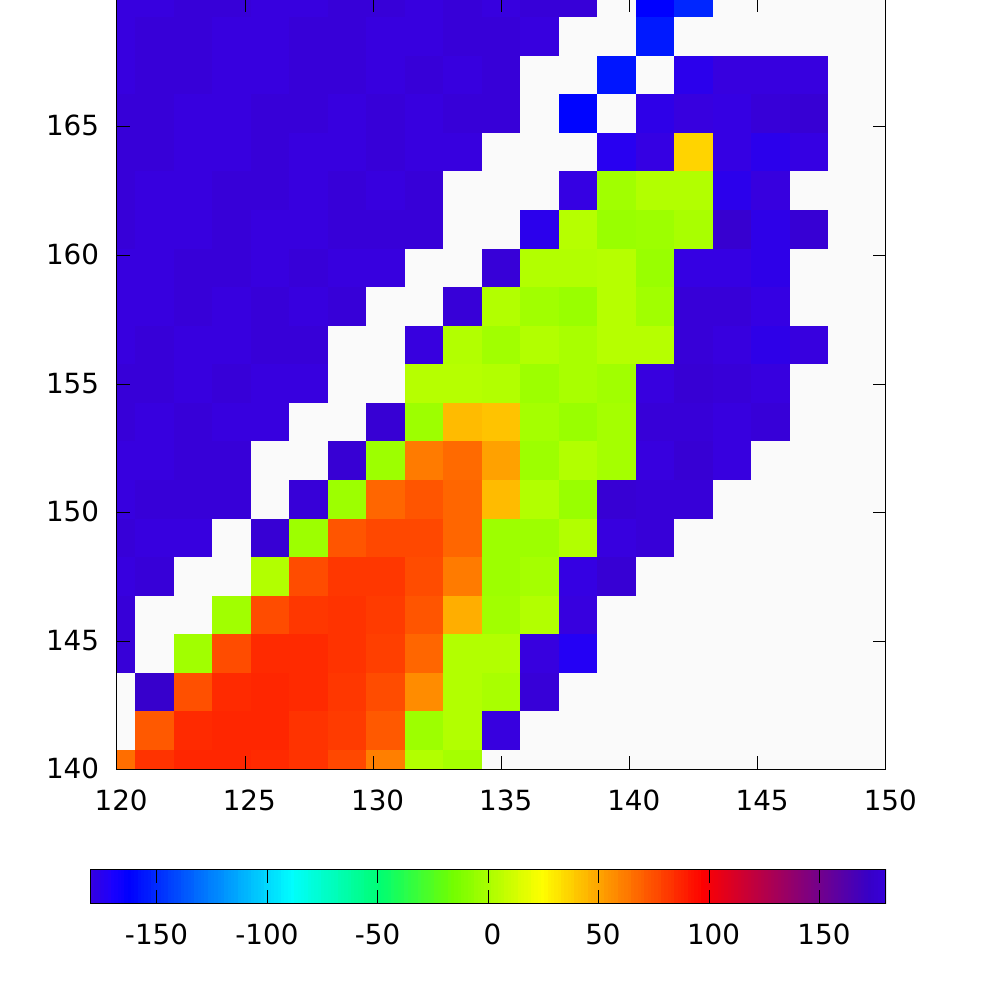}\includegraphics[width=0.5\linewidth]{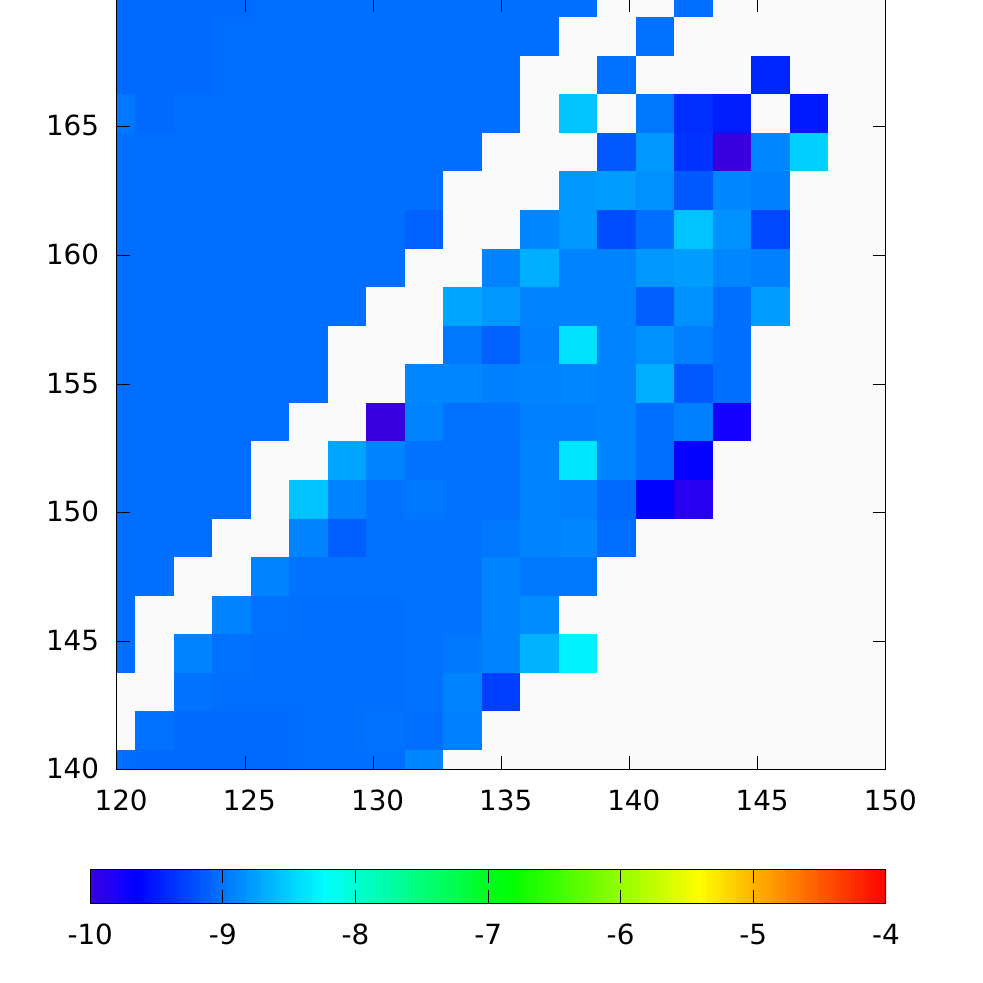}\\
  \setlength{\unitlength}{1cm}
\begin{picture}(.001,0.001)
\put(-7.5,4.5){\rotatebox{90}{$\Dv$}}
\put(0,4.5){\rotatebox{90}{$\Dv$}}
\put(-3.8,1.5){{$\zeta$}}
\put(3.8,1.5){{$\zeta$}}
\put(-4.2,0.5){{moy $\Dv$}}
\put(3.3,0.5){{$\Deta$}}
\end{picture}

\vspace{-0.05cm}

\includegraphics[width=0.5\linewidth]{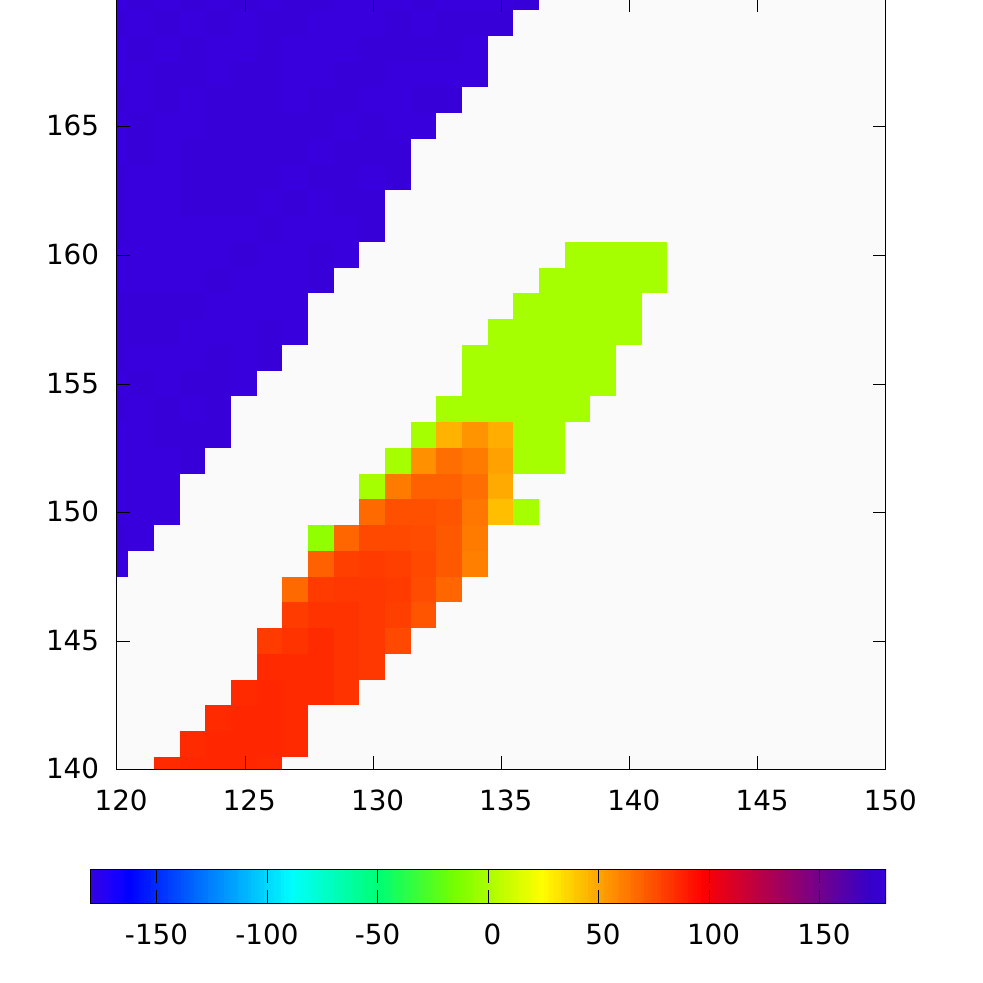}\includegraphics[width=0.5\linewidth]{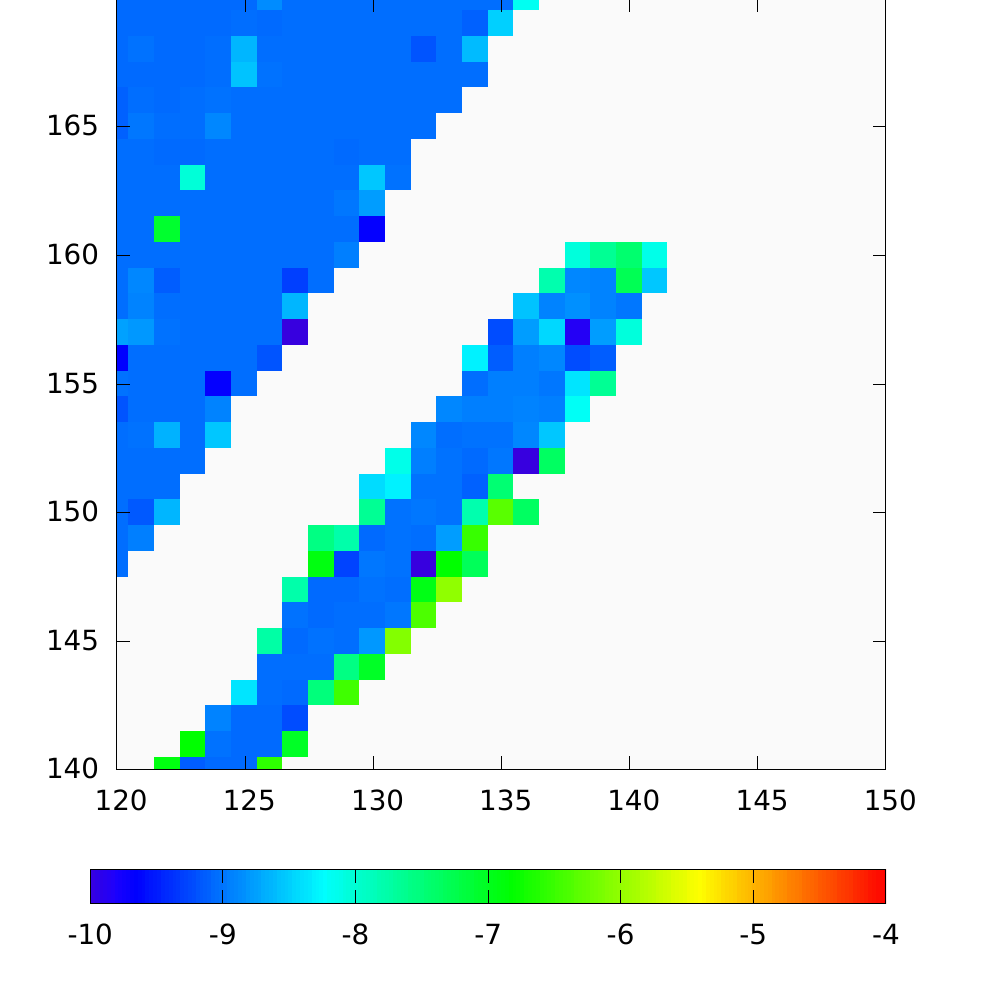}\\
  \setlength{\unitlength}{1cm}
\begin{picture}(.001,0.001)
\put(-7.5,4.5){\rotatebox{90}{$\Dv$}}
\put(0,4.5){\rotatebox{90}{$\Dv$}}
\put(-3.8,1.5){{$\zeta$}}
\put(3.8,1.5){{$\zeta$}}
\put(-4.2,0.5){{moy $\Dv$}}
\put(3.3,0.5){{$\Deta$}}
\end{picture}

\vspace{-0.5cm}

\caption{\label{fig:stab_G} Domaine de stabilité de la configuration $G$ pour $\eps=m_1/m_0=m_2/m_0=10^{-5}$ pour les graphes du haut et $\eps=10^{-4}$ pour les graphes du bas. Intégration sur $100/\eps$ dans les deux cas. Les graphes de gauche représentent la valeur moyenne de $\Dv$ sur la trajectoire pour pouvoir identifier la position de la séparatrice. Les graphes de droite représentent la diffusion de $\eta$ pour chaque trajectoire.}
\end{center}
\end{figure}

Les trajectoires passant proche de $AL_3$ sont stables sur des temps long par rapport à $100/\eps$ pour $\eps=m_1/m_0=m_2/m_0=10^{-5}$. En effet, ces trajectoires (en rouge sur le graphe de gauche figure~\ref{fig:stab_G}) ont une diffusion similaire aux orbites régulières du domaine des quasi-satellites (voir tous les graphes des figures~\ref{fig:compint} et le graphe en haut à droite \ref{fig:stab_G}). Cela reste vrai pour $\eps=10^{-4}$, voir les graphes du bas sur la figure~\ref{fig:stab_G}.\\

Le domaine de la configuration $G$, situé au-dessus de la séparatrice identifiée par l'interface orange/vert sur les graphes de gauche de la figure~\ref{fig:stab_G}, contient également des orbites stables sur des temps longs devant $100/\eps$ pour $\eps=10^{-5}$ et $\eps=10^{-4}$. En effet, pour ces deux valeurs des masses, ce domaine contient des orbites dont la diffusion est comparable aux orbites régulières du système.

 De plus, pour $\eps=10^{-5}$ on remarque que le domaine des configurations $G$, possède la même séparation que dans les autres domaines: des configurations librent autour de $\Dv=0^\circ$ (vert sur le graphe en haut à gauche de la figure \ref{fig:stab_G}) et d'autres autour de $\Dv=180^\circ$ (bleu).\\

\an{
Trajectories passing in the vicinity of $AL_3$ are stable for long durations with respect to $100/\eps$ for $\eps=m_1/m_0=m_2/m_0=10^{-5}$. Indeed, these trajectories (in red in the left-hand graph, figure~\ref{fig:stab_G}) have a diffusion similar to the trajectories of the quasi-satellite domain (see the figure~\ref{fig:compint} and the top-right graph figure~\ref{fig:stab_G}). That still holds for $\eps=10^{-4}$, see the bottom-right graph, figure~\ref{fig:stab_G}.\\}
\an{Trajectories of the $G$ domain, above the separatrix identified by the transition orange/green of the left-hand graphs, are also stable for long durations with respect to $100/\eps$~orbital periods for both $\eps=10^{-5}$ and $10^{-4}$. Moreover, in the $\eps=10^{-5}$ case, the $G$ domain is split into configurations that librate around $\Dv=0^\circ$ (in green on the left-hand graphs) and configurations that librate around $\Dv=180^\circ$ (in blue). The same split was observed in the horseshoe, tadpole and quasi satellite domains.}

\subsubsection{Reconnexion des familles $\ol \cF^g$ }

  \begin{figure}[h!]
\begin{center}
\includegraphics[width=0.5\linewidth]{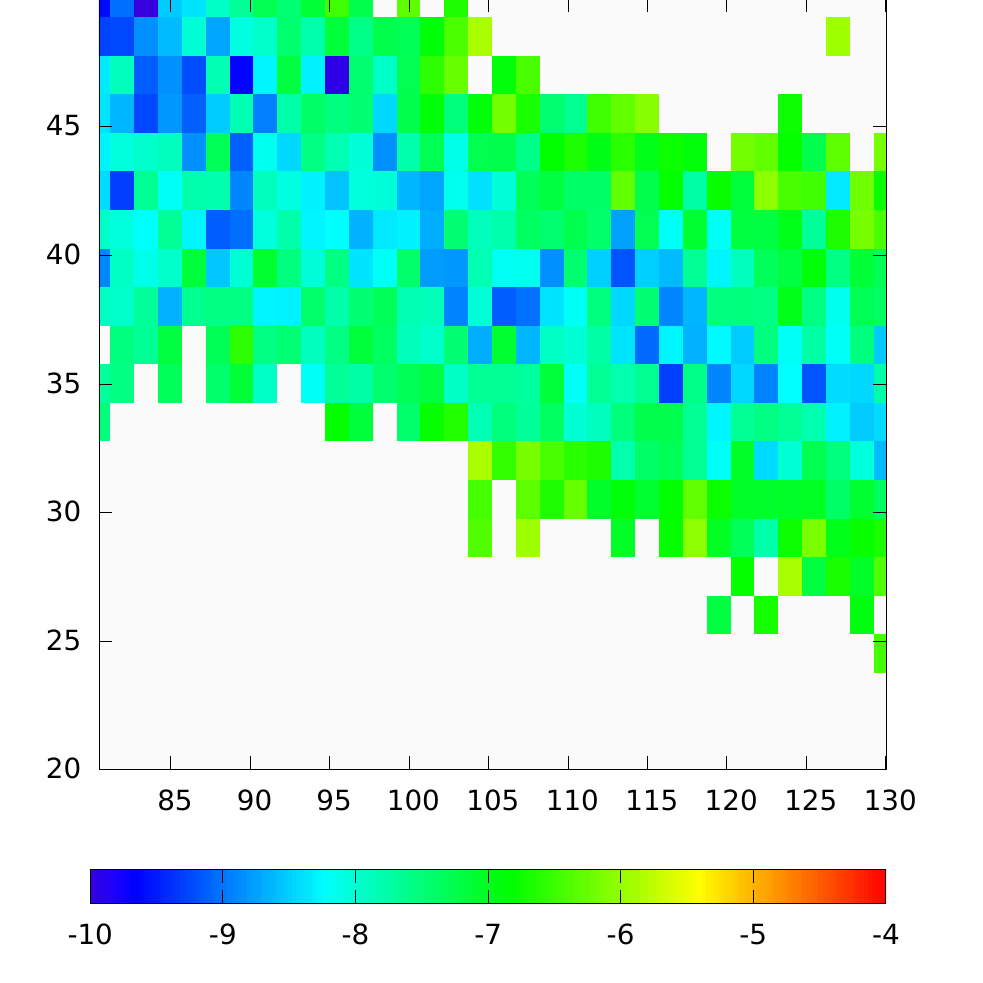}\includegraphics[width=0.5\linewidth]{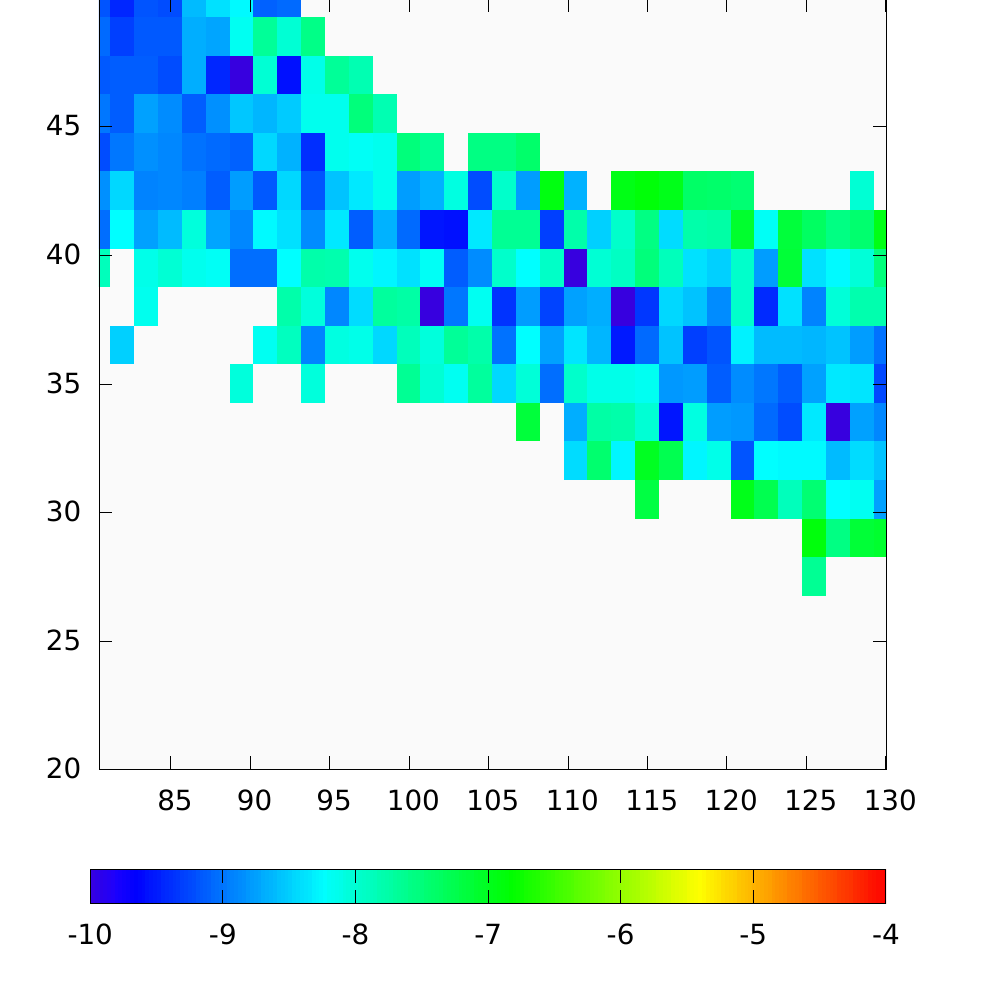}\\
  \setlength{\unitlength}{1cm}
\begin{picture}(.001,0.001)
\put(-7.5,4.5){\rotatebox{90}{$\Dv$}}
\put(0,4.5){\rotatebox{90}{$\Dv$}}
\put(-3.8,1.5){{$\zeta$}}
\put(3.8,1.5){{$\zeta$}}
\put(-4.2,0.5){{$\Deta$}}
\put(3.3,0.5){{$\Deta$}}
\end{picture}

\vspace{-0.05cm}

\includegraphics[width=0.5\linewidth]{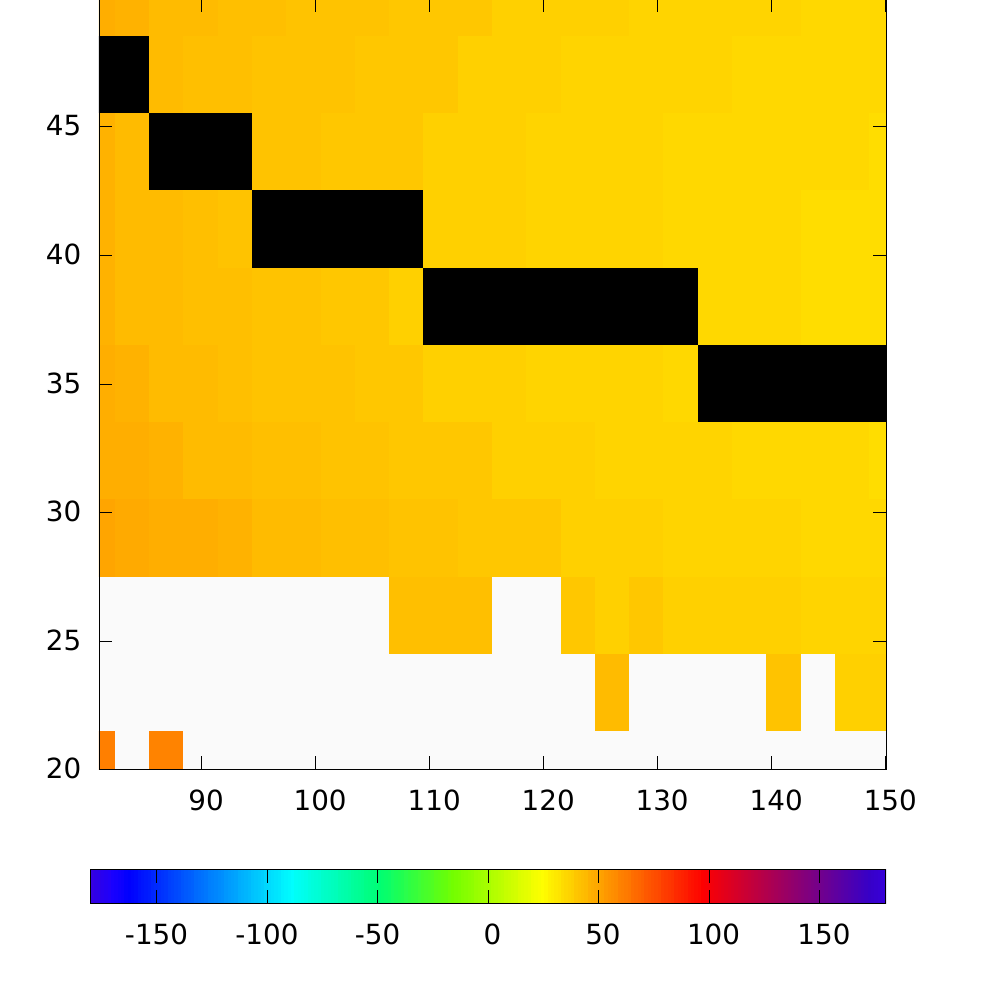}\includegraphics[width=0.5\linewidth]{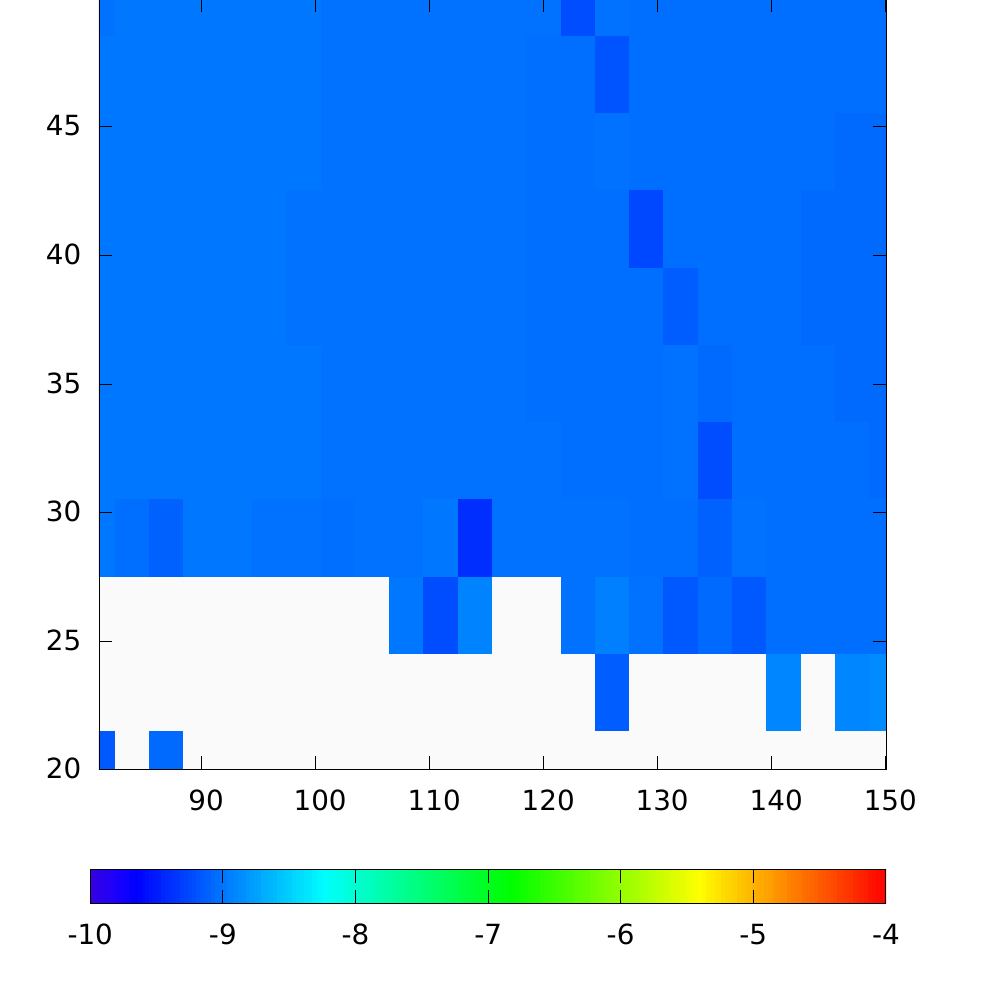}\\
  \setlength{\unitlength}{1cm}
\begin{picture}(.001,0.001)
\put(-7.5,4.5){\rotatebox{90}{$\Dv$}}
\put(0,4.5){\rotatebox{90}{$\Dv$}}
\put(-3.8,1.5){{$\zeta$}}
\put(3.8,1.5){{$\zeta$}}
\put(-4.2,0.5){{moy $\Dv$}}
\put(3.3,0.5){{$\Deta$}}
\end{picture}

\vspace{-0.5cm}

\caption{\label{fig:stab_recon} Stabilité au voisinage de la reconnexion des $\ol \cF^g$ pour $\eps=m_1/m_0=m_2/m_0=10^{-6}$ pour les graphes du bas (en intégrant sur $100/\eps$ périodes orbitales) et $\eps=10^{-5}$ pour les graphes du haut ($10/\eps$~périodes orbitales à gauche et $100\eps$ à droite). Le graphe en bas à gauche représente la valeur moyenne de $\Dv$. Les orbites proches de $\ol \cF^g$, vérifiant la relation (\ref{eq:condFb1}) avec $\epsilon_g= 3^\circ$, sont représentées en noir. Les autres graphes représentent la diffusion de $\eta$ pour chaque trajectoire.}
\end{center}
\end{figure}

Nous avons vu en section \ref{sec:Coo2meg} que les familles $\ol \cF^g_{L_4}$, $\ol \cF^g_{\gH\cS_1}$ et $\ol \cF^g_{\gH\cS_2}$, ainsi que leur équivalent du voisinage de $L_5$ semblent se reconnecter pour des valeurs des excentricités semblables à celles nécessaires à la reconnexion des familles $\ol \cF^\nu$. Dans le problème séculaire, la reconnexion des $\ol \cF^g$ crée une famille continue d'orbites périodiques passant par $L_4$, les domaines $\cA\gH\cS_1$, $\cA\gH\cS_2$, et $L_5$, avant de revenir à $L_4$ en repassant par les domaines $\cA\gH\cS_2$ et $\cA\gH\cS_1$ (voir les pixels noirs sur la figure~\ref{fig:glob_e7_m6}). 

Nous nous demandons si il est effectivement possible de passer continûment (par le biais d'orbites quasi-périodiques stables) des configurations fers à cheval au voisinage de $L_4$ dans le problème complet. Il est important de noter que, comme illustré par les symboles $*$ et $\times$ sur la figure~\ref{fig:glob_e7}, notre variété de conditions initiales intersecte les mêmes domaines de fers à cheval asymétriques en des endroits différents: le domaine $\cA \gH \cS_1$ est intersecté en $\Dv > 0$ et $\cA \gH \cS_2$ est intersecté en $\Dv < 0$. Les orbites qui nous permettraient de passer continûment du voisinage de $L_4$ au fer à cheval asymétriques pour $\zeta>180^\circ$ sont donc, si elles existent, les mêmes  qui nous permettraient de passer continûment du voisinage de $L_4$ au fer à cheval asymétriques pour $\zeta<180^\circ$. il est donc suffisant d'étudier les trajectoires issues du rectangle violet de la figure~\ref{fig:glob_box} pour conclure sur cette reconnexion.\\

Pour $\eps=10^{-5}$ et $e_1=e_2=0.7$, on trouve une zone d'orbites instables au voisinage de $\zeta=100^\circ$, $\Dv=50^\circ$ empêchant cette reconnexion dans le problème complet (voir figure~\ref{fig:glob_e7}). En prenant $e_1=e_2=0.62$, les graphes du haut de la figures~\ref{fig:stab_recon} indiquent que ce passage existe au moins sur des temps de l'ordre de $100/\eps$ périodes orbitales. Cependant, les orbites sur le bord de cette reconnexion ont une diffusion non négligeable, et seule les trajectoires dans le voisinage immédiat de $\ol \cF^g$ semblent stable sur des temps long devant $100/\eps$ périodes orbitales (comparer les intégrations sur $10/\eps$ périodes orbitales à gauche et $100/\eps$ périodes orbitales à droite).

En revanche, pour des masses plus faibles ($\eps=10^{-6}$), cette reconnexion existe sur des temps longs devant $100/\eps$~périodes orbitales pour $e_1=e_2=0.7$ (voir les graphes du bas figure~\ref{fig:stab_recon}).\\

\an{
We saw in section \ref{sec:Coo2meg} that the families $\ol \cF^g_{L_4}$, $\ol \cF^g_{\gH\cS_1}$ and $\ol \cF^g_{\gH\cS_2}$, and their equivalence in the vicinity of $L_5$, seemed to reconnect one another for eccentricities similar to the ones necessary to the reconnection of the $\ol \cF^\nu$ families. In the secular problem, the reconnection of the $\ol \cF^g$ families creates a continuous family of periodic orbit that passes through $L_4$, the $\cA\gH\cS_1$ domain, the $\cA\gH\cS_2$ domain, and $L_5$, before it comes back to $L_4$ passing though the $\cA\gH\cS_2$ and $\cA\gH\cS_1$ domains once again (see black pixels in figure~\ref{fig:glob_e7_m6}).\\
It is enough to check if there is a path of stable quasi periodic orbits allowing to pass continuously from the neighbourhood of $L_4$ to the $\cA\gH\cS_1$ domain. Indeed, as shown by the trajectories identified by the symbols $*$ and $\times$ in the figure~\ref{fig:glob_e7}, our reference manifold crosses the $\cA\gH\cS_2$ and $\cA\gH\cS_1$ domains in two different sections (one for $\zeta<180^{\circ}$ and the other for $\zeta>180^{\circ}$); the stability of the vicinity of $L_5$ is equivalent to the vicinity of $L_4$ by symmetry; and the stability within the $\cA\gH\cS_k$ domains will be confirmed in the next paragraph.
For $\eps=10^{-5}$ and $e_1=e_2=0.7$, we found an unstable area in the vicinity of ($\zeta=100^\circ$, $\Dv=50^\circ$), preventing the reconnection (see figure~\ref{fig:glob_e7}). However, taking  $e_1=e_2=0.62$ (top graphs in figure~\ref{fig:stab_recon}, we found that this reconnection is stable in the 3-body problem for duration of the order of $100/\eps$ orbital periods at least. However, the diffusion of the orbits on the border of this reconnection is not negligible, and only the orbits in the neighbourhood of $\ol \cF^g$ seem to be stable over long durations with respect to  $100/\eps$~orbital periods (compare the $10/\eps$~orbital periods (left) and $100/\eps$~orbital periods (right) in the top graphs in figure \ref{fig:stab_recon}).\\
For lower masses ($\eps=10^{-6}$) this reconnection is stable for long durations with respect to $100/\eps$~orbital periods for $e_1=e_2=0.7$, see the bottom graphs figure~\ref{fig:stab_recon}. 
}

\subsubsection{Configuration $\cA \gH \cS$}

  \begin{figure}[h!]
\begin{center}
\includegraphics[width=0.5\linewidth]{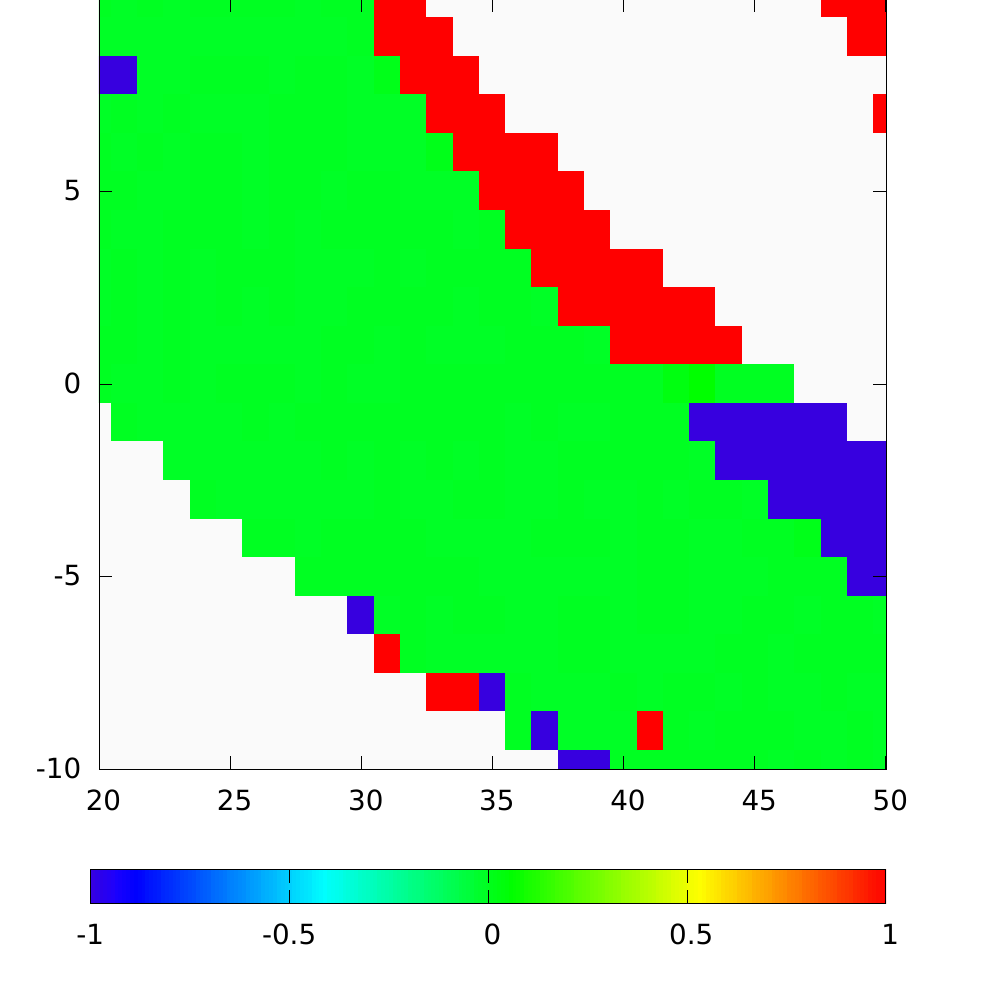}\includegraphics[width=0.5\linewidth]{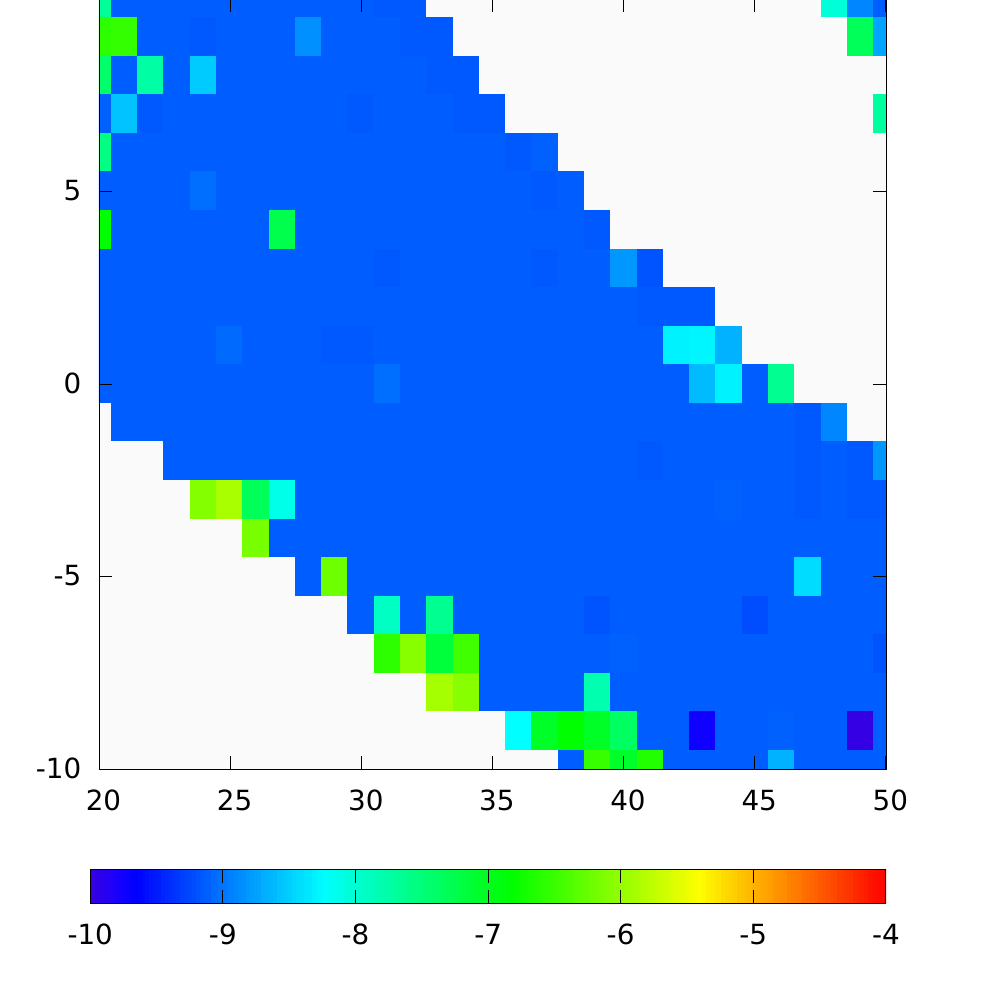}\\
  \setlength{\unitlength}{1cm}
\begin{picture}(.001,0.001)
\put(-7.5,4.5){\rotatebox{90}{$\Dv$}}
\put(0,4.5){\rotatebox{90}{$\Dv$}}
\put(-3.8,1.5){{$\zeta$}}
\put(3.8,1.5){{$\zeta$}}
\put(-4.2,0.5){{moy $\Dv$}}
\put(3.3,0.5){{$\Deta$}}
\end{picture}

\vspace{-0.5cm}

\caption{\label{fig:stab_AHS} Stabilité au voisinage de la reconnexion des $\ol \cF^g$ pour $\eps=m_1/m_0=m_2/m_0=10^{-6}$ pour les graphes du haut et $\eps=10^{-5}$ pour les graphes du bas. Intégration sur $10/\eps$ dans les deux cas. Les graphes de gauche représentent la valeur moyenne de $\Dv$. Les orbites proches de $\ol \cF^g$, vérifiant la relation (\ref{eq:condFb1}) avec $\epsilon_g= 1^\circ$, sont représentées en noir. Les graphes de droite représentent la diffusion de $\eta$ pour chaque trajectoire.}
\end{center}
\end{figure}

En ce qui concerne les configurations en fers à cheval asymétriques ($\cA \gH \cS_k$), on montre figure~\ref{fig:stab_AHS} qu'il existe un domaine stable pour $\cA \gH \cS_1$ (rouge sur le graphe de gauche) et $\cA \gH \cS_2$ (bleu) sur des temps longs devant $100/\eps$ périodes orbitales pour $\eps=10^{-4}$. Comme la taille des zones instable tend à diminuer quand $\eps$ diminue, des orbites stables $\cA \gH \cS_k$ existent pour tout $\eps \leq 10^{-4}$.\\

\an{Regarding the asymmetric horseshoe configurations ($\cA \gH \cS_k$), we show in figure~\ref{fig:stab_AHS} that orbits in the $\cA \gH \cS_1$ (red in the left-hand graph) and $\cA \gH \cS_2$ (blue) configurations are stable over long durations with respect to $100/\eps$~orbital periods for $\eps=10^{-4}$. As the size of the unstable areas tends to shrink as $\eps$ decreases, we conclude that stables orbits in the $\cA \gH \cS_k$ domains exist for $\eps \leq 10^{-4}$. }
%
%
%
%

\section{Étude des coorbitaux excentriques de masse quelconque}
\label{sec:mdif}


Lorsque $m_2 \neq m_1$, les équilibres de Lagrange elliptiques $L_k$ se trouvent dans le plan $e_1=e_2$, alors que les équilibres anti-lagranges $AL_k$ se trouvent en $m_1 e_1=m_2 e_2$ pour une valeur donnée du moment cinétique, au moins pour de faibles excentricités \citep[voir][]{HaVo2011}. De même, comme pour le cas circulaire, ces équilibres ne se situent pas en la même valeur de $Z$ (voir section \ref{sec:pperdi}). Des conditions initiales sur une variété à $Z$ et $\varPi$ constants ne peuvent donc pas être représentatives de l'ensemble des configurations pour une valeur du moment cinétique donné.

Dans un premier temps, on tracera de tels plans afin de comparer certaines propriétés de l'espace des phases quand $m_1 \neq m_2$ avec le cas $m_1=m_2$ vu dans les sections précédentes. Dans un second temps, on décrira un algorithme qui vise à déterminer une variété de référence pour une configuration coorbitale donnée. \\

\an{
When $m_1\neq m_2$, the plane of initial conditions $e_1=e_2$, $a_1=a_2$ does not describe all the possible orbital configurations: we knows for example that the orbits librating around the $L_4$ equilibrium have their eccentricities librating around $e_1=e_2$, while for the orbits librating around $AL_4$ they librate around $m_1 e_1 = m_2 e_2$ (at least for moderate eccentricities). We thus cannot represent all the co-orbital configurations by choosing a manifold of initial conditions for $Z$ and $\varPi$ constant.\\
However, the determination of the position of a representative manifold in the phase space is not trivial (although we did it in the neighbourhood of the circular $L_4$ equilibrium in section \ref{sec:VaReL4}). At first, we will integrate plans of initial conditions with $a_1=a_2$ and $\varPi$ constant to verify that the modifications of the phase space that were observed in the case $m_1=m_2$ also occur in the case $m_1\neq m_2$. Second, we will describe an algorithm to determine the position of the representative manifold in the phase space.}

\subsection{Étude de l'espace des phases}

Comme nous l'avons fait dans le cas $m_1=m_2$, on s'intéresse à des plans de conditions initiales à valeur du moment cinétique total fixé.

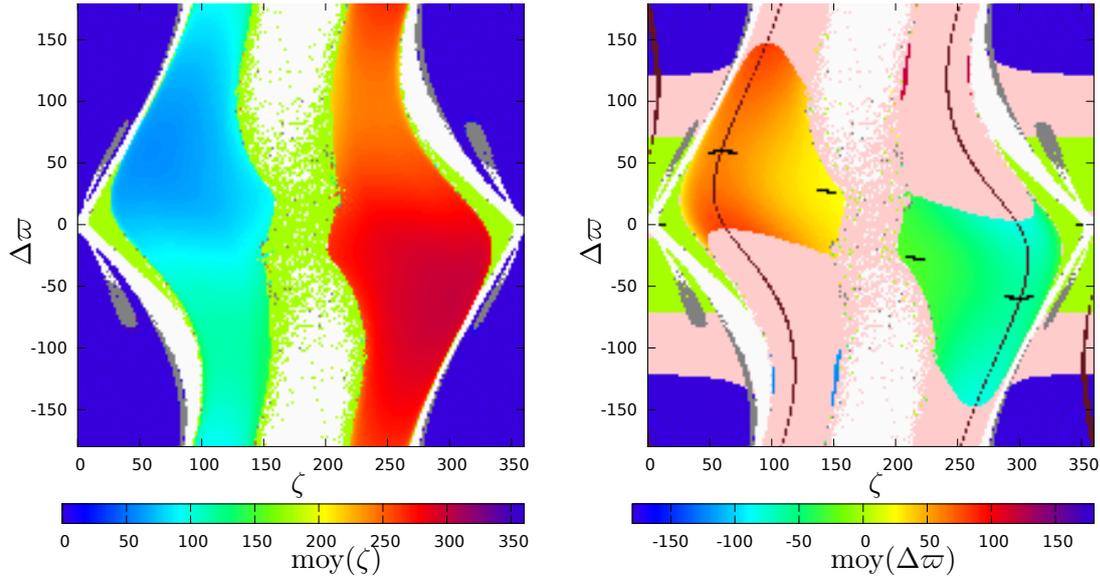
\begin{figure}[h!]
\begin{center}
\includegraphics[width=0.5\linewidth]{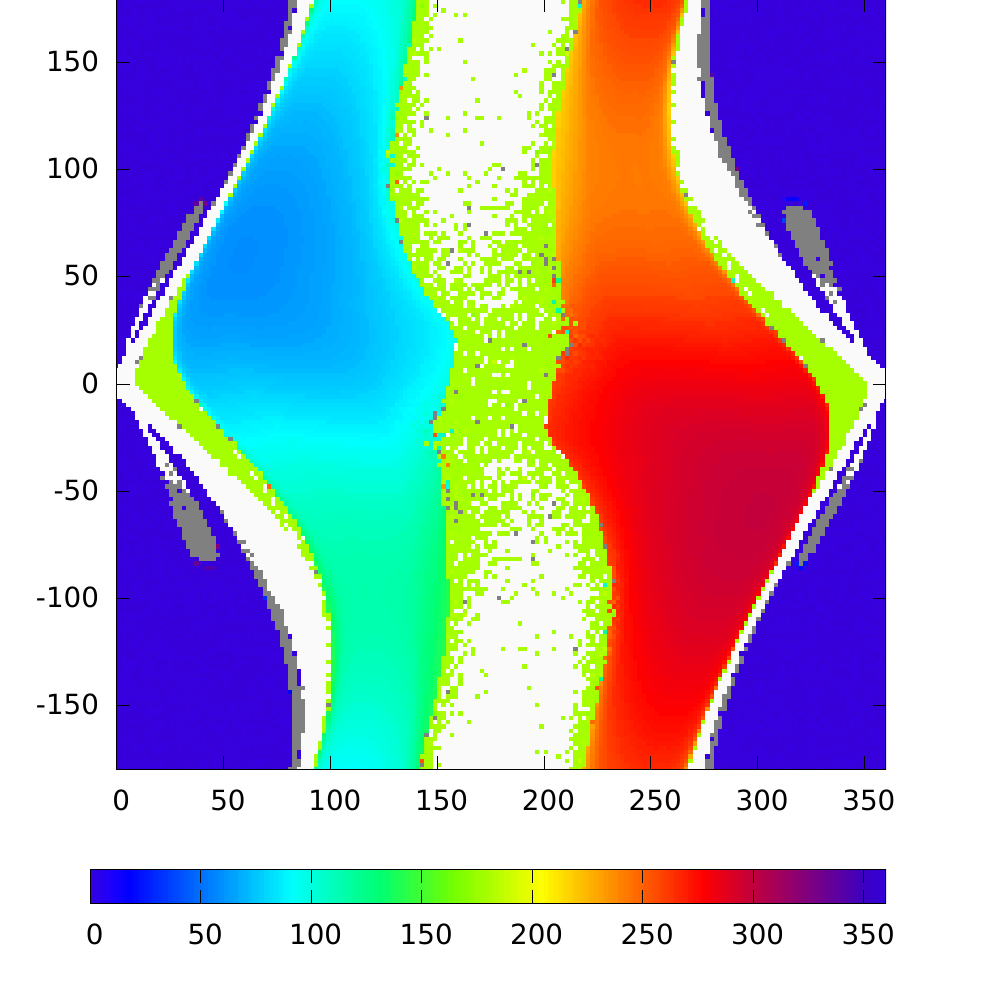}\includegraphics[width=0.5\linewidth]{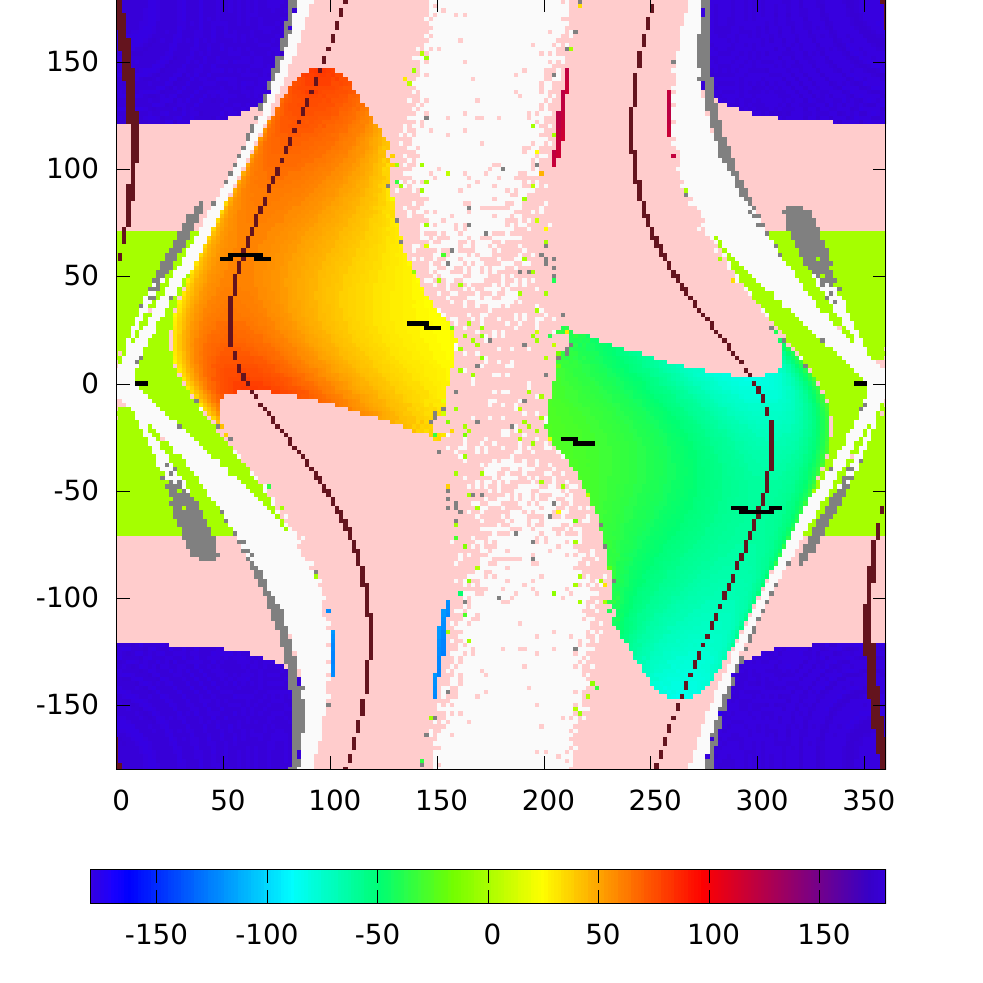}\\
  \setlength{\unitlength}{1cm}
\begin{picture}(.001,0.001)
\put(-7.5,4.5){\rotatebox{90}{$\Dv$}}
\put(0,4.5){\rotatebox{90}{$\Dv$}}
\put(-3.8,1.5){{$\zeta$}}
\put(3.8,1.5){{$\zeta$}}
\put(-3.8,0.5){{moy($\zeta$)}}
\put(3.3,0.5){{moy($\Dv$)}}
\end{picture}
\caption{\label{fig:glob_e4_md} Grille de conditions initiales pour $a_1=a_2=1\,ua$, $m_2=3m_1=1.5\, 10^{-5}$ et $e_1=e_2=0.4$. Sur la figure de gauche, le code couleur correspond à la valeur moyenne de $\zeta$ sur cette orbite. Sur la figure de droite, le code couleur représente la valeur moyenne de $\Dv$ sur chaque orbite, et la couleur saumon est affectée aux orbites pour lesquelles $\Dv$ circule. Les orbites proches de $\ol \cF^\nu$, donc celles vérifiant la relation (\ref{eq:condFb02}) avec $\epsilon_\nu=10^{-3.5}$, sont représentées en marron. Les orbites proches de $\ol \cF^g$, donc celles vérifiant la relation (\ref{eq:condFb1}) avec $\epsilon_g= 3^\circ$ sont représentées en noir. $\epsilon_a=10^{-5.5}$.\\
}
\end{center}
\end{figure}

\begin{figure}[h!]
\begin{center}
\includegraphics[width=0.5\linewidth]{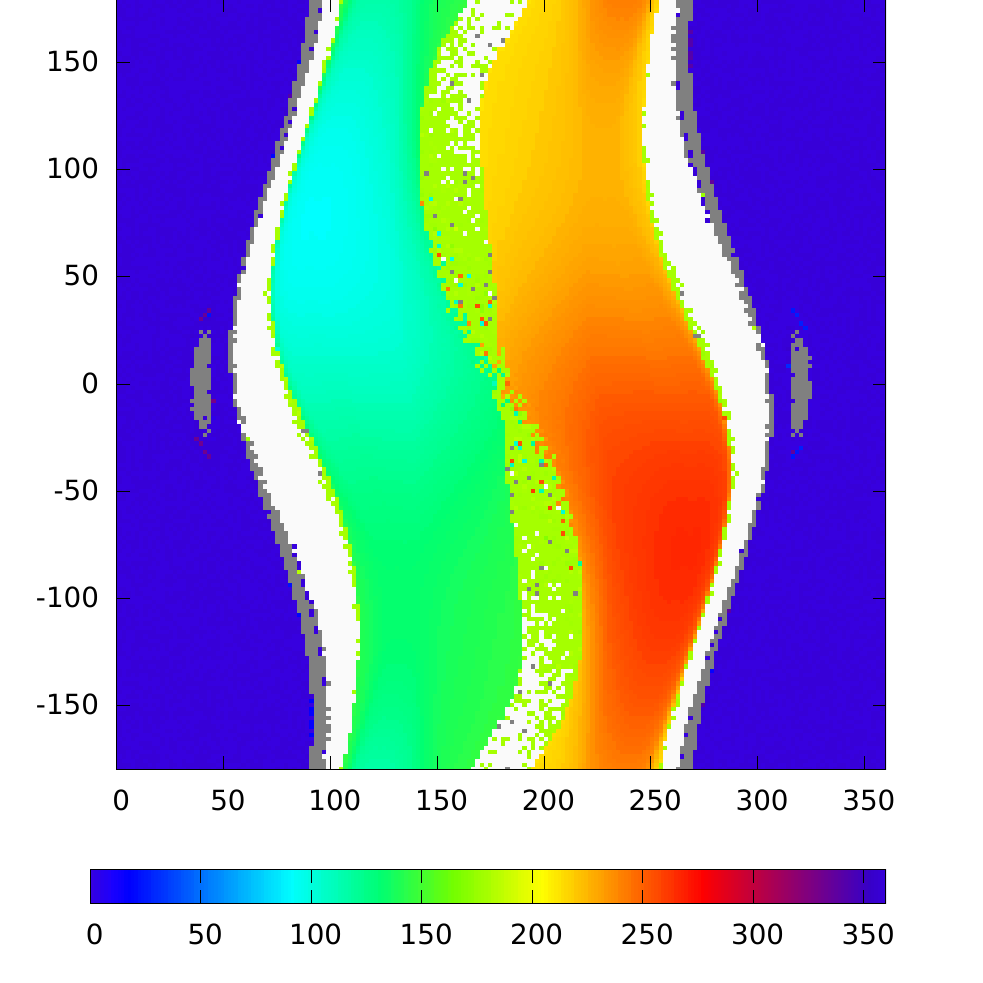}\includegraphics[width=0.5\linewidth]{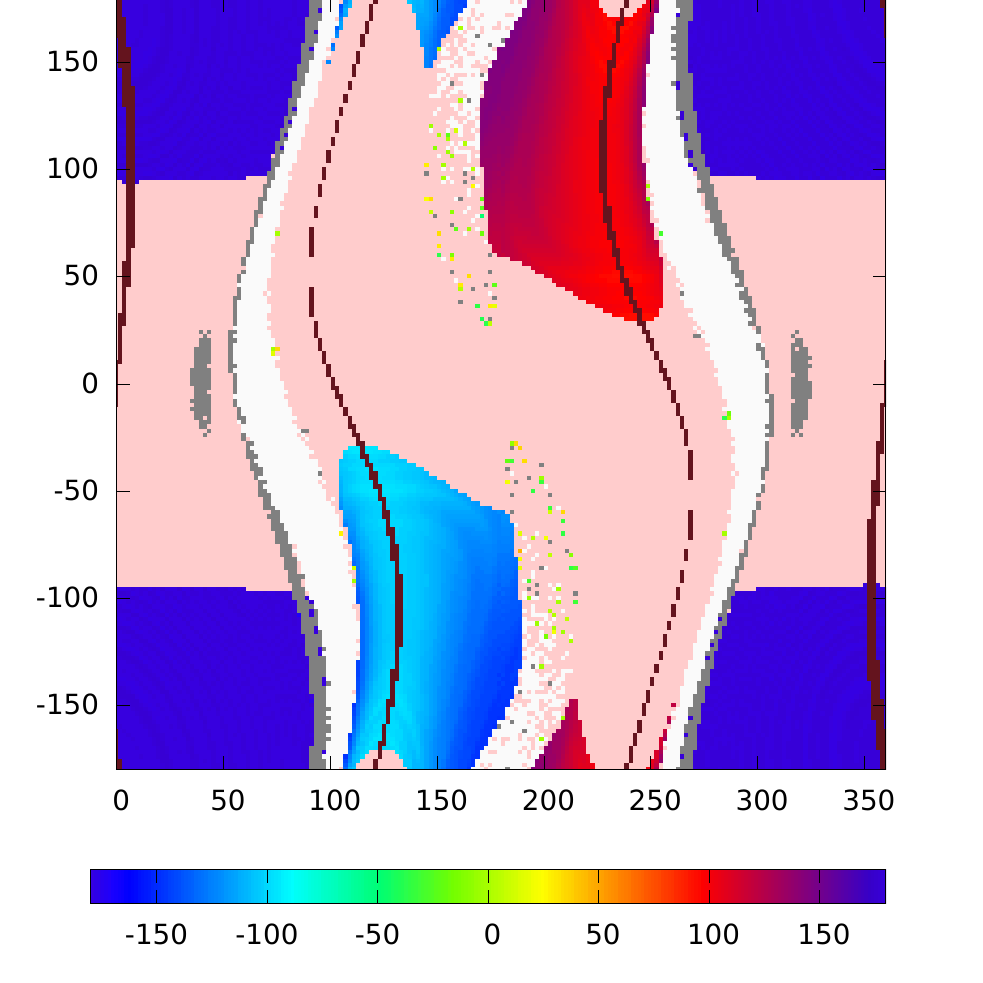}\\
  \setlength{\unitlength}{1cm}
\begin{picture}(.001,0.001)
\put(-7.5,4.5){\rotatebox{90}{$\Dv$}}
\put(0,4.5){\rotatebox{90}{$\Dv$}}
\put(-3.8,1.5){{$\zeta$}}
\put(3.8,1.5){{$\zeta$}}
\put(-3.8,0.5){{moy($\zeta$)}}
\put(3.3,0.5){{moy($\Dv$)}}
\end{picture}
\caption{\label{fig:glob_e4d_md} Grille de conditions initiales pour $a_1=a_2=1\,ua$, $m_2=3m_1=1.5\, 10^{-5}$, $e_1=0.7$ et $e_2\approx0.18$ de manière à ce que la valeur de $J_1$ soit identique à celle de la figure~\ref{fig:glob_e4_md}. Sur la figure de gauche, le code couleur correspond à la valeur moyenne de $\zeta$ sur cette orbite. Sur la figure de droite, le code couleur représente la valeur moyenne de $\Dv$ sur chaque orbite, et la couleur saumon est affectée aux orbites pour lesquelles $\Dv$ circule. Les orbites proches de $\ol \cF^\nu$, donc celles vérifiant la relation (\ref{eq:condFb02}) avec $\epsilon_\nu=10^{-3.5}$, sont représentées en marron. Les orbites proches de $\ol \cF^g$, donc celles vérifiant la relation (\ref{eq:condFb1}) avec $\epsilon_g= 3^\circ$ sont représentées en noir. $\epsilon_a=10^{-5.5}$.\\
}
\end{center}
\end{figure}

\subsubsection{$J_1(e_1=e_2=0.4)$}

Pour commencer, comparons les orbites issues de deux plans de conditions initiales identiques à l'exception de la répartition de la masse entre les coorbitaux: $m_1=m_2$ pour le plan de référence figure~\ref{fig:glob_e4}, et $m_2=3m_1$ pour le plan de conditions initiales figure \ref{fig:glob_e4_md}. La dynamique du degré de liberté ($Z,\zeta$) (graphe de gauche) est similaire sur ces deux plans. Les domaines de stabilité de chaque configuration semblent légèrement moins étendus dans le cas de deux masses différentes, mais il ne s'agit justement pas d'une variété représentative: les extremums des angles $\zeta$ et $\Dv$ peuvent être atteints autre part dans l'espace des phases. La dynamique du degré de liberté $\Dv$ est quant à elle totalement différente sur près de la moitié du plan de condition initiales: pour $m_2=3m_1$, une portion significative des trajectoires issues du plan voit $\Dv$ circuler (ces orbites sont identifiées en saumon sur le graphe de droite, figure~\ref{fig:glob_e4_md}). Sur ce même graphe, on constate que seules quelques orbites des configurations troyennes vérifient la condition (\ref{eq:condFb1}). Cela est cohérent avec les résultats de la section \ref{sec:DFFb} où nous verrons que la variété $\ol \cF^g$ est tangente au plan $e_1=e_2$ en les équilibres elliptiques $L_4$ et $L_5$ et qu'elle s'en éloigne quand on s'éloigne de ces points d'équilibres. La variété $\ol \cF^\nu$ semble quant à elle rester au voisinage du plan $a_1=a_2$ (elle y reste $\eps$ proche au moins dans le cas circulaire, voir section \ref{sec:pperdi}). 

Le plan de conditions initiales représenté en figure \ref{fig:glob_e4d_md} ne diffère de celui de la figure \ref{fig:glob_e4_md} que par la valeur initiale des excentricités ($e_1=0.7$ et $e_2\approx0.18$), mais la valeur du moment cinétique total reste la même. Il s'agit donc d'une coupe différente du même portrait de phase. Cette coupe est significativement plus proche des points $AL_4$ et $AL_5$ (situés en $e_1 \approx 0.67$, $e_2\approx0.22$ selon l'approximation linéaire). Ce plan de conditions initiales intersecte d'ailleurs le domaine des configurations voisines de ces points (zone centrée en $\zeta=140^\circ$, $\Dv=-100^\circ$ pour le voisinage de $AL_4$). Notons que dans le cas de masses différentes il n'est peut-être pas possible de passer directement d'orbites voisines de $L_4$ à des orbites voisines de $AL_4$ car les deux régions semblent toujours séparées par une région où $\Dv$ circule (vérifié pour $e_1 \in \{0, 0.15, 0.3 0.55 ,0.7\}$ et $e_2$ tel que $J_1=J_1(e_1=e_2=0.4)$). \\

\an{
In this section, we consider the case $m_2=3m_1=1.5 \times 10^{-5} m_0$. The figures~\ref{fig:glob_e4_md} to \ref{fig:glob_e4d_md} display the same information than in the previous section. In addition, on the right graphs the salmon color represents the initial conditions of the trajectories for which the angle $\Dv$ circulates (instead of librating in the other cases).\\}  
\an{The figure~\ref{fig:glob_e4_md} gives the mean value of the angles $\zeta$ and $\Dv$ when the initial conditions are taken across the plane $e_1=e_2=0.4$, $a_1=a_2$ (with $m_2=3m_1=1.5 \times 10^{-5} m_0$). 
On the left hand graph (evolution of the mean value of $\zeta$), the dynamics of the degree of freedom ($Z,\zeta$) seems not to change much from the case $m_1=m_2$ for the same value of the total angular momentum (compare with the figure~\ref{fig:glob_e4}). 
The stability domains seems to evolve, but we are not looking at a representative manifold, so we cannot conclude on the range of values that can assumes $\zeta$ in a given domain: the extrema of $\zeta$ and $\Dv$ could be reached somewhere else in the phase space. 
On the right hand graph, we can see that the dynamics of the degree of freedom ($\varPi,\Dv$) is different from the case $m_1=m_2$: $\Dv$ circulates for a large amount of the integrated trajectories (salmon color). 
The family $\ol \cF^\nu$ seems to be close to this plan of initial conditions, but $\ol \cF^g$ departs from it as soon as we depart from the $L_4$ equilibrium (this is consistent with the estimation of the position of $\ol \cF^g$ in the neighbourhood of $L_4$, see section \ref{sec:DFFN}).  }       

\an{
The plan of initial conditions integrated in figure~\ref{fig:glob_e4d_md} represents another section of the same phase space, with $e_1=0.7$ and $e_2 \approx 0.18$. This plane intersects the phase space closer from the trojan domain librating around the $AL_4$ equilibrium ( $e_1 \approx 0.67$, $e_2\approx0.22$ in the linear approximation). The trajectories taking their initial conditions on this plane librate around $AL_4$ (domain centred on $\zeta=130^\circ$, $\Dv=-100^\circ$), but no orbit librating around $L_4$ intersects this plane. Note that for $m_1 \neq m_2$ it may be impossible to pass directly from orbits librating around $L_4$ to orbits librating around $AL_4$ as it seems that these areas are separated by a region where $\Dv$ circulate (checked for $e_1 \in \{0, 0.15, 0.3 0.55 ,0.7\}$ and $e_2$ such that $J_1=J_1(e_1=e_2=0.4)$).
}

 \subsubsection{$J_1(e_1=e_2=0.65)$}
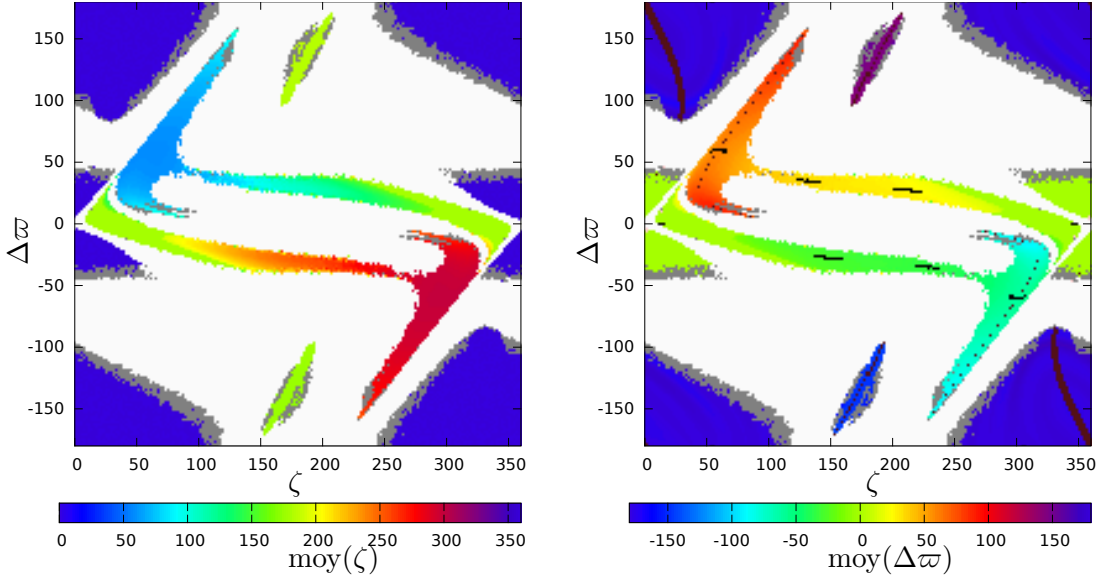
\begin{figure}[h!]
\begin{center}
\includegraphics[width=0.5\linewidth]{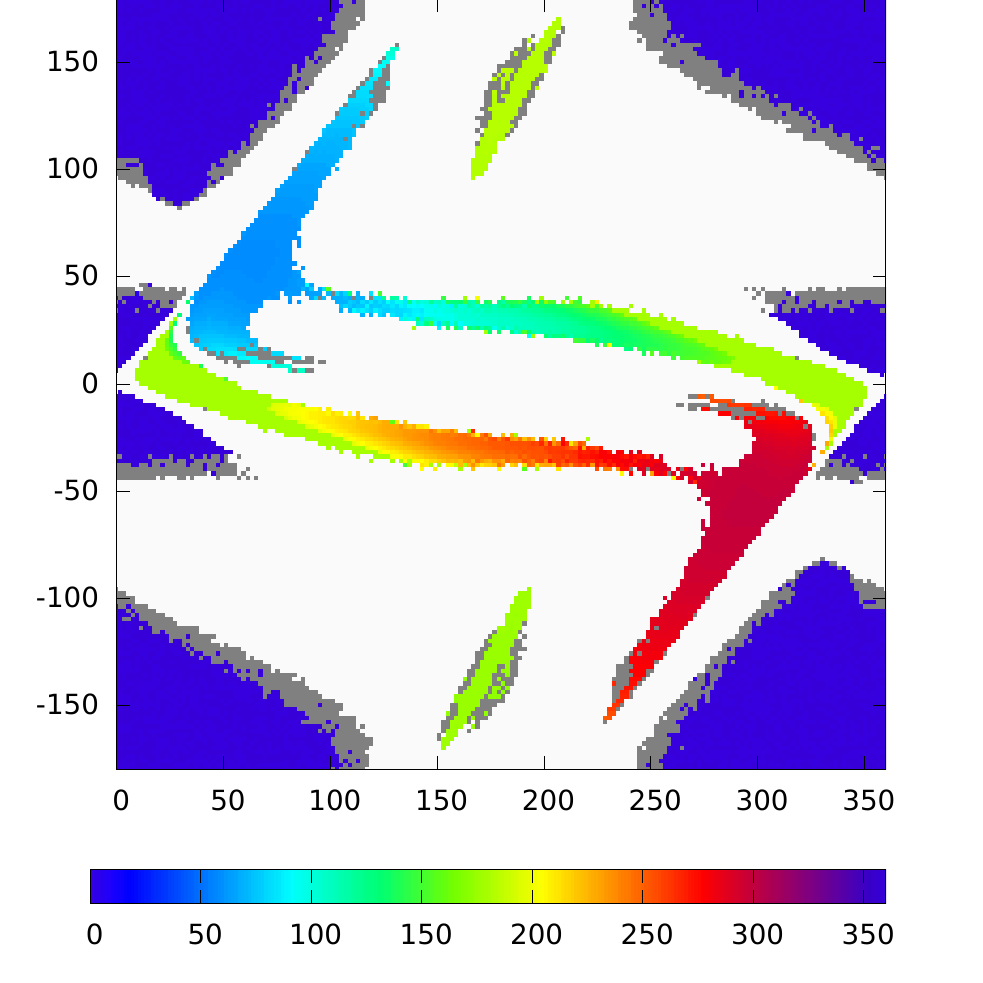}\includegraphics[width=0.5\linewidth]{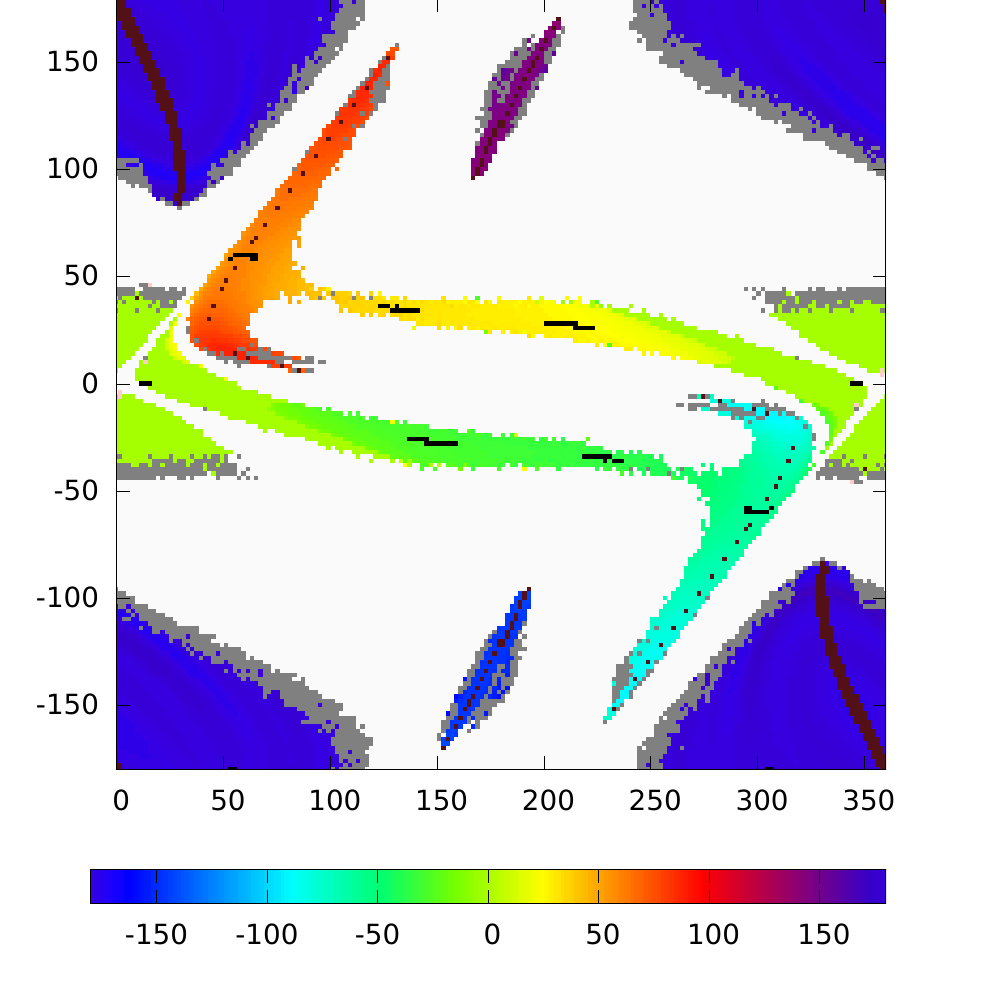}\\
  \setlength{\unitlength}{1cm}
\begin{picture}(.001,0.001)
\put(-7.5,4.5){\rotatebox{90}{$\Dv$}}
\put(0,4.5){\rotatebox{90}{$\Dv$}}
\put(-3.8,1.5){{$\zeta$}}
\put(3.8,1.5){{$\zeta$}}
\put(-3.8,0.5){{moy($\zeta$)}}
\put(3.3,0.5){{moy($\Dv$)}}
\end{picture}
\caption{\label{fig:glob_e7_md} grille de conditions initiales pour $a_1=a_2=1\,ua$, $m_2=3m_1=1.5\, 10^{-5}$ et $e_1=e_2=0.65$. Le plan est défini par $a_1=a_2=1\,ua$, et $e_1=e_2$. Chaque point de ce plan est une condition initiale d'orbite. Sur la figure de gauche, le code couleur correspond à la valeur moyenne de $\zeta$ sur cette orbite. Sur la figure de droite, le code couleur représente la valeur moyenne de $\Dv$ sur chaque orbite. Les orbites proches de $\ol \cF^\nu$, donc celles vérifiant la relation (\ref{eq:condFb02}) avec $\epsilon_\nu=10^{-3.5}$ sont représentées en marron. Les orbites proches de $\ol \cF^g$, donc celles vérifiant la relation (\ref{eq:condFb1}) avec $\epsilon_g= 3^\circ$ sont représentées en noir. Les orbites ayant une diffusion en demi-grand axe supérieur à $\epsilon_a=10^{-5.5}$ ont été représentées en gris.\\
}
\end{center}
\end{figure}

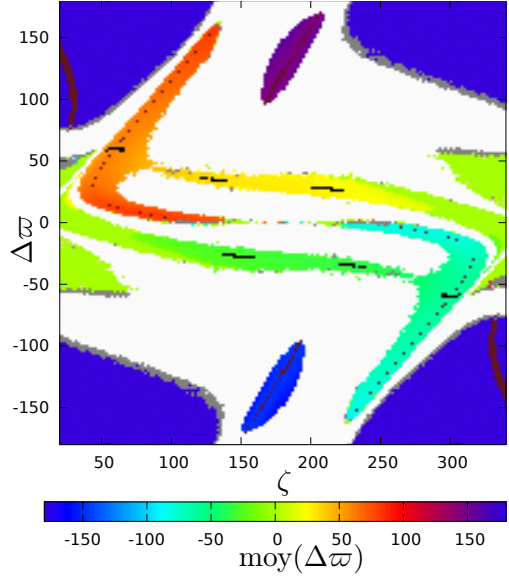
\begin{figure}[h!]
\begin{center}
\includegraphics[width=0.5\linewidth]{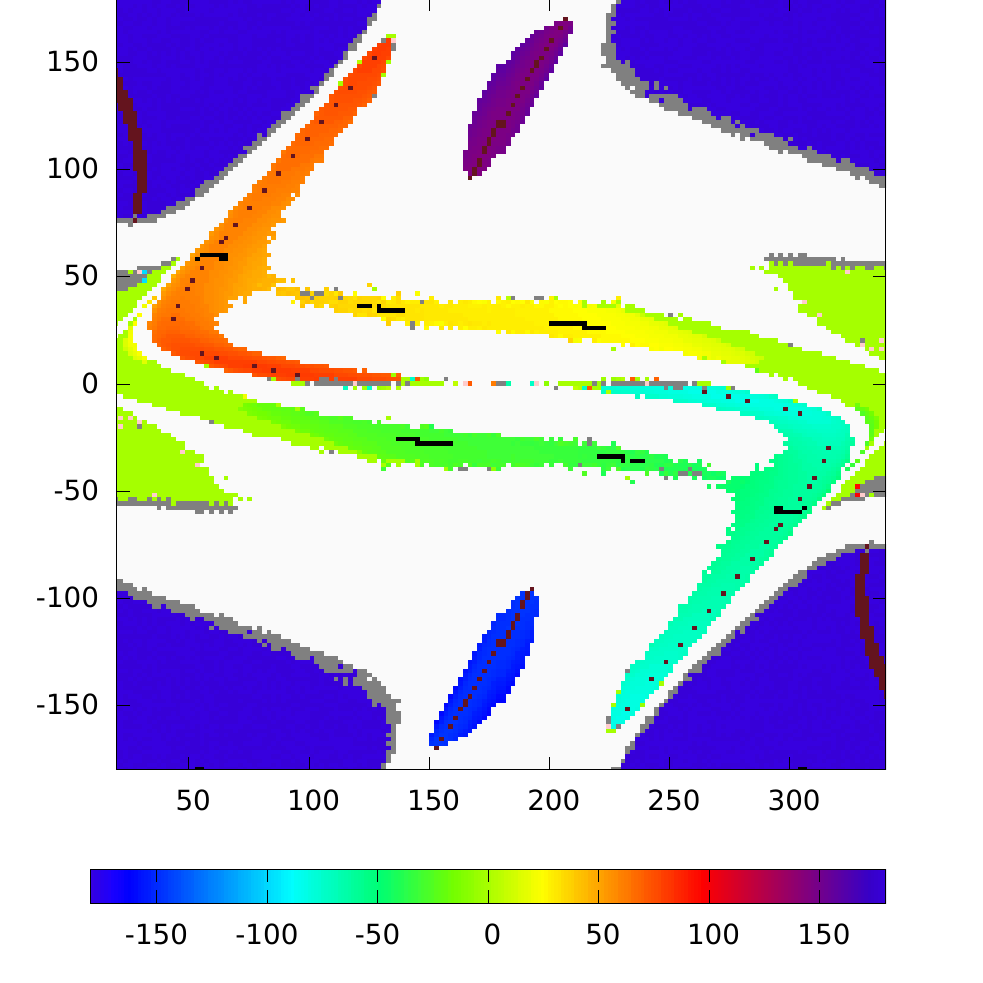}\\
 \setlength{\unitlength}{1cm}
\begin{picture}(.001,0.001)
\put(-3.5,4.5){\rotatebox{90}{$\Dv$}}
\put(0,1.5){{$\zeta$}}
\put(-0.5,0.5){{moy($\Dv$)}}
\end{picture}
\caption{\label{fig:glob_e7_mdt} Conditions initiales, temps d'intégration et code couleur identiques au graphe de droite de la figure~\ref{fig:glob_e7_md} pour $\zeta \in [20^\circ,340^\circ]$. Pas de temps: $dt=0.001$~périodes orbitales.\\
}
\end{center}
\end{figure}

La figure \ref{fig:glob_e7_md} représente l'évolution de la valeur moyenne de $\zeta$ et $\Dv$ sur un plan de conditions initiales $e_1=e_2=0.65$, $a_1=a_2=1$ pour $m_2=3m_1=1.5\, 10^{-5}$. Cette figure est à comparer avec la figure \ref{fig:glob_e65} où les trajectoires ont les mêmes conditions initiales pour $m_1=m_2$. On observe ici le même changement de topologie que dans le cas de masses égales, la reconnexion des familles $\ol \cF$ semble également survenir quand $m_1 \neq m_2$. On constate notamment l'apparition les régions des fers-à-cheval asymétriques. Pour cette valeur des excentricités, aucune des trajectoires issue de ce plan voit l'angle $\Dv$ circuler (il n'est pas exclu que ce soit le cas en prenant des excentricités initiales différentes pour la même valeur de moment cinétique). Notons que pour ces valeurs des masses et des excentricités, le pas de temps de $0.01$ périodes orbitales n'est pas adaptée: des orbites stables sont éjectées (comparer les figures \ref{fig:glob_e7_md} et \ref{fig:glob_e7_mdt}).  \\

\an{
The figure~\ref{fig:glob_e7_md} gives the mean value of the angles $\zeta$ and $\Dv$ when the initial conditions are taken across on the plane $e_1=e_2=0.7$, $a_1=a_2$ (with $m_2=3m_1=1.5 \times 10^{-5} m_0$). The topological change that we described in the previous section occurs in this case as well (compare with the figure~\ref{fig:glob_e65} - $m_1=m_2$). 
No orbit for which $\Dv$ circulate crosses this plane. Note that for these values of the masses and eccentricities, the time step of $0.01$ is not adapted: stable orbits are ejected (compare the figures \ref{fig:glob_e7_md} and \ref{fig:glob_e7_mdt}). 
}

\subsection{variété de référence pour $m_2 \neq m_1$}
\label{sec:VaRemdif}

En section \ref{sec:VaReL4}, nous avons déterminé l'expression d'une variété de référence pour le voisinage de l'équilibre $L_4$ circulaire pour des masses quelconques. Dans cette section, nous cherchons à étendre la détermination de cette variété représentative à l'ensemble du domaine des configurations troyennes librant autour de l'équilibre excentrique $L_4$: $\Sigma_{L_4}$. On simplifie le problème en considérant que toutes les trajectoires de $\Sigma_{L_4}$ passent par le plan $a_1=a_2$, ou de manière équivalente par $Z=Z_\cV=Z(a_1=a_2)$. En pratique nous savons que, au moins pour le cas quasi-circulaire, elles passent par une variété qui reste $\eps$ proche de ce plan. Il nous reste à déterminer $\cV_\varPi$. Nous cherchons donc à déterminer la variété $\Ddv=0$ (section \ref{sec:VaReL4}).\\

\an{
In section \ref{sec:VaReL4}, we determined the expression of the reference manifold in the neighbourhood of the circular $L_4$ equilibrium for arbitrary masses. In this section, we seek to extend the determination of this representative manifold to the whole area of the Trojan configurations librating around the eccentric $L_4$ equilibrium: $ \Sigma_{L_4} $. We simplify the problem by considering that all the trajectories of $ \Sigma_{L_4} $ go through the plan $ a_1 =  a_2$, or equivalently by $Z =  Z_\cV = Z (a_1 = a_2)$. In practice we know that, at least in the vicinity of the circular case, they go through a variety remaining $ \eps $ close to this plan. We are left with the determination of $ \cV_\varPi$. We define this manifold by $ \Ddv = 0$ (see section \ref{sec:VaReL4}).
}


\subsubsection{Description de l'algorithme}
\label{sec:algodes}

Afin d'estimer numériquement la position de la variété $\Ddv=0$ dans l'espace des phases, on implémentera l'algorithme suivant: on prend en condition initiale une carte $(\zeta, \Dv) \in [0^\circ,360^\circ]^2$, $Z=Z_\cV$ et $\varPi=\varPi^0_\cV$ avec $\varPi^0_\cV$ quelconque. On prendra par exemple $\varPi^0_\cV$ tel que $e_1=e_2$ (nous prenons donc en condition initiale de l'algorithme le plan représenté en figure \ref{fig:glob_e4_md}). Pour chacune des trajectoires intégrées depuis ces conditions initiales sur plusieurs périodes de $g$, on enregistre les instants $t^k_{Z}$ où $Z=Z_\cV$ (par interpolation linéaire). Notons $t^{k_{\min{}}}_{Z}$ l'instant où $\Dv(t^k_{Z})$ est minimal et $t^{k_{\max{}}}_{Z}$ l'instant où $\Dv(t^k_{Z})$ est maximal (on discutera le cas où $\Dv$ circule à la fin de cette section). On définit ensuite: 
\begin{equation}
\begin{aligned}
\zeta_{1,1} & = \min{(\zeta(t^{k_{\min{}}}_{Z}),\zeta(t^{k_{\min{}}+1}_{Z}))}\\ 
\zeta_{1,2} & = \max{(\zeta(t^{k_{\min{}}}_{Z}),\zeta(t^{k_{\min{}}+1}_{Z}))}\\ 
\zeta_{2,1} & = \min{(\zeta(t^{k_{\max{}}}_{Z}),\zeta(t^{k_{\max{}}+1}_{Z}))}\\ 
\zeta_{2,2} & = \max{(\zeta(t^{k_{\max{}}}_{Z}),\zeta(t^{k_{\max{}}+1}_{Z}))}\\ 
\end{aligned}
\label{eq:extrezet}
\end{equation}
Pour obtenir un ensemble de $4$ points qui sont dans le voisinage de $\cV$:
\begin{equation}
\begin{aligned}
 (Z=Z_\cV,\zeta_{1,1}, \Dv_1, \varPi_1)\\ 
 (Z=Z_\cV,\zeta_{1,2}, \Dv_1, \varPi_1)\\ 
  (Z=Z_\cV,\zeta_{2,1}, \Dv_2, \varPi_2)\\ 
   (Z=Z_\cV,\zeta_{2,2}, \Dv_2, \varPi_2)\\ 
\end{aligned}
\label{eq:}
\end{equation}
Avec l'indice $_1$ pour la valeur au temps $t^{k_{\min{}}}_{Z}$ et l'indice $2$ pour $t^{k_{\max{}}}_{Z}$. on note $\cV_\varPi^1$ la variété définie par $\cV_\varPi^1(\zeta_{j,k},\Dv_j)=\varPi_j$ pour l'ensemble des orbites issues de la variété de conditions initiales ($\cV_Z$, $\cV_\varPi^0$). Comme nous l'avons rappelé, toutes les orbites ne passent pas dans le voisinage de $e_1=e_2$, la variété définie par $(Z=0,\varPi=\varPi^1_\cV)$ comporte donc des "trous", il est donc nécessaire d'itérer plusieurs fois cet algorithme en interpolant à chaque fois la variété $\cV_\varPi^j$ afin de reconstituer un maillage régulier selon les directions $\zeta$ et $\Dv$ et de converger vers une variété proche de $\cV$ en tout point de l'espace des phases.\\

\an{The algorithm that we use to identify the position of $\Ddv=0$ works as follow: we take a grid of initial conditions $(\zeta, \Dv) \in [0^\circ,360^\circ]^2$, $Z=Z_\cV=Z(a_1=a_2)$ and $\varPi=\varPi^0_\cV$ with an arbitary value for $\varPi^0_\cV$. We then integrate the orbits and identify the instants $t^k_{Z}$ when $Z=Z_\cV$ (linear interpolation). Among all these instants, we identify the one for which $\Dv$ is maximal ($t^{k_{\max{}}}_{Z}$), and the one for which it is minimal ($t^{k_{\min{}}}_{Z}$). We then identify the extrema of $\zeta$ when $\Dv$ is near its own extrema (eq. \ref{eq:extrezet}). We hence obtain 4 points which are in the vicinity of both the manifold $\dot \zeta=0$ and $\Ddv=0$, hence in the vicinity of $\cV$.\\
After one iteration of this algorithm, $\cV$ is not complete: some orbits may not pass near the manifold $\varPi=\varPi^0_\cV$. We hence need to interpolate between the points of the manifold that we identified to create a new grid of initial conditions, and then repeat the algorithm until we obtain a satisfying result.}

\subsubsection{Application de l'algorithme}
\label{sec:resalgo}
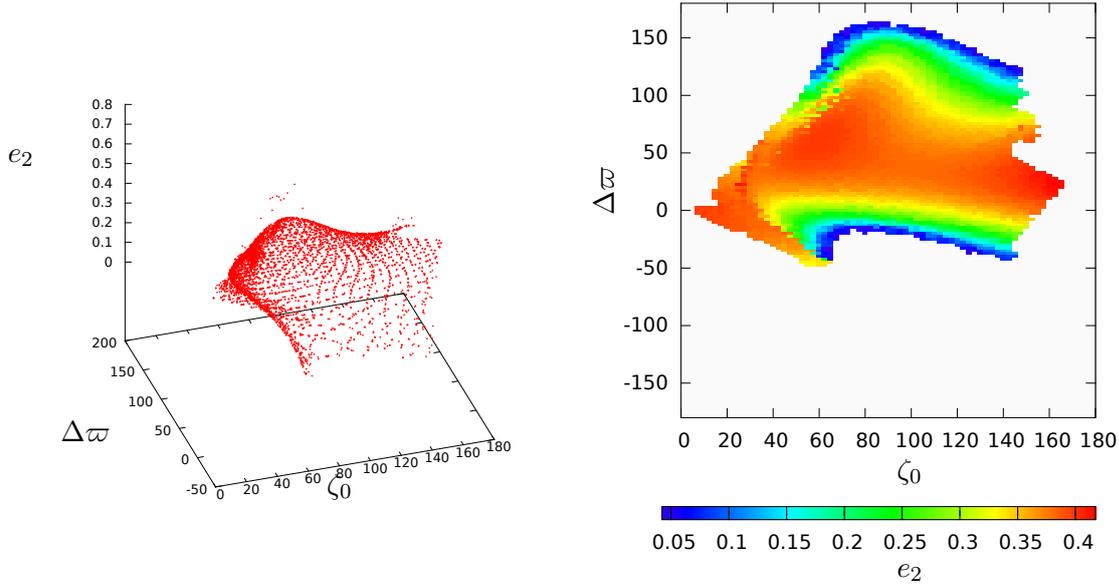
\begin{figure}[h!]
\begin{center} 
\includegraphics[width=0.49\linewidth]{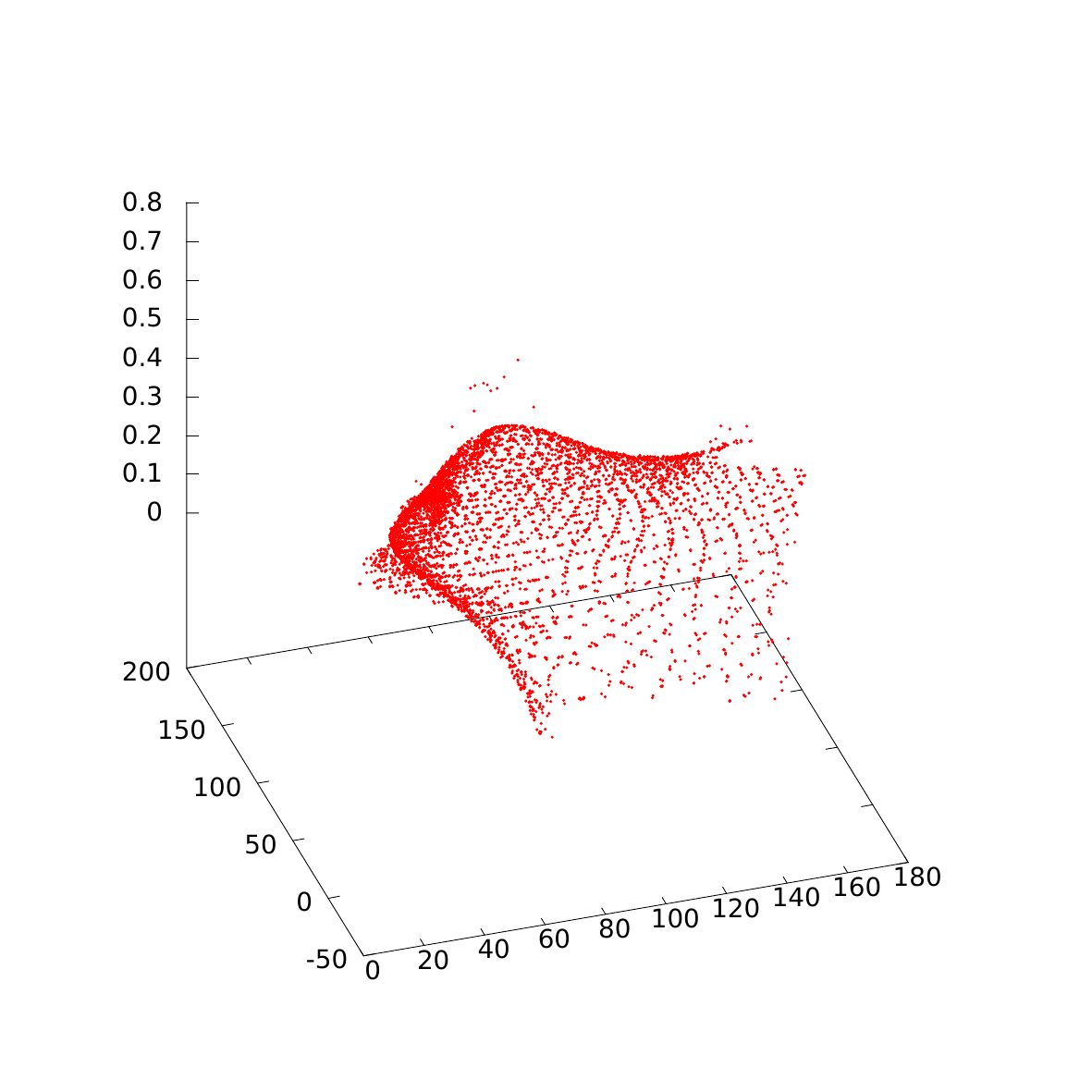}
\includegraphics[width=0.49\linewidth]{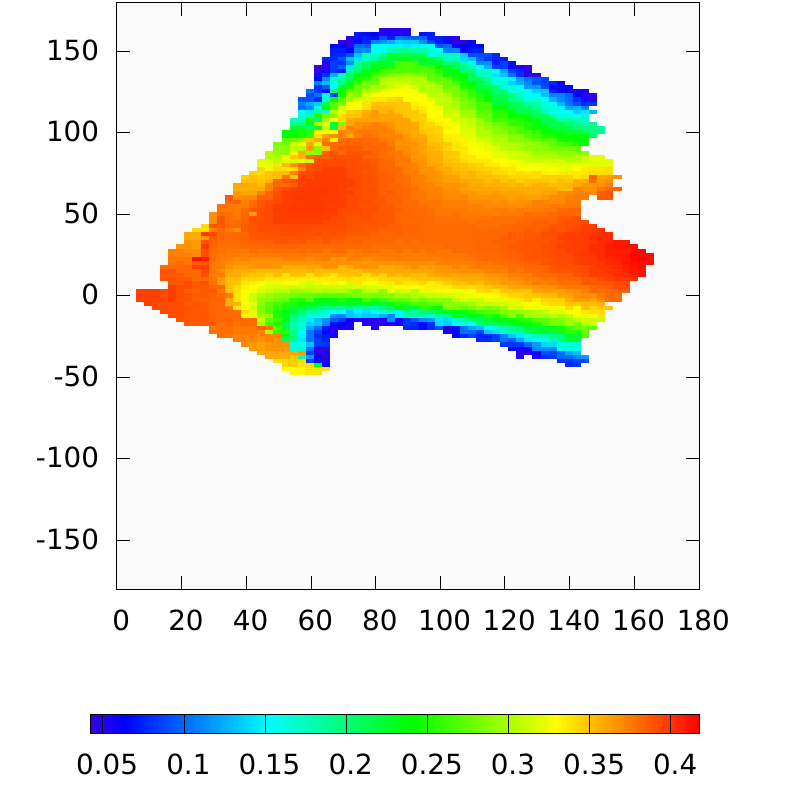}  
 \setlength{\unitlength}{1cm}
\begin{picture}(.001,0.001)
\put(0,5){\rotatebox{90}{$\Dv$}}
\put(-7.7,5.7){$e_2$}
\put(-3.5,1.3){{$\zeta_0$}}
\put(4,1.5){{$\zeta_0$}}
\put(-7,2){{$\Dv$}}
\put(4,0.2){{$e_2$}}
\end{picture}
\caption{\label{fig:VaReMdif} Résultat de l'algorithme détaillé en section \ref{sec:algodes} après 3 itérations, pour $m_1=3m_2$ et une énergie cinétique fixée calculée pour $e_1=e_2=0.4$. Le graphe de gauche représente les points de $\Ddv=0$ identifiés par l'algorithme dans les coordonnées ($\zeta$,$\Dv$,$e_2$). Le graphe de droite représente la valeur de $e_2$ sur cette variété après interpolation. Voir le texte pour plus de détails.\\
\an{
Result of the algorithm explained in section \ref{sec:algodes} after 3 iterations, for $ m_1 = 3m_2$ and a fixed angular momentum equivalent to $ e_1 =e_2 =  0.4$. The left graph shows the points of $ \Ddv = 0$ identified by the algorithm in the coordinates ($ \zeta $, $ \Dv$, $e_2 $). The right graph represents the value of $e_2$ on this manifold after interpolation. See the text for more details.
}} 
\end{center}
\end{figure}

L'algorithme précédemment décrit à été utilisé sur le voisinage de l'équilibre de Lagrange elliptique pour $m_1=3m_2$ et une énergie cinétique calculée pour $e_1=e_2=0.4$ Le résultat de la 3ème itération de l'algorithme est présenté en figure~\ref{fig:VaReMdif} dans les variables $\zeta$, $\Dv$, $e_2$. On rappelle que $Z_\cV=Z(a_1=a_2)$ et que la valeur de $\varPi$ est imposée par celles de $e_2$ et du moment cinétique choisi. On peut voir sur la figure de gauche que la variété $\dot \Dv=0$ semble être lisse et passe bien par l'équilibre de Lagrange excentrique ($\zeta=60^\circ,\Dv=60^\circ$, $e_1=e_2=0.4$). Certains points de la variété $\dot \Dv=0$ ont été identifiés dans le domaine des fers-à-cheval.

Le résultat de l'interpolation (à droite) semble régulier, à l'exception de la séparatrice entre les fers-à-cheval et les configurations troyennes, et les zones de fortes pentes dans le quadrant supérieur gauche par rapport à l'équilibre de Lagrange. Cela peut être amélioré en augmentant le nombre de points dans cette région, en améliorant la détermination de l'instant ou $\dot \Dv=0$ en augmentant par exemple la durée de l'intégration, et en améliorant la fonction d'interpolation (on a utilisé ici une simple interpolation linaire à trois points). \\

\an{
The algorithm described above was used in the vicinity of $L_4$ for $ m_1 = 3m_2$ and a total angular momentum calculated for $e_1 =e_2 = 0.4$. The result of the third iteration of the algorithm is represented in figure~\ref{fig:VaReMdif} in the $\zeta$, $\Dv$, $e_2$ variables. Remember that $ Z_\cV = Z(a_1 = a_2)$ and $\varPi$ is imposed by the value of $e_2$ and the angular momentum. We can see on the left figure that the manifold $ \Ddv =  0$ appears to be smooth and to contain indeed $L_4$ ($ \zeta = 60^\circ, \Dv = 60^\circ $ , $ e_1 = e_2 = 0.4$). Some points of the variety $ \Ddv = 0$ have been identified in the horseshoe area.\\
On the right, the result of the interpolation seems smooth as well, except for the area near the separatrix between the trojan and horseshoe configurations, and the area of steep slope in the upper left quadrant with respect to $L_4$. This could be improved by increasing the number of points in the region, by improving the determination of the instant where $ \Ddv = 0$ (for example by increasing the duration of integration), and by improving the interpolation (we used here a simple linear interpolation).
}

\subsubsection{Exploitation de la variété de référence}

\begin{figure}[h!]
\begin{center} 
\includegraphics[width=0.49\linewidth]{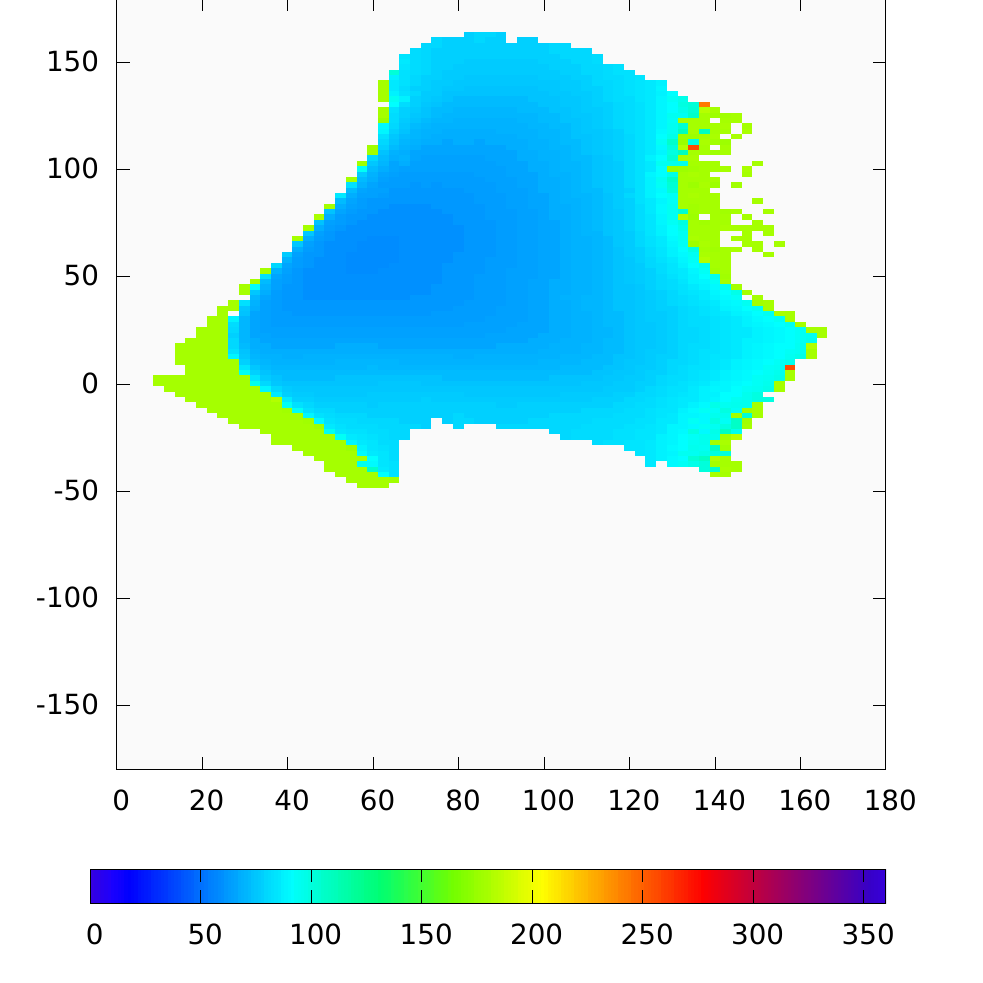}  
\includegraphics[width=0.49\linewidth]{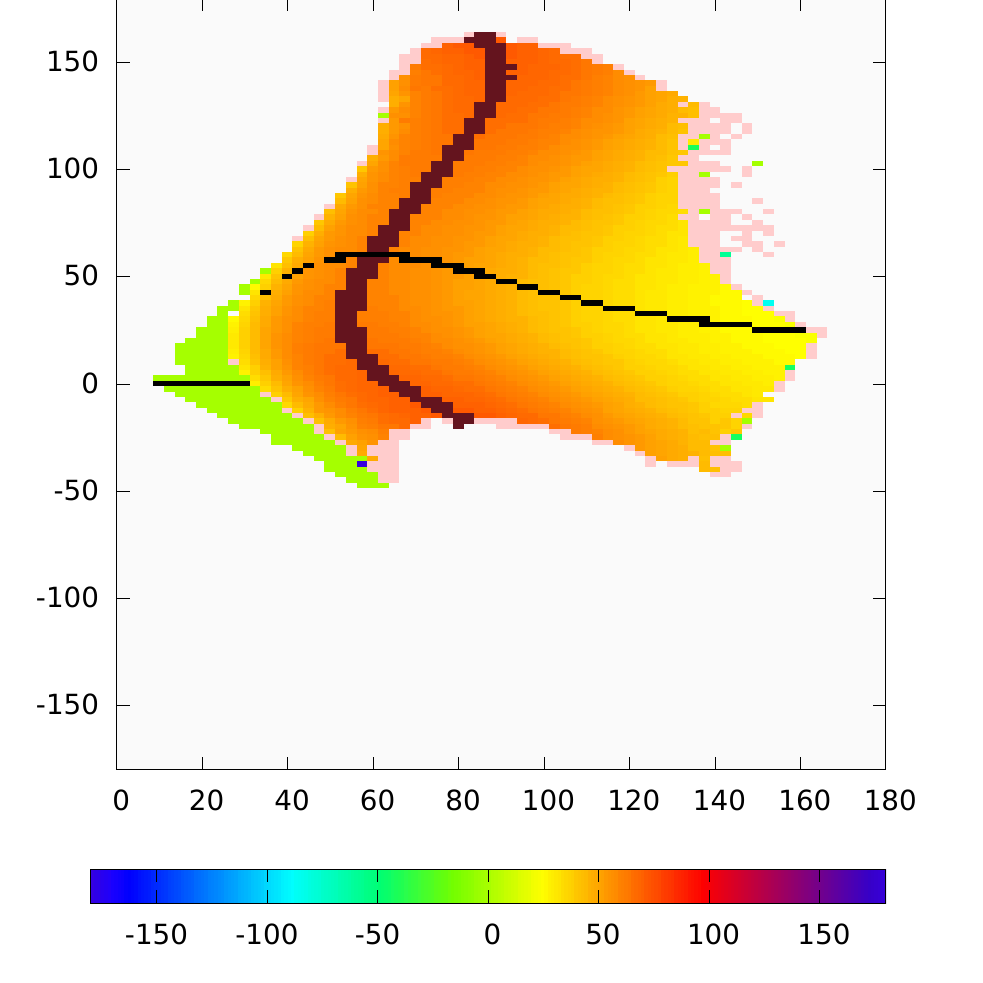}  \\
  \setlength{\unitlength}{1cm}
\begin{picture}(.001,0.001)
\put(-7.5,4.5){\rotatebox{90}{$\Dv$}}
\put(0,4.5){\rotatebox{90}{$\Dv$}}
\put(-3.8,1.5){{$\zeta$}}
\put(3.8,1.5){{$\zeta$}}
\put(-3.8,0.5){{moy($\zeta$)}}
\put(3.3,0.5){{moy($\Dv$)}}
\end{picture}
\caption{\label{fig:VaReMdif} Le code couleur représente la valeur moyenne de $\zeta$ sur le graphe de gauche et $\Dv$ sur le graphe de droite pour chaque orbite en prenant en condition initiale le résultat de l'algorithme représenté sur la figure~\ref{fig:VaReMdif}. La couleur saumon est attribuée sur le graphe de droite quand $\Dv$ circule. Les orbites proches de $\ol \cF^\nu$, donc celles vérifiant la relation (\ref{eq:condFb02}) avec $\epsilon_\nu=10^{-3.5}$ sont représentées en marron sur le graphe de droite. Les orbites proches de $\ol \cF^g$, donc celles vérifiant la relation (\ref{eq:condFb1}) avec $\epsilon_g= 3^\circ$ sont représentées en noir sur le graphe de droite. Les orbites ayant une diffusion en demi-grand axe supérieur à $\epsilon_a=10^{-5.5}$ on été représentées en gris.\\
\an{The color code represents the mean value of $ \zeta$ on the left graph and $\Dv$ on the right graph for each orbit taking as initial condition the result of the algorithm shown in figure~\ref{fig:VaReMdif}. The salmon color is assigned to the right graph when $\Dv$ circulates. The orbits close to $ \ol \cF^\nu $, hence those satisfying the relationship (\ref{eq:condFb02}) with $ \epsilon_\nu = 10^{- 3.5}$ are shown in brown on the right hand graph. The orbits satisfying the relationship (\ref{eq:condFb1}) with $ \epsilon_g = 3^\circ $, hence close to $ \ol \cF^g $, are represented in black on the right hand graph. The orbits having a diffusion of the semi-major axis higher than $ \epsilon_a = 10^{-5.5}$ are displayed in gray.
} } 
\end{center}
\end{figure}

On intègre ici les trajectoires ayant pour conditions initiales le maillage résultant de l'algorithme explicité en section \ref{sec:algodes}. Ces conditions initiales sont donc proches de la variété de référence définie par $Z_\cV=Z(a_1=a_2)$ et $\varPi_\cV$ tel que $\Ddv=0$. L'approximation obtenue de cette variété a été présentée en section \ref{sec:resalgo}. Nous rappelons que pour les conditions initiales choisies, les masses des co-orbitaux, l'énergie cinétique totale, ainsi que l'ensemble des autres conditions initiales sont identiques à celles du plan de conditions initiales représenté en figures \ref{fig:glob_e4_md} et \ref{fig:glob_e4d_md} à l'exception de la valeur de $\varPi$. Nous pouvons donc comparer ces différentes variétés de conditions initiales qui font parties du même espace des phases.

Nous rappelons que notre but est de représenter exhaustivement les orbites troyennes qui librent autour de l'équilibre de Lagrange excentrique situé en $e_1=e_2=0.4$ avec $\zeta=\Dv=60^\circ$. Le plan de conditions initiales figure~\ref{fig:glob_e4d_md} ne génère aucune trajectoire librant autour de $L_4$ excentrique car toutes les trajectoires intersectées par ce plan voient leur angle $\Dv$ circuler. Le plan de conditions initiales figure~\ref{fig:glob_e4_md} contient quant à lui des orbites du voisinage de $L_4$, ainsi que le point $L_4$ lui-même. Cependant, la plupart des orbites de la famille $\ol \cF^g_4$ ne sont pas représentées par ce plan: pour $m_2 \neq m_1$, ces orbites n'ont jamais $e_1=e_2$. A l'inverse, nous pouvons voir que la variété de référence $\cV$ que nous avons construite semble être au voisinage de l'ensemble des membres de la famille $\ol \cF^g_4$ (voir les pixels noirs sur la figure~\ref{fig:VaReMdif}). On note également que la partie de la variété de référence construite dans le domaine des fers-à-cheval intersecte également la famille $\ol \cF_{\gH\cS}$. \\

\an{
Here we integrate the trajectories whose initial conditions are the results of the algorithm explained in section \ref{sec:algodes}. These initial conditions are close to the reference manifold defined by $ Z_\cV = Z (a_1 = a_2) $ and $ \varPi_\cV $ such as $ \Ddv = 0$. We recall that for the chosen initial conditions, the masses of the co-orbitals, the total angular momentum, as well as all other initial conditions are identical to those of the figures~\ref{fig:glob_e4_md} and \ref{fig:glob_e4d_md} except for the value of $\varPi$. We can therefore compare these different manifolds of initial conditions which are part of the same phase space.\\
We recall that our goal is to represent exhaustively the Trojan orbits librating around the $L_4$ equilibrium located at $e_1 = e_2 =0.4 $ with $ \zeta = \Dv = 60^\circ $. The plane of initial conditions in figure~\ref{fig:glob_e4d_md} generates no trajectory librating around $L_4$. On the contrary, the plane of initial conditions (figure~\ref{fig:glob_e4_md}) generates trajectories in the neighborhood of $ L_4 $ and the point $ L_4$ itself. However, most of the orbits of the family $ \ol \cF^g_4 $ are not represented by this plan: for $m_2 \neq m_1$, most of these orbits never reach $e_1  = e_2$. \\
On the contrary, we can see that the reference manifold $\cV$ that we built seems to contain orbits in the vicinity of all members of $ \ol \cF^g_4$ (see the black pixels in figure~\ref{fig:VaReMdif}). Note also that the part of the reference manifold in the horseshoe domain also contains orbits in the vicinity of $ \ol \cF_{\gH \cS}$.
}

\subsubsection{Discussion}

La variété $\cV$ obtenue avec l'algorithme décrit en section \ref{sec:algodes} semble répondre aux attentes. Il resterait à vérifier, avec une méthode similaire à celle utilisée en section \ref{sec:VaRemeq} que l'ensemble des orbites librant autour de $L_4$ passe bien au voisinage de $\cV$.

On peut bien sûr appliquer cet algorithme pour l'ensemble des configurations coorbitales: fer-à-cheval, quasi-satellite, etc. pour une valeur donnée du moment cinétique et des masses. Certains détails techniques sont à prendre en compte:\\
- On ne peut pas espérer, comme nous pouvons le faire pour $m_2=m_1$, représenter l'ensemble des configurations passant par le plan $a_1=a_2$ sur un seul graphe en deux dimensions, en tout cas pas avec les coordonnées $\zeta$ et $\Dv$. En effet, quand $m_2 \neq m_1$, pour un couple donné ($\zeta$, $\Dv$), peut passer la variété représentative de deux configurations différentes, par exemple celle des orbites librant autour de $L_4$ et celle des orbites librant autour de $AL_4$ (pour des $\varPi$ différents).\\
- Les orbites où $\Dv$ circule doivent être traitées différemment. Toutes ces orbites passent par exemple par $\Dv=0$, on peut donc remplacer la variété $\dot \Dv=0$ par $\Dv=0$ et paramétrer la variété de conditions initiales par $\zeta$ et $\varPi$.\\
- Certaines orbites peuvent voir $\dot \Dv$ s'annuler plus de deux fois par période $2\pi/g$ sur leur orbite (c'est le cas par exemple de certaines orbites de la configuration $G$). dans ce cas il faut bien "choisir" les points qui constitueront la variété de référence.\\

\an{
The variety $ \cV$ obtained by the algorithm described in section \ref{sec:algodes} appears to meet our expectations: to represent all the trojan orbits that librate around $L_4$. It remains to check it with a method similar to the one used in \ref{sec:VaRemeq}.\\}

\an{
This algorithm can be applied to all co-orbital configurations: horseshoe, quasi-satellite, etc. for a given value of the angular momentum and given masses. Some technical details are to be considered: \\
- We can not hope to represent all the configurations going through the plane $ a_1 =  a_2$ with one continuous two-dimensional manifold like we did in the $m_1=m_2$ case, at least not parametrised by $ \zeta$ and $ \Dv$. Indeed, when $ m_2 \neq m_1 $, a couple ($ \zeta $, $ \Dv $) can belong to the representative manifold of different configurations for different values of $ \varPi$. \\
- The orbits where $ \Dv$ circulate must be treated differently. All these orbits pass through the plane $\Dv =  0$, we can hence replace the variety $ \Ddv = 0$ by $ \Dv =  0$ in the definition of $\cV$ and parametrise the variety of initial conditions by $ \zeta $ and $ \varPi $ for this configuration. \\
- For some orbits, $ \Ddv $ vanishes more than twice by period of $g$ (this is the case for example of some orbits in the $G$ configuration). In this case we have to chose which point form the reference manifold.
}

\section{Dynamique au voisinage des équilibres elliptiques $\cF^2_{k}$}
\label{sec:DynFl}


Dans cette section, nous allons décrire la dynamique au voisinage des équilibres de Lagrange excentriques émergeant des points fixes circulaires $L_3$, $L_4$ et $L_5$: les familles $\cF^2$. Cette étude sera effectuée analytiquement. On se replace dans le problème moyen (non réduit), et on repart des hamiltoniens $\ol H_{L_k}$ dont la partie quadratique diagonale est explicitée par les équations (\ref{eq:HquadL4}) et (\ref{eq:HquadL3}). \\

Les $\cF^2_{k}$ étant des familles de points fixes du problème moyen (familles des équilibres de Lagrange excentriques), nous avons expliqué en section \ref{sec:genF2} qu'il était possible de translater le hamiltonien le long de celles-ci afin de connaître la dynamique au voisinage d'une orbite donnée de ces familles. Nous allons commencer par calculer l'évolution des fréquences fondamentales le long des familles $\cF^2_{3}$ et $\cF^2_{4}$ au premier ordre en $\eps$ et à un ordre $m$ en $e_1$. On calculera ensuite la position des familles $\ol \cF^j_k$ au voisinage des $\cF^2_k$ afin de confirmer les identifications des familles $\ol \cF^\nu_k$ et $\ol \cF^g_k$ effectuées dans le problème moyen réduit dans les sections précédentes.\\

\an{
In this section we describe the dynamics in the vicinity of the eccentric Lagrangian equilibriums emerging from the circular fixed points $L_3$, $L_4$ and $L_5$: the families $ \cF^2$. This study will be done analytically. We consider once again the averaged problem (unreduced), and we start from the Hamiltonian $\ol H_{L_k}$ whose diagonal quadratic part is given by the equations (\ref{eq:HquadL4}) and (\ref{eq:HquadL3}).\\
The $ \cF^2_{k} $ being fixed point families in the average problem (eccentric Lagrangian equilibriums), we explained in section \ref{sec:genF2} that it is possible to translate the Hamiltonian along them in order to describe the dynamics near a given orbit of these families. We will start by calculating the evolution of the fundamental frequencies along the families $ \cF^2_{3} $ and $ \cF^2_{4} $ at first order in $ \eps $ and at the order $m$ in $e_1$. We will also compute the positions of the $\ol \cF^j_k$ families in the neighbourhood of the $\cF^2_k$ families, in order to confirm the identifications of the $\ol \cF^\nu_k$ and $\ol \cF^g_k$ families in the reduced averaged problem.  
}

\subsection{Hamiltonien au voisinage de la famille $\cF^2_{4}$}

Nous partons du membre connu de cette famille situé en $e_1=0$, le point $L_4$. Le hamiltonien $\ol H_{L_4}$, dont la partie quadratique est exprimée eq. (\ref{eq:HquadL4}), est le hamiltonien au voisinage de ce point d'équilibre. $\cF^2_{4}$ vérifie $z_0=z_1=0$ (voir section \ref{sec:H0pf}). On effectue donc le changement de variable suivant:
\begin{equation}
  \chi_{\cF^2_{4}}(e_1)(z'_0,\tilde{z}'_0,z'_1,\tilde{z}'_1,z'_2,\tilde{z}'_2)= (z_0,\tilde{z}_0,z_1,\tilde{z}_1,z_2,\tilde{z}_2)
\label{eq:cv3f}
\end{equation}
où
\begin{equation}
z_2 = z_2^0(e_1) + z'_2\ , \ \ \ \tilde{z}_2 = -i \overline{ z_2^0(e_1)} + \tilde{z}'_2,
\label{eq:cv3ex}
\end{equation}
et l'identité pour les autres variables. 
%

La valeur de la translation $z_2^0(e_1)$ pour un $e_1$ donné se calcule grâce à l'équation (\ref{eq:L4exz}):
\begin{equation}  
\begin{aligned}
 z^0_2=\frac{\sqrt{m'_1\Lambda_1}+\sqrt{m'_2\Lambda_2 }}{\sqrt{2}\sqrt{m'_1+m'_2}} \sqrt{1-\sqrt{1-e^2_1}} \e^{-i \pi/3}.
\end{aligned}
\label{eq:eel4}
\end{equation}
Après translation, nous obtenons le hamiltonien suivant: 
\begin{equation}
  \ol H_{\cF^2_{4}}(e_1) = \ol H_{L_4}  \circ \chi^{-1}_{\cF^2_{4}}(e_1) = \ol H'_{\cF^2_{4}}(e_1) + \gO(e_1^{m+1})
\label{eq:H3}
\end{equation}
où chaque monôme de $\ol H_{\cF^2_{4}}(e_1)$ est un polynôme en $e_1$ d'ordre maximum $m$.
\begin{rem} Pour obtenir tous les termes de $\ol H_{\cF^2_{4}}(e_1)$ à l'ordre $m$ en $e_1$ et à l'ordre $p$ en les variables ($z'_j,\zt'_j$), il faut calculer le hamiltonien $\ol H_{L_4}$ jusqu'à l'ordre $m+p$ avant d'effectuer le changement de variable $\chi_{\cF^2_{4}}(e_1)$. 
\end{rem}
Pour avoir une idée claire de la dynamique dans le voisinage de la famille $\cF^2_{4}$, nous allons diagonaliser $\cQ(e_1)$, la partie quadratique du hamiltonien  $\ol H_{\cF^2_{4}}(e_1)$. Pour ce faire, nous allons la diagonaliser degré par degré de $e_1$. $\cQ(0)$ est déjà diagonal, car identique à la partie quadratique de $\ol H_{L_4}$. Nous pouvons donc écrire:
\begin{equation}
 \cQ(e_1) = \cQ_d + \sum_{l=1}^m e_1^l \cQ_l \ ,
  \label{eq:Qexp}
\end{equation}
où $\cQ_d$ est diagonal, et où les coefficients des $\cQ_l$ ne dépendent pas de $e_1$. 
%
%
%
%
Cette transformation est effectuée de manière itérative, où le monôme de $\cQ$ en $e^l$ est diagonalisé au pas $l$. Chaque transformation élémentaire est le flot au temps $1$ d'un hamiltonien $\cW_l$. calculé de la manière suivante:
\begin{equation}
  W_l= - \int  \left( \cQ_l^{(l-1)}-  \cQ_{l\,d}^{(l-1)}  \right)  dt \, ,
  \label{eq:Wl}
\end{equation}
\begin{figure}[h!]
\begin{center}
\includegraphics[width=0.5\linewidth]{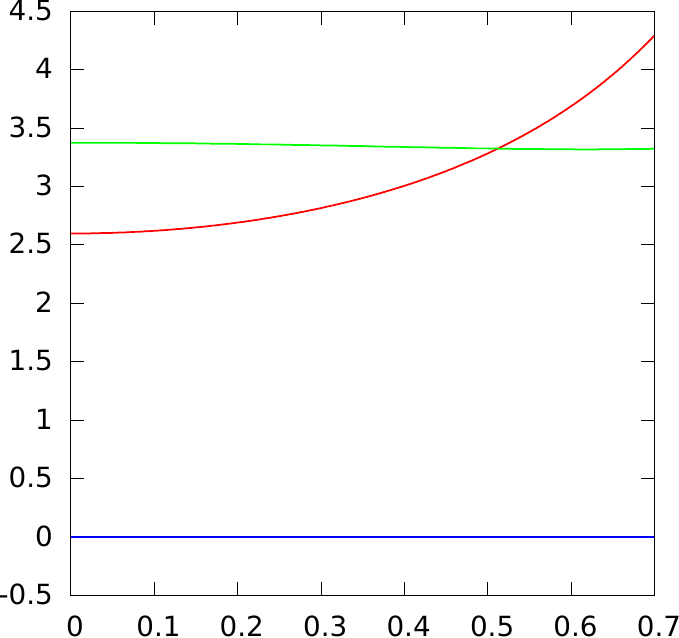}\\
  \setlength{\unitlength}{0.1\linewidth}
\begin{picture}(.001,0.001)
\put(-3,1.75){\rotatebox{90}{fréquence normalisée}}
\put(0,0){{$e_1$}}
\end{picture}
\caption{\label{fig:freqF2L4} évolution des valeurs propres le long de la famille $\cF^2_{4}$. $\nu/(\eta\sqrt{\eps})$ en rouge, $g/(\eta\eps)$ en vert et $g_2/(\eta\eps)$ en bleu. $m=14$.}
\end{center}
\end{figure}
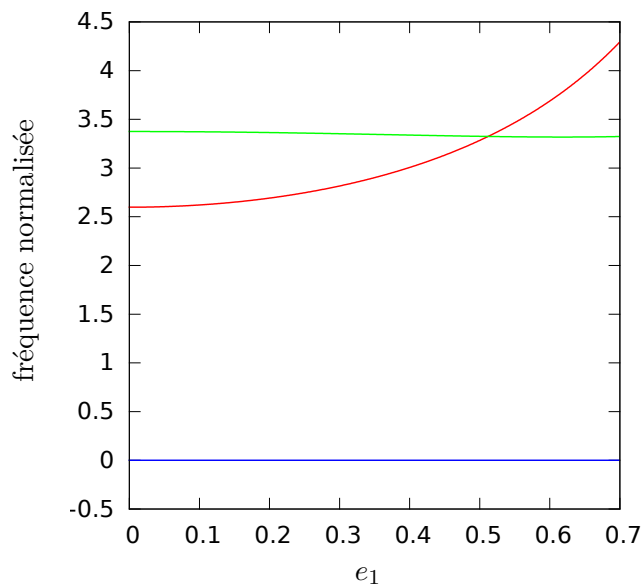
\begin{figure}[h!]
\begin{center}
\includegraphics[width=0.5\linewidth]{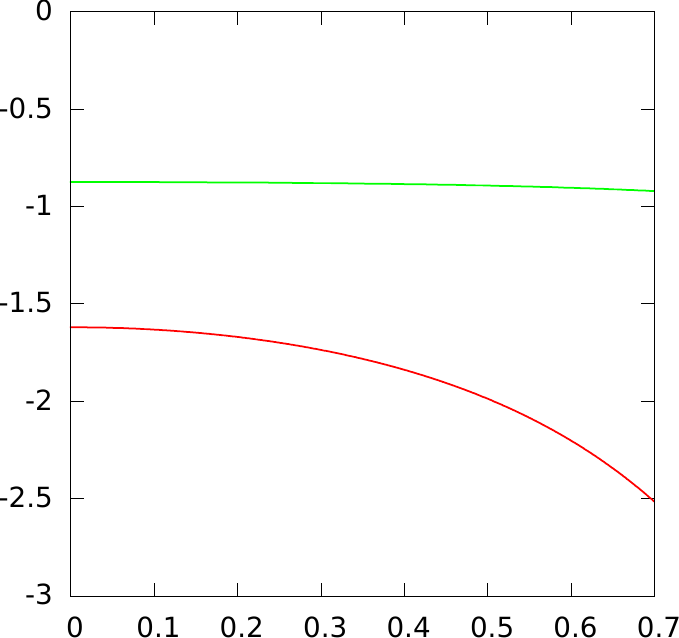}\\
  \setlength{\unitlength}{0.1\linewidth}
\begin{picture}(.001,0.001)
\put(-3,1.5){\rotatebox{90}{valeur propre normalisée}}
\put(0,0){{$e_1$}}
\end{picture}
\caption{\label{fig:freqF2L3} évolution des valeurs propres le long de la famille $\cF^2_{3}$. $|\omega_0|/(\eta\sqrt{\eps})$ en rouge, $|\omega_1|/(\eta\eps)$ en vert et $|\omega_2|/(\eta\eps)$ en bleu. la valeur propre $\omega_0$ est réelle, $\omega_1$ est imaginaire pure et $\omega_2$ est nulle. $m=14$.}
\end{center}
\end{figure}
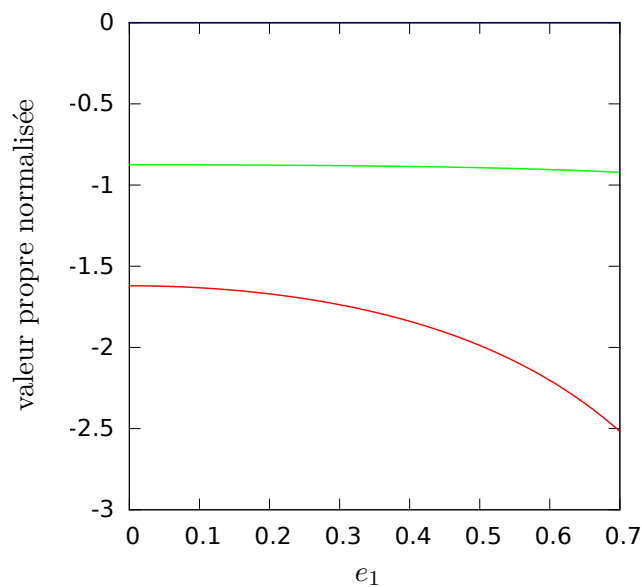
où $\cQ_{l\,d}^{(l-1)}$ est la partie diagonale de $ \cQ_l^{(l-1)}$, en prenant $ \cQ_l^{(0)}=\cQ_l$. Puis on calcule le nouveau hamiltonien  $\cQ^{(l)}$ par application du flot de $e_1^l W_l $ au temps $1$. Par construction, cette opération diagonalise $\cQ^{(l)}$ jusqu'au degré $l$ en $e_1$. Comme $e_1$ est un petit paramètre, chaque changement de variable canonique est proche de l'identité. Celle-ci étant menée jusqu'au degré $m$, nous diagonalisons la partie quadratique du hamiltonien $\cQ^{(m)}$. La composition des flots au temps $1$ des hamiltoniens $e\cW_l$ peut alors être appliquée à l'ensemble du hamiltonien $\ol H_{\cF^2_{4}}(e_1)$. Nous obtenons le hamiltonien $\gH_{\cF^2_{4}}(e_1)$, dont la partie quadratique est diagonale. et le nouveau changement de variable proche de l'identité: 
\begin{equation}
   (z_j,\tilde{z}_j) = \chi_d(Z_j,\tilde{Z}_j)\, .
\label{eq:cv4f}
\end{equation}
$\gH_{\cF^2_{4}}(e_1)$ décrit la dynamique au voisinage de la famille $\cF^2_{4}$ à l'ordre $m$ en $e_1$. La figure \ref{fig:freqF2L4} montre l'évolution des valeurs propres $\omega_j$ le long de cette famille on notera $\omega_0=-i\nu$, $\omega_1=-ig$ et $\omega_2=-i g_2$, où $\nu$, $g$ et $g_2$ sont des réels car toutes les directions sont elliptiques. Pour des excentricités faibles, on retrouve les résultats de \cite{RoPo2013}. \\

De la même manière, on peut calculer le hamiltonien $\gH_{\cF^2_{3}}(e_1)$, décrivant la dynamique dans le voisinage de $\cF^2_{3}$ et dont la partie quadratique est diagonale. La figure \ref{fig:freqF2L4} montre l'évolution des valeurs propres $\omega_j$ le long de cette famille.\\

\an{
To compute the Hamiltonian in the neighbourhood of an orbit of the $\cF^2_4$ family, we start from the know member of this family, located at $e_1=0$: the $L_4$ circular equilibrium. Departing from the Hamiltonian $\ol H_{L_4}$, whose quadratic part is given by the equation (\ref{eq:HquadL4}), we translate it along the $z_2$ direction, using the change of variables (\ref{eq:cv3f}) (the expression of  $z_2^0(e_1)$ is given eq. \ref{eq:eel4}). We obtain the Hamiltonian $\ol H_{\cF^2_{4}}(e_1)$ (\ref{eq:H3}). Each monomial of this Hamiltonian is a polynomial in $e_1$ of maximum order $m$.\\
Remark: to obtain the correct expression of the monomials of $\ol H_{\cF^2_{4}}(e_1)$ at the order $m$ in $e_1$ and $p$ in ($z'_j,\zt'_j$), you need to compute $\ol H_{L_4}$ at the order $m+p$ before the translation.\\
Then, we diagonalise $\cQ(e_1)$, the quadratic part of $\ol H_{\cF^2_{4}}(e_1)$. To do so, we process degree by degree in $e_1$. $\cQ(0)$ is already diagonal (it is the quadratic part of $\ol H_{L_4}$). We can hence expend $\cQ(e_1)$ (see eq. \ref{eq:Qexp}), where $\cQ_d$ is diagonal, and where the coefficients of the $\cQ_l$ do not depend on $e_1$.\\
Each elementary transformation is the time-1 map of the Hamiltonian flows generated by $\cW_l$ (eq. \ref{eq:Wl}), where $\cQ_{l\,d}^{(l-1)}$ is the diagonal part of $ \cQ_l^{(l-1)}$, taking $ \cQ_l^{(0)}=\cQ_l$. Then we compute the new Hamiltonian $\cQ^{(l)}$ by applying the flow of $e_1^l W_l $ at time $1$. By construction, it diagonalises $\cQ^{(l)}$ at the order $l$ in $e_1$. As $e_1$ is a small parameter, all these transformations are close to the identity. Carrying these step up to the order $m$, we diagonalise $\cQ^{(m)}$. The composition of the time-1 map of the Hamiltonian flows generated by the Hamiltonians $e\cW_l$ is then applied to $\ol H_{\cF^2_{4}}(e_1)$. We obtain the Hamiltonian $\gH_{\cF^2_{4}}(e_1)$, which describe the dynamics in the neighbourhood of $\ol \cF^2 (e_1)$ at the order $m$ in $e_1$, and whose quadratic part is diagonal.\\
Figure \ref{fig:freqF2L4} shows the evolution of the eigen values $\omega_j$ along this family. We note $\omega_0=-i\nu$, $\omega_1=-ig$ and $\omega_2=-i g_2$, where $\nu$, $g$ and $g_2$ are reals because all directions are elliptic. For low excentricities, the results coincide with those of \cite{RoPo2013}.\\
We can also compute $\gH_{\cF^2_{3}}(e_1)$, which describe the dynamics in the neighbourhood of $\cF^2_{3}$. The figure \ref{fig:freqF2L4} shows the evolution of the eigenvalues $\omega_j$ along this family.
}

\subsection{Direction des familles $\overline{\cF}^l_{k}$}
\label{sec:DFFb}

\begin{figure}[h!]
\begin{center} 
\includegraphics[width=0.5\linewidth]{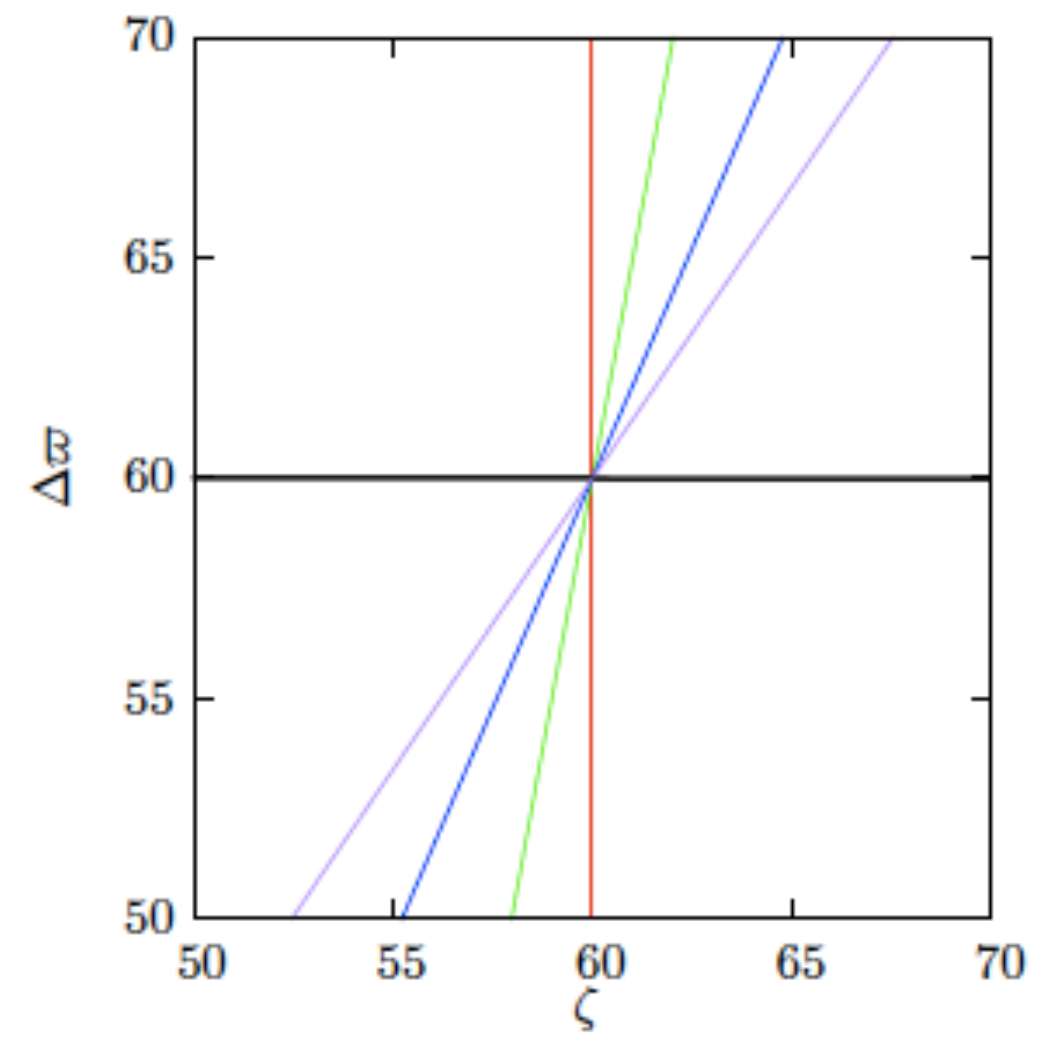}
\caption{\label{fig:FbL4dir} approximation quadratique des familles $\ol \cF^0_{4}$ et $\ol \cF^1_{4}$ pour $m_1=m_2=1\, 10^{-6}m_0$ sur la variété $\cP$ définie par le jeu d'équations (\ref{eq:syseqFb0L4}) au voisinage de $\cF^2_{4}(e_1)$, la famille $\ol \cF^0_{4}$ est représentée en rouge pour $e_1=0.001$, en vert pour $e_1=0.3$, en bleu pour $e_1=0.5$ et en violet pour $e_1=0.7$. La direction de la famille  $\ol \cF^1_{4}$ est constante à l'ordre 2 en $\Re (Z_1)$ et est représentée en noir.  } 
\end{center}
\end{figure}

\begin{figure}[h!]
\begin{center} 
\includegraphics[width=1\linewidth]{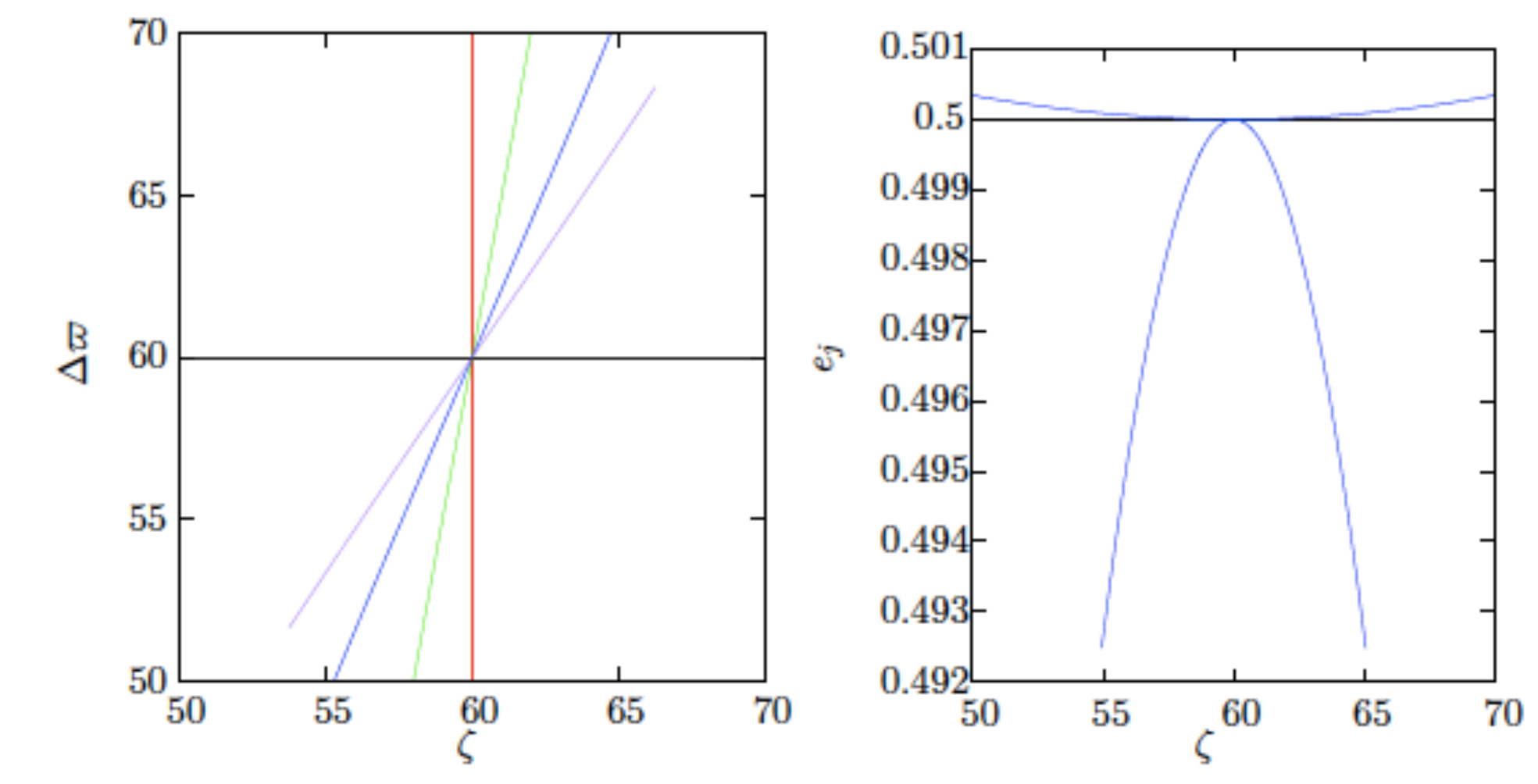}
\caption{\label{fig:FbL4mdifdir} Approximation quadratique des familles $\ol \cF^0_{4}$ et $\ol \cF^1_{4}$ pour $m_1=1\, 10^{-7}m_0 $ et $m_2=19\, 10^{-7}m_0$ sur la variété $\cP$ définie par le jeu d'équations (\ref{eq:syseqFb0L4}) au voisinage de $\cF^2_{4}(e_1)$. Sur la figure de gauche, la famille $\ol \cF^0_{4}$ est représentée en rouge pour $e_1=0.001$ , en vert pour $e_1=0.3$, en bleu pour $e_1=0.5$ et en violet pour $e_1=0.7$. La direction de la famille $\ol \cF^1_{4}$ est constante à l'ordre 2 en $\Re (Z_1)$ et est représentée en noir. La figure de droite représente l'évolution des excentricités des corps $m_1$ et $m_2$ le long de ces coupes des familles $\ol \cF^0_{4}$ en bleu et $\ol \cF^1_{4}$ en noir pour $e_1=0.5$.} 
\end{center}
\end{figure}

Intéressons-nous maintenant aux directions des familles $\overline{\cF}^l_{k}$ en un point donné des $\cF^2_{k}$. Comme précédemment, nous développons la méthode pour les familles émergeant de $L_4$. Les familles émergeant de $L_5$ seront identiques à une symétrie près, et une méthode similaire sera appliquée pour les familles émergeant de $L_3$.\\

\an{We now compute the quadratic approximations of the $\overline{\cF}^l_{k}$ families in a given point of the $\cF^2_{k}$ families.}

\subsubsection{Voisinage de $\cF^2_{4}$}

Reprenons la partie quadratique du hamiltonien $\gH_{\cF^2_{4}}(e_1)$. Comme précédemment évoqué, les directions des familles $\ol \cF^0_{4}$ et $\ol \cF^1_{4}$ en $\cF^2_{4}(e_1)$ sont données respectivement par $z_0=0$ et $z_1=0$. Afin de connaître leur position dans l'espace des phases initial ($Z,z,x_j,\xt_j$) nous devons exprimer ces variables en fonction des ($Z_j,\tilde{Z}_j$). Nous avons la relation suivante:
\begin{equation}
  (Z,z,x_j,\xt_j) =\chi_{L_4} \circ \chi_{\cF^2_{4}} \circ \chi_d(e_1) (Z_j,\tilde{Z}_j) .
\label{eq:varini1}
\end{equation}
Nous obtenons donc l'expression des variables ($Z,z,x_j,\xt_j$) comme des polynômes de degré $2$ en ($Z_j,\tilde{Z}_j$) et $m$ en $e_1$. Nous pouvons par exemple écrire, pour $z$ dans la direction de $\ol \cF^0_{4}$:
\begin{equation}
 z= \sum_{l\in \{1,...,m\},|k_1|+|k_2|+|\tilde{k}_1|+|\tilde{k}_2|=2} c_{l,k_1,k_2,\tilde{k}_1,\tilde{k}_2}  e_1^l Z^{k_1}_1 \tilde{Z}^{\tilde{k}_1}_1 Z^{k_2}_2 \tilde{Z}^{\tilde{k}_2}_2 
\label{eq:varini2}
\end{equation}
Nous obtenons donc une expression des membres la famille $\ol \cF^0_{4}$ au voisinage de $\cF^2_{4}(e_1)$ à l'ordre $m$ en $e_1$ et $2$ en ($Z_j,\tilde{Z}_j$).

 Afin de visualiser l'approximation quadratique de $\ol \cF^0_{4}$ (de dimension $4$) dans le voisinage de $\cF^2_{4}$, nous allons calculer son intersection avec la variété $\cP$ (de dimension $3$):
\begin{equation}
\left \{
\begin{array}{c @{=} c}
J_1(x_j,\xt_j) & J_{1,0}\, , \\
\Im (x_1) & 0\, ,\\
Z & 0\, ,
\end{array}
\right.
\label{eq:syseqFb0L4}
\end{equation}
où $J_1$ est la valeur du moment cinétique total du système:
\begin{equation}
J_1 = (\Lambda^0_1+\Lambda^0_2-i(x_1 \xt_1+x_2 \xt_2))/2
\label{eq:mcin}
\end{equation}
et $J_{1,0}$ sa valeur en $e_2=e_1$. Notons que la définition choisie pour $\cP$ est celle des variétés de référence (pour le cas $m_1=m_2$) que nous avons étudié dans les sections précédentes. L'intersection de ces deux variétés est dans le cas général de dimension $1$. Nous obtenons donc un système de trois équations polynomiales en $Z_1$, $\tilde{Z}_1$, $Z_2$ et $\tilde{Z}_2$. nous rappelons que $\tilde{Z}_j=-i \bar{Z}_j$. Le système (\ref{eq:syseqFb0L4}) permet donc d'obtenir une expression pour l'ensemble de ces variables en fonction d'un seul paramètre. Nous prendrons la partie réelle de $Z_1$, $\Re (Z_1)$, pour la famille $\ol \cF^0_{4}$, et $\Re (Z_0)$ pour la famille $\ol \cF^1_{4}$.
Comme nous sommes au voisinage de la famille $\cF^2_{4}(e_1)$, les variables ($Z_j,\tilde{Z}_j$) sont petites. Nous pouvons donc résoudre le système d'équations (\ref{eq:syseqFb0L4}) degré par degré: on résout d'abord le système en tronquant les polynômes à l'ordre 1, on obtient alors une expression des ($Z_j,\tilde{Z}_j$) du type $Z_j = c_1 \Re (Z_1)$. On injecte ensuite cette solution dans le système (\ref{eq:syseqFb0L4}) tronqué à l'ordre 2 des polynômes en cherchant une modification de l'expression des $Z_j$ de taille $(\Re (Z_1))^2$. On obtient donc $Z_j = c_1 \Re (Z_1) + c_2 (\Re (Z_1))^2$.\\

De cette manière, nous obtenons l'approximation quadratique de l'intersection $\ol \cF^0_{4} \cap \cP$ et $\ol \cF^1_{4}$ à l'ordre 2 en $\Re (Z_1)$. Les figures \ref{fig:FbL4dir} et \ref{fig:FbL4mdifdir} montrent la direction de ces familles au voisinage de $\cF^2_{4}(e_1)$ pour plusieurs valeurs de $e_1$ dans le plan donné par le système (\ref{eq:syseqFb0L4}). La figure \ref{fig:FbL4dir} montre le cas de deux masses égales, alors que $m_2=19 \times m_1$ sur la figure \ref{fig:FbL4mdifdir}. Dans le cas de masses égales, la coupe des familles $\ol \cF^0_{4}$ et $\ol \cF^1_{4}$ effectuée par le système d'équations (\ref{eq:syseqFb0L4}) reste sur la variété $e_1=e_2$, comme c'est le cas pour la famille de points fixes $\cF^2_{4}$. Cette propriété disparait quand les coorbitaux ont des masses différentes, comme l'illustre le graphe de droite de la figure \ref{fig:FbL4mdifdir}. \\

\an{
$\gH_{\cF^2_{4}}(e_1)$ describes the dynamics in the neighbourhood of $\cF^2_{4}(e_1)$. The direction of $\ol \cF^0_{4}$ and $\ol \cF^1_{4}$ in $\cF^2_{4}(e_1)$ is given by $z_0=0$ and $z_1=0$, respectively. The relation (\ref{eq:varini1}) gives their position in the variables ($Z,z,x_j,\xt_j$). For example, the expression of the variable on the quadratic approximation of $\ol \cF^0_{4}$ is of the form (\ref{eq:varini2}).\\
To visualise the quadratic approximation of $\ol \cF^0_{4}$ (4-dimensional manifold) in the neighbourhood of $\cF^2_{4}$, we compute its intersection with the manifold $\cP$ (eq. \ref{eq:syseqFb0L4}). Note that the expression of $\cP$ is similar to the expression of the reference manifold (for $m_1=m_2$) that we used in the previous sections. The intersection of these two manifolds is generally 1-dimensional. As we are in the neighbourhood of $\ol \cF^2$, the variables ($Z_j$,$\tilde{Z}_j$) are smalls. The solutions of the equations (\ref{eq:varini2}) and (\ref{eq:syseqFb0L4}) can hence be expanded as a polynomial in one of the variable (we chose $\Re (Z_1)$ here). We can thus solve the equations (\ref{eq:varini2}) and (\ref{eq:syseqFb0L4}) degree by degree.\\
We obtain the quadratic approximation of $\ol \cF^0_{4} \cap \cP$ and $\ol \cF^1_{4}$ at the order $2$ in $\Re (Z_1)$.  The figures \ref{fig:FbL4dir} and \ref{fig:FbL4mdifdir} shows the results in the vicinity of $\cF^2_{4}(e_1)$. The quadratic approximation of $\ol \cF^1_{4}$, in black, is constant with respect to $e_1$. The quadratic approximation of $\ol \cF^0_{4}$ (colored lines) is given for several values of $e_1$. The figure \ref{fig:FbL4dir} shows the case $m_1=m_2$, while $m_2=19 \times m_1$ in figure \ref{fig:FbL4mdifdir}. When $m_1=m_2$, $\ol \cF^0_{4} \cap \cP$ and $\ol \cF^1_{4} \cap \cP$ stay in the plane $e_1=e_2$, as it is the case for the family $\cF^2_{4}$. This does not hold for $m_1 \neq m_2$, see figure \ref{fig:FbL4mdifdir}. 
}


\subsubsection{Voisinage de $\cF^2_{3}$}

\begin{figure}[h!]
\begin{center} 
\includegraphics[width=0.5\linewidth]{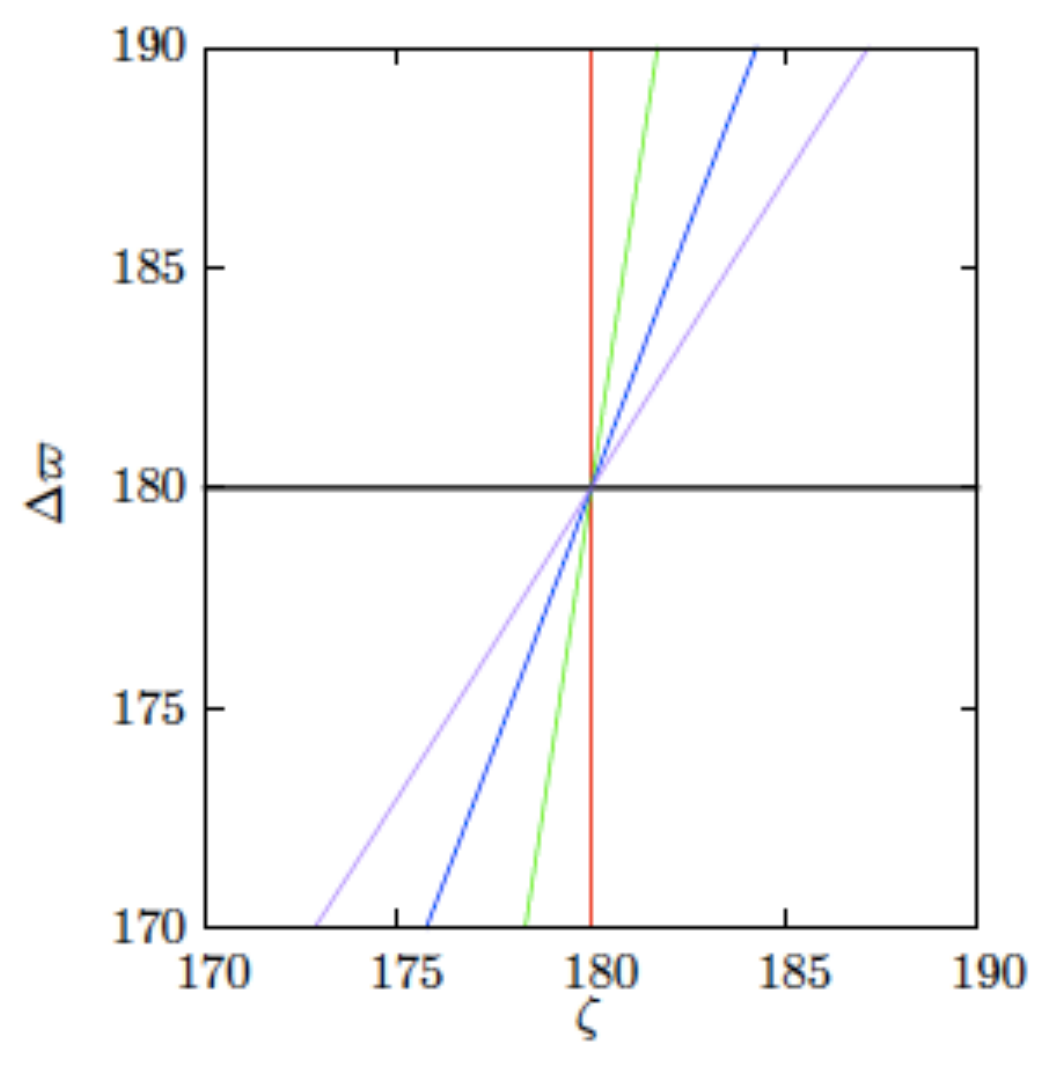}
\caption{\label{fig:FbL3dir}  approximation quadratique des familles $\overline{\cF}^0_{3}$ et $\ol \cF^1_{3}$ pour $m_1=m_2=1\, 10^{-6}m_0$ sur la variété $\cP$ définie par le jeu d'équations (\ref{eq:syseqFb0L4}) au voisinage de $\cF^2_{3}(e_1)$, la famille $\overline{\cF}^0_{3}$ est représentée en rouge pour $e_1=0.001$ , en vert pour $e_1=0.3$, en bleu pour $e_1=0.5$ et en violet pour $e_1=0.7$. La direction de la famille  $\ol \cF^1_{3}$ est constante sur $\cF^2_{3}(e_1)$ à l'ordre 2 en $|Z_1|$ et est représentée en noir.} 
\end{center}
\end{figure}

Une approche similaire peut être effectuée pour le voisinage de $\cF^2_{3}(e_1)$. La seule différence avec celui de $\cF^2_{4}(e_1)$ est que nous n'avons pas la relation $\tilde{Z}_j=-i \bar{Z}_j$. Un algorithme similaire à celui qui vient d'être décrit peut cependant également être utilisé. En pratique le paramètre choisi pour le parcours de ces familles sera $|Z_1|$. La figure \ref{fig:FbL3dir} montre la direction de ces familles au voisinage de $\cF^2_{3}(e_1)$ dans le plan donné par le système (\ref{eq:syseqFb0L4}) pour plusieurs valeurs de $e_1$.\\

Il est possible de calculer l'expression des familles $\overline{\cF}^l_{k}$ dans un voisinage des $\overline{\cF}^l_{k}$ à un ordre plus élevé. Dans le cadre de cette étude, on se limitera au voisinage $\cF^2_{4}$ car les orbites dans le voisinage de $\cF^2_{3}$ sont généralement instables et présentent donc moins d'intérêt.

\an{Similarly, we compute the quadratic approximation of the intersection $\overline{\cF}^0_{3} \cap \cP$ and $\ol \cF^1_{3}\cap \cP$ in the case $m_1=m_2$, see figure \ref{fig:FbL3dir}. The quadratic approximation of $\ol \cF^1_{3}$, in black, is constant with respect to $e_1$. The quadratic approximation of $\ol \cF^0_{3}$ (colored lines) is given for several values of $e_1$. 
}

\subsection{Forme normale au voisinage d'un point de $\cF^2_{4}$}
\label{sec:FNBP}
Repartons du hamiltonien $\gH_{\cF^2_{4}}(e_1)$ (obtenu par le changement de variable \ref{eq:cv4f}) à l'ordre $m$ en $e_1$ et $p$ en les variables ($Z_j$,$\tilde{Z}_j$) dont la partie quadratique est diagonale. Nous sommes ici dans le cas où la fréquence $\omega_2$ est nulle, et la famille  $\cF^2_{4}$ est une famille de points fixes (section \ref{sec:gen}). Nous obtenons donc une forme normale de la forme:
\begin{equation}
\cN_{\cF^2_{4}}(e_1) = \sum_{d=2}^p \left( \sum_{k_0+k_1+k_2+\bar{k}_2=d}  \left( C_{k_0,k_1,k_2,\tilde{k}_2} (Z_0 \tilde{Z}_0)^{k_0} (Z_1 \tilde{Z}_1)^{k_1} Z^{k_2}_2 \tilde{Z}^{\tilde{k}_2}_2\right) \right)\, .
  \label{eq:HFN}
\end{equation}
Notons que les termes en $Z_2$ et $\tilde{Z}_2$ n'apparaissent pas dans l'expression du hamiltonien au voisinage des équilibres circulaires, et proviennent des termes croisés lors de la translation (\ref{eq:transIj}).  


Pour obtenir cette forme normale on élimine degré par degré tous les termes que nous ne voulons pas voir apparaître par le biais d'un changement de variable bien choisi. Comme pour la diagonalisation de la partie quadratique du hamiltonien, on effectue ce changement de variable par composition des flots au temps un de $p$ hamiltoniens $\cW_p$ \citep{MeHa1992}. On note ce changement de variable
\begin{equation}
\begin{aligned}
 (Z_j, \tilde{Z}_j) & = \chi_{\cN}(I_j,\theta_j),\\
                  & = (\sqrt{2I_j} \e^{i \theta_j},\sqrt{2I_j} \e^{-i \theta_j}).
\end{aligned}
\label{eq:cv5}
\end{equation}
La forme (\ref{eq:HFN}) peut, par un dernier changement de variable, être exprimée en variables action-angles. On obtient alors, à l'ordre $p$ en $I_j$ et $m$ en $e_1$:
\begin{equation}
\cN'_{\cF^2_{4}}(e_1) = \sum_{d=2}^p \left( \sum_{2k_1+2k_2+k_3+\bar{k}_3=d}  \left( C_{k_1,k_2,k_3,\bar{k}_3} I_1^{2k_1} I_2^{2k_2} I^{k_3+\bar{k}_3}_3 \e^{i(k_3-\bar{k}_3) \theta_3}\right) \right) .
  \label{eq:H6}
\end{equation}
avec le changement de variable:
\begin{equation}
\begin{aligned}
 (Z_j, \tilde{Z}_j) & = \chi_{\cN'}(I_j,\theta_j),\\
                  & = (\sqrt{2I_j} \e^{i \theta_j},\sqrt{2I_j} \e^{-i \theta_j}).
\end{aligned}
\label{eq:cv6f}
\end{equation}
Cette forme permet d'identifier la position des $\ol \cF^l_{4}$ au voisinage de $\cF^2_{4}$. Une utilisation rigoureuse de cette forme normale nécessite de calculer son rayon de convergence\footnote{En pratique la série diverge, on peut cependant estimer l'ordre optimal (celui qui minimise le reste), voir \cite{Morbidelli02}}. Dans cette étude nous nous contenterons de calculer les premiers termes de la forme normale afin d'approximer les $\ol \cF^l_{4}$ au voisinage de $\cF^2_{4}$. \\

\an{
To obtain a better approximation of $\ol \cF^0_{4} \cap \cP$ and $\ol \cF^1_{4} \cap \cP$, we will compute the normal form of the Hamiltonian $\gH_{\cF^2_{4}}(e_1)$ at the order $p$ in the variables ($Z_j$,$\tilde{Z}_j$). We are here in the case $\omega_2=0$ (see section \ref{sec:gen}).\\
We hence obtain a normal form whose expression is given equation (\ref{eq:HFN}). Note that the terms in $Z_2$ and $\tilde{Z}_2$ do not appear in the circular case, but they come from the translation (\ref{eq:transIj}).\\
To obtain the normal form, we cancel degree by degree all the monomials that are not of the form (\ref{eq:HFN}). Similarly to the transformation to diagonalise the quadratic part of the Hamiltonian, we compute the change of variables  $\chi_{\cN}$ (eq. \ref{eq:cv5}) by composing the time-one flow of $p$ Hamiltonians $\cW_p$ \citep{MeHa1992}. One can then put the Hamiltonian under the form (\ref{eq:H6}) using a last change of variables (\ref{eq:cv6f}).\\
The form (\ref{eq:H6}) allows to identify the approximation of the  $\ol \cF^l_{4}$ in the neighbourhood of $\cF^2_{4}$ at the order $p$ in ($Z_j \tilde{Z}_j$) and $m$ in $e_1$.
}

\subsection{$\ol \cF^l_{4}$ au voisinage de $\cF^2_{4}$} 
\label{sec:DFFN}

\begin{figure}[h!]
\begin{center} 
\includegraphics[width=0.49\linewidth]{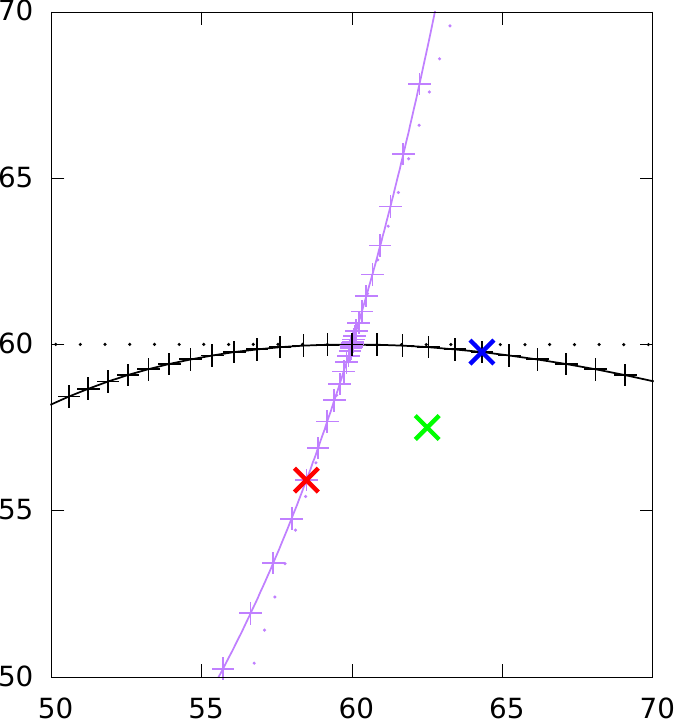}\\
  \setlength{\unitlength}{0.1\linewidth}
\begin{picture}(.001,0.001)
\put(-3,3){$\Dv$}
\put(0,0){{$\zeta$}}
\end{picture}
\vspace{0.5cm}

\includegraphics[width=0.47\linewidth]{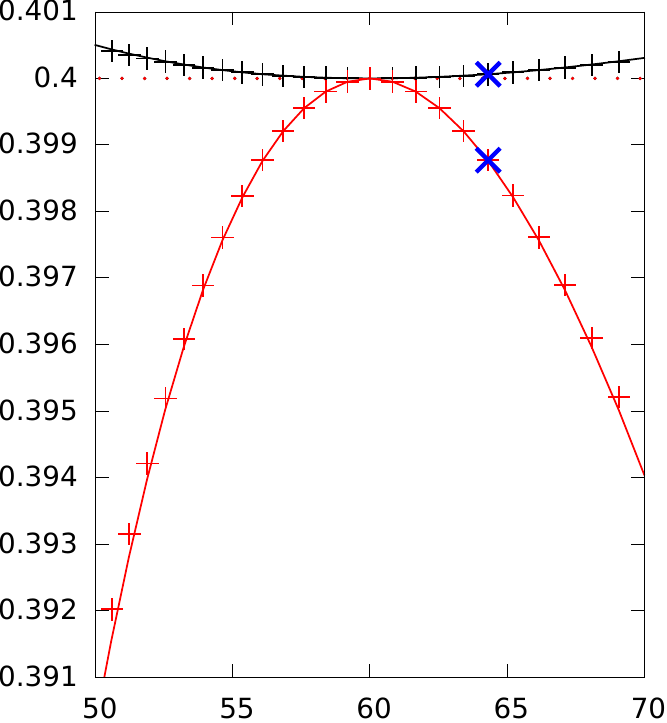} \hfill
\includegraphics[width=0.47\linewidth]{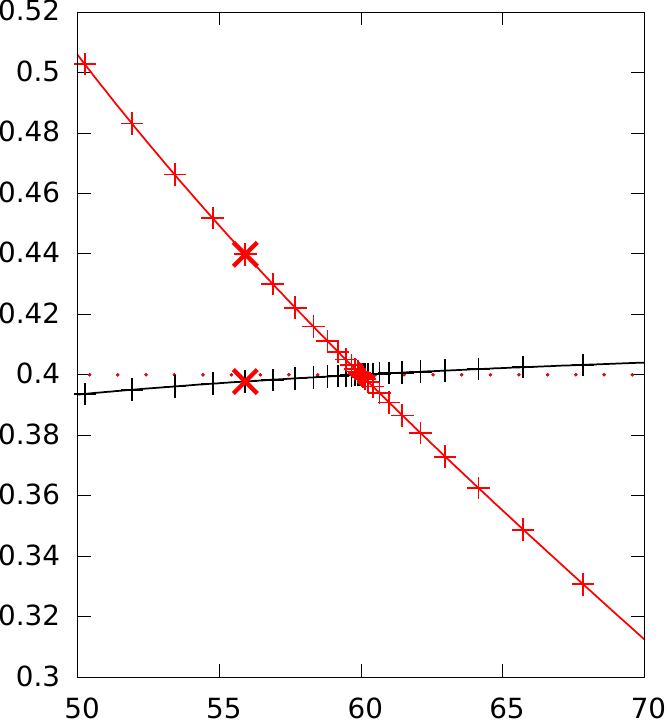} \\
\vspace{0.1cm}
  \setlength{\unitlength}{0.1\linewidth}
\begin{picture}(.001,0.001)
\put(0,3){$e_j$}
\put(-5.5,3){$e_j$}
\put(-2.5,0){{$\zeta$}}
\put(2.5,0){{$\Dv$}}
\end{picture}
\caption{\label{fig:FbL4ana} approximation des familles $\ol \cF^l_{4}$ vérifiant les équations (\ref{eq:syseqFb0L4}) et paramétrées par $\Re (Z_1)$, pour $m_2=19m_1$. Ces familles sont représentées pour $p=1$ (dots), $p=3$ (line) and $p=5$ ($+$). Sur le graphe du haut: $\ol \cF^0_{4}$ en violet et $\ol \cF^1_{4}$ en noir. Sur les graphes du bas: $e_1$ en rouge et $e_2$ en noir, le long de $\ol \cF^1_{4}$ à gauche et $\ol \cF^0_{4}$ à droite. Les '$\times$' représentent les conditions initiales pour les intégrations de la figure \ref{fig:FbL4orb}.} 
\end{center}
\end{figure}

\begin{figure}[h!]
\begin{center} 
\includegraphics[width=0.47\linewidth]{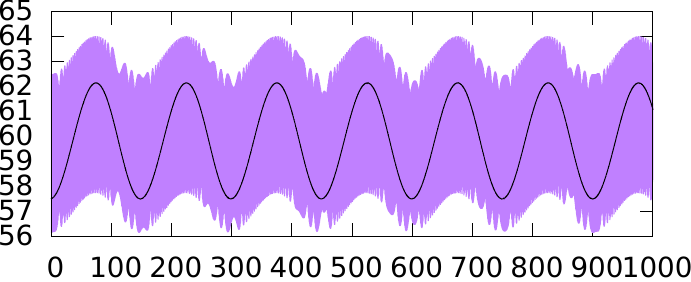} \hfill
\includegraphics[width=0.47\linewidth]{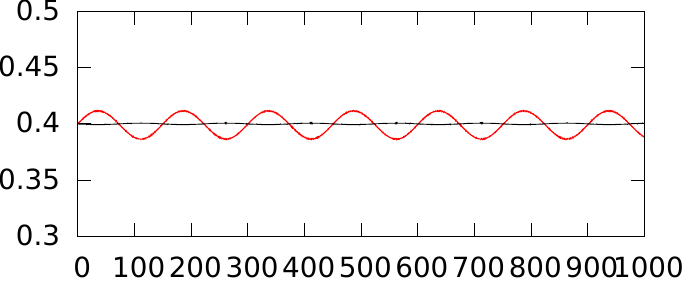}  \\
\includegraphics[width=0.47\linewidth]{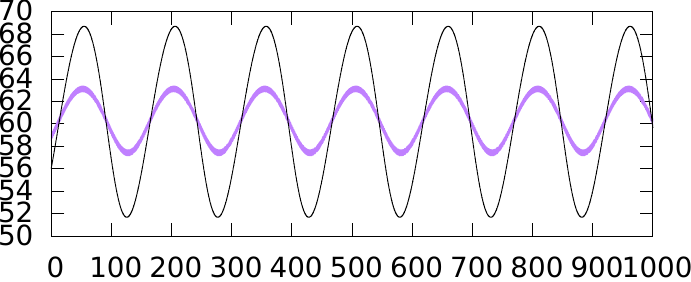} \hfill
\includegraphics[width=0.47\linewidth]{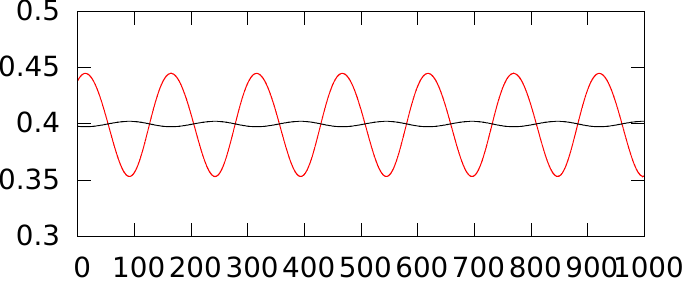}  \\
\includegraphics[width=0.47\linewidth]{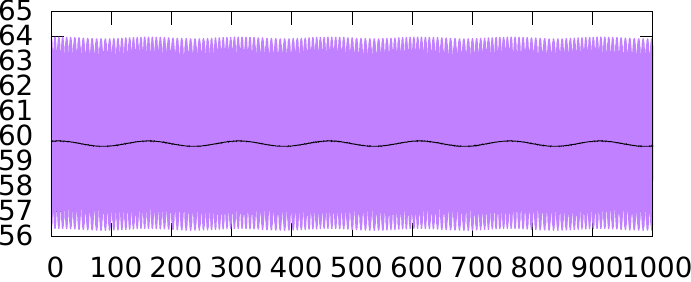} \hfill
\includegraphics[width=0.47\linewidth]{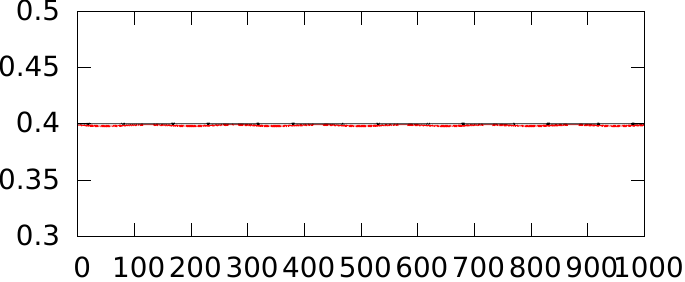} \\
\vspace{0.1cm}
  \setlength{\unitlength}{0.1\linewidth}
\begin{picture}(.001,0.001)
\put(0,3.2){$e_j$}
\put(-5.5,3.2){$[^\circ]$}
\put(-3,0){{$t[\times 10^{4} ]$~year}}
\put(2,0){{$t[\times 10^{4} ]$~year}}
\end{picture}
\caption{\label{fig:FbL4orb} Intégration des trajectoires dont les conditions initiales sont représentées par des '$\times$' sur la figure \ref{fig:FbL4ana}. Sur la ligne du haut, la trajectoire identifiée par les croix vertes, avec $e_1=e_2=0.4$ en conditions initiales. Sur la ligne du centre, les croix bleues. Cette trajectoire est dans le voisinage de $\ol \cF^1_{4}$. Et sur la ligne du bas, la trajectoire identifiée par les croix rouges. Cette trajectoire est dans le voisinage de $\ol \cF^0_{4}$. La colonne de gauche représente l'évolution des angles $\Delta \varpi$ (noir) et $\zeta$ (violet) la colonne de droite représente l'évolution de $e_1$ (rouge) et de $e_2$ (noir). les abscisses représentent le temps en dizaine de millier de périodes orbitales. pour ces trois orbites $m_2=19 m_1 = 19\times10^{-7}m_0$. } 
\end{center}
\end{figure}

Nous avons vu en section \ref{sec:DFFb} comment déterminer l'intersection des familles $\ol \cF^l_{4}$ avec la variété $\cP$ dans les variables elliptiques. Pour le hamiltonien à l'ordre $p$ en les variables $(Z_j, \tilde{Z}_j)$, c'est sous la forme normale (\ref{eq:HFN}) que ces directions apparaissent de manière triviale. Nous procédons donc ici au même algorithme que en section \ref{sec:DFFb} en ajoutant le changement de variable $\chi_{\cN}$.\\

 La figure \ref{fig:FbL4ana} montre l'évolution des familles $\ol \cF^l_{4}$ vérifiant les équations (\ref{eq:syseqFb0L4}) et paramétrées par $\Re (Z_1)$. Ces familles sont représentées pour $p=1$ (dots), $p=3$ (line) and $p=5$ ($+$). Nous pouvons voir que dans le cas de deux coorbitaux de masse différente, ces familles sortent de la variété $e_1=e_2$ dès que l'on quitte l'équilibre de Lagrange. Sur cette figure, 3 points de l'espace des phases sont identifiés par le symbole '$\times$'. Ces points sont pris comme conditions initiales (avec le jeu d'équations \ref{eq:syseqFb0L4}) pour les intégrations du problème à trois corps dont l'évolution temporelle est présentée sur la figure \ref{fig:FbL4orb}. 
 
 Rappelons que $g=\gO(\eps)$ et $\nu=\gO(\sqrt{\eps})$. Puisque ici $\eps\approx 2\, 10^{-6}$, les échelles de temps semi-rapide et séculaire sont bien distinctes, avec $P_g/P_\nu\approx 700$. La trajectoire ayant pour condition initiale la croix verte (représentée en haut de la figure \ref{fig:FbL4ana}, pour laquelle on prendra en condition initiale $e_1=e_2=0.4$) est sur une orbite à priori quelconque du voisinage de $\cF^2_{4}(0.4)$. Sur cette trajectoire, $\Delta \varpi$ et les excentricités évoluent principalement avec la fréquence $g$, alors que $\zeta$ est impactée à la fois par les échelles de temps semi-rapide et séculaire. La trajectoire bleue (représentée au centre de la figure \ref{fig:FbL4ana}), quant à elle, fait partie de l'expression de la famille $\ol \cF^0_{4}$ à l'ordre $p=5$ de la forme normale. On s'attend donc à ce que l'ensemble des paramètres représentés n'évolue que sur l'échelle de temps séculaire. A l'inverse, les conditions initiales de la trajectoire rouge (représentée en bas de la figure \ref{fig:FbL4ana}), fait partie de l'expression de la famille $\ol \cF^1_{4}$ à l'ordre $p=5$ de la forme normale. On s'attend donc à ce que l'ensemble des paramètres représentés n'évolue que sur l'échelle de temps semi-rapide.\\
 
 La détermination des familles  $\ol \cF^l_{4}$ par cette méthode est approximative: la trajectoire bleue comporte de faibles variations semi-rapides dans l'évolution de $\zeta$, et la trajectoire rouge comporte de faibles variations de $\Delta\varpi$ et des $e_j$ sur une échelle de temps séculaire. Néanmoins, ces variations sont faibles par rapport à une orbite générique, et peuvent être dues en partie à la différence (de taille $\eps$) entre la valeur des conditions initiales dans le problème moyen (là ou elles ont été estimées) et leur valeur dans le problème complet (là où elle ont été intégrées). On considèrera donc que les familles déterminées par la forme normale sont au voisinage de leur position exacte dans l'espace des phases et que leur position déterminée ici ne reste valable que très localement. L'expression de ces familles $\ol \cF^l_k$ permet cependant de confirmer que les familles mises en avant par les méthodes semi-analytiques et numériques décrites en section \ref{sec:SAIF} sont bien les mêmes objets que les familles définies au voisinage des équilibres de Lagrange circulaires dans la section \ref{sec:H0pf}.\\
 
\an{
In this section we compute the approximation of $\ol \cF^l_{4} \cap \cP$ in the neighbourhood of $\cF^2_{4}$ at the order $p$ in ($Z_j \tilde{Z}_j$) and $m$ in $e_1$. As we did in the section \ref{sec:DFFb}, we solve the equations (\ref{eq:varini2}) and (\ref{eq:syseqFb0L4}) degree by degree (with $Z_0=\tilde{Z}_0=0$ for $\ol \cF^0_{4}$ and $Z_1=\tilde{Z}_1=0$ for $\ol \cF^1_{4}$), adding the change of variables $\chi_{\cN}$.\\  
 The figure \ref{fig:FbL4ana} shows the approximation of $\ol \cF^l_{4} \cap \cP$. These approximations are plotted at the order $p=1$ (dots), $p=3$ (line) and $p=5$ ($+$) in ($Z_j \tilde{Z}_j$). For $m_1 \neq m_2$, these families depart from the plane $e_1=e_2$ as soon as they depart from the Lagrangian equilibrium. In this figure, 3 points of the phase space are identified by the symbol '$\times$'. These points are taken as initial conditions for 3-body problem integrations whose temporal evolutions are represented in figure~\ref{fig:FbL4orb}.\\
The black lines on the left column represent the evolution of $\Dv$, the purple lines represent the evolution of $\zeta$. The curves on the right column represent the evolution of $e_1$ (red) and $e_2$ (black). Note that, since $\eps\approx 2\, 10^{-6}$, we have $P_g/P_\nu\approx 700$. The green $\times$ is a generic orbit of the neighbourhood of $\cF^2$. Its evolution is represented on the top row. We can see that $\Dv$ and $e_j$ evolve mainly with a period $2\pi/g$, while $\zeta$ is similarly impacted by both periods. The middle row take as initial conditions the blue $\times$, in the approximation of $\ol \cF^0$, the dynamics seems to be indeed dominated by the period $2\pi/g$. Finally, the bottom row take the red $\times$ as initial conditions. The impact of the period $2\pi/g$ is negligible with respect to its impact on a generic orbit.\\}

\an{In this section we computed the positions of the $\ol \cF^j_k$ at the order $m$ in $e_1$ and at the order $p$ in $z$, $Z$, $x_j$ and $\xt_j$ in the neighbourhood of the $\cF^2_k$ families in the averaged problem. With these results, we can confirm the position of the $\ol \cF^\nu_k$ and $\ol \cF^g_k$ families of the averaged reduced problem that we identified with numerical criterion in the previous sections.}

%
%
%
%
%
%
%
%
%

  \chapter{Co-orbitaux circulaires inclinés}
\label{chap:coi}

Dans ce chapitre, nous étudions l'évolution des différentes configurations coorbitales du cas circulaire quand l'inclinaison mutuelle entre les deux planètes augmente. Cela a un intérêt notable pour la détection d'exoplanètes coorbitales: si l'inclinaison mutuelle est non négligeable, il est probable qu'une planète transite mais pas son compagnon troyen. Il est donc intéressant de pouvoir contraindre, grâce a un argument de stabilité, les valeurs possibles des paramètres orbitaux d'un éventuel compagnon troyen.  

Nous avons montré en section \ref{sec:ivcp} que la variété $e_1=e_2=0$ est stable par le flot du Hamiltonien moyen. Nous pouvons donc nous intéresser à la dynamique des coorbitaux circulaires inclinés. Ce problème moyen comporte 3 degrés de libertés représentés par les variables canoniques conjuguées ($Z,\zeta$), ($y_1,\tilde{y}_1$) et ($y_2,\tilde{y}_2$).\\

\an{
In this chapter we study the evolution of the circular coorbital configurations when the mutual inclination between the co-orbital increases. It provides useful information to detect co-orbital exoplanets: if the mutual inclination is not too small, the co-orbital companion of a transiting exoplanet might not transit. However we may be able to constrain, for stability reason, the orbital parameters of the eventual co-orbital companion.\\ 
We will study the circular inclined case. We showed in section \ref{sec:ivcp} that the manifold $e_1=e_2=0$ is stable by the flow of the averaged Hamiltonian. The averaged circular problem has 3 degrees of freedom, associated to the canonical conjugated variables ($Z,\zeta$), ($y_1,\tilde{y}_1$) and ($y_2,\tilde{y}_2$).
}

\section{Réduction du problème}
\label{sec:Rdpi}
Suivant \cite{Poi1892} et \cite{Ro1995}, nous allons effectuer la réduction de Jacobi. Celle-ci utilise l'invariance du moment cinétique afin de réduire le problème de deux degrés de liberté supplémentaires en s'affranchissant de la longitude des n\oe uds et des inclinaisons et en introduisant un paramètre supplémentaire: le moment cinétique total du système. Cette réduction consiste à nous placer dans un référentiel tournant (non uniformément) dont le plan $Oxy$ coïncide avec le plan invariant perpendiculaire au moment cinétique total et qui tourne avec la ligne des n\oe uds qui est commune aux deux corps dans le plan invariant (dans ce plan, $\Omega_2 = \Omega_1+\pi$). Si nous utilisons les variables canoniques de Delaunay définies dans le repère lié au plan invariant par:
\begin{equation}
\begin{aligned}
L_j&=\sqrt{\mu_ja_j}\, , \vspace{2cm}\ &l_j=M_j\, , \\
\Theta_j&=L_j\cos i_j\, , \vspace{2cm}\ & \theta_j=\Omega_j\, , \\
\end{aligned}
\label{eq:Delaunay}
\end{equation}
dans le cas où $e_1=e_2=0$, le moment cinétique total s'écrit:
\begin{equation}
\bold{C}= 
   \begin{pmatrix}
   \beta_1 L_1 \sin i_1 \sin \theta_1 + \beta_2 L_2 \sin i_2 \sin \theta_2 \\
   -\beta_1 L_1 \sin i_1 \cos \theta_1 - \beta_2 L_2 \sin i_2 \cos \theta_2 \\	 
   	\beta_1 \Theta_1 + \beta_2 \Theta_2
    \end{pmatrix}   
    =
       \begin{pmatrix}
   0\\0\\ C
    \end{pmatrix}   
\label{eq:P2}
\end{equation} 
 Or dans le repère lié au plan invariant, le moment cinétique est porté par la direction $Oz$, nous obtenons donc le jeu d'équations suivant \cite{Ro1995}:
 \begin{equation}
\left \{
\begin{array}{c @{=} c}
\beta_1 L_1 \sin \theta_1 - \beta_2 L_2 \sin \theta_2 & 0\\
	\beta_1 \Theta_1 + \beta_2 \Theta_2 & C\\
	\theta_1 - \theta_2 & \pi
\end{array}
\right.
\label{eq:syseqPI}
\end{equation}
 Donc, en effectuant le changement de variable canonique suivant:
  \begin{equation}
\begin{aligned}
\Psi_1&=\beta_1 \Theta_1 + \beta_2 \Theta_2\, , \vspace{2cm}\ &\psi_1=(\theta_1+\theta_2)/2\, , \\
\Psi_2&=\beta_1 \Theta_1 - \beta_2 \Theta_2\, , \vspace{2cm}\ &\psi_2=(\theta_1-\theta_2)/2\, , \\
\end{aligned}
\label{eq:varPI}
\end{equation}
 où les variables $\Psi_1=C$ et $\psi_2= \pi/2$ sont constantes dans le repère lié au plan invariant. En écrivant $\chi_{PI}$ le changement de variable canonique entre les coordonnées de Poincaré et les variables ($\Psi_j, \psi_j$), nous pouvons déduire que le Hamiltonien moyen $\ol \gH_{PI}$ défini par:
\begin{equation}
\begin{aligned}
\ol \gH_{PI} = \ol \gH \circ \chi_{PI}\, ,
\end{aligned}
\label{eq:HPI}
\end{equation}
ne dépend pas des variables $\psi_1$ et $\Psi_2$. En effet:
\begin{equation}
\begin{aligned}
\frac{\partial \ol \gH_{PI}}{\partial  \psi_1} & = - \dot{\Psi}_1 &= 0\, , \\
\frac{\partial \ol \gH_{PI}}{\partial  \Psi_2} & = \dot{\psi}_2 &= 0\, . \\
\end{aligned}
\label{eq:varinvar}
\end{equation}
$\ol \gH_{PI}$ ne dépend donc que des variables ($Z,\zeta,\psi_2,\Psi_1$). Or les variables $\psi_2$ et $\Psi_1$ sont ici ignorables car leur variable conjuguée n'apparaît pas dans l'expression du Hamiltonien. Le Hamiltonien $\ol \gH_{PI}$ est donc intégrable et est paramétré par la valeur de $C$ ($\psi_2$ valant $\pi/2$ dans tout les cas). Par analogie au cas circulaire, on peut également s'attendre à ce que le paramètre de masse $\eps$ n'agisse que comme un facteur d'échelle de temps dans le cas du problème moyen réduit. Ceci ne sera pas démontré ici mais les diverses simulations numériques tendent à confirmer cette hypothèse.

 Les variations de $\Psi_2$ dépendent quant à elles de la dynamique du degré de liberté ($Z,\zeta$). Les relations (\ref{eq:syseqPI}) et (\ref{eq:varPI}) permettent d'écrire \citep{Ro1995}:
 \begin{equation}
\left \{
\begin{array}{c @{=} c}
	\beta_1 \Theta_1 & \frac{C}{2} + \frac{\beta_1^2 L_1^2 - \beta_2^2 L_2^2}{2C} \\
    \beta_2 \Theta_2 & \frac{C}{2} - \frac{\beta_1^2 L_1^2 - \beta_2^2 L_2^2}{2C}
\end{array}
\right.
\label{eq:syseqPI}
\end{equation}
La quantité $\cos i_j$ varie donc comme $\frac{\beta_1^2 L_1^2 - \beta_2^2 L_2^2}{2C\beta_jL_j}$. Nous pouvons estimer la taille des variations des $\cos i_j$ en calculant ce terme à l'ordre $1$ en $Z$:
 \begin{equation}
\frac{\beta_1^2 L_1^2 - \beta_2^2 L_2^2}{2C\beta_jL_j} = \frac{{\Lambda^0_1}^2-{\Lambda^0_2}^2}{2 C {\Lambda^0_1}} + Z \frac{({\Lambda^0_1}+{\Lambda^0_2}) (2-({\Lambda^0_1}-{\Lambda^0_2})/{\Lambda^0_1})}{ 2 C {\Lambda^0_1}}
\label{eq:evocosi}
\end{equation}
 Ce terme comporte donc une partie constante et un terme linéaire en $Z$. En considérant que $m_1 \leq m_2$, le terme linéaire est borné par:
 \begin{equation}
\left|\frac{({\Lambda^0_1}+{\Lambda^0_2}) (2-({\Lambda^0_1}-{\Lambda^0_2})/{\Lambda^0_1})}{ 2 C {\Lambda^0_1}}\right| < 2 \frac{\left(\frac{\Lambda^0_2}{\Lambda^0_1}+\left(\frac{\Lambda^0_2}{\Lambda^0_1}\right)^2\right)}{C}
\label{eq:bornZ}
\end{equation}
On ne peut conclure ici pour des conditions initiales quelconques: en effet le moment cinétique peut être nul, par exemple en prenant $i_1=i_2=90^\circ$ et $m_1=m_2$. Intéressons-nous à des inclinaisons modérées: en prenant par exemple $\max (i_1,i_2)< \pi/3$ Nous avons l'inégalité suivante:
 \begin{equation}
C > (\Lambda^0_1+\Lambda^0_2)/2
\label{eq:Cmin}
\end{equation}
et nous pouvons donc conclure que:
\begin{equation}
\left|\frac{({\Lambda^0_1}+{\Lambda^0_2}) (2-({\Lambda^0_1}-{\Lambda^0_2})/{\Lambda^0_1})}{ 2 C {\Lambda^0_1}}\right| < \frac{1}{\Lambda^0_1} \left( 1 + \frac{\Lambda^0_2}{\Lambda^0_1} \right)\, .
\label{eq:varimax}
\end{equation}

Pour des inclinaison modérées et des masses comparables, $\cos i_j$ subit des variations de taille $Z$.

D'autre part, on rappelle que l'inclinaison mutuelle $J$ des deux co-orbitaux est définie par:
\begin{equation}
\cos J= \cos i_1 \cos i_2 + \sin i_1 \sin i_2 \cos (2 \psi_2)
\label{eq:cosJ}
\end{equation}
donc, dans le plan invariant, nous avons simplement $J= i_1+i_2$. De plus nous avons, pour le moment cinétique total \citep{Ro1995}:
\begin{equation}
C^2 =\Lambda^2_1 + \Lambda^2_2 + 2\Lambda_1 \Lambda_2 \cos J
\label{eq:C2}
\end{equation}

\an{
Following \cite{Poi1892} and \cite{Ro1995}, we will perform the Jacobi's reduction of the node. This reduction uses the invariance of the total angular momentum of the system in order to obtain a system with 2 degrees of freedom. We place ourselves in a rotating frame whose plane $Oxy$ coincides with the invariant plane orthogonal to the total angular momentum and which rotates with the line of the nodes. The line of the nodes is common to both co-orbitals in the invariant plane (in this plane, $\Omega_2 = \Omega_1+\pi$). We use the canonical variables of Delaunay in the invariant plane (equation \ref{eq:Delaunay}) when $e_1=e_2$. The total angular momentum is given by the expression (\ref{eq:P2}).\\}

\an{In the invariant plane, the total angular momentum is along the direction $Oz$, we hence obtain the set of equations (\ref{eq:syseqPI}). Performing the canonical variables change (\ref{eq:varPI}), where $\Psi_1=C$ and $\psi_2= \pi/2$ with our choice of coordinates, and writing $\chi_{PI}$ the canonical change of variables between the Poincaré variables and the variables ($\Psi_j, \psi_j$), we obtain the averaged Hamiltonian $\ol \gH_{PI}$ (eq. \ref{eq:HPI}. This Hamiltonian does not depend on the variables $\psi_1$ and $\Psi_2$ (see equation \ref{eq:varinvar}), it is hence a one-degree of freedom Hamiltonian parametrised by $C$ (since $\psi_2=\pi/2$).\\}

\an{By analogy to the circular case, we can expect that the mass parameter $\eps$ only acts as a time and space scaling factor of the dynamics in the stable areas of the phase space ($\eps$ highly impact the shape of the stability domain). We will not demonstrate this result here, but the numerical simulations appear to confirm this hypothesis.\\}

\an{The evolution of the inclinations is given by the equation (\ref{eq:syseqPI}) (deduced from the equations \ref{eq:syseqPI} and \ref{eq:varPI}). The quantity $\cos i_j$ evolves as $\frac{\beta_1^2 L_1^2 - \beta_2^2 L_2^2}{2C\beta_jL_j}$. Its expression at order $1$ in $Z$ is given by the equation (\ref{eq:evocosi}). For moderate initial inclinations, using the expressions (\ref{eq:bornZ}) and (\ref{eq:Cmin}), we obtain the inequality (\ref{eq:varimax}). For low inclination and comparable masses, $\cos i_j$ undergoes variations of the order of $\gO(Z)$.\\}

\an{In the invariant plane, the mutual inclination is simply given by $J=i_1+i_2$ (eq. \ref{eq:cosJ}), and we have $C^2 =\Lambda^2_1 + \Lambda^2_2 + 2\Lambda_1 \Lambda_2 \cos J$.}

\section{Familles émergeant des équilibres de Lagrange circulaire}

\begin{figure}[h!]
\begin{center} 
\includegraphics[width=0.49\linewidth]{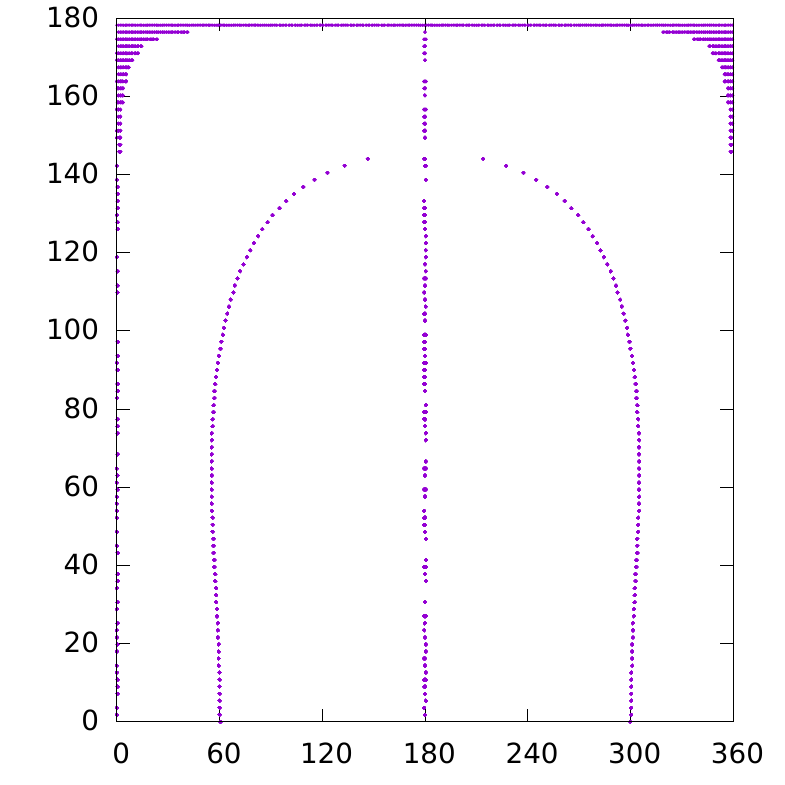}\\
 \setlength{\unitlength}{1cm}
\begin{picture}(.001,0.001)
\put(-4.5,5){$J$}
\put(0.5,0.1){{$\zeta$}}
\end{picture}
\caption{\label{fig:Fpi} Trace des familles de points fixes du problème moyen réduit émergeant des équilibres de Lagrange circulaires. On identifie les endroits de la variété $a_1=a_2$ et $e_1=e_2=0$ où $\frac{\partial }{\partial \zeta} \gH_{\cR\cM}=0$ pour $m_1=m_2$.} 
\end{center}
\end{figure}
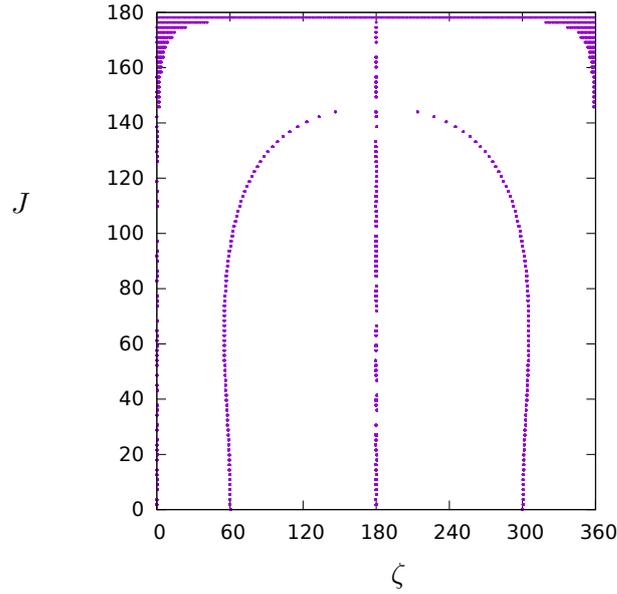

De manière similaire à l'étude semi-analytique effectuée dans le cas des co-orbitaux coplanaires, nous pouvons explorer l'espace des phases à la recherche des familles émergeant des points $L_3$, $L_4$ et $L_5$ dans la direction verticale. Le Hamiltonien du problème moyen dans le plan invariant $\ol \gH_{PI}$ comprenant un unique degré de liberté associé aux variables conjuguées $Z$ et $\zeta$, nous pouvons aisément définir une variété représentative pour une valeur donnée du moment cinétique total. Pour cette valeur du moment cinétique, les points fixes du problème moyen réduit sont donnés par l'équation:

\begin{equation}
\frac{\partial }{\partial Z}\gH_{PI} = \frac{\partial }{\partial \zeta} \gH_{PI}= 0\, . 
\label{eq:condFpi}
\end{equation}

A l'instar du cas coplanaire, pour $m_1=m_2$ on fait l'hypothèse suivante: $\frac{\partial }{\partial Z}\gH_{PI} =0$ partout sur la variété $Z=0$. Nous avons donc à vérifier où $\frac{\partial }{\partial \zeta}\gH_{PI} = 0$ sur la variété $Z=0$. Cette vérification sera faite sur une approximation numérique du Hamiltonien moyenné, comme en section \ref{sec:SAIF}. La variété $Z=0$ étant de dimension $1$ pour le problème moyen réduit, nous pouvons représenter l'évolution de ces familles de points fixes en fonction du moment cinétique total (ou, de manière équivalente pour des masses données, en fonction de l'inclinaison mutuelle $J$). Les points où $\frac{\partial }{\partial \zeta}\gH_{PI}$ s'annule sont représentés sur la figure \ref{fig:Fpi}. 

La coupe $J=0$ représente le cas circulaire plan: on retrouve les équilibres de Lagrange $L_3$, $L_4$ et $L_5$ en $180^\circ$, $60^\circ$ et $300^\circ$. Quand $J$ augmente, la famille issue de $L_3$ reste à $\zeta$ constant, alors que la position des familles issues de $L_4$ et $L_5$ évolue en $\zeta$, et semble tendre vers la famille émergente de $L_3$ vers $J \approx145^\circ$. Ces résultats sont similaires à ceux obtenu par \cite{Marchal2009} dans le cas du problème restreint circulaire. \\ 

\an{
We search for the position of the families of fixed points of the reduced averaged problem emanating from the circular coplanar equilibriums $L_3$, $L_4$ and $L_5$. We use the same semi-analytical method as in the coplanar case (section \ref{sec:SAIF}).\\
The fixed points verify the equation (\ref{eq:condFpi}). Let us suppose that, as it is the case in the coplanar circular case, $\dot \zeta= 0$ on the line $a_1=a_2$ when $m_1=m_2$. We then only need to find the points on this line that verify $\frac{\partial }{\partial \zeta}\gH_{PI} = 0$. This is done on a grid of initial conditions, where we estimate numerically the value of $\ol \gH_{PI}$ (see section \ref{sec:SAIF}) and compute its derivative. Since the manifold $Z=0$ is a line (1-dimensional), we can represent in a 2-D graph the evolution of the position of these families as the total angular momentum increases (or, equivalently, as a function of $J$).\\
The points satisfying the relation (\ref{eq:condFpi}) are represented in figure~\ref{fig:Fpi}. For $J=0^\circ$, we have the coplanar case: $L_3$, $L_4$ and $L_5$ are located at $180^\circ$, $60^\circ$ and $300^\circ$ respectively, and the collision manifold is located at $\zeta=0^\circ$. As $J$ increases, the position of the family emanating from $L_3$ and the collision manifold stay at a constant value of $\zeta$. However, the families emanating from $L_4$ and $L_5$ evolve, and seem to tend to the family emanating from $L_3$ when $J\approx 145^\circ$. These results agree with those of \cite{Marchal2009} in the restricted circular case.}

\section{Dynamique et stabilité des co-orbitaux circulaires inclinés}

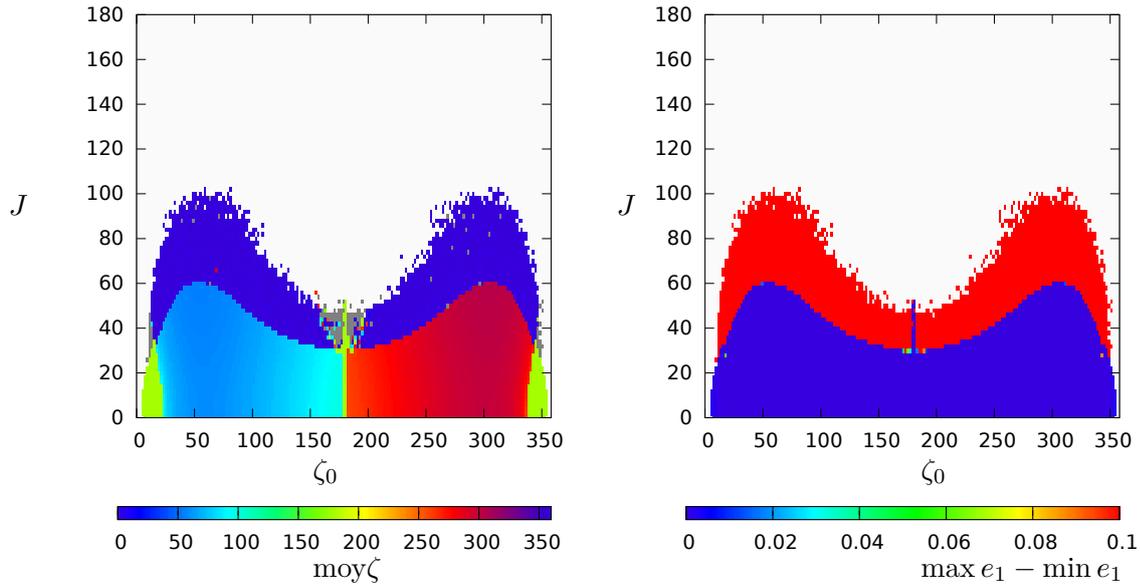
\begin{figure}[h!]
\begin{center} 
\includegraphics[width=0.49\linewidth]{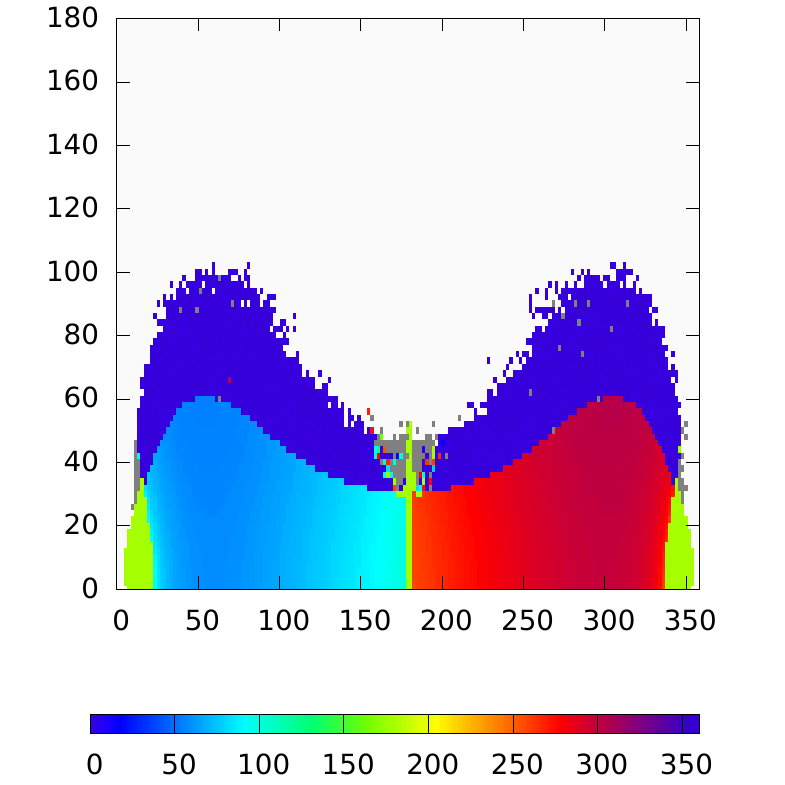}
\includegraphics[width=0.49\linewidth]{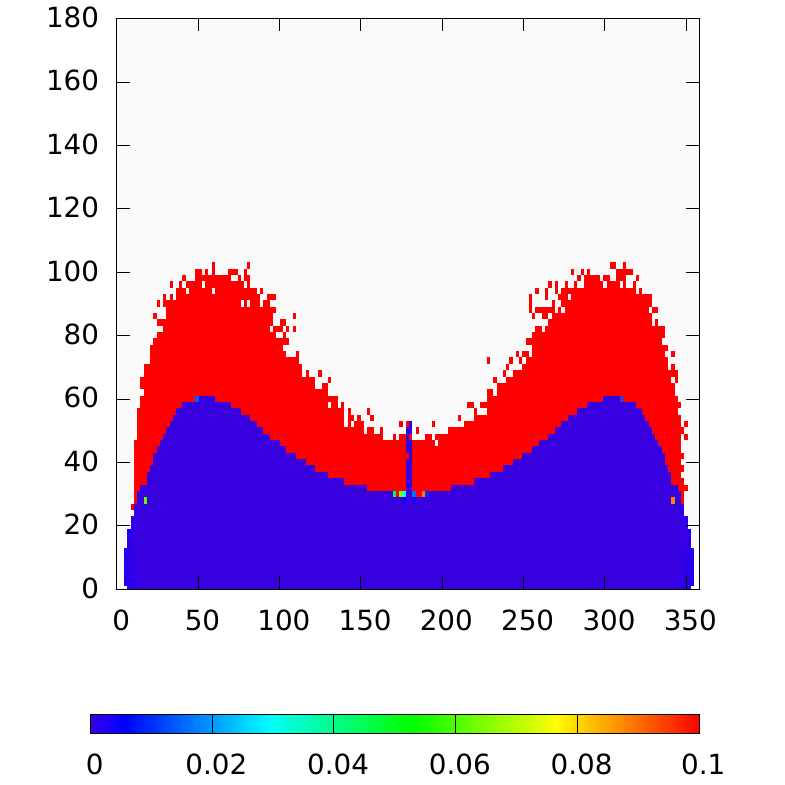}  \\
 \setlength{\unitlength}{1cm}
\begin{picture}(.001,0.001)
\put(0,5){$J$}
\put(-8,5){$J$}
\put(-4,1.5){{$\zeta_0$}}
\put(4,1.5){{$\zeta_0$}}
\put(-4,0.2){{$\text{moy} \zeta$}}
\put(4,0.2){{$\max e_1 - \min e_1$}}
\end{picture}
\caption{\label{fig:VaReJm4} Co-orbitaux inclinés ayant pour conditions initiales $e_1=e_2=0$, $a_1=a_2=1$, $\lambda_1=\Omega_1=0$ et $i_1$, $i_2$ et $\Omega_2$ de manière à être dans le plan invariant. $m_1=m_2=10^{-5} m_0$.} 
\end{center}
\end{figure}

Nous allons intégrer les trajectoires issues de plans de conditions initiales afin d'avoir un aperçu de l'espace des phase (voir la section \ref{sec:stabcirc} pour les informations concernant l'intégrateur).
Le Hamiltonien moyen réduit $\ol \gH_{PI}$ ne possède qu'un degré de liberté représenté par les variables ($Z,\zeta$). Comme exposé en section \ref{sec:pperdi} et \ref{sec:VaRe}, l'ensemble de ces orbites peut être représenté par les trajectoires issue de la variété $\dot \zeta=0$. A l'instar du cas coplanaire, on fait l'approximation que cette variété coïncide avec le plan $Z=0$. On exclut donc toute orbite ne passant pas par le plan $Z=0$ (par exemple l'équilibre de Lagrange circulaire lorsque $m_1 \neq m_2$, mais on en reste $\eps$ proche). Puisque parcourir l'ensemble des valeurs de $\zeta$ possible sur la droite $Z=0$ suffit à représenter l'ensemble des configurations pour une valeur du moment cinétique donné, on représente en figure~\ref{fig:VaReJ} la valeur initiale de $\zeta$ en abscisse et celle de $J$ en ordonnée. Pour le reste des conditions initiales, on prend $a_1=a_2=1$~ua, $\lambda_1=0^\circ$, $e_1=e_2=0$, $\Omega_1=0^\circ$, $\Omega_2=180^\circ$, et on calcule $i_1$ et $i_2$ grâce aux équations (\ref{eq:syseqPI}) afin de se placer dans le plan invariant. Pour la colonne de gauche de la figure~\ref{fig:VaReJ} nous avons $m_2=m_1=1\times10^{-4}m_0$ et pour la colonne de droite $m_2= 19 m_1=19\times10^{-5}m_0$ de manière à ce que $\mu=2\times 10^{-4}$ dans les deux cas. Selon la ligne, le code couleur représente la moyenne de l'angle $\zeta$, l'amplitude de variation maximale $\Delta i_1 = \max i_1 - \min i_1$ et $\Delta e_1= \max e_1 - \min e_1$. 
La trajectoire partant de chaque condition initiale a été intégrée pendant $1\times10^{6}$~an avec un pas de $0.01$~an.

Sur chaque graphe pour $J=0$ se trouve le cas circulaire plan: la dynamique est bien connue et expliquée en section \ref{sec:pperdi}. Cette ligne nous permet d'identifier aisément les différentes configurations qui existent à faible inclinaison mutuelle: les régions troyennes, pour lesquelles $\zeta$ libre autour de $60^\circ$ ou $300^\circ$, et les fers-à-cheval librant autour de $\zeta=180^\circ$. On traverse le domaine des fers-à-cheval en des zones centrées en $\zeta \approx 20^\circ$ et $\approx 340^\circ$, mais aussi dans le voisinage l'équilibre $L_3$ circulaire.

Quand $J$ augmente, la taille respective de ces domaines évolue, et la coupe du domaine des fers-à-cheval n'est plus d'épaisseur discernable avant que l'inclinaison mutuelle n'atteigne $40^\circ$. Le domaine de stabilité des configurations troyennes est quant à lui délimité par une séparatrice qui sera décrite ci-après. Au-delà de cette séparatrice, dans le domaine prograde ($J<90^\circ$), une région d'orbites chaotiques apparaît (voir figures~\ref{fig:orbJ}, ligne du haut). Ces orbites diffusent lentement et ne semblent pas stable à très long termes. Dans le cas de planètes à masse différentes, une autre région stable apparait pour des orbites rétrogrades ($J>90^\circ$). Ces orbites, centrées en $\zeta=60^\circ$ ou $300^\circ$ et $J=150^\circ$ sont mises en évidence dans les travaux de \cite{MoNa2016} pour le problème restreint (voir figure~\ref{fig:orbJ}, ligne du bas). Entre ces deux régions se trouve une autre configuration d'orbites chaotiques qui diffuse également lentement (voir figure~\ref{fig:orbJ}, ligne du milieu).

Dans les deux zones chaotiques, les hypothèses utilisées en section \ref{sec:Rdpi} ne sont pas valables car le mouvement n'est pas quasi-périodique. La dynamique sur la variété $e_1=e_2=0$ du problème moyen n'est donc pas représentative de la dynamique sur cette même variété dans le problème complet. Comme nous pouvons le voir sur les figures~\ref{fig:VaReJ} et \ref{fig:orbJ}, l'excentricité peut être excitée.\\


\an{
As previously stated, the reduced averaged Hamiltonian $\ol \gH_{PI}$ has only one degree of freedom. Therefore, as we explained in sections \ref{sec:pperdi} and \ref{sec:VaRe}, all orbits can be represented by trajectories taking the line $\dot \zeta=0$ as initial condition. As we did in the coplanar case, we approximate the position of $\dot \zeta=0$ by the position of $Z=0$. Doing so, we do not represent the orbits that do not cross $Z=0$, such as the exact equilibrium $L_4$ when $m_1 \neq m_2$, but this point is $\eps$-near from the line $Z=0$.\\
Since the representative set of initial conditions is 1-dimensional, we can display on a 2-D graph the effect of the evolution of the total angular momentum as well (we prefer to show the equivalent $J$ instead, see figures~\ref{fig:VaReJm4} and \ref{fig:VaReJ}). We take as initial conditions $a_1=a_2=1$~au, $\lambda_1=0^\circ$, $e_1=e_2=0$, $\Omega_1=0^\circ$, $\Omega_2=180^\circ$. $i_1$ and $i_2$ are given by the equation (\ref{eq:syseqPI}) in order to remain in the invariant plane. More details regarding the integration can be found in section \ref{sec:stabcirc}. Figure~\ref{fig:VaReJm4} shows the mean value of $\zeta$ and the maximum value of $e_1-e_2$ for trajectories taking their initial condition on the plane described above. In this case, we have $m_1=m_2=10^{-5}$.}
\an{In Figure~\ref{fig:VaReJ} we show simulations for $m_2=m_1=1\times10^{-4}m_0$ on the left column and $m_2= 19 m_1=19\times10^{-5}m_0$ on the right ($\eps \approx \mu=2\times 10^{-4}$ in both cases). Comparing the left pictures with the figure~\ref{fig:VaReJm4}, it seems that the value of $\eps$ does not impact much the dynamics.\\}

\an{
In each graph, $J=0^\circ$ represents the well known coplanar circular case, with the tadpole domain where the orbits librate around $L_4$ and $L_5$ and the horseshoe domain (crossed near $\zeta \approx 20^\circ$ $\approx 180^\circ$ and $\approx 340^\circ$). 
As $J$ increases, the stability domain of these configuration evolves, and the horseshoe domain seems to vanish before $J$ reaches $40^\circ$. The trojan domain is delimited by a separatrix that we will describe in the next section. Beyond this separatrix, in the prograde domain ($J < 90^\circ$) we are left only with chaotic orbits (see for example the trajectory at the top of the figure~\ref{fig:orbJ}). These orbits are slowly diffusing and seem to be unstable on the long term.\\
In the case $m_1 \neq m_2$ (right column in fig.~\ref{fig:VaReJ}), another seemingly stable domain appears for retrograde orbits ($J>90^\circ$). This domain crosses the reference plane in two areas centred at $J=150^\circ$ and $\zeta=60^\circ$ or $300^\circ$. These orbits were introduced by \cite{MoNa2016} (see an example of trajectory figures~\ref{fig:orbJ}, bottom). Between these two areas lies another chaotic region, where trajectories are also slowly diffusing, see an example on the middle raw, figure~\ref{fig:orbJ}.\\
In both chaotic areas, the hypothesis used in section \ref{sec:Rdpi} are not valid because the motion is not quasi-periodic. the dynamics on the manifold $e_1=e_2=0$ in the averaged case is hence not representative of the dynamics on this same manifold in the full 3-body problem. As we can see in the figures~\ref{fig:VaReJ} and \ref{fig:orbJ}, the eccentricity of the planets can be excited by angular momentum transfer from the mutual inclination.} 

\subsection{Limite du domaine troyen}

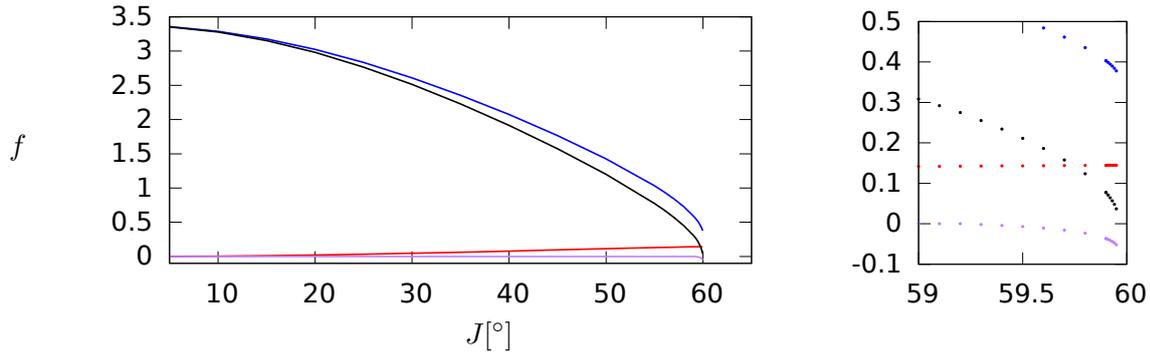
\begin{figure}[h!]
\begin{center} 
\includegraphics[width=0.65\linewidth]{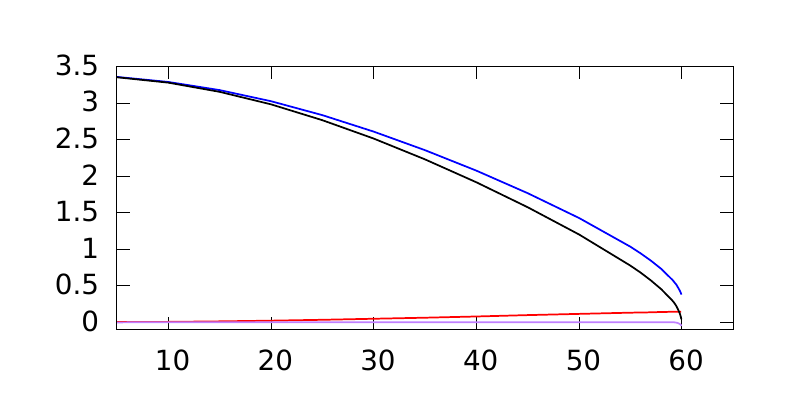}
\includegraphics[width=0.32\linewidth]{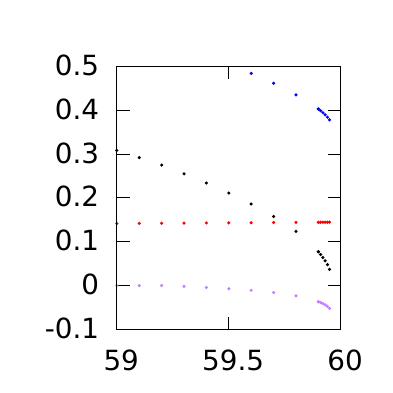}\\  
 \setlength{\unitlength}{1cm}
\begin{picture}(.001,0.001)
\put(-8,2.7){$f$}
\put(-2,0.2){{$J[^\circ]$}}
\end{picture}
\caption{\label{fig:freqJ} Evolution des fréquences pour des conditions initiales où seul $J$ évolue. Le graphe de droite représente un zoom pour $J\in [59^\circ,60^\circ]$. $m_2=m_1=10^{-4}m_0$, $\zeta=60^\circ$, les autres conditions initiales sont identiques à celles des figures~\ref{fig:VaReJ}. les fréquences représentées sont $s$ (rouge), $g_1$ (bleu) et $g_2$ (violet). Toutes les fréqunces sont normalisées par $\eps n$. En noir est représentée la combinaison de caractéristique nulle $(2s -g_1-g_2)/(\eps n)$. Notons qu'on retrouve, quand $J$ tend vers $0$, $g_1/(n\eps)\approx27/8$ et $g_2=s=0$ comme c'est le cas dans de voisinage le l'équilibre $L_4$ circulaire.} 
\end{center}
\end{figure}

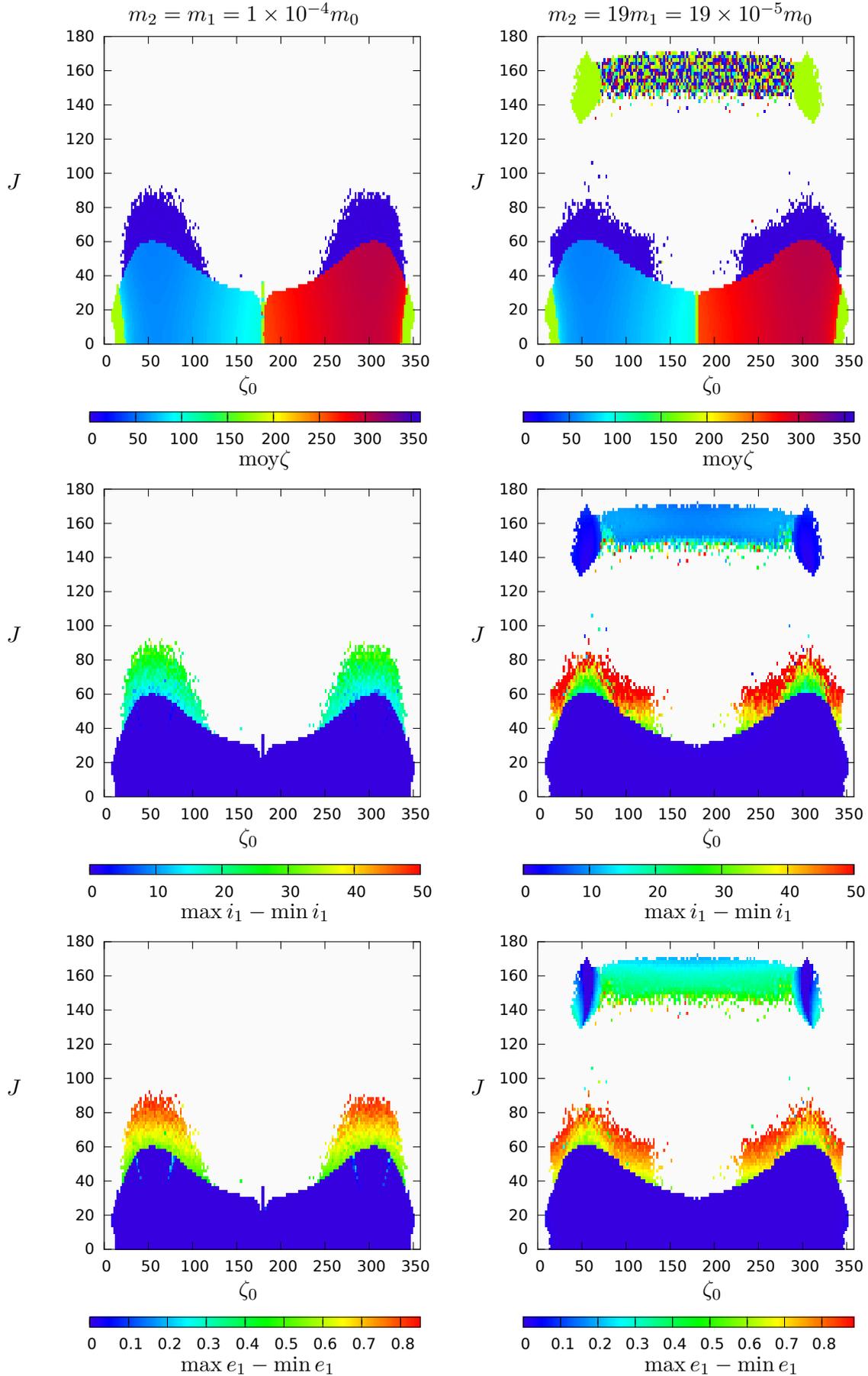
\begin{figure}[h!]
\begin{center} 
$m_2=m_1=1\times10^{-4}m_0$ \hspace{3cm} $m_2= 19 m_1=19\times10^{-5}m_0$
\includegraphics[width=0.49\linewidth]{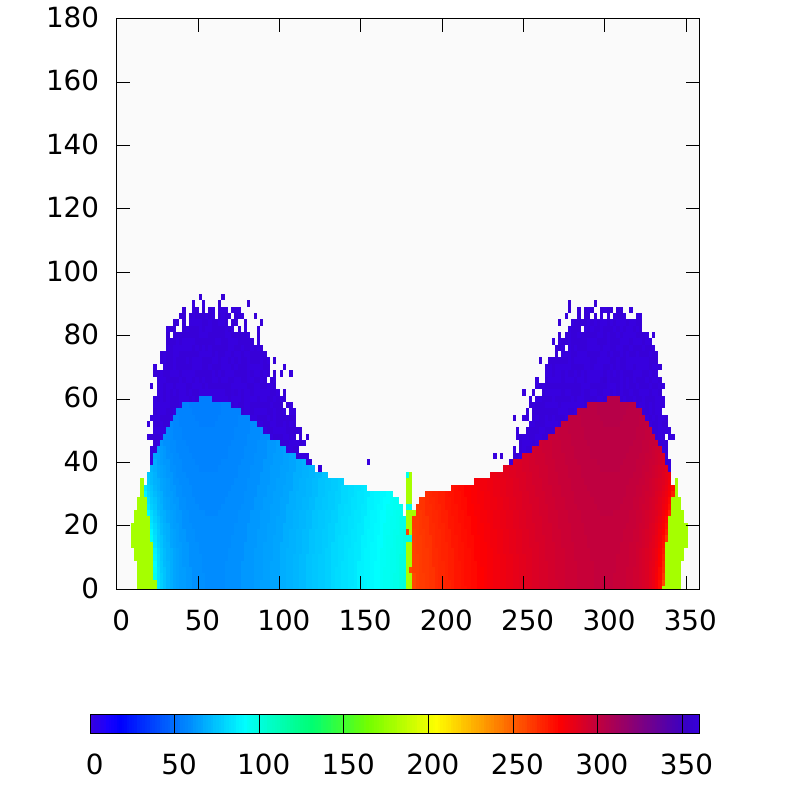}
\includegraphics[width=0.49\linewidth]{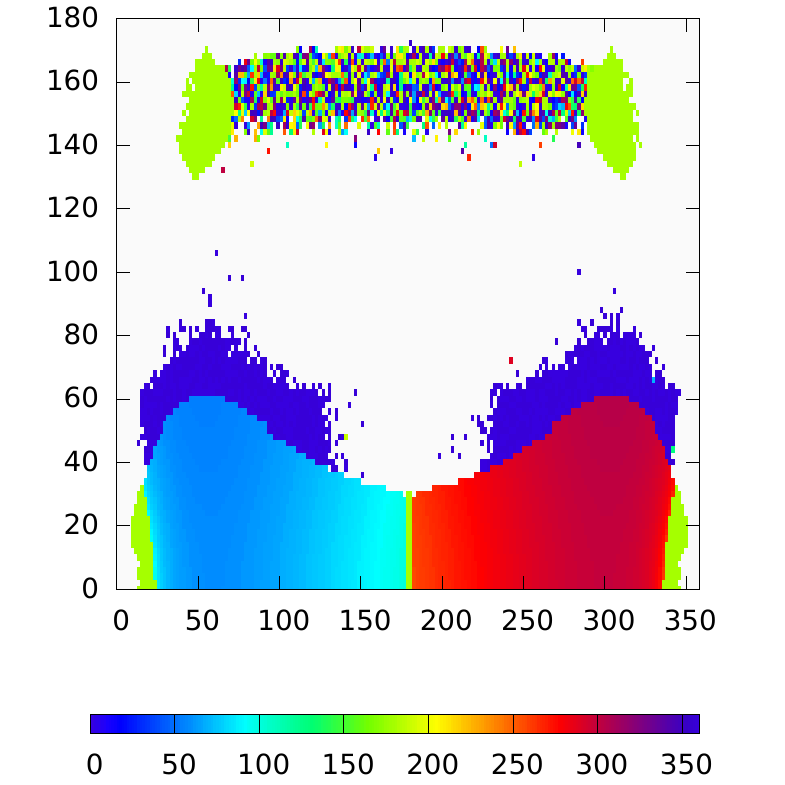}  \\
 \setlength{\unitlength}{1cm}
\begin{picture}(.001,0.001)
\put(0,5){$J$}
\put(-8,5){$J$}
\put(-4,1.5){{$\zeta_0$}}
\put(4,1.5){{$\zeta_0$}}
\put(4,0.2){{$\text{moy} \zeta$}}
\put(-4,0.2){{$\text{moy} \zeta$}}
\end{picture}\\
\includegraphics[width=0.49\linewidth]{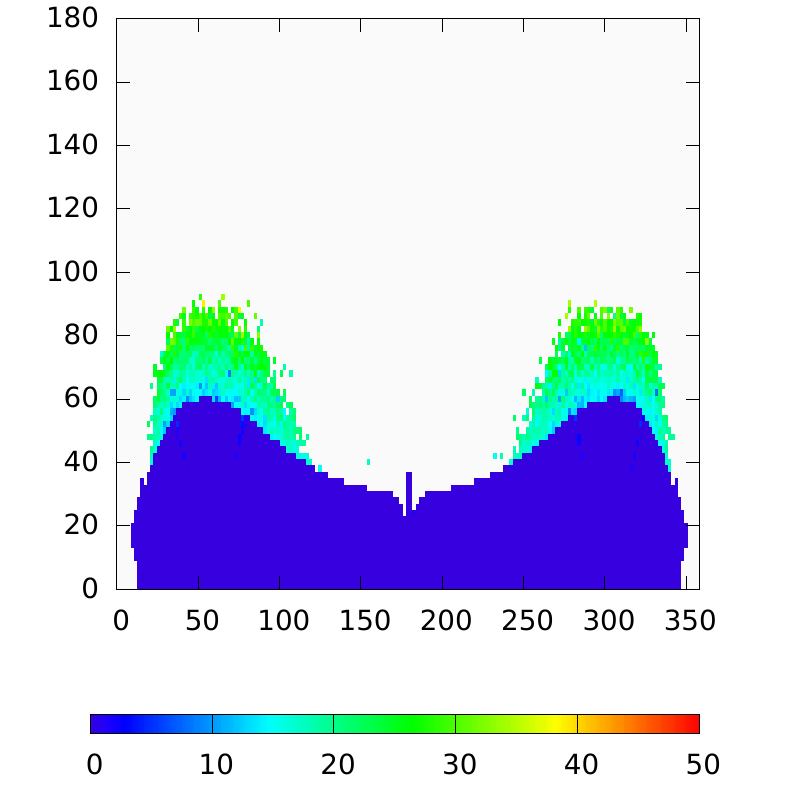}
\includegraphics[width=0.49\linewidth]{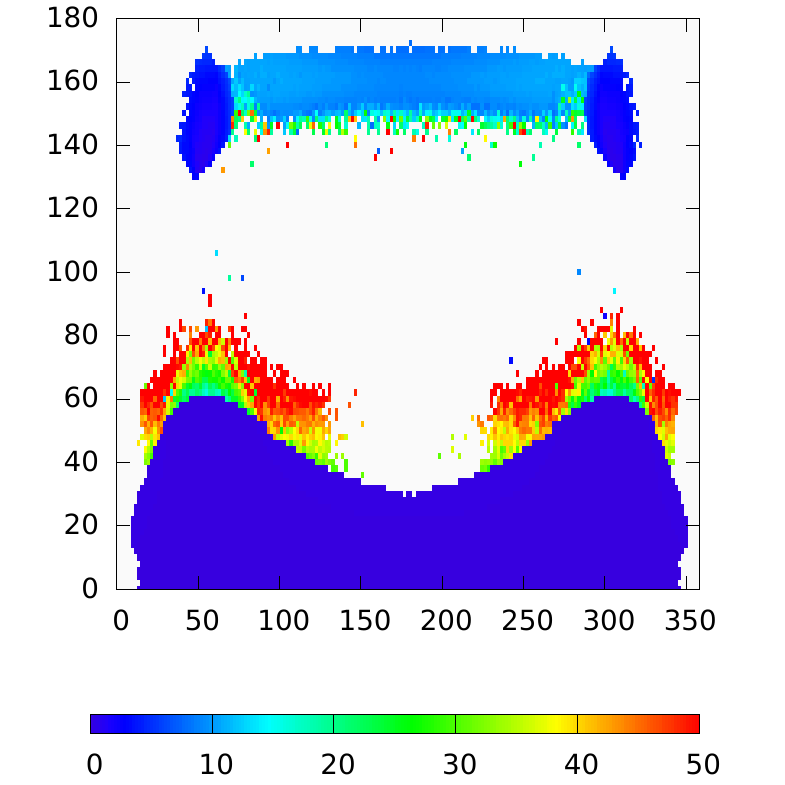} \\
 \setlength{\unitlength}{1cm}
\begin{picture}(.001,0.001)
\put(0,5){$J$}
\put(-8,5){$J$}
\put(-4,1.5){{$\zeta_0$}}
\put(4,1.5){{$\zeta_0$}}
\put(3,0.2){{$\max i_1 - \min i_1$}}
\put(-5,0.2){{$\max i_1 - \min i_1$}}
\end{picture}\\
\includegraphics[width=0.49\linewidth]{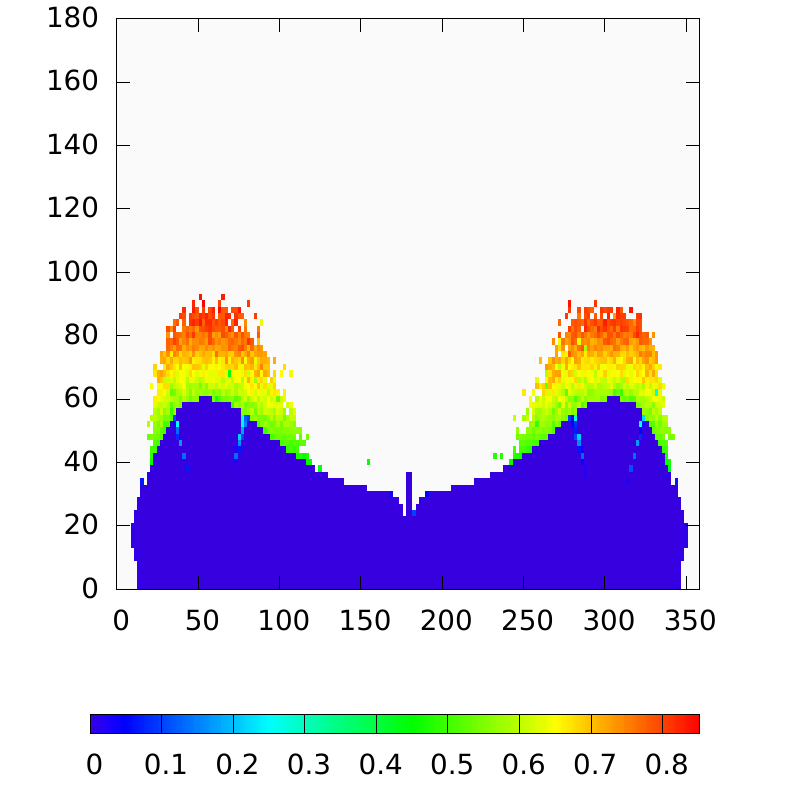}
\includegraphics[width=0.49\linewidth]{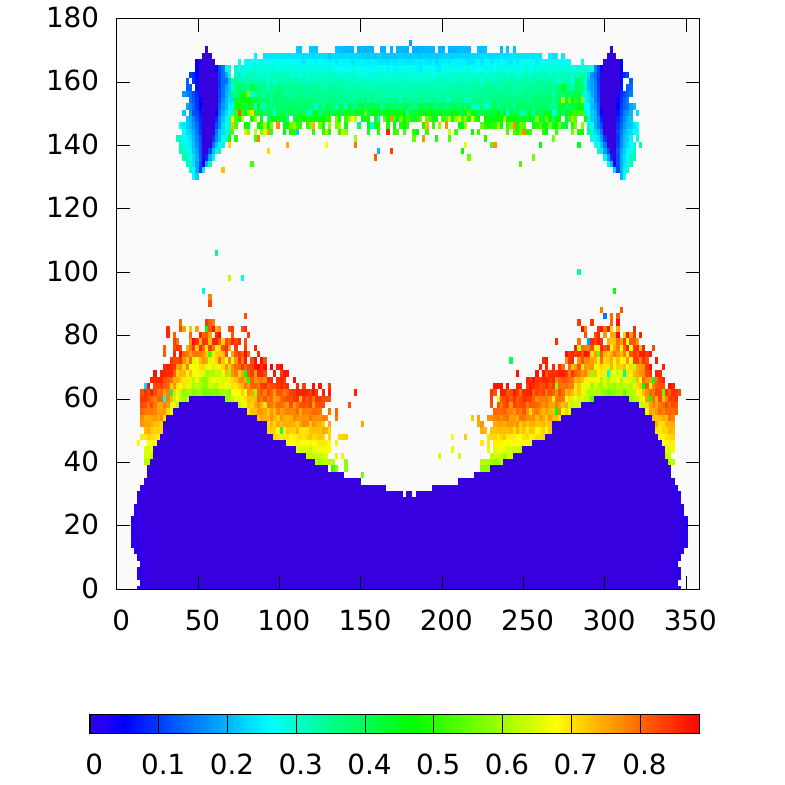}   
 \setlength{\unitlength}{1cm}
\begin{picture}(.001,0.001)
\put(0,5){$J$}
\put(-8,5){$J$}
\put(-4,1.5){{$\zeta_0$}}
\put(4,1.5){{$\zeta_0$}}
\put(3,0.2){{$\max e_1 - \min e_1$}}
\put(-5,0.2){{$\max e_1 - \min e_1$}}
\end{picture}
\caption{\label{fig:VaReJ} Co-orbitaux inclinés ayant pour conditions initiales $e_1=e_2=0$, $a_1=a_2=1$, $\lambda_1=\Omega_1=0$ et $i_1$, $i_2$ et $\Omega_2$ de manière à être dans le plan invariant.} 
\end{center}
\end{figure}

Dans cette section, nous cherchons à identifier ce qui entraîne la transition du domaine troyen au domaine chaotique visible sur les figures~\ref{fig:VaReJ} quand $J$ augmente. Rappelons que ces intégrations numériques sont celles du problème à trois corps complet et que par conséquent, il s'agit d'un problème à 6 degrés de liberté, que nous réduisons a 5 en choisissant des conditions initiales telles que le repère inertiel possède deux axes dans le plan invariant du problème. La dynamique de problème est donc régie par $5$ fréquences. Le moyen mouvement $n$ et la fréquence de libration semi-rapide $\nu$ s'identifient aisément en effectuant l'analyse en fréquence des quantités $\exp^{i\lambda_1}$ et $\zeta$. La fréquence de circulation des n\oe uds ascendants $s$ s'identifie également aisément car il s'agit de la fréquence de plus haute amplitude dans la décomposition de la variable de Poincaré $y_1$.

Les fréquences $g_1$ et $g_2$ associées au mouvement des périhélies et à l'évolution des excentricités sont plus difficiles à identifier, car les excentricités sont prises nulles en conditions initiales pour l'ensemble des trajectoire étudiées dans ce chapitre et n'évoluent que très faiblement dans les zones quasi-périodiques (voir l'amplitude de l'évolution de $e_1$, figure~\ref{fig:VaReJ}). Pour identifier ces fréquences, on procède de la manière suivante: on effectue l'analyse en fréquence de la variable de Poincaré $x_1$ pour un jeu de conditions initiales donné. Cette variable est de caractéristique $1$, elle n'admet donc que des termes de caractéristique $1$ dans son développement. $n$ et $s$ sont de caractéristique $1$ alors que $\nu$ est de caractéristique $0$. On choisit de définir $g_1$ et $g_2$ comme des fréquences de caractéristique $1$ également. La décomposition de $x_1$ se fait donc avec des termes de la forme $C\exp^{k_0\nu + k_1 n + k_2 s + k_3 g_1 + k_4 g_2}$ avec $k_0$ quelconque, $k_2$ pair et $ k_1 + k_2 + k_3  + k_4$ impaire. On choisit d'identifier $g_1$ et $g_2$ de manière à ce que $\sum |k_j|$ soit minimal et que $|g_1|>|g_2|$. On prolonge la détermination de ces fréquences par continuité aux conditions conditions initiales proches.

La transition entre le domaine troyen et le domaine chaotique ne dépend pas de l'orientation du repère choisi, l'éventuelle résonance qui entraine ce changement implique donc des combinaisons de fréquences de caractéristiques nulles. Sur la figure~\ref{fig:freqJ} sont représenté l'évolution des fréquences $s$, $g_1$ et $g_2$ quand $J$ augmente pour $\zeta=60^\circ$ fixé. ces fréquences tendent vers $2s=g_1+g_2$ quand $J$ approche de la séparatrice mise en évidence sur les figures~\ref{fig:VaReJ}. Cette séparatrice peut être liée à la résonance de Kozai ($2s=2g$ dans le problème restreint).\\

\an{We try to understand what leads to the transition from trojan to chaotic orbits in the figures \ref{fig:VaReJm4} and \ref{fig:VaReJ} when $J$ increases. These figures represent integration of the 3-body problem, hence a 12-dimensional phase space. Since we chose our initial conditions to remain in the invariant plane, there are $5$ fundamental frequencies (we do not consider the precession of the nodes). In the regular (quasi-periodic orbits) area of the phase space, $n$ and $\nu$ can be identified by frequency analysis of the quantities $\exp^{i\lambda_1}$ and $\zeta$. The precession frequency of the nodes $s$ is the frequency of the dominant term in the expansion of the Poincaré variable $y_1$. \\
$g_1$ and $g_2$, associated with the eccentricities, are harder to identify because we take $e_1=e_2=0$ as initial condition and they remain close to this value in the regular area of the phase space. In order to identify these frequencies, we process to the expansion of the Poincaré variable $x_1$ of a given trajectory in the regular areas of the phase space. $x_1$ being of characteristic $1$, each term of its expansion is also of characteristic $1$. $n$ and $s$ are of characteristic $1$, and $\nu$ is of characteristic $0$. We define $g_1$ and $g_2$ as characteristic $1$ as well. The expansion of $x_1$ hence writes: $C\exp^{k_0\nu + k_1 n + k_2 s + k_3 g_1 + k_4 g_2}$, with no constrain on  $k_0$, but with $k_2$ even and $ k_1 + k_2 + k_3  + k_4$ odd. We define $g_1$ et $g_2$ such that $\sum |k_j|$ is minimal, and $|g_1|>|g_2|$. We then identify these frequencies for trajectories for nearby initial conditions, considering that the frequencies evolve continuously in the regular areas of the phase space. We check that the value of $g_1$, $g_2$ and $s$ are those expected near the Lagrangian equilibrium when $J$ tends to $0$: $g_1/(n\eps)\approx27/8$ and $g_2=s=0$.\\
The transition from the trojan domain to the chaotic area does not depend on the choice of the reference frame. If this transition is due to a resonance, this resonance is due to combinations of characteristic $0$ of the fundamental frequencies. This combination is unlikely to contain terms in $n$ and $\nu$ as we have $n \gg \nu \gg (s,g_1,g_2)$. In figure~\ref{fig:freqJ} we show the evolution of the frequencies $s/(\eps n)$ (red), $g_1/(\eps n)$ (blue) and $g_2/(\eps n)$ (purple) when $J$ increases for a fixed value $\zeta=60^\circ$. These frequencies tend to the relation $2s=g_1+g_2$ ($(2s -g_1-g_2)/(\eps n)$ plotted in black) when $J$ get close from the transition. This separatrix may have a link with the Kozai resonance  ($2s=2g$ in the restricted case).}

\chapter{Détection des coorbitaux quasi-circulaires}

\an{
This chapter is dedicated to the detection methods adapted to co-orbital exoplanets. An introduction to these method can be found in the paper \cite{LeRoCo2015} (annex \ref{an:VR}). We discuss here the radial velocity method, the astrometry, the transit of both planets, the TTVs of one them, and the combination of transit and radial velocity. Our work regarding the radial velocity technique and the astrometry is explained in \cite{LeRoCo2015}. Methods to detect both co-orbitals by transit can be found in \cite{Ja2013}. \cite{VoNe2014} explain how co-orbitals can be found in TTV measurements. Finally, a letter in preparation (annex \ref{an:VRT}), based on the method developed by \cite{FoGa2006}, explain how the combination of the time of primary and secondary transits and radial velocity measurements allows the detection of co-orbital exoplanets.\\}     

Ce chapitre est dédié aux méthodes de détection des exoplanètes en résonance coorbitale. Depuis la découverte de la première exoplanète \citep{WoFr1992}, une grande diversité de systèmes a été découverte, dont certains sont en résonance de moyen mouvement (MMR). Certaines de ces configurations, comme la MMR $2/1$, sont très présentes dans les systèmes connus, alors qu'aucun système en configuration coorbitale n'a été découvert jusqu'à maintenant. \citet{LauCha2002} ont introduit deux processus pouvant former ces exo-planètes coorbitales: (i) la diffusion due aux interactions entre les planètes et (ii) l'accrétion \textit{in situ} au voisinage des points $L_4$ et $L_5$ d'une planète déjà existante.  

Dans le cas (i), le profil de la densité du disque de gaz semble impacter le rapport de masse entre les coorbitaux: dans leurs simulations, \cite{GiuBe2012} forment principalement des coorbitaux de masses comparables, alors que \cite{CreNe2008,CreNe2009} obtiennent une grande diversité de rapport de masse. Dans cette dernière étude, les auteurs forment des coorbitaux dans $\approx 30\%$ des systèmes planétaires générés par leurs simulations. Ces coorbitaux se forment généralement en configuration fer-à-cheval, puis l'amplitude de libration est amortie par le disque, ce qui entraîne une transition en configuration troyenne. L'accrétion de gaz semble renforcer les écarts de masses entre les coorbitaux, le plus massif des deux atteignant une masse de l'ordre de celle de Jupiter alors que le moins massif reste en dessous de 70 masses terrestres. Les coorbitaux formés dans leur modèle ont généralement une inclinaison et une excentricité assez faible ($e<0.02$). 

Dans le cas (ii), les simulations de \cite{LyJo2009} forment des troyens allant jusqu'à 5-20 masses terrestres au voisinage des points $L_4$ et $L_5$ de jupiters.

Les modèles qui génèrent des configurations coorbitales dues à la dissipation dans le disque entraînent généralement la migration des coorbitaux vers le centre du système. Cette migration tend à augmenter l'amplitude de la libration de l'angle résonant. L'effet de la dissipation dû au gaz et aux effets de marée sur les configurations coorbitales est mal connu. \citet{PiRa2014} constatent que dans leur modèle les coorbitaux de masses égales (super-terre jusqu'à saturne) sont fortement perturbés en cas de longue phase de migration, et \citet{RoGiMi2013} concluent sous certaines hypothèses que les effets de marrée déstabilisent les coorbitaux de masses comparables proches de l'étoile. Les coorbitaux peuvent également être perturbés par la présence d'autres planètes, en particulier quand ils traversent une résonance en moyen mouvement avec une planète massive lors des phases migratoires du système planétaire \citep{MoLeTsiGo2005, RoBo2009}. De plus, les configurations fer-à-cheval sont davantage susceptibles d'être perturbées par des résonances que les orbites troyennes car la fréquence fondamentale de libration peut être très différente en fonction des conditions initiales choisies (voir figure \ref{fig:zr}).\\

Les études de formation et d'évolution de systèmes planétaires montrent donc que des planètes en configuration coorbitale peuvent exister, mais n'apportent pour l'instant pas de contrainte sur l'orbite ou sur la masse de ces coorbitaux indépendamment d'hypothèses faites sur les conditions initiales du système. Dans ce chapitre, on se limitera donc aux contraintes imposées par la stabilité des configurations dans le problème à trois corps non dissipatif.\\

Les exo-planètes coorbitales induisent un signal souvent semblable à celui d'autres configurations orbitales pour les méthodes basiques de détection indirecte (vitesse radiale - RV, transit, astrométrie, variation du temps de transit - TTV). Des méthodes spécifiques peuvent cependant être développées afin d'extraire la signature de coorbitaux pour chacune de ces techniques. Ces méthodes s'appuient sur des aspects de la dynamique coorbitale et peuvent être séparées en deux groupes: les méthodes observant le mouvement de l'étoile (RV, astrométrie) et les méthodes mesurant la diminution du flux de lumière qui nous provient de l'étoile quand une des planètes passe devant (transit et TTV). Les méthodes RV et astrométrie dépendent donc du mouvement du barycentre des coorbitaux alors que transit et TTV dépendent de la position de la planète observée.

\section{Mouvement du barycentre}
\label{sec:MBar}

\begin{figure}[h!]
\includegraphics[width=0.49\linewidth]{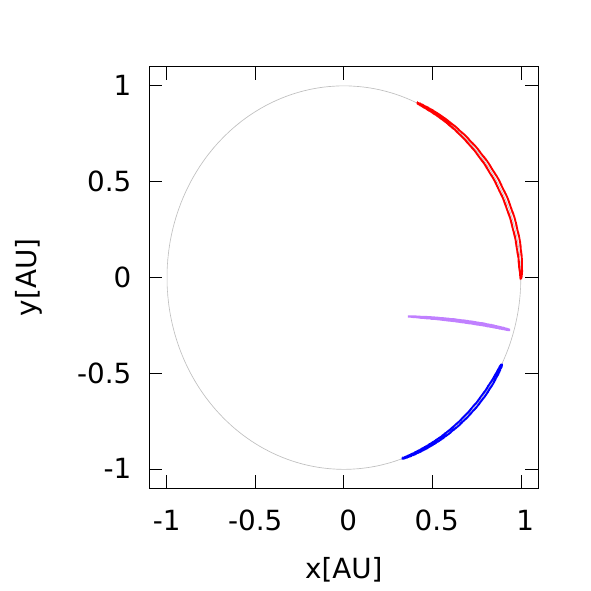}
\includegraphics[width=0.49\linewidth]{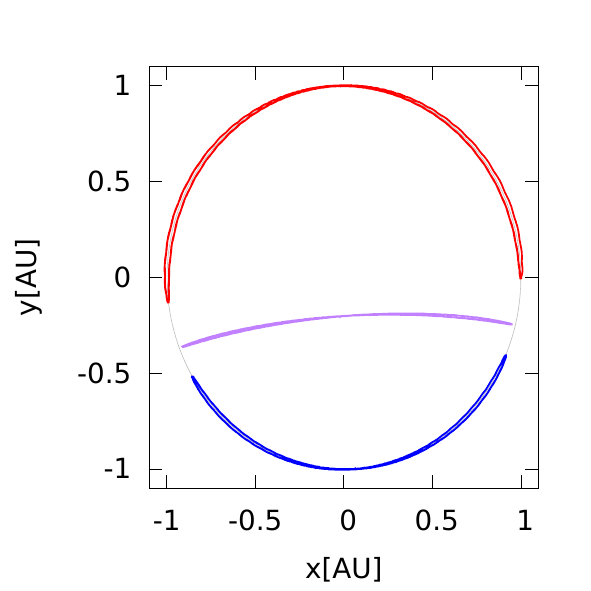}
\caption{Motion of the two co-orbital bodies (red and blue) and their barycenter (purple) in a co-rotating frame  with frequency $n$. Tadpole (left) and horseshoe (right). $\delta=0.6$. Here $\mu=2\,10^{-4}$ and the planets are located at $1\,AU$ from the star. Getting rid of the influence of $n$, one can see the long term motion of the barycenter of the planets. $P_\nu$ is the period of the periodic trajectories represented by the coloured lines. See the text for more details.}
\label{fig:schema_sym}
\end{figure}
%
%
%
%

C'est par le biais du mouvement de leur barycentre que les coorbitaux influent sur les mesures de vitesse radiale ou d'astrométrie d'une étoile. La position de l'étoile par rapport au barycentre du système est donnée par ($\bold{r_j} =$x$_j+i $y$_j$):

\begin{equation}
\bold{r_0}= -\mu [(1-\delta)\bold{r_1}+ \delta \bold{r_2}]\ ,
\label{eq:r0_v1}
\end{equation}
avec $\mu=\frac{m_1+m_2}{m_0+m_1+m_2}$ et $\bold{r_j}= r_j \exp^{f_j}$ où les $r_j$ et les $f_j$ sont donnés par les équations (\ref{eq:polar}). Dans le cas circulaire, à l'ordre $0$ en $\mu$ et en choisissant l'origine du temps telle que y$_1(0)=0$, la projection de la position $\bold{r_0}$ dans la direction $x$ s'écrit:
\begin{equation}
\Re(\bold{r_0})= -a\mu [(1-\delta) \cos (nt+\delta(\zeta(t)-\zeta(0)))+ \delta \cos (nt+(\delta-1)\zeta(t) - \delta \zeta(0)) ] \, ,
\label{eq:r0_v2}
\end{equation}
Sur un temps court devant la fréquence fondamentale de libration $\nu$, $\zeta$ peut être considéré constant. En faisant cette hypothèse, l'équation (\ref{eq:r0_v2}) est équivalente à:
\begin{equation}
\Re(\bold{r_0})= C \cos (nt + \phi)\, ,
\label{eq:r0_v3}
\end{equation}
où $C=a \frac{m_{eq}}{m_0+m_{eq}}$ est l'amplitude des variations de position et
\begin{equation}
\begin{aligned}
m_{eq}  & = \sqrt{m_1^2+m_2^2 + 2 m_1 m_2 \cos \zeta}\\
 & = (m_1+m_2) \sqrt{(1-\delta)^2+\delta^2 + 2 (1-\delta)\delta \cos \zeta}\,
\end{aligned}
\label{eq:meq}
\end{equation}
est la masse équivalente d'une planète seule en orbite keplerienne circulaire et de même demi-grand axe $a$ autour de l'étoile observée. Pour obtenir la vitesse de l'étoile dans la direction x, on dérive l'équation (\ref{eq:r0_v3}) par rapport au temps:
\begin{equation}
\Re(\bold{\dot{r}_0})= K \sin (nt + \phi)\, ,
\label{eq:dr0_v1}
\end{equation}
où $K=a n \frac{m_{eq}}{m_0+m_{eq}}$. Si on prend en compte l'inclinaison $I$ de la ligne de visée par rapport au plan de l'orbite, $K=a n \frac{m_{eq}}{m_0+m_{eq}} \sin I$. 

Si la libration des coorbitaux n'est pas perceptible à cause d'une trop faible amplitude ou d'une fréquence trop lente par rapport à l'étendue temporelle des observations, le mouvement qu'ils induisent sur l'étoile est indiscernable de celui induit par une seule planète sur une orbite circulaire. On peut vérifier que cela reste vrai à l'ordre $1$ en excentricité, voir la lettre en annexe \ref{an:VRT}.

 Les mesures de vitesses radiales et astrométriques d'un tel système sont donc identiques à celles d'une seule planète fictive sur une orbite circulaire dont la masse dépend de la valeur instantanée de l'angle $\zeta$ et du rapport $\delta$ et est toujours inférieure à la somme des masses des coorbitaux. Dans le cas extrême ou $\zeta=\pi$ et $\delta=1/2$ la masse de la planète équivalente est nulle.
 
 \begin{figure}[h!]
\includegraphics[width=1\linewidth]{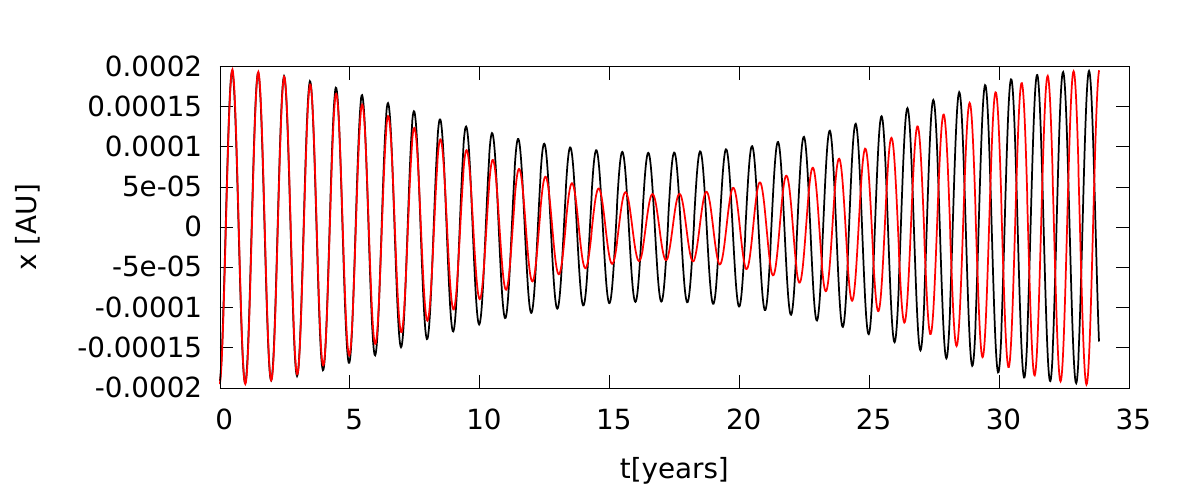}\\
\includegraphics[width=1\linewidth]{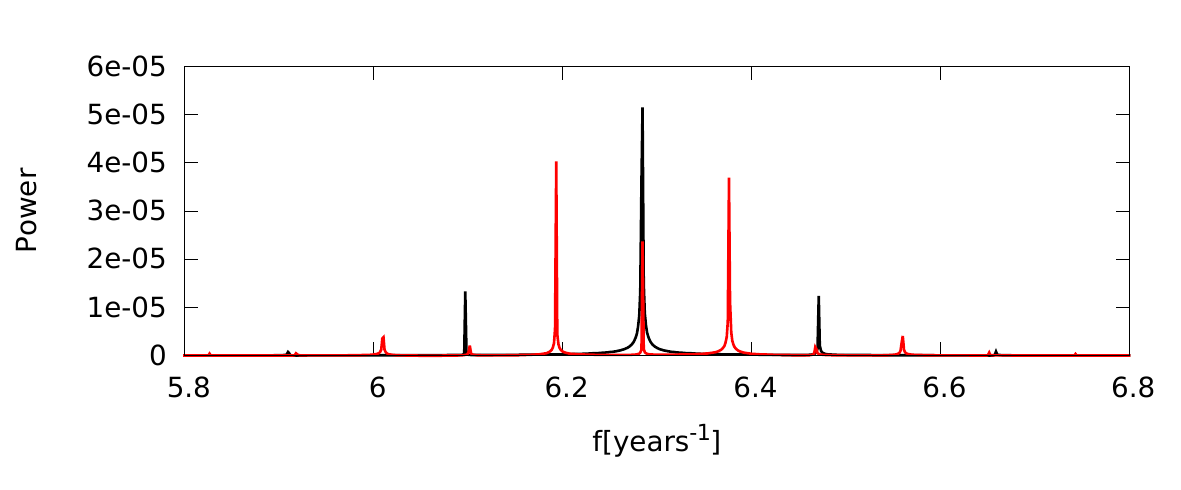}
\caption{Motion of the star in the configurations of Fig. \ref{fig:schema_sym} in the direction x in the inertial frame. In black is the tadpole orbit and in red the horseshoe. The top graph represents the evolution of the position of the star over time and the bottom one is its spectrum. In those examples, the libration period of the horseshoe orbits is about twice the period of the tadpole orbits. See the text for more details.}
\label{fig:VR}
\end{figure}

Considérons maintenant un temps suffisamment long pour voir l'évolution de $\zeta$. \cite{LauCha2002} on remarqué que le signal de vitesse radiale induit par des coorbitaux circulaires est semblable à une sinusoïde modulée en amplitude. Ceci est équivalent à une variation sur une période $P_\nu$ de la masse de la planète fictive précédemment décrite (\ref{eq:meq}). La figure \ref{fig:schema_sym} représente le mouvement du barycentre des coorbitaux pour une configuration troyenne et une configuration en fer-à-cheval. Quand le barycentre se rapproche de l'étoile, l'amplitude des variations de position de l'étoile diminue. L'amplitude de la modulation du signal dépend principalement de $\delta$ et de l'amplitude de libration de $\zeta$ (eq. \ref{eq:meq}). La position de l'étoile dans les cas de la figure \ref{fig:schema_sym} est tracée sur la figure \ref{fig:VR}. En première approximation, la FFT d'un signal modulé se compose d'un pic central à la fréquence rapide (ici le moyen mouvement moyen $\bar{n}$) et de deux pics symétriques séparés du pic central par la fréquence plus lente de la modulatrice (ici $\nu$).

Cette signature particulière est présente dans les signaux de vitesse radiale et d'astrométrie. Elle a été étudiée dans l'article \cite{LeRoCo2015} disponible en Annexe \ref{an:VR}. Il y est décrit un algorithme permettant d'extraire la signature de la modulation d'un signal. Pour identifier cette signature dans le cas d'exoplanètes coorbitales, il faut observer le système sur des temps de l'ordre de la période de libration. Il faut également que l'amplitude de libration, ainsi que la masse du plus petit des coorbitaux, soient suffisamment importantes. De plus, nous montrons que les caractéristiques principales du signal peuvent être inversées afin d'obtenir les paramètres du système coorbital. L'inversion est partielle dans le cas de fer-à-cheval et complète dans le cas d'une configuration troyenne, où nous pouvons également déterminer l'inclinaison des coorbitaux par rapport à la ligne de visée, et donc leur masse. L'article se termine par une illustration de la méthode proposée sur des exemples réalistes. 

\section{Transit}

La détectabilité par transit d'une planète sur une orbite circulaire autour d'une étoile de rayon $R_*$ avec un instrument donné dépend de son rayon $R$, de son demi-grand axe $a$ et de son inclinaison à la ligne de visée $I$. On dira qu'une planète transite si la projection de son centre de masse dans un plan orthogonal à la ligne de visée a une intersection non nulle avec la projection de l'étoile sur ce même plan. Par défaut on parlera du transit primaire où la planète est entre l'observateur et l'étoile, sauf si le contraire est précisé.   

\subsection{transit des deux coorbitaux}
 
Si deux coorbitaux sont sur des orbites coplanaires, soit les deux planètes transitent, soit aucune ne transite. Les deux coorbitaux peuvent également transiter dans le cas où leur inclinaison mutuelle est non nulle, auquel cas le double transit a plus de chance de se produire si l'inclinaison mutuelle est faible, si les planètes sont proches de l'étoile et si leur ligne des n\oe uds dans le plan invariant est proche de la ligne de visée. Dans ce cas, une configuration coorbitale peut être détectée en mesurant des transits aux instants ($k \in \mathbb{N}$):
\begin{equation}
t_{1,k}=t_i + k P_n\ + \delta\frac{\zeta(t_{1,k}) P_n}{2\pi}\ \text{et}\ t_{2,k}=t_i + k P_n + (\delta-1)\frac{\zeta(t_{2,k}) P_n}{2\pi}\, .
\label{eq:ttrans}
\end{equation}
où $t_i$ est la phase du premier transit et $P_n$ la période orbitale. Notons que si rien ne permet de différencier les transits des deux planètes (profondeur du transit, durée ou formes d'entrée/sortie) et que $\zeta=\pi$ (en configuration fer-à-cheval), alors les deux coorbitaux peuvent être confondus avec une planète seule avec une période de $P_n/2$. 

\begin{figure}[h!]
\begin{center}
\includegraphics[width=0.65\linewidth]{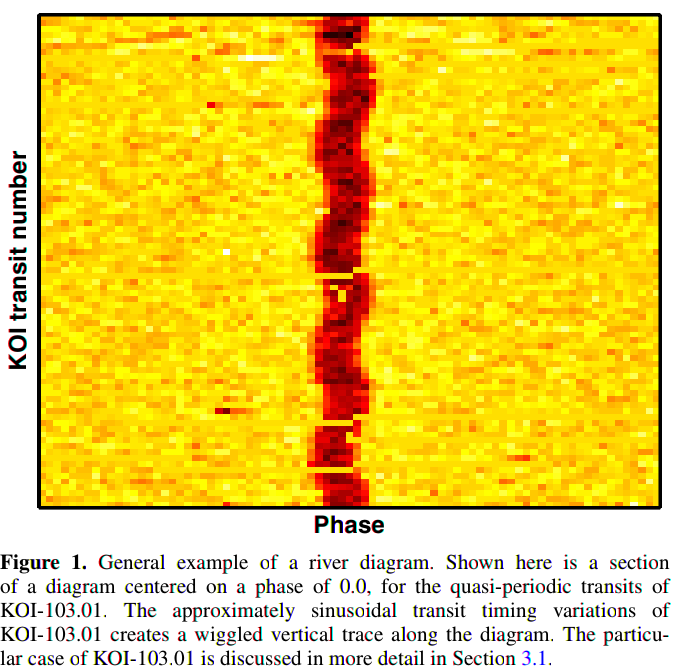}
\caption{\label{fig:river} Diagramme en rivière proposé par \cite{Ja2013}. Chaque ligne correspond à une période orbitale centrée sur l'instant moyen du transit et la couleur indique le flux lumineux reçu de l'étoile. On voit ici une oscillation de l'instant du transit primaire (TTV). Si un compagnon troyen transite et a un rayon suffisamment important, une courbe sombre oscillant autour de $\pm60^\circ$ devrait apparaître sur ce graphe.}
\end{center}
\end{figure}

Dans le cas où une planète est détectée et l'autre est à la limite d'être détectable, \cite{Ja2013} propose une méthode basée sur des diagrammes en rivière qui permettent de faire ressortir le signal de coorbitaux à la limite de la sensibilité instrumentale, voir Fig. \ref{fig:river}.

\subsection{transit d'un des coorbitaux}

Considérons maintenant que nous observons le transit d'une seule planète, et que nous voulons savoir si celle-ci possède un compagnon coorbital. L'absence de détection par transit du compagnon peut provenir de son trop faible rayon ou d'une inclinaison mutuelle non nulle avec la planète qui transite. En effet, nous avons vu au Chapitre 3 qu'il existait des configurations coorbitales stables pour d'importantes valeurs de l'inclinaison mutuelle.  


Dans le cas où une seule des planètes transite, les instants des transits sont donnés par exemple par $t_{1,k}$, équation (\ref{eq:ttrans}). Deux transits successifs de la même planète sont séparés de:
\begin{equation}
t_{k+1}-t_{k}=P_n + \delta (\zeta(t_{k+1})-\zeta(t_{k})) \frac{ P_n}{2\pi}\\
\label{eq:Dtuk}
\end{equation}
On ne connaît pas à priori $P_n$, la période de l'orbite de la planète observée. Si on n'observe que deux transits successifs, on fait une erreur de $\delta (\zeta(t_{1,k+1})-\zeta(t_{1,k})) \frac{ P_n}{2\pi}$ sur la période orbitale de la planète. La détection de davantage de transits permet d'étudier les TTV (variation du temps de transit). Si on mesure trois transits successifs, l'écart de temps entre les transits $k$ et $k+1$ et les transits $k+1$ et $k+2$ s'écrit:
\begin{equation}
t_{k+2} + t_k -2 t_{k+1} = \delta (\zeta(t_{k+2})+\zeta(t_{k})-2\zeta(t_{k+1})) \frac{ P_n}{2\pi}\, .
\label{eq:ttrans2}
\end{equation}
On rappelle que l'évolution de $\zeta$ est lente devant la période orbitale ($\nu \propto \sqrt{\mu}n$). Il est donc préférable, après avoir estimé $P_n$ à $\delta\sqrt{\mu}$ près, d'attendre plusieurs périodes orbitales et de comparer l'instant effectif du transit avec l'instant prédit pour une orbite keplerienne. Au maximum, on peut mesurer un écart temporel de:
\begin{equation}
\max{t_k} - \min{t_k} = \delta (\zeta_{\max{}}-\zeta_{\min{}}) \frac{P_n}{2\pi}\, ,
\label{eq:ttrans2}
\end{equation}
si les transits sont mesurés aux extremums de la libration (donc des mesures séparées de $\approx P_n/\sqrt{\mu}$). Cette méthode ne fonctionne que si l'amplitude de libration de $\zeta$ est non nulle. Le signal tend vers $0$ avec $m_1/m_2$ si on observe $m_2$.  
Par ailleurs, cette mesure seule ne permet pas de distinguer l'amplitude de libration  $\zeta_{\max{}}-\zeta_{\min{}}$ de la répartition des masses $\delta$. On définit l'application:
\begin{equation}
t(k) = t_k-k P_n\, ,
\label{eq:ttrans2}
\end{equation}
où $P_n$ peut être approximé par l'équation (\ref{eq:Dtuk}), ou déterminé comme la moyenne des intervalles entre deux transits. \cite{VoNe2014} ont montré que l'étude de cette fonction permet d'identifier la configuration coorbitale du système observé, notamment dans le cas de fer-à-cheval.

\section{Transit et vitesse radiale}

\begin{figure}[h!]
\begin{center}
\includegraphics[width=0.65\linewidth]{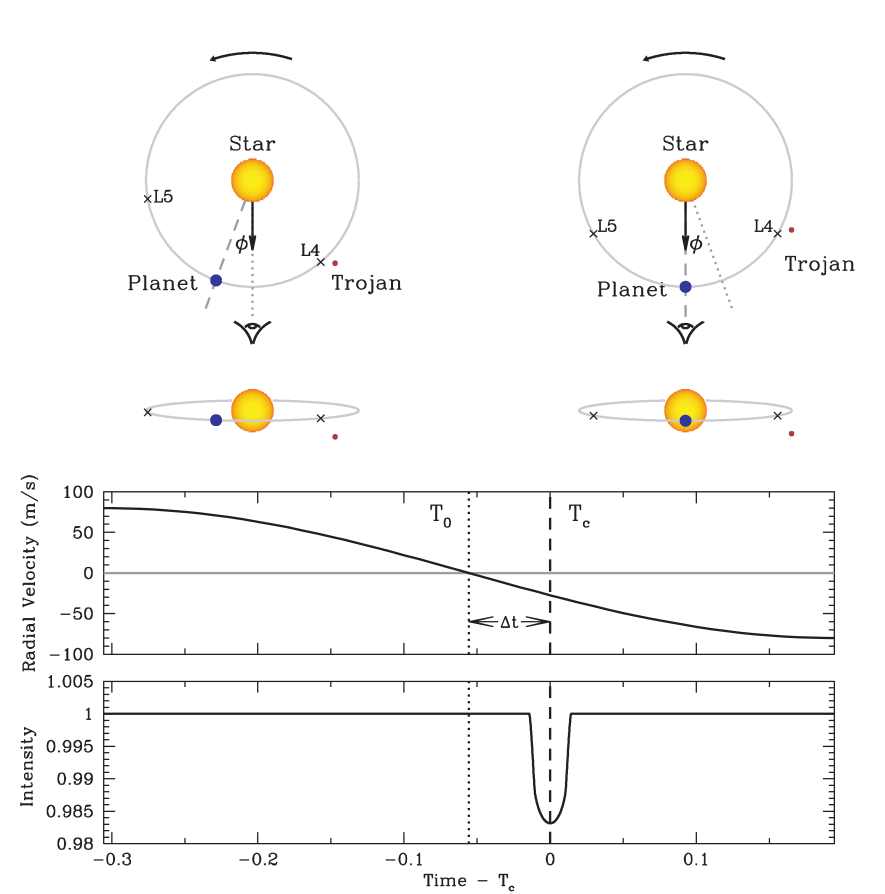}
\caption{\label{fig:ford} Illustration de la méthode de \cite{FoGa2006}. Le $\phi$ représenté ici est identique à la phase de l'équation (\ref{eq:r0_v3}). Les graphes A et B représentent l'instant $T_0$ où les vitesses radiales valent leur valeur moyenne et où la planète fictive passe par la ligne de visée. Les graphes C et D représentent l'instant $T_e$ de transit effectif, où la planète $m_1$ passe par la ligne de visée.}
\end{center}
\end{figure}


Dans le cas où seule la planète $m_1$ transite, la combinaison avec des mesures de vitesse radiale peut permettre d'identifier des systèmes coorbitaux, même avec des mesures réparties sur un temps court devant la période de libration de l'angle résonant.

Cette méthode, proposée par \cite{FoGa2006} dans le cas de coorbitaux en configuration équilatérale circulaire, confronte la date effective du transit d'une planète avec la date prédite par les mesures de vitesses radiales. Prenons par exemple une planète seule sur une orbite circulaire autour de l'étoile. Si celle-ci transite, la date du milieu du transit coïncide avec l'instant où la vitesse radiale de l'étoile vaut sa valeur moyenne.

D'autre part, si la planète qui transite a un compagnon coorbital, alors la vitesse radiale mesurée atteint sa valeur moyenne à l'instant $T_{0}$, lorsque la planète équivalente $m_{eq}$ décrite en section \ref{sec:MBar} passe dans la ligne de visée. La planète qui transite, quant à elle, passera dans la ligne de visée avant ou après cet instant à la date $T_{e}$. On mesure donc l'écart de temps $\Delta t$: 
\begin{equation}
\Delta t = |T_{e}-T_{0}|= \frac{\phi(\zeta,\delta)}{2\pi}P_n\, .
\label{eq:Dt}
\end{equation}
où $\phi$, l'angle entre $m_1$ et la position du barycentre des deux planètes, est représenté sur la figure~\ref{fig:ford}.
%
%

\subsubsection{Au voisinage de l'équilibre de Lagrange circulaire coplanaire}

\begin{figure}[h!]
\begin{center}
\includegraphics[width=0.49\linewidth]{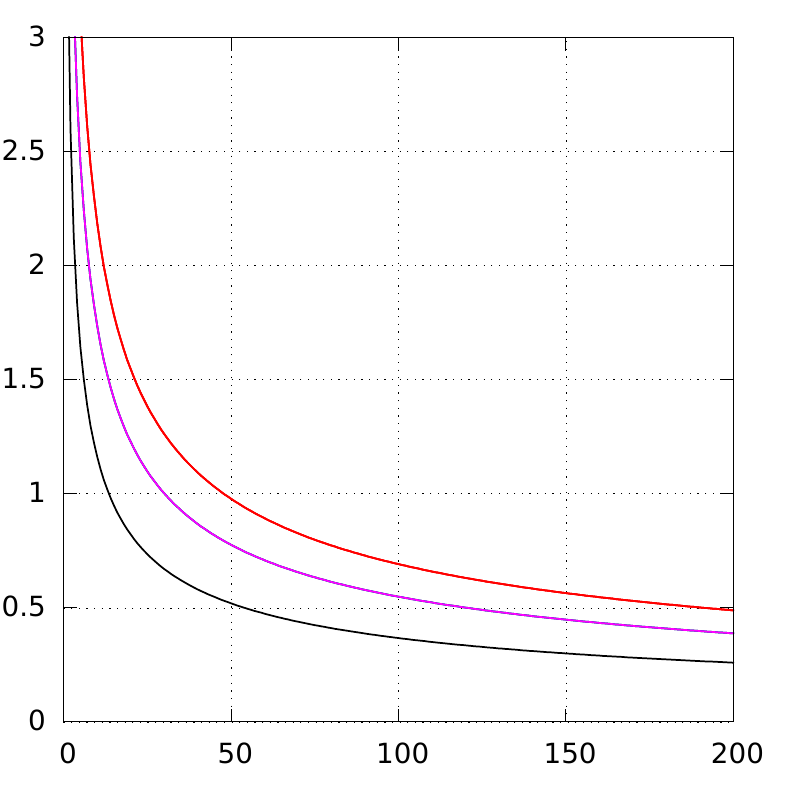}\\
\setlength{\unitlength}{0.05\linewidth}
\begin{picture}(.001,0.001)
\put(0,0){$N_{RV}$}
\put(-6,5){\rotatebox{90}{$\sigma_{m_2}[m_\oplus]$}}
\put(1,9.5){{$P=3$~day}}
\put(1,8.5){{\color{purple}$P=10$~day}}
\put(1,7.5){{\color{red}$P=20$~day}}
\end{picture}
\caption{\label{fig:sigm2} Écart-type de la détermination de la masse d'un compagnon coorbital $m_2<<m_1$ situé en $\zeta=\pm 60^\circ$ en fonction du nombre de point de vitesse radiale ($N_{VR}$). On a pris ici $\sigma_{VR}=1$~m/s. }
\end{center}
\end{figure}
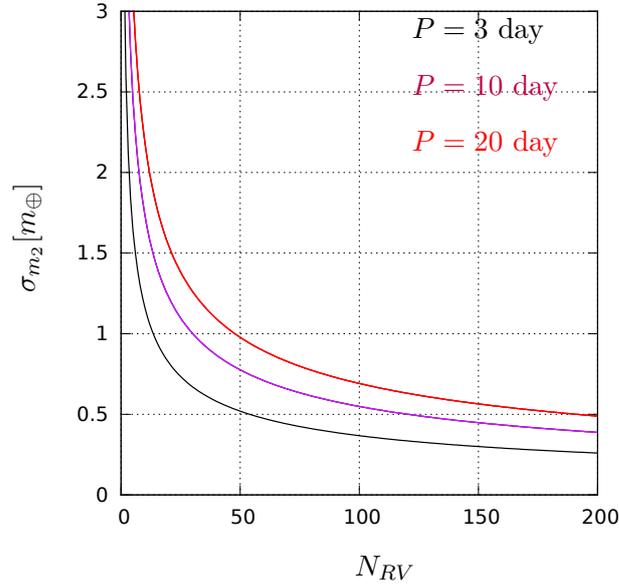

On rappelle qu'on observe le transit de la planète $m_1$. Si $m_2 > m_1$ ou $m_2 \approx m_1$, alors $\Delta t \geq P_n/12$ est facilement détectable. 

Considérons maintenant que $m_2 << m_1$ ($\delta << 1$) et que les coorbitaux sont proches de la configuration équilatérale. $\phi$ devient \citep{FoGa2006}:
\begin{equation}
\Delta t = \pm \frac{\sqrt{3}\delta P_n}{4 \pi}\, .
\label{eq:DtL4}
\end{equation}
On cherche à identifier la limite inférieure de détection, donc l'écart type sur la mesure de la masse du compagnon troyen $m_2$. De l'équation (\ref{eq:DtL4}) nous obtenons:
\begin{equation}
\sigma_{m_2}= 6 (m_1+m_2)  \frac{\sigma_{\Delta t}}{P}\, ,
\label{eq:sigm1}
\end{equation}
Cette expression fait intervenir l'imprécision sur la mesure de $\Delta t$. On considère que cette imprécision est la somme de l'imprécision sur la mesure $T_e$ et $T_0$, et que le transit est mesuré par une série de mesures photométriques continues prises avec un pas $\Gamma$ et une incertitude gaussienne $\sigma_{ph}$, alors l'incertitude sur l'instant du milieu d'un transit est donnée par \citep{FoGa2006}:
\begin{equation}
\sigma_{T_e}=\left(\frac{t_e}{2\Gamma}\right)^{1/2} \sigma_{ph}\left(\frac{R_*}{R}\right)^2\, ,
\label{eq:sigTe}
\end{equation}
où $R_*$ et $R$ sont les rayons de l'étoile et de la planète, et $t_e$ le temps d'entrée/sortie de la projection de la planète dans la projection de l'étoile. \cite{FoGa2006} ont considéré que $\sigma_{T_e}$ était négligeable devant $\sigma_{T_0}$ ($\sigma_{T_e} \approx 10\, s$ pour les précisions instrumentales de 2006).

D'autre part, si on considère $N_{RV}$ mesures de vitesses radiales réparties uniformément sur la phase orbitale avec une erreur gaussienne d'écart type $\sigma_{RV}$, l'incertitude sur le temps $T_0$ tend vers \citep{FoGa2006}: 

\begin{equation}
\sigma_{T_0}=\left(\frac{1}{2\pi^2 N_{RV}}\right)^{1/2} \frac{P_n \sigma_{RV}}{K}\, ,
\label{eq:sigDt}
\end{equation}
où $P$ est la période orbitale et $K\approx 2\pi m_{eq}/m_0 P_n^{1/3}$ (à l'ordre $1$ en $m_{eq}$, équation \ref{eq:meq}). Une bonne détermination de $T_0$, donc de $\Delta t$ et de $m_2$ demande donc l'acquisition de nombreux points de vitesse radiale. En remplaçant $\sigma_{\Delta t} \approx \sigma_{T_0}$ par l'expression (\ref{eq:sigDt}), on obtient:
\begin{equation}
\sigma_{m_2}= \frac{\sigma_{RV}}{N_{RV}^{1/2}} P_n^{1/3} \frac{3}{ \sqrt{2} \pi^2} \frac{m_1+m_2}{m_{eq}} m_0\, .
\label{eq:sigm12}
\end{equation}
Or, dans le cas $m_1 \gg m_2$, la masse équivalente tend vers $m_1$. nous avons donc:
\begin{equation}
\sigma_{m_2}=  \frac{3}{ \sqrt{2} \pi^2} \frac{\sigma_{RV}}{N_{RV}^{1/2}} P_n^{1/3} m_0 + \gO(m2/m_1)\, .
\label{eq:sigm12}
\end{equation}

Cet écart type donne une information sur l'erreur commise sur la masse de $m_2$ (si cette erreur est gaussienne, mais on ne rentrera pas dans les détails ici). Elle donne donc également la masse limite de détection d'un compagnon troyen pour la planète observée: si un décalage $\Delta t$ supérieur à $3$ sigma est observé, alors le test est positif avec une certitude de $99.7\%$. De la même manière, pour un jeu de mesures donné, on pourra détecter avec une certitude de $99.7\%$ la présence ou l'absence d'un compagnon coorbital de masse $3*\sigma _{m_2}$ (en supposant une configuration circulaire équilatérale).

\begin{rem}
$\sigma_{m_2}$ ne dépend pas de la masse des coorbitaux. Cela peut paraître contre-intuitif car $\Delta t$ augmente quand $m_1$ diminue à $m_2$ fixé (voir équation \ref{eq:DtL4}). Cependant, l'erreur sur la détermination de $T_0$, et donc de $\Delta t$, augmente quand $m_1$ diminue.
\end{rem}

La figure~\ref{fig:sigm2} donne l'évolution de l'écart type $\sigma_{m_2}$ en masse terrestre fonction du nombre de points de vitesse radiale ($N_{RV}$), pour trois valeurs de $P_n$ et une précision sur les vitesses radiales de $1$~m/s. Pour des périodes orbitales de quelques jours, une précision de l'ordre d'une masse terrestre est atteignable avec un nombre de points de vitesses radiales raisonnable ($N_{VR} \approx 100$).\\

\subsubsection{Discussion}

La mesure d'un décalage entre le transit prédit par les vitesses radiales et le transit effectif permet donc d'identifier des systèmes contenant potentiellement des configurations coorbitales. Cependant, ce décalage peut également être induit par l'excentricité mal déterminée de la planète qui transite.\\

Nous abordons ce problème dans une lettre (annexe \ref{an:VRT}). Nous commençons par généraliser la méthode de \cite{FoGa2006} à l'ordre $1$ en excentricité, pour toutes valeurs des masses des coorbitaux et toute valeur de l'angle résonant $\zeta$. Nous montrons qu'à l'ordre $1$ en excentricité, les vitesses radiales induites par des coorbitaux sont équivalentes à celles induites par une planète seule sur une orbite keplerienne (nous montrons également que cela n'est plus valable à l'ordre $2$). Nous discutons la dégénérescence entre le décalage induit par un compagnon coorbital de faible masse et l'excentricité de la planète qui transite et nous donnons, en fonction de l'erreur de mesure sur l'excentricité, la masse maximale que peut avoir un compagnon coorbital non détecté dans le cas incliné et faiblement excentrique.

\chapter{Résonances spin-orbites dans le problème planétaire à trois corps plan}

L'objet de ce chapitre est l'étude de la rotation d'un corps asymétrique indéformable dans le problème planétaire à trois corps plan. On y développe une méthode visant à déterminer la position et la taille des différentes résonances spin-orbites présentes dans l'espace des phases, permettant de prédire la position des zones de libration, circulation, ou d'évolution chaotique du spin d'un corps. La taille relative des îles de résonance donne une indication sur la probabilité de capture d'un corps en résonance spin orbite en cas de dissipation.  

On s'intéresse au cas simplifié où le corps tourne autour de son axe principal d'inertie qui est perpendiculaire au plan de son orbite. Cela représente le cas où l'obliquité du corps a déjà évolué vers son état d'équilibre, que nous supposons être nul. Comme nous pouvons le voir dans l'article \cite{CoLeRaRo2015} la prise en compte de l'évolution de l'obliquité peut grandement augmenter la taille des régions chaotiques, même si la position et la taille relative des îles de résonance spin-orbite restent inchangées dans la plupart des cas. \\



\an{In this chapter we study the rotation of an asymmetric rigid body in the coplanar 3-body planetary problem. We develop a method to determine the position and the size of the main spin-orbit resonances in the phase space, in order to predict the areas where the rotation angle is either circulating, librating, or in a chaotic state. The relative width of the resonant islands gives also an indication regarding the probability for the rotating body to be trapped in a given resonance if we consider dissipative forces. We consider the simplified case where the body rotates around its principal axis of inertia, which is held perpendicular to the orbital plane.}

\section{Formulation Hamiltonienne de l'équation de la rotation plane}

Sous l'hypothèse de rotation autour de l'axe d'inertie maintenu perpendiculaire au plan de l'orbite, la rotation du corps peut être décrite par un angle unique, que nous appelons $\theta$. L'équation du mouvement de  $\theta$ est donnée par les équations canoniques associées au Hamiltonien $\gH$ données par \cite{1962DJ}:
\begin{equation}
\gH=T +\frac{I^2}{2} - \frac{3}{4} \frac{B-A}{C} \cG \sum_{i=0,2} m_i \frac{ \cos\,2(\theta-\Psi_i)}{||\bold{r_1}-\bold{r_i}||^3},
\label{eq:danby} 
\end{equation}
où $\Psi_i$ est l'angle entre la direction x et la droite reliant les centres de gravité du corps tournant et du corps $i$, et $I$ (resp. $T$) est la variable canoniquement associée à $\theta$ (resp. au temps $t$). $A < B < C$ sont les moments d'inertie du corps tournant. Ceux-ci sont considérés constants dans cette étude (corps rigide). Pour plus de commodité, l'équation (\ref{eq:danby}) peut s'écrire:
\begin{equation}
\gH=T +\frac{I^2}{2} - \frac{3}{4} \frac{B-A}{C} \cG\, \Re \left[ \sum_{i=0,2} \left( m_i \frac{ \operatorname{e}^{-i2\Psi_i}}{||\bold{r_1}-\bold{r_i}||^3} \right) \operatorname{e}^{i2\theta} \right].
\label{eq:danby2} 
\end{equation}
Tout d'abord, il est clair que si le mouvement des trois corps considéré est chaotique, les équations canoniques associées au Hamiltonien (\ref{eq:danby2}) ne décriront que des solutions chaotiques. Nous considèrerons donc par la suite que le mouvement des trois corps est quasi-périodique, et donc que la quantité complexe
\begin{equation}
\sum_{i=0,2} \left( m_i \frac{ \operatorname{e}^{-i2\Psi_i}}{||\bold{r_1}-\bold{r_i}||^3} \right)
\label{eq:danby3} 
\end{equation}
peut se décomposer sur la base des fréquences fondamentales associées au mouvement du problème des trois corps: $\varsigma \in  \mathbb{R}^{n_f}_+$, avec $n_f$ le nombre de fréquences fondamentales du problème. Nous pouvons donc écrire: 

\be 
  \sum_{i=0,2} \left( m_i \frac{ \operatorname{e}^{-i2\Psi_i}}{||\bold{r_1}-\bold{r_i}||^3} \right) = \sum_{j \geq 0 } \ \hat{\rho}^2_{\eta_j} \ \operatorname{e}^{i( 2 \langle \eta_j, \varsigma \rangle t+ \phi_j )},
\label{eq:danbydec}
\ee
avec $2\eta_j \in \mathbb{Z}^{n_f}$, $ \langle \eta_j, \varsigma \rangle$ le produit scalaire usuel entre les vecteurs $\eta_j$ et $\varsigma$ et $\hat{\rho}_{\eta_j} \in \mathbb{R}_+$, où:   
\be
\hat{\rho}^2_{\eta_j} \operatorname{e}^{i \phi_j} = \frac{1}{\pi} \int_0^{\pi}   \sum_{i=0,2} \left( \frac{ m_i \operatorname{e}^{i 2 (-\Psi_i - \langle \eta_j, \varsigma \rangle t)}}{||\bold{r_1}-\bold{r_i}||^3} \right)  \ d t
\ .
\label{eq:rhoexp}
\ee 
En injectant l'expression (\ref{eq:danbydec}) dans le Hamiltonien (\ref{eq:danby2}) et en notant
\be 
\rho_{\eta_j}\equiv \frac{3}{2} \cG \sqrt{\frac{B-A}{C}}\hat{\rho}_{\eta_j}, 
\label{eq:rhoeq}
\ee
nous obtenons le Hamiltonien suivant:

\begin{equation}
\label{eq:theppqp} 
\gH=T +\frac{I^2}{2} -  \sum_{j \geq 0 }\ \frac{\rho^2_{\eta_j}}{2} \cos\ (2\theta + 2 \langle \eta_j, \varsigma \rangle t+ \phi_j).
\end{equation}
%


$\gH$ peut être vu comme le Hamiltonien d'une somme de pendules centrés en $\dot{\theta}=\langle \eta_j, \varsigma \rangle$, et de demi-largeur $\rho_{\eta_j}$. Dans le cas d'une orbite keplerienne excentrique, la seule fréquence présente est le moyen mouvement $n$. Dans ce cas, il y a une famille de résonances spin-orbites centrées en $\dot{\theta}=pn$, avec $2p \in \mathbb{Z}$ et $n$ le moyen mouvement de ce corps \citep{1965C,1966GP}. Dans le cas général d'une orbite quasi-périodique, nous avons potentiellement beaucoup plus d'îles de résonance dans l'espace des phases.  \\

\an{
We define the angle $\theta$ which parametrises the rotation of the rotating body around its principal axis of inertia. The equations of motion of $\theta$ are given by the canonical equations associated with the Hamiltonian $\gH$ (\ref{eq:danby}) \citep{1962DJ}, where $\Psi_i$ is the angle between the x direction and the line between the center of gravity of the rotating body and the body $i$. $I$ (resp. $T$) is the variable canonically associated with $\theta$ (resp. to the time $t$). $A < B < C$ are the moments of inertia of the rotating body. From now, let us consider quasi-periodic orbits (in the chaotic case, the solution of the equations associated with the Hamiltonian (\ref{eq:danby}) would be chaotic as well). Rewriting the Hamiltonian under the form (\ref{eq:danby2}), the quantity (\ref{eq:danby3}) can be expanded as Fourier series on the base of the fundamental frequencies associated with the 3-body motion: $\varsigma \in  \mathbb{R}^{n_f}_+$, with $n_f$ the number of fundamental frequency of the problem. We can hence obtain the expression (\ref{eq:danbydec}), where $2\eta_j \in \mathbb{Z}^{n_f}$, $ \langle \eta_j, \varsigma \rangle$ is the usual scalar product between $\eta_j$ and $\varsigma$, and$\hat{\rho}_{\eta_j} \in \mathbb{R}_+$. $\hat{\rho}^2_{\eta_j} \operatorname{e}^{i \phi_j}$ is given equation (\ref{eq:rhoexp}).   \\
Injecting the expression (\ref{eq:danbydec}) in the Hamiltonian (\ref{eq:danby2}), and noting $\rho_{\eta_j}$ the quantity (\ref{eq:rhoeq}), we obtain the Hamiltonian (\ref{eq:theppqp}).\\
Under the form (\ref{eq:theppqp}), $\gH$ can be seen as the Hamiltonian of a sum of pendulum centred at $\dot{\theta}=\langle \eta_j, \varsigma \rangle$, and of half-width $\rho_{\eta_j}$. In the case of a keplerian orbit, $n_f=1$ and $\varsigma$ contain only the mean motion of the rotating body. In this case, there is a family of spin-orbit resonance centred at $\dot{\theta}=pn$, with $2p \in \mathbb{Z}$ and $n$ the mean motion of the rotating body \citep{1965C,1966GP}. In the generic case of a quasi-periodic orbit, there may be many more resonant islands in the phase space.
}

\section{Description de l'espace des phases}

On définit trois états possibles pour la rotation du corps: la libration, lorsque le corps évolue en résonance spin-orbite (on définit la résonance exacte comme un cas particulier de libration), la rotation chaotique, lorsque le spin est imprévisible, et la circulation, lorsque le corps est en dehors des deux états précédents. La nature dynamique d'une région donnée de l'espace des phases ($\theta, \dot{\theta}$) dépend de la proximité des résonances voisines et de leur épaisseur respective \citep{1979CB}. On définit la distance entre le centre de deux résonances de la manière suivante:
\be
 \epsilon^k_j=|\langle \eta_j-\eta_k, \varsigma \rangle|.
  \label{eq:epsjk}
\ee
Selon le critère de recouvrement de Chirikov, l'interaction entre deux résonances voisines peut être de trois sortes:\\

(a) quand $(\rho_{\eta_j}+\rho_{\eta_k}) \ll \epsilon^k_j$, les deux îles sont bien séparées, et seules leurs séparatrices sont entourées d'une fine zone chaotique. La plupart des trajectoires au voisinage de ces îles sont donc quasi-periodiques. Si on considère une faible dissipation, le spin peut être capturé dans une de ces résonances. Plus l'île est large, plus la probabilité d'y être capturé est importante \citep{1966GP}.\\   

(b) quand $(\rho_{\eta_j}+\rho_{\eta_k}) \gg \epsilon^k_j$, les deux îles se recouvrent totalement. La dynamique résultante est celle d'un pendule modulé \citep[voir][page 224]{Morbidelli02}. Adaptons l'étude menée par \cite{Morbidelli02} à l'interaction de deux résonances voisines, indicées $j$ et $k$. La restriction du Hamiltonien (\ref{eq:theppqp}) à ces résonances s'écrit:
\begin{equation}
\label{eq:morb2r} 
\gH=T +\frac{I^2}{2} -  \frac{\rho^2_{\eta_j}}{2} \cos\ (2\theta + 2 \langle \eta_j, \varsigma \rangle t+ \phi_j) -  \frac{\rho^2_{\eta_k}}{2} \cos\ (2\theta + 2 \langle \eta_k, \varsigma \rangle t+ \phi_k) .
\end{equation}
Ce qui, après quelques transformations trigonométriques, est équivalent à:
\begin{equation}
\label{eq:morb2r2} 
\gH=T +\frac{I^2}{2} -  A(t)   \cos\ (2\theta + 2 \langle \eta_j, \varsigma \rangle t+ \phi').
\end{equation}
où $\phi'$ dépend de l'ensemble des paramètres, et:
\begin{equation}
\label{eq:morb2rA} 
A(t)= \sqrt{  \frac{\rho^4_{\eta_j}}{4} + \frac{\rho^4_{\eta_k}}{4}  + \frac{\rho^2_{\eta_j}\rho^2_{\eta_k}}{2} \cos\ (\epsilon^k_jt +\phi_k-\phi_j) } .
\end{equation}
 De manière générale, l'expression (\ref{eq:morb2r2}) est moins limpide que (\ref{eq:morb2r}). Cependant, lorsque $\epsilon^k_j$ est petit devant la fréquence de libration dans l'île de résonance, c'est à dire:
 \begin{equation}
\label{eq:morb2rcond} 
 \epsilon^k_j \ll \min (\sqrt{A(t)}) = \sqrt{\left|\frac{\rho^2_{\eta_j}}{2}-\frac{\rho^2_{\eta_k}}{2}\right|} \ , 
  \end{equation}
  nous pouvons faire l'approximation d'invariance adiabatique \citep{He1982,Morbidelli02} pour $A(t)$. Dans ce cas, l'expression (\ref{eq:morb2r2}) est celle d'un pendule modulé, dont l'épaisseur de l'île de libration varie sur la période $P_{\epsilon^k_j} = 2\pi/\epsilon^k_j$ d'une amplitude égale à $\max (A(t)) - \min (A(t)) = \min(\rho^2_{\eta_j},\rho^2_{\eta_k})$. Sur une échelle de temps courte par rapport à la période $P_{\epsilon^k_j}$, nous pouvons considérer une valeur instantanée pour $A(t)$, afin de déterminer si notre système se trouve dans une zone de circulation, de libration, ou au voisinage de la séparatrice de l'île de résonance. Cependant, en considérant la rotation sur un temps plus long, l'ensemble de l'espace des phases balayé par la séparatrice est chaotique, comme chaque point de cette région de l'espace des phases est à tour de rôle dans un état de circulation, de libration, ou bien au voisinage de la séparatrice.\\

(c) quand la distance entre le centre de deux résonances est comparable avec la somme de leurs deux demi-épaisseurs non perturbées ($\epsilon^k_j\approx \rho_{\eta_j}+\rho_{\eta_k}$), il y a apparition d'une zone chaotique d'épaisseur significative \citep{1979CB,1999ssd}, comme c'est le cas pour la rotation d'Hyperion, dans le cas d'une orbite Keplerienne excentrique \citep{1984W}.\\


Décrire l'espace des phases du problème considéré s'avère donc difficile, car il faut connaître l'ensemble des $\rho_{\eta_j}$ qui sont de manière générale de nombre infini. Cependant, dans la plupart des cas, l'amplitude des termes de la décomposition (\ref{eq:danbydec}) décroît. Cette décroissance peut être plus ou moins lente, mais elle permet néanmoins de ne considérer qu'un nombre fini de termes. En effet, à partir d'un certain rang les termes sont soit d'amplitude négligeable par rapport à leur séparation en fréquence (cas b), mais avec des îles de résonances très petites, avec des probabilités de capture proportionnelles à leur largeur; soit ils ne font que perturber légèrement la taille de la zone chaotique autour de la séparatrice d'îles de largeur plus élevée, en créant une zone chaotique de taille comparable à leur propre largeur (cas c). On peut donc avoir un bon aperçu de la dynamique au voisinage d'une résonance spin-orbite excentrique (Keplerienne, de la forme $\dot \theta = k \eta/2$) en ne considérant que les îles de largeur comparable avec l'île la plus large présente, étant le terme de plus grande amplitude de la décomposition (\ref{eq:danbydec}). \\

Pour que des perturbations orbitales aient un impact significatif sur l'espace des phases ($\theta$, $\dot{\theta}$), il faut donc que la décomposition (\ref{eq:danbydec}) possède des termes supplémentaires par rapport au cas keplerien de taille significative et une séparation fréquentielle non négligeable par rapport à la taille des îles. Ce dernier critère est généralement relatif à l'échelle de temps considérée: sur un temps de l'ordre de la periode orbitale, l'évolution de l'excentricité n'a aucun effet sur l'espace des phases, mais celle-ci peut générer d'importantes zones chaotiques sur des temps séculaires \citep[][en annexe]{LeRoCo2016}. Reprenant l'expression (\ref{eq:danby2}), on étudiera deux cas pouvant faire apparaître des fréquences comparables avec la taille des résonances spin-orbites kepleriennes: Le cas d'une étoile et de deux planètes en résonance en moyen mouvement, où $m_0 >> \max (m_1,m_2)$ (section \ref{sec:csp}), et le cas de deux étoiles et d'une planète $(m_0,m_2) >> m_1$ (section \ref{sec:cb}).\\

\an{
We define three possible states for the rotation of the body: the circulation, the libration, and the chaotic motion. To know which one of these states dominates in a given area of the phase space, we compare the quantity $\epsilon^k_j$ (eq. \ref{eq:epsjk}), which is the departure between the center of the resonant islands $j$ and $k$, with the quantity $\rho_{\eta_j}+\rho_{\eta_k}$, which is the sum of the half width of these two islands.\\
 From Chirikov's criterion, we know that if $(\rho_{\eta_j}+\rho_{\eta_k}) \ll \epsilon^k_j$ (case a), the $k^{th}$ and $j^{th}$ islands are separated, and most of the trajectories in their vicinity are quasi-periodic (either librating inside one of these resonance or circulating). If we consider dissipative forces, the spin can get locked in one of these resonances. The probability to stay in that resonance increases with its width \citep{1966GP}.\\
If $(\rho_{\eta_j}+\rho_{\eta_k}) \gg \epsilon^k_j$ (case b), the islands overlap. The resulting dynamics is equivalent to the one of a modulated pendulum, see the expression of the restriction of $\gH$ to these 2 islands \citep[see the equations \ref{eq:morb2r} to \ref{eq:morb2rA}, and][page 224]{Morbidelli02}.\\
If $\epsilon^k_j\approx \rho_{\eta_j}+\rho_{\eta_k}$, a chaotic area of significant width appears \citep{1979CB,1999ssd}.    
\\}

\an{
Generally, there is an infinite number of therm in the expansion (\ref{eq:danbydec}). However, the size of these terms generally decreases. We will discard all terms whose coefficient $\rho_{\eta_j}$ is negligible with respect to the largest value of $\rho_{\eta_j}$ in the studied area.\\
The importance of a given term may depend on the considered time scale: for example the evolution of the eccentricity due to planetary interactions do not impact the rotational phase space on a short time scale, but it can create wide chaotic area on a secular time scale \citep[][annex]{LeRoCo2016}. 
To perturb significantly the rotational phase space of a body on a quasi-keplerian orbit on a relatively short time scale hence requires to generate terms of significant width, and significant frequency separation, with respect to the Keplerian ones.  
We identified two cases that may generate such perturbations:\\
the case of two planets in mean motion resonance around a star, $m_0 >> \max (m_1,m_2)$, where the frequency is generally of the order of $\sqrt{(m_1+m_2)/m_0}n$ \citep[][in annex]{LeRoCo2016}.\\
And the case of a planet orbiting around binaries, $(m_0,m_2) >> m_1$. In this case, the mean motion of the binaries can be similar to the mean motion of the planet, see \cite{CoLeRaRo2015} in annex \ref{an:rot_cb}.
}

\subsection{Système planétaire}
\label{sec:csp}
Considérons un système comprenant un corps significativement plus massif que les autres, tel une étoile $m_0$, et deux corps gravitant autour sur des orbites perturbées stables sans être en résonance en moyen mouvement, $m_1$ et $m_2$. Tant que les deux planètes restent suffisamment éloignées (en dehors de la sphère de Hill), l'impact du potentiel gravitationnel induit directement par la planète $2$ sur la rotation de la planète $1$ est négligeable devant le potentiel de l'étoile. Cependant, la dynamique de l'angle $\theta$ est perturbée indirectement par la planète $m_2$: elle perturbe l'orbite du corps tournant, ce qui modifie l'expression du terme $i=0$ dans la somme (\ref{eq:danby3}).

Dans le cas non résonant, les fréquences impliquées dans la dynamique orbitale de chacun des corps sont les moyens mouvements $n_j=\gO(1)$, et fréquences séculaires associées à l'évolution de leur excentricité et de leur inclinaison $g_j=\gO(\eps)$ et $s_j=\gO(\eps)$, où $\eps$ est de taille $\frac{m_1+m_2}{m_0}$. Nous avons donc $(n_1,n_2) \gg (g_1,g_2,s_1,s_2)$. La séparation fréquentielle des îles de résonance induites par ces fréquences ($\epsilon^k_j=\gO(\eps)$) est dans la plupart des cas négligeable par rapport à la largeur des îles de résonance induites par les termes kepleriens du développement (ça n'est pas le cas quand $\rho_{\eta_j} = \gO(\eps)$ mais cela implique que la largeur des îles de résonance considérées sont très faibles).

Dans le cas de fortes variations de l'excentricité ou de l'inclinaison avec des fréquences $g_j$ et $s_j$, la séparation des échelles de temps impliquées dans la dynamique nous permet de considérer l'excentricité, l'inclinaison et leurs angles associés comme des invariants adiabatiques. Si on s'intéresse à la rotation du corps sur un temps court devant des temps séculaires, on peut négliger l'évolution de ces variables. L'impact de l'évolution des invariants adiabatiques peut ensuite être étudié: les séparatrices balayent la région de l'espace des phases qui se situe entre leurs valeurs extrémales.\\

Dans le cas planétaire résonant, en plus des échelles de temps évoquées précédemment intervient celle de l'angle résonant de fréquence fondamentale $\nu=\gO(\sqrt \eps )$. Cette libration orbitale crée des îles de résonance spin-orbites centrées en $\dot \theta = kn/2+ p\nu/2$ avec $(k,p) \in \mathbb{Z}^2$ qui viennent s'ajouter aux familles de résonances spin-orbites kepleriennes $\dot \theta = kn/2$ ($\epsilon^k_j=\gO(\sqrt \eps)$). Cette séparation angulaire est trop faible pour qu'il y ait une modification significative de l'espace des phases ($\theta,\dot \theta$) pour des corps à très forte asymétrie tels que les satellites co-orbitaux de Saturne, mais les modifications de l'espace des phases deviennent très importantes pour des co-orbitaux qui auraient des caractéristiques proches de la terre.

 L'article \cite{LeRoCo2016} (annexe \ref{an:rot_coorb}) détaille l'espace des phases de l'angle de rotation $\theta$ d'une planète ayant un compagnon co-orbital dans le cas plan excentrique. Nous y montrons l'apparition des familles de résonances en $n$, $\nu$ et $g$ et étudions l'impact de l'amplitude de libration et de l'excentricité sur la position et la taille des îles. Nous montrons notamment que pour de larges amplitudes de libration, les îles de résonances centrées en $\dot \theta = kn/2+ p\nu/2$ peuvent devenir prépondérantes par rapport aux résonances spin-orbites kepleriennes. \\

\an{In the case of two planets in a mean motion resonance, the fundamental frequency $\nu$ associated with the libration is generally proportional to $\sqrt{\eps}n$. If the planets don't get too close to one another, we can neglect the direct effect of the planet on the rotational phase space of the rotating body. However, since the orbit of the rotation body is perturbed by the planet, resonant islands located at $\dot \theta = kn/2+ p\nu/2$ with $(k,p) \in \mathbb{Z}^2$ appear in addition to the Keplerian resonant islands. In some cases, these new islands can become preponderant with respect to the Keplerian ones. Moreover, on secular time scales, the evolution of the eccentricities can induce large chaotic areas \citep[see][in annex \ref{an:rot_coorb}]{LeRoCo2016}.}

\subsection{Cas binaire}
\label{sec:cb}

Dans le cas $(m_0,m_2) \gg m_1$, à l'exception du cas où $||\bold{r_1}-\bold{r_0}|| \ll ||\bold{r_1}-\bold{r_2}||$, le potentiel des deux étoiles doit être pris en compte pour l'étude de la rotation en plus de leurs effets sur l'orbite du corps. Cependant, si l'orbite de la planète $1$ reste suffisamment proche d'une orbite keplerienne (on rappelle que la méthode développée ici est valable dans le cas d'orbites quasi-périodiques), comme par exemple dans le cas circumbinaire où le demi grand-axe de la planète est grand devant celui de la binaire, ou dans le cas où la planète tourne autour d'une des étoiles, on peut dans un premier temps négliger l'effet des perturbations orbitales sur la rotation: l'effet dominant est la somme des potentiels gravitationnels des deux étoiles. 

Considérons le cas circumbinaire, où le corps tournant a une orbite proche d'une orbite keplerienne. notons $n$ le moyen mouvement du corps tournant et $n_b$ le moyen mouvement de la binaire. Les résonances spin-orbites apparaissent en $\dot \theta = kn/2 + p(n_b-n)/2$ avec $(k,p) \in \mathbb{Z}^2$ \citep{CoLeRaRo2015} (Annexe \ref{an:rot_cb}). Dans le cas où $n_b$ est comparable à $n$, de nouvelles îles apparaissent par rapport au cas keplerien. La largeur de ces îles augmente quand $n_b$ se rapproche de $n$, ce qui modifie complètement l'espace des phases. L'article \cite{CoLeRaRo2015} traite par exemple le cas des satellites circumbinaires du système Pluton-Charon. Dans cette lettre, on montre que le modèle analytique considérant le potentiel des deux corps massifs et des orbites circulaires donne des résultats similaires à l'intégration numérique du problème à trois corps et des équations de la rotation. On verra également que selon sa valeur initiale, l'obliquité peut fortement augmenter la taille des zones chaotiques dans l'espace des phases $(\theta, \dot \theta$).\\

\an{In the case of a planet orbiting binaries, the mean motion of the binaries, $n_b$ can be commensurable to the mean motion of the rotating body. The difference of the frequencies $n_b-n$ appears in the expansion (\ref{eq:danbydec}), there are hence resonant islands located at $\dot \theta = kn/2 + p(n_b-n)/2$, with $(k,p) \in \mathbb{Z}^2$ \citep{CoLeRaRo2015} (Annexe \ref{an:rot_cb}). }



  
\chapterstar{Résumé et discussion}

Après le rappel de résultats concernant la dynamique coorbitale dans le problème circulaire plan dans le chapitre 1, nous avons étudié le problème excentrique plan dans le chapitre 2, puis le problème circulaire incliné dans le chapitre 3.\\

Dans le cas excentrique plan nous nous sommes intéressé à l'évolution de l'espace des phases à mesure que l'excentricité des coorbitaux augmentait. Dans les premières sections, nous avons introduit les familles d'orbites quasi-périodiques de dimension non maximale $\cF$ et $\ol \cF$ qui émergent des équilibres de Lagrange circulaires. Nous avons ensuite réduit le problème à deux degrés de liberté: un degré associé à la libration de l'angle résonant $\zeta=\lambda_1-\lambda_2$, et un degré associé à l'évolution séculaire de la différence des périhélies $\Dv$ et des excentricités. Dans cet espace moyen réduit, nous avons introduit les variétés représentatives: des variétés de conditions initiales de dimension 2, choisies de manière à ce que les trajectoires qui en sont issues représentent un volume significatif de l'espace des phases de dimension 4. Nous montrons notamment que dans le cas où les coorbitaux sont de masses égales, le plan défini par $e_1=e_2$ et $a_1=a_2$ représente effectivement une part significative des orbites coorbitales pour une valeur fixée du moment cinétique. Puis nous proposons des critères numériques et semi-analytiques permettant d'identifier la position des familles $\ol \cF$ (et donc $\cF$ à l'intersection de celles-ci) dans l'ensemble de l'espace des phases. L'établissement de ces critères porte sur la séparation entre l'échelle de temps semi-rapide (associée à $\zeta$), et l'échelle de temps séculaire (associée à $\Dv$). Nous utilisons ensuite ces critères dans le cas $m_1=m_2$. Grâce à ceux-ci, nous montrons que les changements de topologie intervenant lorsque l'excentricité des coorbitaux augmente sont lié à l'évolution des familles $\ol \cF$. De plus, nous montrons que mis à part dans le domaine quasi-satellite, à mesure que l'excentricité et les masses des coorbitaux augmentent, les régions stables semblent se limiter au voisinage des familles $\ol \cF$. Les changements majeurs mis en évidence dans le cas de coorbitaux à masses égales apparaissent également pour des coorbitaux de masse différentes. Cependant, en plus des différentes configurations présentes dans le cas $m_1=m_2$, une partie significative de l'espace des phases est occupée par des orbites ou l'angle $\Dv$ circule. Enfin, nous calculons analytiquement la position des $\ol \cF$ dans le voisinage des familles $\cF$ afin de valider leur mise en évidence numérique.
        
Dans le chapitre $3$, nous effectuons la réduction de Jacobi afin de représenter la dynamique des coorbitaux circulaires inclinés par un hamiltonien à 1 degré de liberté. Nous appliquons ensuite des méthodes similaires à celles utilisées dans le chapitre 2. Nous étudions notamment l'évolution des familles émergeant des équilibres de Lagrange circulaires et la limite de stabilité des configurations troyennes lorsque l'inclinaison mutuelle des deux coorbitaux augmente.  

Il reste encore beaucoup à faire dans les deux cas étudiés (le cas excentrique plan et le cas circulaire incliné). Dans le cas circulaire incliné, la réduction à un degré de liberté permet l'étude exhaustive de l'espace des phases dans le problème moyen réduit. Dans le cas excentrique plan, pour $m_1=m_2$, il reste à affiner la vérification des variétés représentatives: une vérification plus approfondie dans la direction de la variable $\varPi$ (associées aux excentricités), mais également dans la direction de la variable $Z$ (associée aux demi-grand axes): vérifier qu'il n'existe pas de configurations où $a_1 \neq a_2$ sur toute l'orbite. Ces vérifications, dans un espace à deux paramètres, doivent être possible afin de se rapprocher de l'exhaustivité pour des valeurs fixées des masses et du moment cinétique.

Dans le cas $m_1 \neq m_2$, nous donnons en section \ref{sec:VaRemdif} un algorithme permettant de construire une variété de référence pour une configuration orbitale donnée. Cet algorithme a été utilisé pour l'ensemble des configurations troyennes, fer à cheval et quasi-satellite pour une valeur du moment cinétique $J_1=J_1(e_1=e_2=0.4)$. Il reste à adapter l'algorithme pour les orbites où $\Dv$ circule, puis à vérifier (également dans l'espace à deux paramètres $\varPi$ et $Z$) que chaque orbite de l'espace des phases passe bien par la variété de référence d'une des configurations connues.    

S'attaquer au problème excentrique incliné avec cette méthode est envisageable, mais nécessiterait de considérer des variétés représentatives de dimension $3$ (la moitié du nombre de dimension de l'espace moyen réduit), on perdrait donc l'avantage de pouvoir représenter sur des graphes en deux dimensions la dynamique d'un espace de dimension supérieure. Mais cela n'empêche pas l'étude des familles $\cF$ et $\ol \cF$ dans ce cas.

Nous pouvons également utiliser cette approche pour d'autres problème. Par exemple, identifier de manière semi-analytique ou numérique les familles d'orbites quasi-périodiques de dimension non maximale émergeant des points d'équilibres d'autres résonances en moyen mouvement afin d'observer leur évolution quand les paramètres orbitaux des planètes changent. Cela peut permettre, dans certains cas, de faire le lien avec des modifications de l'espace des phases.  \\

Dans le chapitre 4, nous abordons la détection des exoplanètes coorbitales quasi-circulaires. Nous rappelons que les études sur leur formation et leur évolution ne permettent pas à ce jour de contraindre les éléments orbitaux ou la masse de telles exoplanètes. Nous rappelons ensuite les différentes méthodes de détection adaptées au cas coorbital. On développe particulièrement le cas des vitesses radiales, ainsi que leur combinaison avec des mesures de transit. Nos travaux sur ces deux méthodes sont fournis en annexe. La détection de coorbitaux par vitesse radiale, astrométrie, ou variation du temps de transit nécessite d'identifier la signature de la libration de $\zeta$ dans le signal. Dans le cas contraire, le signal est équivalent à celui d'une planète seule (cela reste vrai à l'ordre 1 en excentricité pour les vitesses radiales, voir la lettre pour plus de détails). Pour pouvoir identifier cette signature, il faut observer sur des temps de l'ordre de la période de libration. Il faut également que l'amplitude de libration, ainsi que la masse du plus petit des deux coorbitaux, soient suffisamment importantes. L'observation de la libration n'est plus nécessaire si on peut combiner des mesures de vitesses radiales avec les mesures de transits: l'instant du transit prédit par les vitesses radiales est différent de l'instant effectif de transit de la planète si le signal de vitesses radiales à été induit par des coorbitaux. Cependant ce déphasage peut également être dû à une erreur dans la prédiction de l'instant transit à cause d'une mauvaise détermination de l'excentricité de la planète. Nous abordons ce problème dans la lettre en annexe.

La prochaine étape est évidement l'application de ces méthodes sur des systèmes concrets. Cela est déjà en cours dans le cadre du projet TROY\footnote{http\!\! ://www.sc.eso.org/\textasciitilde jlillobo/troy/index.html}. L'amélioration de ces techniques, quant à elle, nécessite d'approfondir la connaissance de l'impacte de l'excentricité et de l'inclinaison sur le signal, ainsi que de développer des techniques fines de traitement du signal et de `fit' pour faire ressortir la signature coorbitale. Mais ce n'est pas tout: il est nécessaire de comprendre comment nous pouvons confirmer la détection d'exoplanètes coorbitales, notamment en combinant différentes techniques entre elles et en étudiant quelles autres configurations peuvent induire des signatures semblables à celles des configurations coorbitales. D'autre part, une étude de stabilité adaptée aux voisinage des paramètres orbitaux et de la masse estimée d'une planète détectée peut grandement contraindre les paramètre orbitaux d'un éventuel compagnon troyen de cette planète.\\

Pour finir, on décrit dans le chapitre 5 une méthode permettant d'étudier l'effet de perturbations orbitales sur les résonances spin-orbites d'un corps indéformable dans le problème planétaire à trois corps plan. Nous appliquons cette méthode dans deux cas: le cas coorbital excentrique, et le cas circumbinaire. Ces travaux sont disponibles en annexe. 
Dans ces deux cas, la dynamique du système comporte des fréquences fondamentales qui peuvent être comparables au moyen mouvement du corps tournant. Cela entraîne l'apparition de nouvelles résonances spin-orbites qui viennent s'ajouter à celles du cas keplerien. Dans certains cas, ces nouvelles résonances sont de taille comparable à celle des résonances képleriennes. Selon les paramètres orbitaux et la forme du corps tournant, cela entraine soit l'apparition de larges zones chaotiques, soit de nouvelles configurations d'équilibre dans lesquelles le spin du corps tournant peut être capturé en présence de forces dissipatives.

D'autre part, nous avons vu que dans le cas circumbinaire l'ajout d'une obliquité initiale non nulle entrainait une très forte augmentation de la taille des zones chaotiques de l'espace des phases. Il serait intéressant de quantifier cela, et de l'expliquer à partir d'un formalisme adapté.\\    

\an{After we recalled results regarding the circular coplanar co-orbital case in chapter 1, we studied the coplanar eccentric problem in Chapter 2, then the inclined circular problem in chapter 3.\\} 

\an{In the coplanar eccentric case, we followed the evolution of the phase space as the eccentricity of the co-orbitals increased. In the first sections, we introduced the families of non-maximal quasi-periodic orbits $ \cF $ and $ \ol \cF$, which emerge from the circular Lagrangian equilibriums. We then reduced the problem to two degrees of freedom: a degree associated to the libration of the resonant angle $ \zeta = \lambda_1- \lambda_2 $, and a degree associated to the secular evolution of the difference of the periastron $\Dv$ and eccentricities. In this averaged reduced problem, we introduced the representative manifolds: 2-dimensional manifolds of initial conditions, chosen such that the trajectories taking their initial conditions on it would represent a significant volume of the 4-dimensional phase space. We give the expression of such manifold in the case where co-orbitals are of equal masses: we show that the plane $ e_1 = e_2$ and $ a_1 = a_2$ represents a significant part of the co-orbital orbits for a fixed value of the total angular momentum. Then we give numerical and semi-analytical criteria to identify the position of the $ \ol \cF $ families (thus the $ \cF $ families as well at the intersection of the $\ol \cF$) throughout the phase space. The establishment of these criteria is based on the separation of semi-fast time scale (associated with $ \zeta$), and secular time scale (associated with $ \Dv$). We then use these criteria in the case $ m_1 = m_2 $. Thanks to them, we show that the topological changes, occurring when the eccentricity increases, are linked to the evolution of the families $ \ol \cF$. Furthermore we show that, apart from the quasi-satellite configuration, as the eccentricity and the masses of the co-orbitals rise, stable areas shrink down to the vicinity of the families $ \ol \cF$. The major changes highlighted in the case of equal masses co-orbitals also appear for different-mass co-orbitals. However, in addition to the various configurations existing in the case $ m_1 = m_2 $, a significant part of the phase space is occupied by orbits for which $ \Dv $ circulates. Finally, we compute analytically the position of the $ \ol \cF $ in the neighbourhood of the $\cF$ families to validate their numerical identification.\\        
In Chapter 3, we use the Jacobi's reduction to represent the dynamics of the inclined circular co-orbitals by a 1-degree of freedom Hamiltonian. We then apply similar methods to those used in Chapter 2. We study in particular the evolution of the families emerging from the circular Lagrangian equilibriums, and the stability limit of the trojan configurations when the mutual inclination between the co-orbitals increases.\\
There is still much to do in both cases studied (coplanar eccentric case and inclined circular case). In the inclined circular case, the reduction to a 1-degree of freedom system makes possible the exhaustive study of the phase space (in the reduced averaged problem). In the coplanar eccentric case, for $ m_1 = m_2 $, it remains to refine the verification of the representative manifolds: a more thorough check in the direction of the variable $ \varPi $ (associated with eccentricities), but also in the direction of the variable $ Z $ (associated with semi-major axes): to check that there are no configurations where $ a_1 \neq a_2 $ throughout the orbit. These checks in a two-parameters space should be possible to approach completeness for given values of the masses and total angular momentum.\\
If $ m_1 \neq m_2 $ we give in section \ref{sec:VaRemdif} an algorithm to build a reference manifold for a given co-orbital configuration. This algorithm has been used for all the trojan, horseshoe and quasi-satellite configurations for the angular momentum equal to $ J_1 = J_1 (e_1 = e_2 = 0.4)$. It remains to adapt the algorithm to the orbits where $ \Dv $ circulates, then check (also in the two-parameters space $ \varPi$ and $ Z $) if each orbit of the phase space goes through the reference manifold of a known configurations.\\
To tackle the inclined eccentric problem with this method is feasible, but requires to consider 3-dimensional representative manifolds (half the number of dimensions of the reduced averaged phase space), we would lose the advantage of representing in 2-dimensional graphs the dynamics of a higher dimensional phase space. But this does not prevent the study of the families $ \cF $ and $ \ol \cF $ in this case.\\
We can also use this approach for other problems. For example, we can identify the non-maximal quasi-periodic families emerging from the equilibrium points of another mean motion resonance to study their evolution when the orbital parameters of the planets are changing. In some cases, it may allow to better understand the changes in the phase space.\\} 

\an{
In Chapter 4, we discuss the detection of quasi-circular co-orbital exoplanets. We recall that, to this day, the studies on their formation and evolution do not allow to constrain the orbital elements or masses of such exoplanets. We then recall the various detection methods adapted to the co-orbital case. We stress on the radial velocity technique, as well as its combination with transit measures. Our work on these two methods are provided in annex. The detection of co-orbitals by radial velocity, astrometry, or TTVs, requires to identify the signature of the libration of $\zeta$ in the signal, otherwise the signal is equivalent to the signal induced by a single planet on a keplerian orbit (this remains true to order 1 in eccentricity for the radial velocities, see letter for details). To be able to detect this signature, the time span of the observation must be of the order of the period of liberation. It also requires that the amplitude of libration, and the mass of the smaller of the two co-orbitals, are sufficiently large. The observation of the libration is not necessary if we can combine radial velocity measurements with transits measures: The time of transit predicted by the radial velocity is different from the actual transit of the planet if the radial velocity signal is induced by co-orbitals. However, this phase difference may also be due to a mistake in the prediction of the time of transit because of an improper determination of the eccentricity of the planet. We address this problem in a letter (in annex).\\
The next step is obviously to apply these methods on real systems. This is already under way in the TROY project (http\!\! ://www.sc.eso.org/\textasciitilde jlillobo/troy/index.html).\\
Improving the detection techniques requires a deeper understanding of the impact of the eccentricity and inclination on the signal, as well as to develop adapted signal processing techniques and `fitting' algorithm to bring out the co-orbital signature. But that's not all: it is necessary to understand how we can confirm the detection of co-orbital exoplanets, for example by combining different detection techniques together and by studying which other configurations can induce similar signatures to the co-orbital one. Moreover, a stability study adapted to the vicinity of the orbital parameters and the estimated mass of a detected planet can greatly constrain the orbital parameters of a possible trojan companion.\\}

\an{
Finally, in chapter 5 we described a method to study the effect of the orbital perturbations on the spin-orbit resonances of a rigid body in the planetary coplanar 3-body problem. We apply this method in two cases: the eccentric co-orbital case and circumbinary case. These works can be found in the appendix. In both cases, the dynamics includes fundamental frequencies that can be comparable with the mean motion of the rotating body. This causes the appearance of new spin-orbit resonances in addition to those of the Keplerian case. In some cases, these new resonances are comparable in size to the Keplerian's ones. Depending on the orbital parameters and the shape of the rotating body, this causes either the appearance of large chaotic areas or new equilibrium configurations in which the spin of the rotating body may be trapped in the presence of dissipative forces.\\
Beside that, we saw in \cite{CoLeRaRo2015} that adding a nonzero initial obliquity caused a sharp increase in the size of the chaotic areas of the phase space. It would be interesting to quantify this phenomenon, and explain it from a suitable formalism.
}

  \thispagestyle{empty}

\epigraph{Outside of a dog, a book is a man's best friend.\\
	Inside of a dog it's too dark to read.}{Groucho Marx}
	
\bibliographystyle{apalike-fr}
\bibliography{astro,mabiblio}

  \appendix

\chapter{Matrices de Passage}
\label{an:MP}

Afin de s'affranchir de la partie linéaire en $Z$, on effectue le changement de variable suivant:
\begin{equation}
  \chi^{(1)}(z,Z^{(1)},x_j,\bar{x}_j) = (z,Z,x_j,\bar{x}_j)
\label{eq:cv1f}
\end{equation}
où
\begin{equation}
Z^{(1)} = Z + h_{Z}/(2 h_{Z^2}).
\label{eq:cv1ex}
\end{equation}
Pour la suite, on reprendra la notation $Z$ pour $Z^{(1)}$:. Nous obtenons donc $h_{Z}=0$. Le changement de variable suivant, $\chi_{L_k}$, a pour but de diagonaliser la partie quadratique du Hamiltonien au voisinage de l'équilibre circulaire $L_k$. On remarque que les variables ($z$,$Z$) et ($x_j$,$\bar{x}_j$) sont découplées, la matrice de passage du changement de variable suivant prend donc la forme:
\begin{equation}
 P^{(2)}= 
   \begin{pmatrix}
    A^{(2)}& 0_{2 \times 4} \\
 	^t 0_{2 \times 4} &  B_j^{(2)} 
    \end{pmatrix}   
\label{eq:P2}
\end{equation} 
avec
\begin{equation}
 A= 
   \begin{pmatrix}
    -i\frac{1}{\alpha \sqrt{2}} &  -\frac{1}{\alpha \sqrt{2}} \\
    \frac{\alpha}{\sqrt{2}}& i\frac{\alpha}{\sqrt{2}} 
    \end{pmatrix}   
\label{eq:P2A2}
\end{equation} 
avec $\alpha=(h_{z}/h_{Z})^{1/4}$ et 
\begin{equation}
 B_4^{(2)}= \frac{1}{\sqrt{\Lambda^0_1+\Lambda^0_2}}
   \begin{pmatrix}
    \sqrt{m_2}\exp^{i\pi/3} & 0 & \sqrt{m_1}\exp^{i\pi/3} & 0 \\
   0 & \sqrt{m_2}\exp^{-i\pi/3} & 0 & \sqrt{m_1}\exp^{-i\pi/3}  \\
    -\sqrt{m_1} & 0 & \sqrt{m_2} & 0 \\
     0 &    -\sqrt{m_1} & 0 & \sqrt{m_2} 
    \end{pmatrix}   
\label{eq:P2B2L4}
\end{equation} 
et pour $L_3$:
\begin{equation}
 B_3^{(2)}= \frac{1}{\sqrt{\Lambda_1+\Lambda_2}}
   \begin{pmatrix}
    \sqrt{\Lambda_2} & 0 & \sqrt{\Lambda_1} & 0 \\
   0 & \sqrt{\Lambda_2} & 0 & \sqrt{\Lambda_1}  \\
    \sqrt{\Lambda_1} & 0 & -\sqrt{\Lambda_2} & 0 \\
     0 &    \sqrt{\Lambda_1} & 0 & -\sqrt{\Lambda_2}
    \end{pmatrix}   
\label{eq:P2B2L3}
\end{equation} 
On donne par ailleurs:
\begin{equation}
 A^{-1}= 
   \begin{pmatrix}
    i\frac{\alpha}{\sqrt{2}}  & \frac{1}{\alpha \sqrt{2}}  \\
 	-\frac{\alpha}{\sqrt{2}}  & -i\frac{1}{\alpha \sqrt{2}}   
    \end{pmatrix}   
\label{eq:P2A2}
\end{equation} 

$\chi_{L_k}$ s'écrit donc:
\begin{equation}
  \chi_{L_k}(z_0,\tilde{z}_0,z_1,\tilde{z}_1,z_2,\tilde{z}_2)= (z,Z,x_1,\tilde{x}_1,x_2,\tilde{x}_2)
\label{eq:cv2f}
\end{equation}
où
\begin{equation}
^t (z,Z,x_1,\bar{x}_1,x_2,\bar{x}_2) =  P^{(2)} \,  ^t(z_0,\tilde{z}_0,z_1,\tilde{z}_1,z_2,\tilde{z}_2).
\label{eq:cv2ex}
\end{equation}

\chapter{Exemples d'orbites}

\begin{figure}[h!]
\begin{center}
 \includegraphics[scale=0.9]{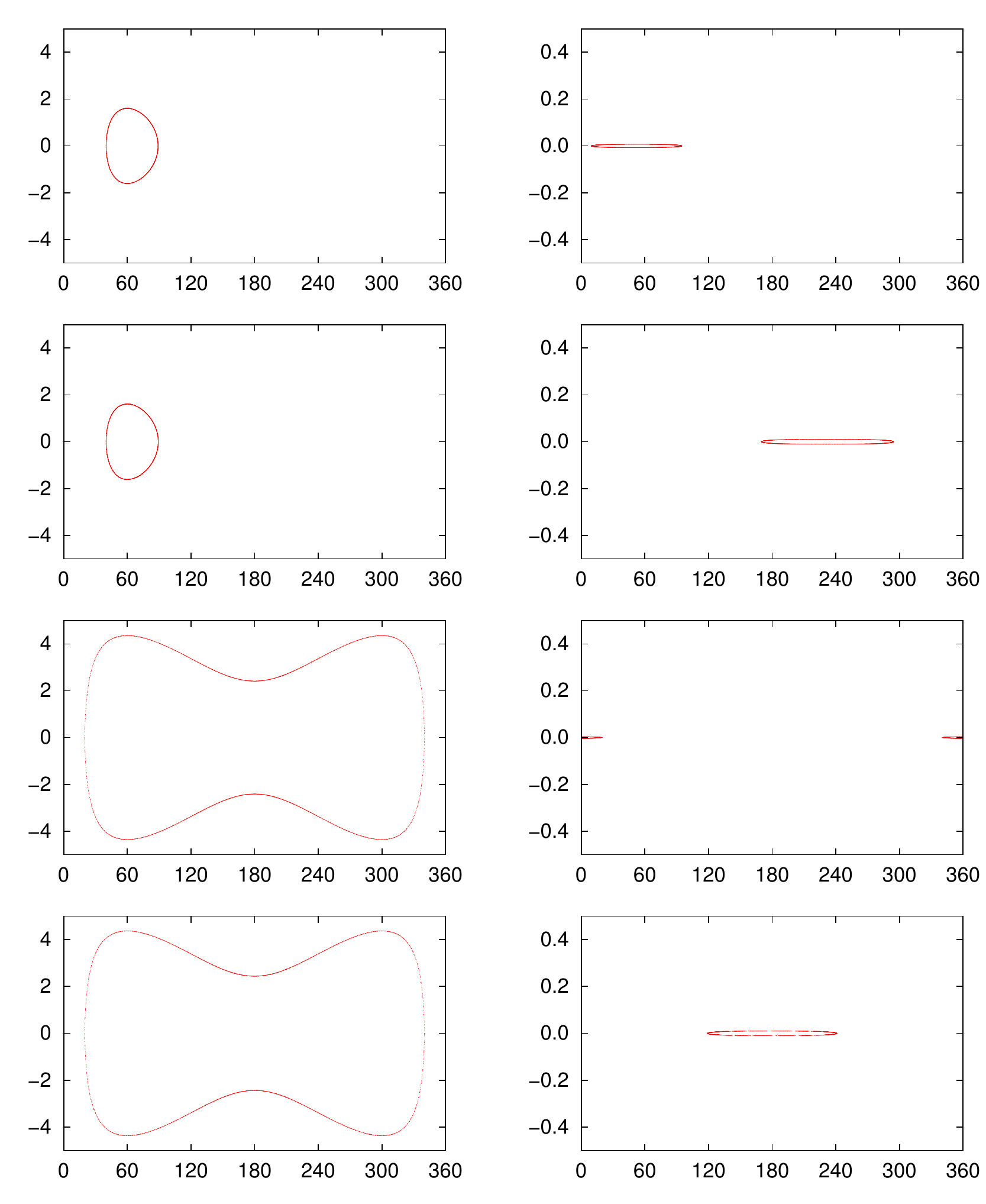}
   \setlength{\unitlength}{1cm}
\begin{picture}(1,0.001)
\put(0,9){\rotatebox{90}{$e_1-e_2$}}
\put(-7.8,9){\rotatebox{90}{$Z/\eps^{3/2}$}}
\put(-3.5,.1){{$\zeta$}}
\put(4,.1){{$\Delta \varpi$}}
\put(-6.3,18){{(a)}}
\put(-6.3,13.5){{(b)}}
\put(-6.3,9){{(c)}}
\put(-6.3,4.5){{(d)}}
\end{picture} 
\hspace{-1cm}
 \caption{\label{fig:orbe01} Trajectoires issues du plan de conditions initiales de la figure~\ref{fig:glob_e01}. (a): $\zeta=40^\circ$, $\Dv=100^\circ$; (b): $\zeta=40^\circ$, $\Dv=170^\circ$; (c): $\zeta=20^\circ$, $\Dv=20^\circ$; (d): $\zeta=20^\circ$, $\Dv=120^\circ$}
 \end{center}
 \end{figure}

\begin{figure}[h!]
\begin{center}
 \includegraphics[scale=0.9]{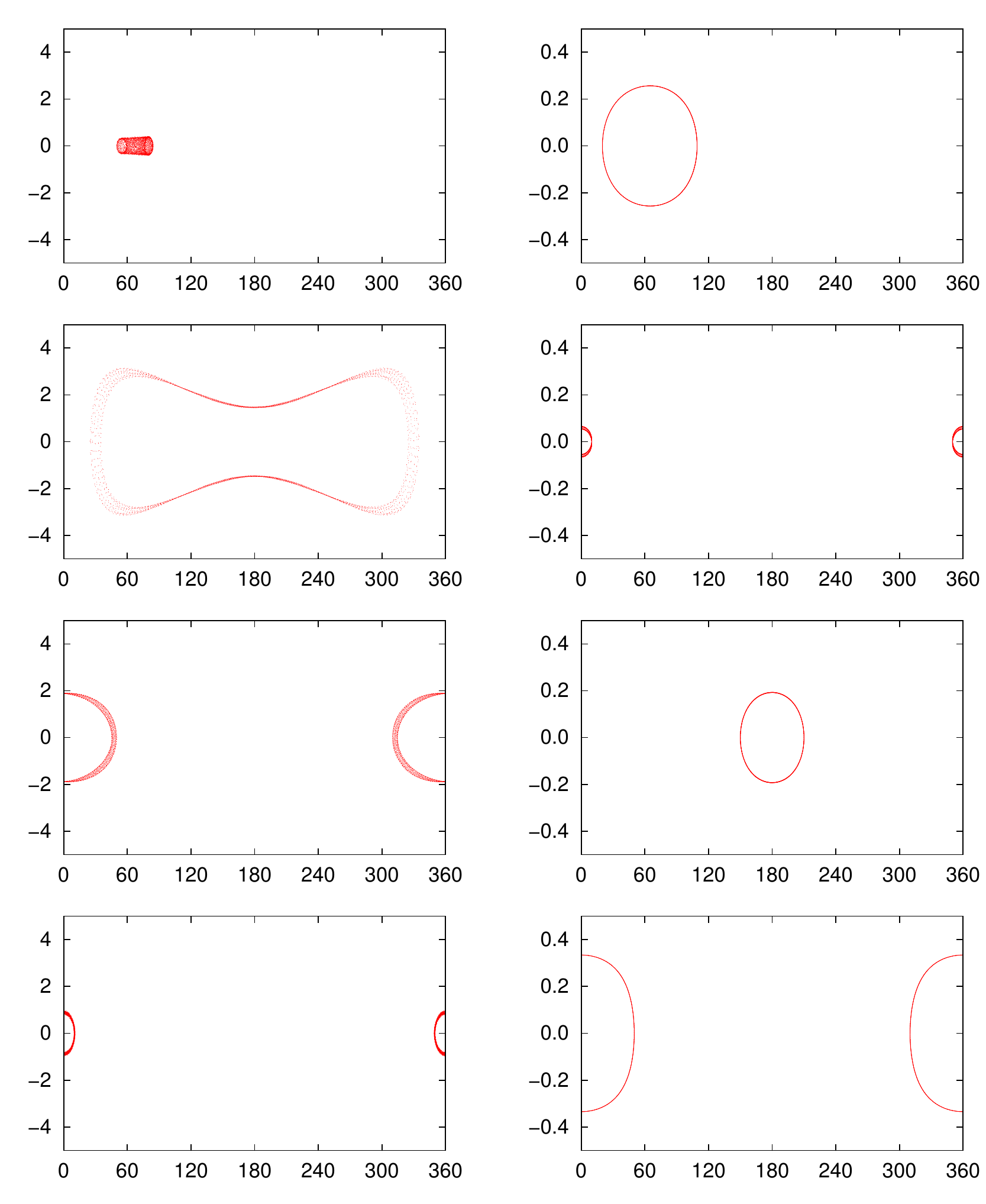}
   \setlength{\unitlength}{1cm}
\begin{picture}(1,0.001)
\put(0,9){\rotatebox{90}{$e_1-e_2$}}
\put(-7.8,9){\rotatebox{90}{$Z/\eps^{3/2}$}}
\put(-3.5,.1){{$\zeta$}}
\put(4,.1){{$\Delta \varpi$}}
\put(-6.3,18){{(a)}}
\put(-6.3,13.5){{(b)}}
\put(-6.3,9){{(c)}}
\put(-6.3,4.5){{(d)}}
\end{picture} 
\hspace{-1cm}
 \caption{\label{fig:orbe4} Trajectoires issues du plan de conditions initiales de la figure~\ref{fig:glob_e4}. (a): $\zeta=55^\circ$, $\Dv=20^\circ$; (b): $\zeta=20^\circ$, $\Dv=10^\circ$; (c): $\zeta=50^\circ$, $\Dv=160^\circ$; (d): $\zeta=10^\circ$, $\Dv=50^\circ$}
 \end{center}
 \end{figure}
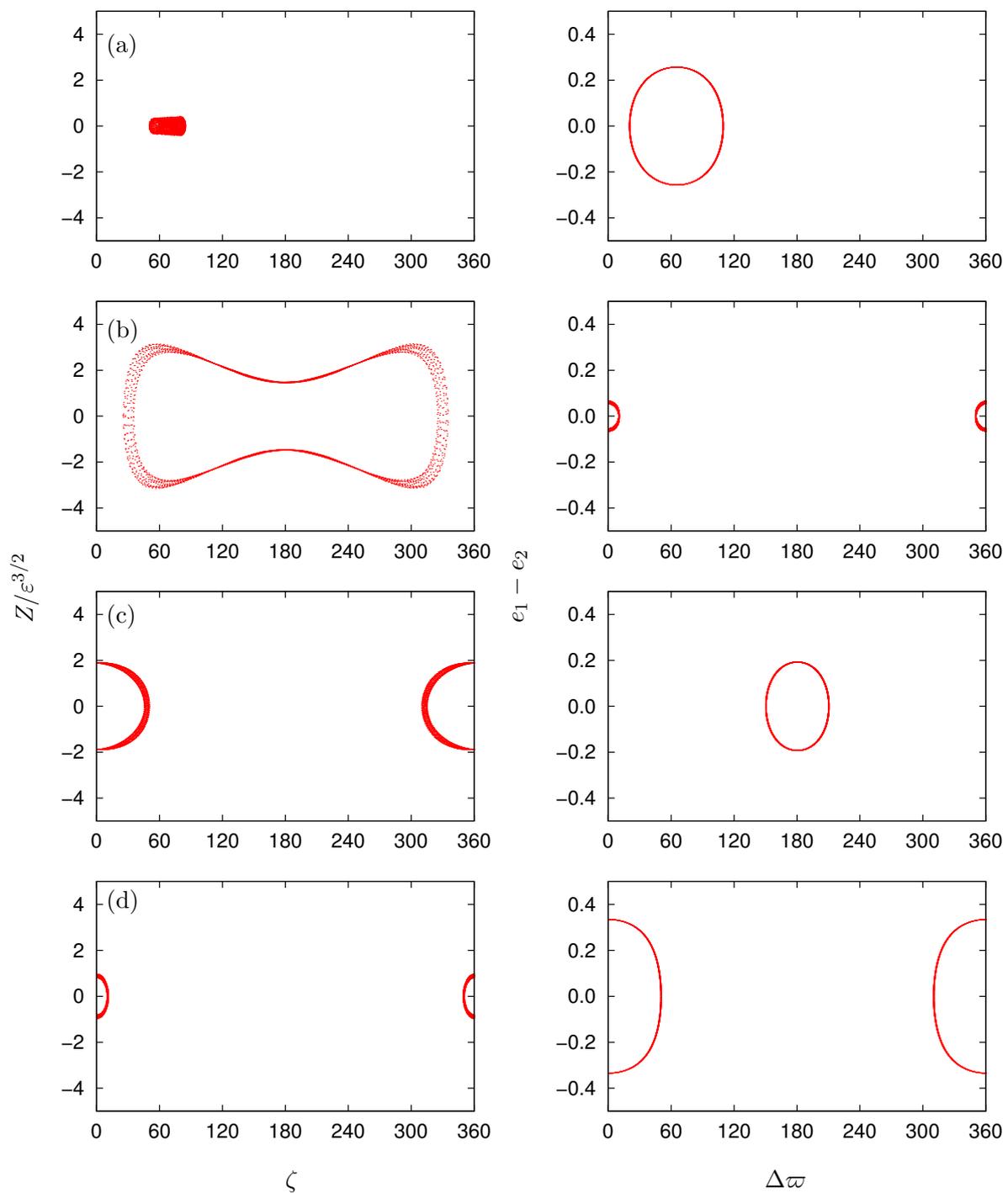

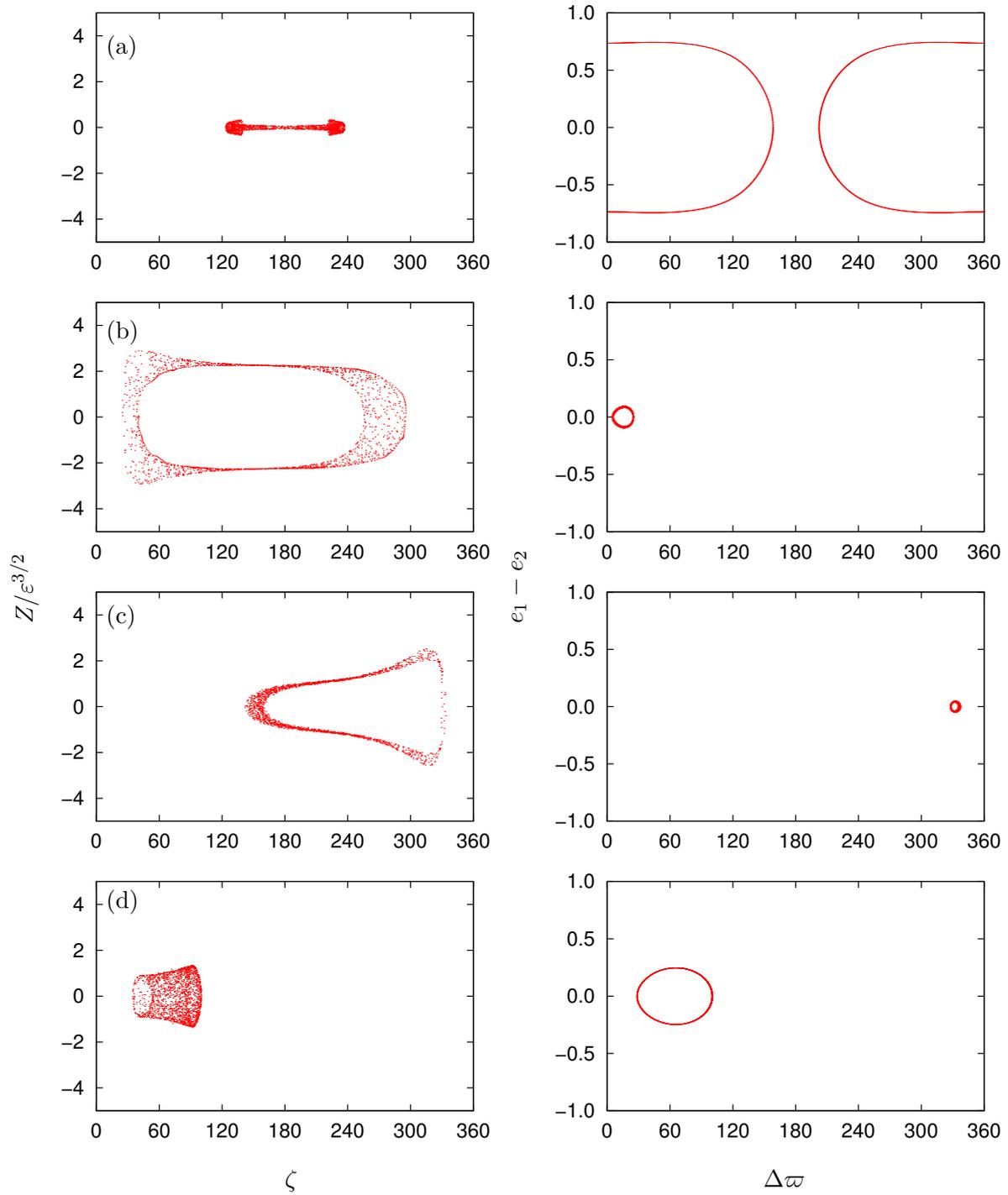
\begin{figure}[h!]
\begin{center}
 \includegraphics[scale=0.9]{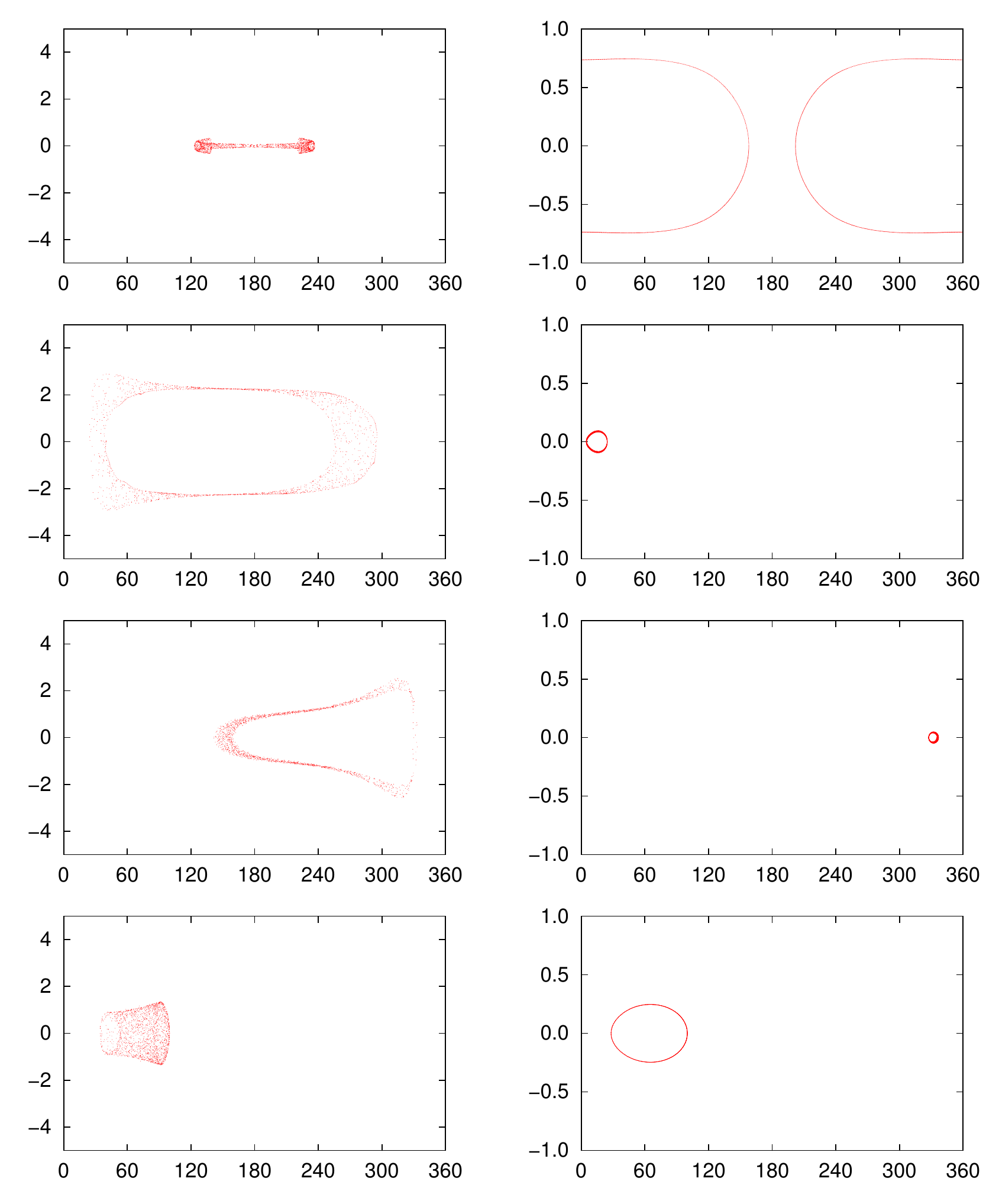}
   \setlength{\unitlength}{1cm}
\begin{picture}(1,0.001)
\put(0,9){\rotatebox{90}{$e_1-e_2$}}
\put(-7.8,9){\rotatebox{90}{$Z/\eps^{3/2}$}}
\put(-3.5,.1){{$\zeta$}}
\put(4,.1){{$\Delta \varpi$}}
\put(-6.3,18){{(a)}}
\put(-6.3,13.5){{(b)}}
\put(-6.3,9){{(c)}}
\put(-6.3,4.5){{(d)}}
\end{picture} 
\hspace{-1cm}
 \caption{\label{fig:orbe7} Trajectoires issues du plan de conditions initiales de la figure~\ref{fig:glob_e7}. (a): $\zeta=136^\circ$, $\Dv=158^\circ$; (b): $\zeta=40^\circ$, $\Dv=5^\circ$; (c): $\zeta=150^\circ$, $\Dv=-25^\circ$; (d): $\zeta=100^\circ$, $\Dv=100^\circ$ }
 \end{center}
 \end{figure}

\begin{figure}[h!]
\begin{center}
\includegraphics[width=0.32\linewidth]{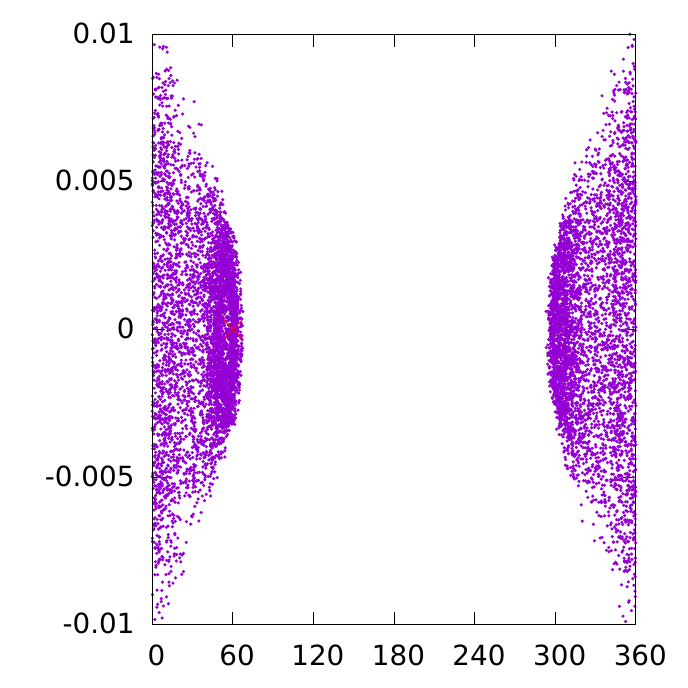} \includegraphics[width=0.32\linewidth]{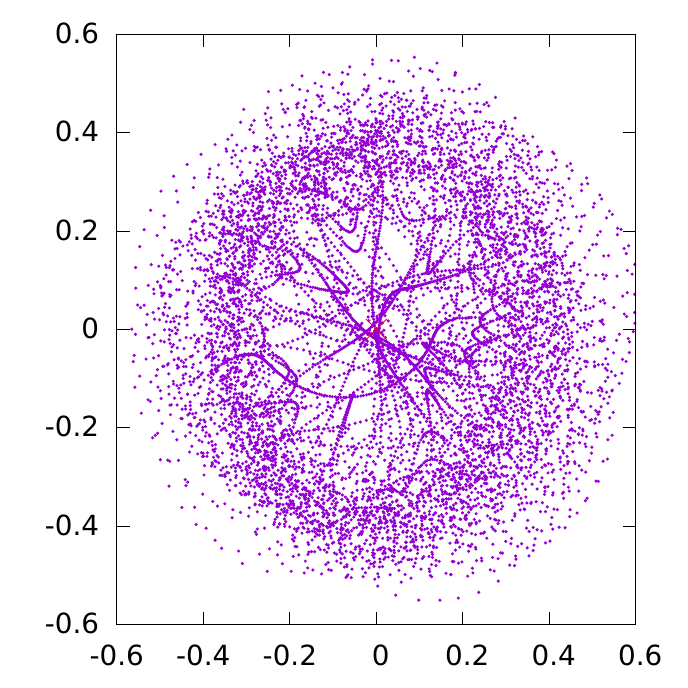} \includegraphics[width=0.32\linewidth]{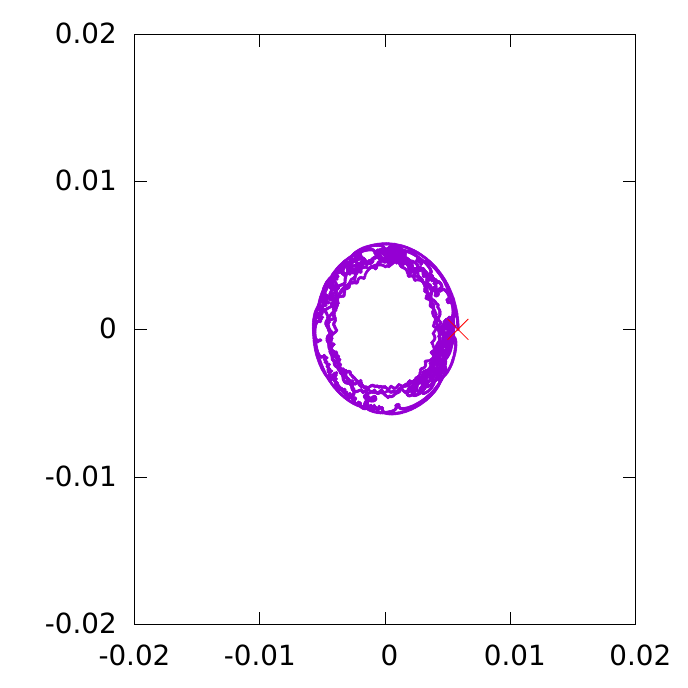}\\
\includegraphics[width=0.32\linewidth]{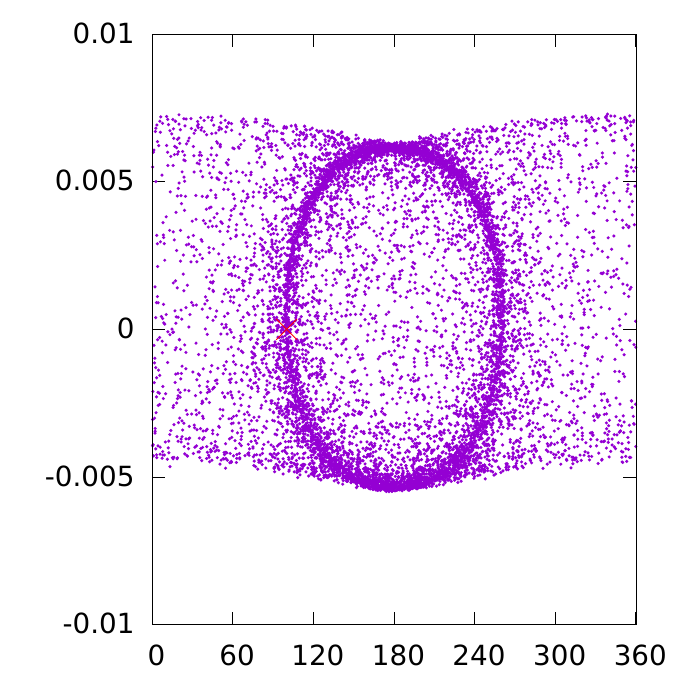} \includegraphics[width=0.32\linewidth]{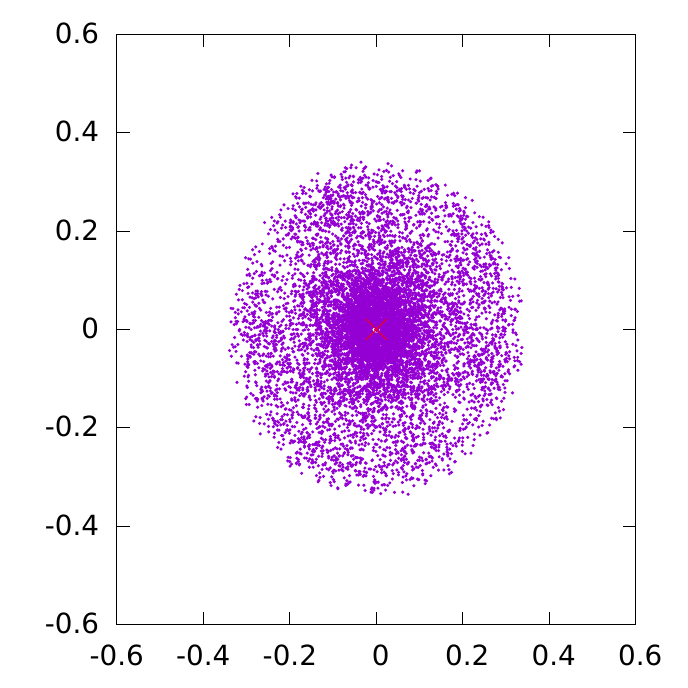} \includegraphics[width=0.32\linewidth]{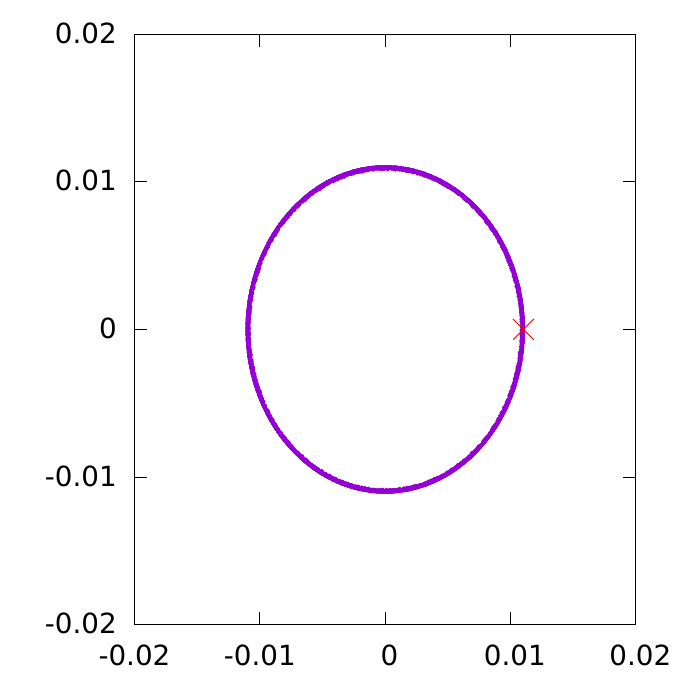}\\
\includegraphics[width=0.32\linewidth]{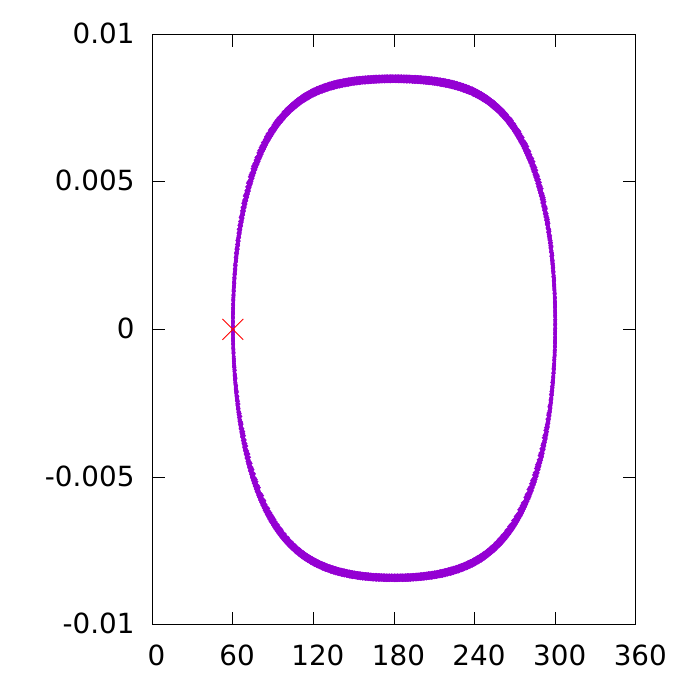} \includegraphics[width=0.32\linewidth]{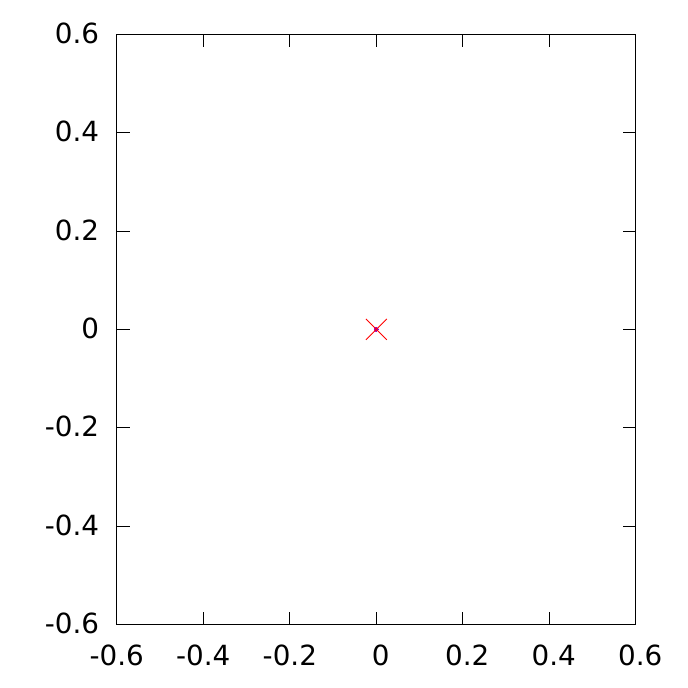} \includegraphics[width=0.32\linewidth]{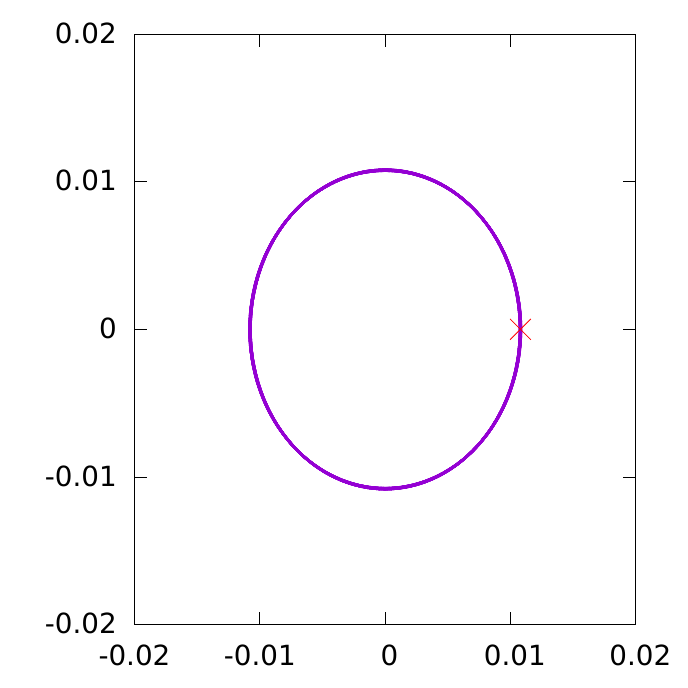}\\
   \setlength{\unitlength}{1cm}
\begin{picture}(1,0.001)
\put(-2.6,7){\rotatebox{90}{$\Im (e_1\exp^{i\varpi_1})$}}
\put(2.6,7){\rotatebox{90}{$\Im (y_1)$}}
\put(-7.8,7){\rotatebox{90}{$a_1-a_2$}}
\put(-4.5,.1){{$\zeta$}}
\put(5,.1){{$\Re (y_1)$}}
\put(0,.1){{$\Re (e_1\exp^{i\varpi_1})$}}
\end{picture} 
\hspace{-1cm}
 \caption{\label{fig:orbJ} Orbites ayant pour condition initiale dans le plan invariant $e_1=e_2=0$, $a_1=a_2=1$, pour la ligne du haut: $\zeta=60^\circ$, $J=65^\circ$; pour la ligne du milieu $\zeta=100^\circ$, $J=160^\circ$; et pour la ligne du bas $\zeta=60^\circ$, $J=150^\circ$. La condition initiale est indiquée par une croix rouge et le reste de la trajectoire par des points violets tous les $50$~an. la durée de chaque intégration est de $500000$~an.}
 \end{center}
 \end{figure}

    \chapter{Publications}
\label{an:Pub}

\section{D\'etection de coorbitaux par vitesse radiale} 
\label{an:VR}

https://arxiv.org/abs/1509.02276
	
	\section{D\'etection de coorbitaux par vitesse radiale et transit} 
\label{an:VRT}

	submitted

	\section{Rotation de coorbitaux excentriques} 
\label{an:rot_coorb}

	https://arxiv.org/abs/1510.09165

\section{Rotation d'un corps circumbinaire} 
\label{an:rot_cb}

	https://arxiv.org/abs/1506.06733

 
%
%
%
%




  \cleardoublepage
  \renewcommand{\indexname}{Index des notations}
  \printindex

\end{document}